\documentclass{elsart}
\usepackage{graphicx}
\usepackage{amssymb}
\usepackage{mathrsfs}
\begin{document}
\ifx\href\undefined\else\hypersetup{linktocpage=true}\fi
\begin{frontmatter}
\title{Photon-assisted transport in semiconductor nanostructures}
\author{Gloria Platero and Ram\'on Aguado}
\address{Departamento de Teor\'{\i}a de la Materia Condensada,
Instituto de Ciencia de Materiales de Madrid (CSIC), Cantoblanco, 28049
Madrid, Spain.}
\begin{abstract}
In this review we focus on electronic transport through semiconductor nanostructures
which are driven by ac fields.
Along the review we describe the available experimental 
information on different nanostructures, like resonant tunneling diodes, superlattices or quantum dots,
together with the theoretical tools needed to describe the observed features. 
These theoretical tools such as, for instance, the Floquet formalism,
the non-equilibrium Green's function technique or
the density matrix technique, are suitable for tackling with photon-assisted transport problems where
the interplay of different aspects like nonequilibrium, nonlinearity, quantum confinement 
or electron-electron interactions
gives rise to many intriguing new phenomena. Along the review we give many examples which 
demonstrate the possibility of using appropriate ac fields to control/manipulate coherent quantum 
states in semiconductor nanostructures.
\end{abstract}
\end{frontmatter}
\tableofcontents
\section{Introduction
\label{introduction}}
Interaction with external time-dependent fields in low-dimensional systems
leads in many cases to completely new ways of electronic transport.
In this review we shall focus on electronic transport through semiconductor nanostructures
like resonant tunneling diodes, superlattices or quantum dots where
the peculiar synergism between ac fields and quantum confinement
gives rise to many novel phenomena. Among them we can mention
ac-induced absolute negative conductance and the so-called dynamical localization
phenomenon observed in superlattices, electron pumps realized in different nanostructures
or the very recent microwaves studies demonstrating quantum coherence in double quantum dots.

Historically, the first experiments of ac-driven tunneling date back to the early sixties when
Dayem and Martin \cite{da} studied photon-assisted-tunneling (PAT) 
in superconductor-insulator-superconductor hybrid
structures. Soon afterwards, Tien and Gordon proposed a simple model of PAT (Section \ref{Tien-Gordon}) in terms
of ac-induced side-bands \cite{tien}.
During the last few decades, this Tien-Gordon model, based on the Bardeen Hamiltonian, has been shown
to grasp the main qualitative physics for PAT through different nanostructures and mesoscopic devices.
In the first part of this review
we shall discuss different theoretical techniques
which allow to address PAT in semiconductor nanostructures where a treatment beyond the
simple Tien-Gordon model is called for. In this part we shall describe
the Floquet approach (Section \ref{Floquet}), various methods based on the Scattering formalism
(Section \ref{Scattering}), and two
methods based on nonequilibrium Green's functions (Section \ref{Keldysh}).

After these four sections devoted to theoretical techniques for the study of PAT,
we elucidate the physics of PAT, both from the experimental and theoretical
points of view, in different semiconductor nanostructures. Here, we chose to divide this part of the review
according to the nanostructure described. 
This division of the review begins with
Section \ref{Double barriers} where we describe PAT in resonant tunneling diodes.
After a short description of the experiments of Chitta et al \cite{ChittaJPC(94)} where the far infrared response of
double barrier structures was analyzed, we elaborate on the importance of studying PAT in these
systems with models
including mixing of electronic states due to the external field:    
when the resonant states in the quantum well
are strongly coupled to reservoirs a description in terms of extended states is called for.
We discuss how mixing of electronic states can be incorporated into the Transfer Matrix  
(subsection \ref{Iñarrea1}) and the Transfer Hamiltonian (subsection \ref{Aguado1}) 
methods.
This Section is completed with discussions about the effects of external magnetic fields
(subsection \ref{Iñarrea3}) and
charge accumulation effects (subsections \ref{bistability} and \ref{Dynamical-self}) on PAT.

The influence of time-dependent fields on transport through semiconductor superlattices is discussed
in Section \ref{Superlattices}. We start this Section by discussing the intringuing phenomenon of absolute
negative conductance observed in THz irradiated superlattices \cite{KeayPRL(95)a} in the linear 
transport regime (subsection \ref{SLlinear}).
Next, we discuss the nonlinear transport regime. In this regime, semiconductor superlattices
exhibit strongly nonlinear behavior
due to the combined action of tunneling and Coulomb interactions.
In particular, weakly coupled superlattices have been shown to exhibit electric-field domain formation, 
self-sustained oscillations and driven and undriven chaos.
Perturbing the system with an ac field brings about
a great deal of new transport phenomena.
We divide this part into four subsections where the statics and dynamics of undriven
(subsections \ref{SLnonlinear-undriven1} and \ref{SLnonlinear-undriven2})
and ac-driven (subsections \ref{SLnonlinear-driven1} and \ref{SLnonlinear-driven2})
superlattices are elucidated. Finally, this portion of the review ends with a
subsection devoted to strongly coupled superlattices (subsection \ref{Bloch}).

During the last few years, many of the new developments in the field of PAT have been realized in quantum dots\footnote{A review on this subject, with a more limited scope, has been recently published in Ref.~\cite{PATreview-VanderWiel}.}. 
We elaborate on different aspects of these new developments in Sections \ref{quantumdotsI}-\ref{quantumdotsII}. 
Beginning with Section \ref{quantumdotsI} we discuss PAT in the Coulomb blockade regime. 
In this Section, key concepts like PAT spectroscopy through zero-dimensional states are introduced.
Section \ref{doubledots} is devoted to double quantum dots where exciting new experiments 
studying the influence of microwaves on the transport properties of these devices have spurred a great deal 
of theoretical activity. Of special interest here is the regime where the effective Hilbert space of the double quantum dot can 
be reduced to just a few levels. Understanding the interplay between electron correlations and the
driving field in these cases is of outmost importance, both from the fundamental and applied points of view. In particular, 
the ability to
rapidly control electrons using ac fields has
immediate applications to quantum metrology and quantum information processing.
Finally, the last Section of this part devoted to quantum dots focuses on strongly 
correlated quantum dots (Section \ref{quantumdotsII}). Here, we analyse the influence which an external ac field
has on quantum dots in the Kondo regime (subsection \ref{Kondo}), in one-dimensional quantum dots (subsection \ref{1D}) and in 
quantum dots in the Wigner molecule regime (subsection \ref{Wigner}). 

The review is completed with three thematic Sections. The first one (Section \ref{MicrowaveHall}) gives
a short account of recent experiments showing 
microwave-induced zero resistance in two-dimensional electron gases at low magnetic fields and 
their explanation in terms of photon-assisted excitations to higher Landau levels. 
In the second Section we give a brief introduction to electron pumps 
(Section \ref{pumps}), a rapidly 
evolving area of research which, surely, would deserve a review of its own. 
For completeness, we also include a short 
description about photon-assisted shot noise (Section \ref{shot-noise}) another fascinating area of research which has 
rapidly developed during the last few years\footnote{Here we urge the interested reader to consult the excelent review
on shot noise by Blanter and Buttiker in Ref.~\cite{BlaPR(00)}.}.

\section{Tien-Gordon model \label{Tien-Gordon}}
Motivated by the experimental microwave studies in superconductor-insulator-superconductor
tunnel junctions of Dayem and Martin \cite{da},
Tien and Gordon \cite{tien}
presented a theoretical model which, in spite of its simplicity, has proven
to be very successful in describing qualitatively transport in ac-driven nanostructures.
The reason for this success is that Tien and Gordon's simple model already contains the main physical 
ingredient of
photon assisted tunneling:
the idea that a time dependent potential $V_{ac}cos\omega t$ can induce {\it inelastic tunnel events} when the electrons exchange energy quanta, i.e. photons,
with the oscillating field.

The first configuration discussed by Tien and Gordon consists of an electric
field applied normal to the surfaces of the superconducting films.
The electric field sets up a potential difference $V_{ac} cos \omega t$ between
the films. Neglecting the interaction of the microwave field with the
insulating barrier and considering one of the metallic films as a reference
(left region),
the effect of the microwave field is to add a potential $V_{ac} cos \omega t$
to the other metallic film (right region). Importantly, within this simple model the effect of the
external field is accounted for by adding a time-dependent, {\it but spatially constant}, 
potential in the right region which is described by
a local Hamiltonian:
\begin{equation}
{\mathcal H}^{R}={\mathcal H}_0^{R}+ eV_{ac}cos \omega t
\label{eq-TG1}
\end{equation}
It is obvious that the time-dependent potential does not vary the
spatial distribution of the electronic wave function within each
region. Solving the time-dependent Schr\"odinger equation, the
electronic wave function for the right region can be written as:
\begin{eqnarray}
\Psi^R(x,y,z,t) &=& \Psi_{0}^R(x,y,z,t)e^{-i\frac{eV_{ac}}{\hbar\omega}sin
\omega t}\nonumber\\
&=&\Psi_{0}^R(x,y,z,t)\sum_{m=-\infty}^{\infty}
J_{m}(\frac{eV_{ac}}{\hbar\omega})e^{-im\omega t}
\end{eqnarray}
where the relationship:
\begin{equation}
e^{-i\frac{eV_{ac}}{\hbar\omega}sin\omega t}\equiv\sum_{m=-\infty}^{\infty
} J_{m}(\frac{eV_{ac}}{\hbar\omega})e^{-im\omega t}.
\label{Bessel}
\end{equation}
with $J_{m}$ being the Bessel function of $m$-th order, has been used.
 From the previous expression of the wave function, it can be observed that
tunneling between the superconducting films through the insulating
barrier can happen from states of energy $E$ in the left region to states of
energy $E\pm m\hbar\omega$ in the right region, namely through inelastic tunneling. The
time-averaged spectral density, $\langle A\rangle$, for the right region can be written in terms of the
density of states without external potential, $A$, as:
\begin{equation}
\langle A_R(E)\rangle= \sum_{m=-\infty}^{\infty} J_{m}^2(\frac{eV_{ac}}{\hbar\omega}) A_R (E+m\hbar \omega).
\label{tien-gordon-dos}
\end{equation}
Eq.~(\ref{tien-gordon-dos}) can be interpreted physically as follows: photon absorption ($m>0$) and emission ($m<0$)
can be viewed as creating an effective electron density of states at energies $E\pm m\hbar\omega$
with a probability
given by $J_{m}^2(\frac{eV_{ac}}{\hbar\omega})$.

The tunneling current between the
superconducting films can be obtained by means of the Transfer
Hamiltonian (TH) method
\footnote{The Transfer Hamiltonian method, also called Bardeen Hamiltonian method, 
considers the coupling
between the different parts of the system only to lowest order in perturbation theory \cite{BardPRL(61)}.}.
Assuming that the transmission coefficient does
not depend on energy, the tunneling current can be expressed as
\cite{BardPRL(61),tucker1,tucker2}:
\begin{eqnarray}
I_{dc}=  T_{LR} \int_{-\infty}
 ^{+\infty} dE [f(E)-f(E+eV_{dc})] \langle A_{L}(E)\rangle \langle A_{R}(E+eV_{dc})\rangle
\end{eqnarray}
where $T_{LR}$ is the transmission coefficient, which is assumed constant,
$f(E)=1/[1+\exp{(E-E_{F})/k_B T}]$ is the Fermi-Dirac distribution 
function and $V_{dc}$ the applied dc voltage.
\begin{figure}
\begin{center}
\includegraphics[width=0.75\columnwidth,angle=0]{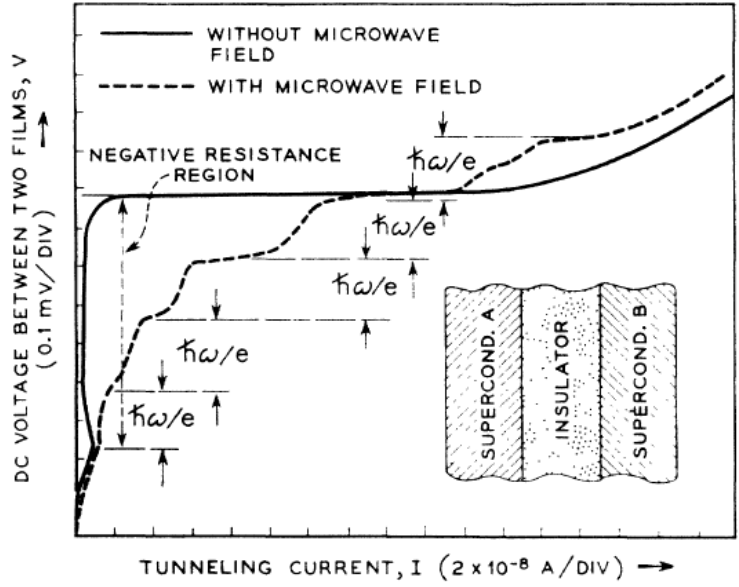}
\end{center}
\caption{Bias voltage vs tunneling current of a superconducting
 $ Al-Al_2 O_3 - In $ diode as measured by Dayem and Martin
 with and without
microwave field. $ \hbar w $ = 0.16 meV. Reprinted with permission from \cite{tien}.
\copyright 1963 American Physical Society.}
\label{TienGordonPR_63__fig1}
\end{figure}
In the presence of microwaves and considering the explicit
expressions for the spectral densities (\ref{tien-gordon-dos}), the
dc current becomes:
\begin{eqnarray}
I_{dc}&=&T_{LR}\sum_{n=-\infty}^{\infty}
J_{m}^{2}(\frac{eV_{ac}}{\hbar\omega})\int_{-\infty}
 ^{+\infty} dE A_{L}(E)
 A_{R}(E+eV_{dc}+m\hbar\omega)\nonumber\\
&\times&[f(E)-f(E+eV_{dc}+m\hbar\omega)]
\end{eqnarray}
Importantly, the dc currents with and without microwaves can be related as:
\begin{eqnarray}
I_{dc}=\sum_{m=-\infty}^{\infty}
J_{m}^{2}(\frac{eV_{ac}}{\hbar\omega}) I_{dc}^0(V_{dc}+\frac{m\hbar\omega}{e}).
\label{Tien-Gordon-rectif}
\end{eqnarray}
where $I_{dc}^0$ is the current without radiation. Namely,
the rectified current of a system biased with a voltage
$V(t)=V_{dc} + V_{ac}\cos\omega t$ is given as a sum of
dc-currents $I_{dc}^0$ {\it without ac driving} evaluated
at voltages shifted by integer multiples of photon energies.  
\section{Floquet theory \label{Floquet}}
In quantum mechanics, symmetry is expressed by an operator S which leaves the
Schr\"odinger equation invariant, i.e., it commutes with the
operator $H(t)-i\hbar\delta_{t}$. Thus, the solutions of the
Schr\"odinger equation are, besides a time-dependent phase factor,
also eigenfunctions of the symmetry operator. For a Hamiltonian
with $\mathcal{T}$-periodic time dependence,

\begin{equation}
H(t) = H(t + n \mathcal{T})
\end{equation}
the related symmetry operation is a discrete time translation by
one period of the driving, $S_\mathcal{T} :t\rightarrow
t+\mathcal{T}$. As symmetry operations have to conserve the norm
of any wavefunction, the eigenvalues of $S_\mathcal{T}$ are pure
phase factors and one may assume for an eigenfunction $|\psi
(t)\rangle$ the eigenvalue exp(-i$\theta$): $S_\mathcal{T} |\psi
(t)\rangle$=$|\psi(t+\mathcal{T})\rangle=
e^{-i\theta}|\psi(t)\rangle$ By inserting the wavefunction
$|\psi(t)\rangle=e^{-iet}|\phi(t)\rangle$
($e=\hbar\theta/\mathcal{T}$) in the Schr\"odinger equation, one
obtains $|\phi(t)\rangle=|\phi(t+\mathcal{T})\rangle$. Thus for a
system which obeys discrete time-translational symmetry, there
exists a complete set {$\Phi(t)$} of solutions of the Schr\"odinger
equation which have Floquet structure,i.e., they are of the form:
$|\psi(t)\rangle
=e^{-ie_{\alpha}t/\hbar}|\phi_{\alpha}(t)\rangle)$ where
$|\phi_{\alpha}(t)\rangle=|\phi_{\alpha}(t+\mathcal{T})\rangle$.

We consider a general quantum system driven by a periodic electric
field, described by a time-dependent Hamiltonian which we can
divide in the following way:
\begin{equation}
H(t) = H_{t} + H_I + H_{ac}(t), \quad H_{ac}(t) = H_{ac}(t + n T)
. \label{divide}
\end{equation}
Here $H_{t}$ holds the tunneling terms, $H_I$ holds the
electron-electron interaction terms and $H_{ac}(t)$ describes the
interaction of the system with the $T$-periodic driving field. The
periodicity of the driving field allows us to use the Floquet
theorem to write solutions of the Schr\"odinger equation as
$\psi(t) = \exp[-i \epsilon_j t] \phi_j(t)$ where $\epsilon_j$ is
called the quasi-energy, and $\phi_j(t)$ is a function with the
same period as the driving field, called the Floquet state. This
type of expression is familiar in the context of solid-state
physics, where {\em spatial} periodicity permits an analogous
rewriting of the spatial wavefunction in terms of quasi-momenta
and Bloch states (Bloch's theorem).

The Floquet states provide a
complete basis, and thus the time-evolution of a general state may
be written as a superposition of Floquet states:
\begin{equation}
|\Psi(t) \rangle= \sum_j \ \left( c_j \mbox{e}^{-i \epsilon_j t}
\right) |\phi_j(t) \rangle ,
\label{floq_exp}
\end{equation}
which is formally analogous to the standard expansion in the
eigenvectors of a time-independent Hamiltonian. Indeed, in the
adiabatic limit, $T = 2 \pi / \omega \rightarrow \infty$, the
quasi-energies evolve to the eigenenergies, and the Floquet states
to the eigenstates. It is important to note that in this expansion
both the basis vectors (the Floquet states) and the expansion
coefficients explicitly depend on time. The nature of this
time-dependence is very different however, and the superposition
of the $T$-periodicity of the Floquet states with the phase
factors arising from the quasi-energies produces a highly
complicated, quasi-periodic time-dependence in general.

As the
Floquet states have the same period as the driving field, they are
only able to produce structure in the time-dependence on short
time-scales. Consequently, the dynamics of the system on
time-scales much larger than $T$ is essentially determined by {\em
just} the quasi-energies, and hence evaluating the quasi-energies
provides a simple and direct way of investigating this behavior. A
number of different methods can be used to numerically calculate
the quasi-energies of a quantum system, and a detailed description
of them is given in Ref.~\cite{GrifPR(98)}.

One commonly used method is
to evaluate the unitary time-evolution operator
for one period of the driving field $U(t+T,t)$, and then to
diagonalize it. It may be easily shown that the eigenvectors of
this operator are equal to the Floquet states, and its eigenvalues
are related to the
quasi-energies via $\lambda_j = \exp[-i \epsilon_j T]$. \\
When two quasi-energies approach degeneracy the time-scale for
tunneling between the states diverges, producing the phenomenon of
coherent destruction of tunneling (CDT) \cite{GrosPRL(91)}. The
time scale for localization is the inverse of the energy
separation of the quasienergies.
For instance, in a two level system
driven by a term $ H_{ac}(t)=eV_{ac} cos\omega t$, 
CDT can be understood from the
renormalization of the level splitting:
\begin{equation}
\Delta\rightarrow \Delta_{eff}\equiv
J_0(\frac{eV_{ac}}{\hbar\omega})\Delta,
\label{CDT}
\end{equation}
where $\Delta$ is the interlevel coupling.
At the first zero of
$J_0$, namely when $eV_{ac}/\hbar\omega=2.4048...$, the {\it
effective tunnel splitting vanishes} leading to a complete
localization of the particle in the initial state. 
This phenomenon is also known as Dynamical Localization in the PAT literature
\footnote{Note that a more general phenomenon is also called Dynamical Localization in
chaotic dynamical systems: In this context, Dynamical Localization is the phenomenon by which 
destructive interference effects
supress difussion in the relevant phase space. See Ref.~\cite{ChiriPhysD(88)}.}.
The expression
for the renormalization of the hopping in Eq.~(\ref{CDT}) is obtained from
first-order perturbation theory in the tunneling, see subsection \ref{Holt-two-lev}. 

As we shall see for some specific quantum systems considered in this review, it
is frequently the case that the total Hamiltonian is invariant
under the generalized parity operation: $x \rightarrow -x; t
\rightarrow t + T/2$. As a result the Floquet states can also be
classified into parity classes, depending  whether they are odd or
even under this parity operation. Quasi-energies belonging to
different parity classes may cross as an external parameter (such
as the field strength) is varied, but if they belong to the same
class the von Neumann-Wigner \cite{vonneum} theorem forbids this,
and the closest approaches possible are avoided crossings.
Identifying the presence of crossings and avoided crossings in the
quasi-energy spectrum thus provides a necessary (though not
sufficient) condition for CDT to occur.

An interesting effect shown by Stockburger \cite{StoPRE(99)} is that
the condition of degenerate Floquet levels, required for localization
in a symmetric system, can be substantially relaxed for tunneling
systems with broken symmetry. He found that the localization
regime was substantially extended due to a synergistic effect of dynamic and static asymmetry.
He called this phenomena "Non-degenerate Coherent
Destruction of Tunneling" (NCDT). His results will decrease the
difficulty to measure coherent destruction of tunneling, i.e, to
control accurately the driving amplitude.
\subsection{Perturbation theory for Floquet states \label{method}}
Although the quasi-energies are extremely useful for
interpretation of the time-dependence of a quantum system, they
are usually difficult to calculate and numerical methods must be
employed. When the driving field dominates the dynamics, however,
it is possible to use a form of perturbation theory introduced by
Holthaus \cite{holtzpb(92)}, in which the time-dependent part of
the problem is solved exactly, and the tunneling part of the
Hamiltonian, $H_t$, acts as the perturbation. Using this method,
Holthaus \cite{holtzpb(92)} and Hanggi \cite{GrifPR(98),hangepl(92)}
have studied the two level system. For high fields,
perturbation theory for Floquet states allows for an analytic
description of the quasienergies and the field parameters where
dynamical localization takes place.

This was generalized to treat interacting systems in
Refs.~\cite{CrefPRB(02)a,CrefPRB(02)b,CreSV03} and was found to be very
successful in the high-frequency regime, where $\hbar \omega$ is
the dominant energy-scale. We now give a brief outline of this
method. Inserting \ref{floq_exp} in the Schr\"odinger equation, the
Floquet states and their quasi-energies may be conveniently
obtained from the eigenvalue equation:
\begin{equation}
\left( H(t) - i \hbar \frac{\partial}{\partial t}  \right) |
\phi_j(t) \rangle = \epsilon_j | \phi_j(t)  \rangle \label{floqeq}
\end{equation}
where the hermitian Floquet operator $\left[ H(t) - i \hbar
\partial / \partial t \right]$ operates in an {\em extended}
Hilbert space of $T$-periodic functions \cite{SambePRA(73)}. The
procedure consists of dividing the Hamiltonian as in
Eq.~(\ref{divide}), and finding the eigensystem of the operator
$\left[ H_{I} + H_{ac}(t) - i \hbar \partial / \partial t
\right]$, while regarding the tunneling Hamiltonian $H_{t}$ as
acting as a perturbation. Standard Rayleigh-Schr\"odinger
perturbation theory can now be used to evaluate the order-by-order
corrections to this result, requiring only that we define an
appropriate inner product for the extended Hilbert space:
\begin{equation}
\langle \langle \phi_m | \phi_n \rangle \rangle_T = \frac{1}{T}
\int_0^{T} \langle \phi_m(t') | \phi_n(t') \rangle dt' .
\label{inner}
\end{equation}
Here $\langle \cdot | \cdot \rangle$ denotes the usual scalar
product for the spatial component of the wavevectors, and
$\langle \cdot | \cdot \rangle_T$ is the integration over the
compact time coordinate. We shall show in later sections how this
method can be used to obtain analytical forms which accurately
describe the behavior of the quasi-energies for the systems we
study.
\subsection{ac-driven two-level Systems \label{Holt-two-lev}}
Holthaus \cite{holtzpb(92)} derived explicitly the quasienergies
for a two-level system in the limit of strong electric
field-electron coupling when the interlevel coupling, $\Delta=E_2-E_1$, can be
treated perturbatively. The starting point of the derivation is the Schr\"odinger equation 
for the two level system:
 \begin{equation}
 i \hbar \frac{\partial}{\partial t} \phi(t)=\left(H_0+ H_I(t)\right) \phi (t)
\label{holt-two-level1}
\end{equation}    
with $\phi(t)=(\phi_1(t),\phi_2(t))^T$, and
\begin{equation}
H_0=
\left(\begin{array}{cccc}
\frac{1}{2}\Delta&0\\
0&-\frac{1}{2}\Delta
\end{array}\right),\quad\quad
H_I(t)=
\left(\begin{array}{cccc}
0&eFd_{0}sin \omega t\\
eFd_{0}sin \omega t&0
\end{array}\right).
\end{equation}   
$F$ is the intensity of the ac field and $d_0$ is the dipole matrix element between the
lowest two double
well eigenstates. In the strong coupling limit $eFd_0>>\Delta$,
the time independent part of the Hamiltonian can be treated as a perturbation
and Eq.~(\ref{holt-two-level1}) reduces to:
 \begin{equation}
i \hbar \frac{\partial}{\partial t} 
\left(\begin{array}{cccc}
\phi_1(t)\\
\phi_2(t)
\end{array}\right)=eFd_{0}sin \omega t
\left(\begin{array}{cccc}
\phi_2(t)\\
\phi_1(t)
\end{array}\right)
\label{flo}
\end{equation}
A fundamental system of solutions of Eq.~(\ref{flo}) 
is:
\begin{equation}
\Phi_1(t)=
\left(\begin{array}{cccc}
cos(\frac {eFd_0}{\omega} cos \omega t)\\
i sin(\frac {eFd_0}{\omega} cos \omega t)
\end{array}\right)\quad\quad
\Phi_2(t)=
\left(\begin{array}{cccc}
i sin(\frac {eFd_0}{\omega} cos \omega t)\\
cos (\frac {eFd_0}{\omega} cos \omega t)
\end{array}\right)
\end{equation}     
$\Phi_1$ and $ \Phi_2$ are time-periodic states (Floquet states)
and the quasienergies are degenerate and equal to zero (modulo
$\omega$). The interlevel part of the Hamiltonian, $H_0$,
is now treated as a
perturbation exploiting the fact that the Floquet states are stationary states 
in the extended Hilbert space of time-periodic functions, see Eq.~(\ref{inner}),
such that the quasienergies, $\epsilon_{\pm}=\langle\langle\Phi_\pm| H_0|\Phi_\pm\rangle\rangle$ 
(with $\Phi_+=\Phi_1$ and $\Phi_-=\Phi_2$) are:
\begin{equation}
\epsilon_{\pm}=\pm
\frac{\Delta}{2}J_0(\frac{2eFd_0}{\omega}). 
\end{equation}
Now, if the absolute value of the dipole matrix
element $d_0$ is approximated by that of the
positions of the center of the wells $d_0\approx d/2$, one gets
\begin{equation}
\epsilon_{\pm}=\pm
\frac{\Delta}{2}J_0(\frac{eFd}{\omega}).
\end{equation} 
This result shows that, in the presence of an ac field of intensity $V_{ac}=Fd$, 
the tunnel coupling is renormalized  by the zero-order Bessel
function, as stated in Eq.~(\ref{CDT}). 

Although sinusoidal driving is considered more frequently when discussing CDT 
(simply because it is the natural form of electromagnetic radiation), 
some work has been also devoted to analyze the degree of localization induced
by ac potentials with different shape. Bavli and Metiu showed that a semi
infinite laser pulse is able to localize an electron in one of the wells
in a double quantum well structure \cite{BavPRL92}.
Holthaus investigated the effect of pulse shaping to enhance 
the rate of tunneling \cite{HoltPRLb(92)}.
Square-wave driving has been considered to a lesser extent.
It was shown in Refs.~\cite{ZhaJPCM94,DigPRL02}
that total CDT can only be produced in a superlattice
if the crossings of the quasienergies are equally spaced.
They showed that this only occurs if the field has discontinuities, and the square-wave is the simplest
example of this type. A comparision of the degree
of localization in two-level systems
for square-wave, sinusoidal and triangular driving
was theoretically performed by Creffield \cite{CrePRB03}.
He analyzed the high frequency regime, where perturbation
theory in the interlevel coupling works well and the low frequency regime where the
crossings move away from the values predicted by perturbation theory.
The position of the crossings can fitted by the function \cite{CrePRB03}:
\begin{equation}
(\frac{eV_{ac}/\hbar\omega}{y_n})^2+ (\frac{\Delta/\hbar\omega}{2n})^2=1
\label{cha}
\end{equation}
where $y_n$ is the n-root of $J_0(y)$, $\Delta$ is the splitting between
the two levels and $\omega$ the frequency of the driving field. Creffield
found that the positions of the crossings at low frequencies follow (\ref{cha})
exactly for square-wave driving and with small deviations for other shapes
of the driving field. Such a general behavior allows the positions of
the quasienergy crossings to be accurately located in all regimes of
driving. In spite of the reduction of the degree of localization observed
at low frequencies, the accurate control of the crossings positions opens
new posibilities for experimental configurations.  

When the
hopping increases, higher order terms should be included and CDT
does not occur at the zeros of the zero-order Bessel function.
Barata et al \cite{BarPRL(00)} have shown that the second order
contribution to the renormalized hopping is identically zero and
the third order contribution is different from zero
just at the zeros of $J_0$
and is given by:
\begin{equation}
\frac{\Delta^3}{4\omega^2}
\sum_{n_1,n_2 =-\infty}^\infty
\frac{J_{2n_1+1}(\frac{eFd}{\omega})J_{2n_2+1}(\frac{eFd}{\omega})J_{2(n_1+n_2+1)}(\frac{eFd}{\omega})}
{(2n_1+1)(2n_2+1)}
\label{Barata}
\end{equation}    
Recently, the perturbation series of a two-level system driven a a sinusoidal field 
(in the strong coupling regime) has been analyzed in detail by Frasca in Ref.~\cite{FrascaPRB(03)}.

The combined effect of radiation and magnetic fields in two level systems
was analyzed by Villas-Boas et al. \cite{VilPRB02}. They did an
analysis of the quasienergy spectrum for the case where the magnetic field is
oriented perpendicular to the quantum wells interfaces and also in
the case where a finite component parallel to the interfaces is applied.
The analysis was performed based in the parity properties for both
configurations and for all ranges of frequency and intensity of the radiation.
They found that at low frequencies the Dynamical Localization
points shift to lower $eFd/\hbar \omega $ ratios yielding to poorer
localization
by the ac field.      

The two-level system has been exhaustively studied in the
literature. However, in many mesoscopic systems the configuration
is such that more than two levels interact and the analysis is
more complicated. The next simplest case corresponds to a three
level system. As we will describe in Section \ref{ac-driven isolated double quantum dots},
a double quantum
dot with two interacting electrons and driven by an ac field can be described by an effective six
dimensional Hubbard Hamiltonian. If spin-flip
processes which can arise by electron scattering with nuclei or
spin-orbit interaction are not included, the spin singlet and
triplet are decoupled and the effective Hamiltonian describing
each sector is three dimensional.
We shall describe in Section \ref{ac-driven isolated double quantum dots} how to obtain
the Floquet spectrum by
using the procedure described above, i.e, considering the interdot
coupling as a perturbation in the limit of strong electric
field-electron coupling.
\subsection{Floquet theory for spatially periodic systems \label{miniband-collapse}}
A superlattice consists on N identical quantum wells coupled by
finite barriers. For strong inter-well coupling the discrete
states of each isolated quantum well hybridize with those of the
neighbor wells and finite width minibands are formed.

Holthaus
studied the spectrum of a superlattice under ac
radiation \cite{holtzpb(92),HoltPRL(92),HoltPRB(93)}. In these systems,
periodicity in time leads to a formulation in terms of quasienergy
eigenvalues and spatial periodicity implies that the quasienergies for the
allowed quantum states group
together in minibands.

The Hamiltonian describing this system is:
\begin{equation}
 H(x,t)= -  \frac{\partial ^2}{{\partial x}^2} + V_{SL}(x) -eFx sin \omega t,
\end{equation}
where $V_{SL}(x)$ is the electrostatic potential of the superlattice.
As $H(x,t)$ is periodic in time, there is a complete set of Floquet
wave functions as solutions of the Schr\"odinger equation. For large
number of quantum wells N, the energies $E_n$ of the lowest
unperturbed miniband of width $\Delta$ are given by:
\begin{equation}
 E_n=\epsilon_0-\frac {\Delta}{2} cos(\frac{n\pi}{N+1}),
 n=1,......N
\end{equation}
where $\epsilon_0$ is the center of the unperturbed
miniband. It can be shown from a quantum mechanical calculation
which neglects finite size effects (i.e., for large N) that the
expression for the quasienergies originated from them is:
\begin{equation}
\epsilon_n= \epsilon_0-\frac{\Delta}{2}J_0(\frac {eFd}{\hbar\omega})
 cos(\frac{n\pi}{N+1}), mod(\omega )
\end{equation}
This result implies that the
width of the quasienergy miniband becomes zero at zeros of $J_0$
\cite{HoltPRL(92)}.

A similar result is found for the average
electron velocity using semiclassical arguments \cite{IgnatovPSS(76),IgnADP94}:
Considering the dispersion relation $E(k)=
\epsilon_0-\frac{\Delta cos(kd)}{2} $ for the undriven superlattice and
a time dependent electric field: $E(t)=F sin \omega t$, the group
velocity of a wave packet centered around $k_0$ and $t=T/4$ is given
by:
\begin{equation}
v(t)=\frac{\Delta d}{2} sin(k_0d+\frac{eFd}{\hbar\omega} cos \omega t)
\end{equation}
and the velocity average over one period of the electromagnetic
field is:
\begin{equation}
 v_{average} =\frac{ \Delta d}{2} sin(k_0d)J_0(\frac
{eFd}{\hbar\omega})
\end{equation}
Hence if the ratio of the Bloch frequency $\omega_B=\frac{eFd}{\hbar}$ and
the external frequency is equal to a zero of $J_0$, namely $J_0(\frac{\omega_B}{\omega})\rightarrow 0$, 
the average
electron velocity is zero and the wave packet becomes localized.
This localization induced by ac field (Dynamical Localization)
is dubbed miniband collapse
by radiation in the context of superlattices.

We conclude this part by mentioning that the Floquet theory described in this section applies to {\it closed systems}.
We will describe along the review how one can combine the Floquet theory with other powerful techniques, 
like the nonequilibrium Green's functions technique or the density matrix technique, in order to treat {\it open systems}.
\section{Scattering approach \label{Scattering}}
\subsection{Transfer matrix approach \label{TM}}
Here we briefly review the transfer matrix approach proposed by Coon and Liu
\cite{CooSSC(85),CooJAP(85),LiuPRB(91)}
to solve a general potential profile in the presence of both dc and
ac signals. The approach is based on the following asumption:
given an arbitrary potential profile, one can always approximate to an
arbitrary accuracy the actual profile by a series of steps,
namely by dividing the space into regions of constant potentials.
Within this approach, one can solve the time-dependent Schr\"odinger
equation for constant $V_{dc}$
and $V_{ac}$ in terms of plane waves:
\begin{equation}
i\hbar\frac{\partial\phi}{\partial t}=\frac{\hbar^{2}\partial^{2}\phi}
{2m\partial z^2}+(v_{dc}+v_{ac}cos\omega t)\phi
\end{equation}
One can easily verify that
\begin{equation}
\phi_k=e^{ikz-iEt/\hbar-iv_{ac}/\hbar\omega sin\omega t}
\end{equation}
is a solution
with $E-v_{dc}=\hbar^{2}k^{2}/2m$.
Then, a general solution can be built up
with energy components $E-n\hbar\omega$ ($n=0,\pm 1,\pm 2\dots,\pm\infty$):
\begin{equation}
\phi=\sum_{n}(a_n\phi_{k_{n}}+b_n\phi_{-k_{n}})
\end{equation}
where $E+n\hbar\omega-v_{dc}=\hbar^2{k_n}^2/2m$.
If the solution in the next constant potential is :
\begin{equation}
\phi'=\sum_{n}(c_n\phi_{k'_{n}}+d_n\phi_{-k'_{n}})
\end{equation}
a general transfer matrix M is required so that: $(c,d)^T=M(a,b)^T$
where $a=(\dots, a_2, a_1,a_0,a_{-1},a_{-2},\dots)$ and similarly for
b, c and d, where the superscript T means transposing the row matrix
\footnote{In order to get the matrix M one needs also to include boundary conditions.
The simplest ones are to impose the continuity of the wave function and its derivative
\cite{LiuPRB(91)}.}.
By using a similar reasoning for all potential steps, one gets the complete transfer matrix
which relates the constants $a$ and $b$ on one side of the structure to $c$ and $d$
on the other side. Transmission and reflection amplitudes fon an incident electron with
energy $E$ are found by setting $a=(\dots,0,0,1,0,0,\dots)$,
$b=(\dots,r_2,r_1,r_0,r_{-1},r_{-2},\dots)$, $c=(\dots,t_2,t_1,t_0,t_{-1},t_{-2},\dots)$
and $d=(\dots,0,0,0,0,0,\dots)$. Then, a multichannel scattering state consists of the
incident wave and the scattered (transmitted and reflected) waves.

As an example of how transmission $t_n$ and refletion $r_n$ are determined,
let us consider a situation in which
just one ac discontinuity occurs, namely $v_{dc}=v'_{dc}$ and $\Delta v_{ac}\equiv v_{ac}-v'_{ac}\neq 0$,
and only the first order side bands ($n=0,\pm 1$) are included.
The transfer matrix becomes $6\times 6$ in this case and ($E>>\hbar\omega$):
\begin{equation}
\mathbf{M}\approx
\left(\begin{array}{cccccc}
J_0&J_1&0&0&0&0\\
J_{-1}&J_0&J_1&0&0&0\\
0&J_{-1}&J_0&J_1&0&0\\
0&0&J_{-1}&J_0&J_1&0\\
0&0&0&J_{-1}&J_0&J_1\\
0&0&0&0&J_{-1}&J_0
\end{array}\right)
\end{equation}
where the Bessel functions have the argument $\Delta v_{ac}/\hbar\omega$.

Including side bands up to first order, the transmitted wave can be written as:
\begin{equation}
\phi=(t_0e^{ik_0z}+t_{+1}e^{ik_1 z-i\omega t}+t_{-1}e^{ik_{-1} z+i\omega t})
e^{-\frac{iEt}{\hbar}}e^{\frac{iV_{ac}sin(\omega t)}{\hbar\omega}},
\end{equation}
such that the electron tunneling current (for a given energy $E$):
\begin{equation}
j=\frac{e\hbar}{2im}
(\phi^*\frac{d\phi}{dz}-\phi\frac{d\phi^*}
{dz})
\end{equation}
reads ($z=0$):
\begin{eqnarray}
j&=&(\frac{e\hbar k_0}{m}|t_0|^2+\frac{e\hbar}{m} Im[ik_0t_0(t_{+1}^*e^{i\omega t}+
t_{-1}^*e^{-i\omega t})\nonumber\\
&-&i{t^*}_{0}(k_{+1}t_{+1}e^{-i\omega t}
+k_{-1}t_{-1}e^{i\omega t})]
\label{current_TM}
\end{eqnarray}
where $\hbar^{2}k_0^{2}/2m=E+V_{dc}$ and $\hbar^{2}k_{\pm 1}^2/2m
=E+V_{dc}\pm \hbar\omega$.
Using Eq.~(\ref{current_TM}) one can split the total current into dc and ac components as:
\begin{eqnarray}
j&=&j_{dc}+j_{ac}\nonumber\\
j_{dc}&\equiv&\frac{e\hbar k_0}{m}|t_0|^2\nonumber\\
j_{ac}&\equiv&\frac{e\hbar}{m}Re[t_0(t_{+1}^*e^{i\omega t}+t_{-1}^*e^{-i\omega t})].
\end{eqnarray}
As an example, the ac current through a resonant tunneling diode reads \cite{LiuPRB(91)}:
\begin{eqnarray}
j_{ac}&=&\frac{e\hbar k}{m}\frac{V_{ac}}{\hbar\omega}T_0
[(\frac{(E-E_{R}+\hbar\omega/2)(E-E_{R}+\hbar\omega)
+\Gamma^2}
{(E-E_R+\hbar\omega)^2+\Gamma^2}\nonumber\\
&-&
\frac{(E-E_{R}-\hbar\omega/2)(E-E_{R}-\hbar\omega)+\Gamma^2}
{(E-E_R-\hbar\omega)^2+\Gamma^2})cos\omega t\nonumber\\
&+&(\Gamma\hbar\omega/2)
(\frac{1}{(E-E_R+\hbar\omega)^2+\Gamma^2}
-\frac{1}{(E-E_R-\hbar\omega)^2+\Gamma^2})sin\omega t].
\end{eqnarray}
Where $T_0\approx T_{0,max}\frac{\Gamma^2}{(E-E_R)^2+\Gamma^2}$
is the transmission coefficient through the double barrier in the absence of ac driving,
and only one resonant level of energy $E_R$ and width $\Gamma$ has been considered.
After doing a Taylor expansion and keeping only the leading order term in $\hbar\omega$
one can write the total current as:
\footnote{Integrating over the emitter Fermi sea, one gets the total current density
($z=0$) at zero temperature: $J=\frac{1}{4\pi^2}\int_0^{k_F}dk (k_F^2-k^2)j=
\frac{m^2}{2\pi^2\hbar^4}\int_0^{E_F}dE(E_F-E)j/k$. Where $E_F=\hbar^2k_F^2/2m$
is the Fermi energy in the emitter.}
\begin{equation}
J_{ac}\approx \frac{\partial J_{dc}}{\partial V_{dc}}
V_{ac}cos\omega t- \hbar\omega
\Gamma^3\frac{\partial}{\partial(\Gamma^2)}
(\frac{\partial J_{dc}}{\Gamma^2\partial
V_{dc}})V_{ac}sin\omega t.
\end{equation}
The first term corresponds
to the classical low-frequency expression while the
second one represents the leading order high frequency correction.
From this expression one can obtain the device admittance.

The above transfer matrix description in terms of piecewise constant potentials was later
extended by Wagner in a series of papers
\cite{WagnerPRB(94),WagnerPRA(95),WagnerPRL(96),WagPRB(97)}
to analyze tunneling through single and
double barriers. In particular, Wagner describes in Ref.~\cite{WagnerPRL(96)}
the possibility of finding
{\it analytical} solutions of the driven problem as a starting point for performing
numerics using the transfer matrix approach.
For example, by considering a quantum well sandwiched between infinitely high barriers
and strongly driven by an external field $eFzcos\omega t$, namely:
\begin{equation}
H(t)=-\frac{\hbar^2}{2m}\frac{\partial^2}{\partial z^2}+eFzcos\omega t,\quad
\textrm{for }-d/2<z<d/2,
\label{Wagner_Hamiltonian}
\end{equation}
the folllowing analytical solution for the lowest Floquet state is proposed:
\begin{eqnarray}
\Phi(z,t,E)&=&exp\left[-i\left(E+\frac{e^2F^2}{4m\omega^2}\right)\frac{t}{\hbar}\right]
\sum_{l=-\infty}^{\infty}A_l \{ exp\left[ik_l
\left(z-\frac{eFcos\omega t}{m\omega^2}\right)\right]\nonumber\\
&+&(-1)^lexp\left[-ik_l\left(z-\frac{eFcos\omega t}{m\omega^2}\right)\right]\}\nonumber\\
&\times&exp\left(-il\omega t-\frac{ieFz sin\omega t}{\hbar\omega}+
\frac{ie^2F^2 sin 2\omega t}{8\hbar m\omega^3}\right),
\label{Wagner_eq1}
\end{eqnarray}
where $\hbar k_l=\sqrt{2m(E+l\hbar\omega)}$. Note that $\Phi(z,t,E)$ is a Floquet state
of the form $\Phi(z,t,E)=exp(-i\epsilon t/\hbar) u(t)$ with $u(t)=u(t+2\pi/\omega)$ and
$\epsilon=E+e^2F^2/4m\omega^2$. The coefficients $A_l$ can be obtained from
the boundary conditions which, for a quantum well sandwiched between two infinitely
high walls at $z=\pm d/2$, read:
\begin{eqnarray}
0=\sum_{l=-\infty}^{\infty}(-1)^l A_l\left[e^{\frac{ik_ld}{2}}+(-1)^n
e^{\frac{-ik_ld}{2}}\right]J_{n+l}(\frac{k_leF}{m\omega^2}),\quad\textrm{for all } n.
\label{Wagner_eq2}
\end{eqnarray}
Eqs.~(\ref{Wagner_eq1}-\ref{Wagner_eq2}) are the starting point for a numerical
implementation of the transfer matrix method for studying the transmission characteristics
of double-barrier diodes with finite barrier heights. Interestingly, the spectral
weights of the photon side bands exhibit strong quenching close to the roots of the Bessel
functions $J_{n}(\frac{k_0eF}{m\omega^2})$, where $n$ is the side band index
and $k_0$ is the wave vector of the centerband resonance.
The $\omega^{-2}$ scaling behavior of the roots is qualitatively different
from the $\omega^{-1}$ dependence found within
Tien-Gordon like models. This is of importance when describing transport in
double-barrier resonant tunneling devices, as we shall describe in section
\ref{Double barriers}.

Eq.~(\ref{Wagner_eq1}) is obtained by making an ansatz which uses all possible particular
solutions of the Hamiltonian in Eq.~(\ref{Wagner_Hamiltonian}) provided that the appropriate
boundary conditions can be satisfied and the symmetries of the problem are properly
considered.
The method for constructing {\it exact} solutions for Hamiltonians like the one in
Eq.~(\ref{Wagner_Hamiltonian}) was developed by Truscott in Ref.~\cite{TrusPRL(93)}.
In this work, Truscott demonstrates that
the solutions to the time-dependent
Schr\"odinger equation for a particle
in a spatially uniform time-dependent field and some potentials
of arbitrary form are like the time-independent eigenfunctions for an identical
static potential
\footnote{A similar approach was developed in the sixties by Henneberger who
proposed a perturbation method for atoms under
intense light beams \cite{HennPRL(68)}. The problem of interaction
of atoms with intense light was formulated via a time-dependent unitary
transformation.
By means of this transformation, which essentially consists of a transformation to an
accelerated frame of reference
in the dipole approximation,
an effective intensity-dependent potential that binds the electrons
can be found.}.
As an application, he considered a rectangular barrier
modulated by a time-dependent field to study the traversal time
for tunneling \cite{BLPRL(82),HauRMP(89)}.

The basic idea of this approach
is to eliminate
the time-dependent field in the Schr\"odinger equation by a
coordinate transformation.
The proof starts by considering a Schr\"odinger equation in which the potential explicitely
includes a spatially uniform field that is an arbitrary function of
time, $V(z,t)-zf(t)$, and a solution $\Psi (z,t)$:
\begin{equation}
-\frac{\hbar^2}{2m}\frac{\partial^2\Psi}{\partial z^2}
+[V(z,t)-zf(t)]\Psi =
i\hbar\frac{\partial \Psi}{\partial t}
\label{Truscot_eq1}
\end{equation}
This is transformed to a new coordinate system ($\xi, t$) where $\xi =z-q(t)$, with
the displacement $q(t)=m^{-1}\int^t p(t')dt'$ and $p(t)=\int^t f(t')dt'$,
by substituting
the product $\phi (\xi,t)\chi (z,t)$ for $\Psi (z,t)$ with:
\begin{equation}
\chi(z,t)=exp[-iE t/\hbar +izp(t)/\hbar
-\int^t \frac{ip^2(t')dt'}{2\hbar m}].
\end{equation}
After division by $\chi(z,t)$ and substracting:
\begin{equation}
[zf(t)+\frac{p^2(t)}{2m}]\phi(\xi,t)+\frac{i\hbar p(t)}{m}\frac{\partial\phi(\xi,t)}
{\partial\xi}
\end{equation}
from both sides, Eq.~(\ref{Truscot_eq1}) becomes:
\begin{equation}
\left[-\frac{\hbar^2}{2m}\frac{\partial^2}{\partial\xi^2}+U(\xi,t)-E\right]\phi(\xi,t)=
i\hbar\frac{\partial\phi (\xi,t)}{\partial t}
\end{equation}
where $U([z-q(t)],t)=V(z,t)$.
For example, for an harmonic time dependent field, $U(\xi,t)-zFcos\omega t$
such that $\xi=z+Fcos(\omega t)/m\omega^2$ and thus:
\begin{eqnarray}
\chi(z,t)=exp[-iEt/\hbar+iFzsin(\omega t)/\hbar\omega-
iF^2[2\omega t-sin(2\omega t)]/8\hbar m \omega^3].\nonumber\\
\label{trus}
\end{eqnarray}
These kind of solutions are used when constructing the ansatz that leads to
Eq.~(\ref{Wagner_eq1}).
\subsection{General formulation \label{Scattering approach}}
We have seen in the previous subsection that
the basic idea of the scattering approach is to relate
transport properties with transmission and reflection
probabilities for electrons incident on a sample. The key
assumption is that the phase of the carrier is preserved over the
entire sample and inelastic scattering is restricted to occur only
in the reservoirs. Here, we describe a more general formulation of PAT using the scattering
approach presented by Pedersen and B\"uttiker in Ref.~\cite{PedPRB(98)}.

The starting point is the current
operator for current incident in contact $\alpha$ in a mesoscopic
system which can be written as \cite{ButtPRB(92)}
\begin{eqnarray}
    \hat{I}_\alpha(t) &=& \frac{e}{h} \int dE \int dE' \left[
        \hat{\bf a}_\alpha^\dagger(E) \hat{\bf a}_\alpha(E') \right.
         - \left.
        \hat{\bf b}^\dagger_\alpha(E) \hat{\bf b}_\alpha(E') \right]
        e^{i\frac{E-E'}{\hbar} t}
\label{current_op}
\end{eqnarray}
where $\hat{\bf a}_\alpha$ and $\hat{\bf b}_\alpha$ are vectors
of operators with components $\hat{a}_{\alpha m}$ and $\hat{b}_{\alpha m}$.
Here $\hat{a}_{\alpha n}$ annihilates an incoming carrier
in channel $m$ in lead $\alpha$ and $\hat{b}_{\alpha m}$ annihilates
an outgoing carrier in channel $m$ in lead $\alpha$.

The incoming and outgoing waves are related by the scattering
matrix ${\bf s}_{\alpha\beta}$ via,
$\hat{\bf b}_\alpha=\sum_\beta {\bf s}_{\alpha\beta} \hat{\bf a}_\beta$ \cite{ButtPRB(92)}.
In a multichannel conductor the s-matrix has dimensions
$N\times M$, where $N$ and $M$ denote the number of channels of lead $\alpha$ and $\beta$ respectively.

Pedersen and B\"uttiker in Ref.~\cite{PedPRB(98)} assumed that a
time dependent field is applied to reservoir $\alpha$. 
The potential is
$eU_\alpha(t)=eV_\alpha(\omega) \cos\omega t$, where
$V_\alpha(\omega)$ is the modulation amplitude.
With this potential
the solution to the single-particle Schr\"odinger equation
at energy $E$ in $\alpha$ is
\begin{equation}
    \psi_{\alpha ,m} (x,t;E) = \phi_{\alpha , m} (x;E) e^{-i Et/\hbar}
        \sum_{l =-\infty}^\infty J_l
        \left( \frac{eV_\alpha}{\hbar\omega}\right)
        e^{-il\omega t}
\label{res}
\end{equation}
where $\phi_{\alpha, m} (x;E) $ is the wave function describing an
incoming (or outgoing) carrier in contact $\alpha$ in channel $m$
in the absence of a modulation potential, and $J_l$ is the $l$'th
order Bessel function.
Thus the potential modulation leads for each state with
central energy $E$ to side bands at energy $E+ l\hbar \omega$
describing carriers which have absorbed $l > 0$ modulation
quanta or have emitted $l < 0$ modulation quanta $\hbar \omega$ \cite{BLPRL(82)}.

Within the scattering approach one assumes that the modulation potential exists only far away
from the conductor and that the time dependent potential vanishes as
one approaches the conductor such that
the annihilation operator of an incoming state close to the conductor
is
\begin{equation}
    \hat{\bf a}_{\alpha,m} (E) = \sum_l
    \hat{\bf a}'_{\alpha , m} (E-l\hbar\omega)
        J_l \left(\frac{eV_\alpha}{\hbar\omega}\right) . \label{lead}
\end{equation}
up to corrections of the order of $\hbar \omega /E_F$ (where $E_F$ is the Fermi energy).
 Using Eq.~(\ref{lead}) one can write the
current operator in terms of the reservoir states $\hat{\bf
a}'_{\alpha , m}$:
\begin{eqnarray}
    \hat{I}_\alpha(t) &=& \frac{e}{h} \int dE \int dE' \sum_{\gamma\delta }
        \sum_{l=-\infty}^\infty \sum_{l=-\infty}^\infty J_l\left(\frac{eV_\gamma}{\hbar\omega}
        \right) J_k\left(\frac{eV_\delta}{\hbar\omega}\right)
        (\hat{\bf a}')_\gamma^\dagger(E-l\hbar\omega) \nonumber \\
        && {\bf A}_{\gamma\delta}(\alpha,E,E')
        \hat{\bf a}'_\delta(E'-k\hbar\omega) e^{i(E-E')t/\hbar}
\label{currentscattering}
\end{eqnarray}
where we have introduced the operator \cite{ButtPRB(92)}
\begin{equation}
    {\bf A}_{\delta\gamma}(\alpha,E,E^{\prime}) = \delta_{\alpha\delta}
    \delta_{\alpha\gamma} {\bf 1}_\alpha -
    {\bf s}_{\alpha\delta}^\dagger(E) {\bf s}_{\alpha\gamma}(E^{\prime}).
\label{a-operator-scattering}
\end{equation}

Assuming that the quantum statistical averages of the reservoir operators are the equilibrium ones
one finds,
\begin{eqnarray}
    I_\alpha(t) &=& \frac{e}{h} \int dE \sum_{\gamma,lk} \label{newcurrent}
        \mbox{Tr} {\bf A}_{\gamma\gamma}(\alpha,E,E+(k-l)\hbar\omega)
        \times \\
        && J_l\left(\frac{eV_\gamma}{\hbar\omega}\right)
        J_k\left(\frac{eV_\gamma}{\hbar\omega}\right)
        e^{-i(k-l)\omega t} f_\gamma(E-l\hbar\omega) . \nonumber
\end{eqnarray}
where $f_\gamma(E)=f(E-\mu_\gamma)$
is the Fermi distribution function for contact $\gamma$.
Here $\mu_\gamma$ is the electrochemical potential of reservoir
$\gamma$.
In Eq.~(\ref{newcurrent}) the trace is over all channels in lead $\alpha$.

Only terms with $l = k$ contribute to the dc-current such that
\begin{eqnarray}
    I_\alpha^{dc} = \frac{e}{h} \int dE \sum_{\gamma,l}
        \mbox{Tr} {\bf A}_{\gamma\gamma}(\alpha,E,E)
        J^{2}_l\left(\frac{eV_\gamma}{\hbar\omega}\right)
        f_\gamma(E-l\hbar\omega) .
\label{cur0a}
\end{eqnarray}

The trace of the operator ${\bf A}$ at equal energy arguments and
equal lower lead indices are just transmission
and reflection probabilities. In particular, $T_{\alpha\gamma}(E)
= - \mbox{Tr} {\bf A}_{\gamma\gamma}(\alpha,E,E)$. For unequal
indices $\alpha$ and $\gamma$ this is the transmission probability
for electrons incident in lead $\gamma$ to be transmitted into
contact $\alpha$. If also $\alpha  = \gamma$ the trace of ${\bf
A}$ is equal to  the probability $R_{\alpha\alpha}$ of electrons
incident in lead $\alpha$ to be reflected back into lead $\alpha$,
minus the number of quantum channels $N$ at energy $E$. Particle
conservation in the scattering process is expressed by the sum
rule $\sum_{\gamma} T_{\alpha\gamma} = 0$. The dc-current thus
read
\begin{eqnarray}
    I_\alpha^{dc} = - \frac{e}{h} \int dE \sum_{\gamma,l}
        T_{\alpha\gamma}(E)
        J^{2}_l\left(\frac{eV_\gamma}{\hbar\omega}\right)
        f_\gamma(E-l\hbar\omega) .
\label{cur0b}
\end{eqnarray}

Let us consider now a two-terminal conductor which consists of
a tunneling barrier between two contacts.
One of the contact potentials is oscillating and the other is kept fixed, 
$V_1(\omega)=V(\omega)$ and $V_2(\omega)=0$. 
This is the geometry considered by Tien and Gordon \cite{tien,tucker1,tucker2}, see Section \ref{Tien-Gordon}.
Assuming that the scattering matrix has been
diagonalized such that transmission through
the barrier is described by a transmission probability $T_m(E)$
and a reflection probability $R_m(E)$ for the m-th eigen channel.
Using Eq.~(\ref{cur0b}) and using the sum rule for Bessel functions,
$\sum_l J_{l+k}(x) J_l(x)=\delta_{k0}$,
one finds
\begin{eqnarray}
I_1^{dc} &=& - \frac{e}{h} \sum_{m}\sum_{l=-\infty}^{\infty}
J_l^2\left(\frac{eV(\omega)}{\hbar\omega}\right)\int dE
T_m(E) [ f_1(E+l\hbar\omega) - f_2(E)]. \label{tiengordon}
\end{eqnarray}
As pointed out by Pedersen and B\"uttiker
\cite{PedPRB(98)}, this simple configuration already suffers from
an important drawback: Eq.~(\ref{tiengordon}) is explicitly {\it
not gauge invariant}\footnote{For instance, 
$V_1(\omega)=V(\omega)$ and
$V_2(\omega)=0$
should be equivalent to $V_1(\omega)=V(\omega)/2$ and
$V_2(\omega)=-V(\omega)/2$. The non-interacting theory yields, however, 
different result in both cases. Another example where this absence of gauge invariance leads 
to completely unphysical results is the situation where
$V_1(\omega)=V_2(\omega)=V(\omega)/2$ which should
give zero current but gives the same as for
$V_1(\omega)=V(\omega)/2$ and $V_2(\omega)=-V(\omega)/2$.}.  
A selfconsistent treatment, like the ones
presented in Refs.~\cite{PedPRB(98),ButtPRL(93),PedPRB(98)b},
beyond the single-particle formulation presented above is needed
in order to achieve both charge and current conservation and
restore gauge invariance. Eq.~\ref{tiengordon} is similar to that
obtained by Tien and Gordon \cite{tien} within Bardeen's Transfer
Hamiltonian approach, see Section \ref{Tien-Gordon}. Although Bardeen's approach
does allow for a gauge-invariant interpretation
\cite{But-GerPLA(68)} it is clear from the
above treatment that one should be careful when using
noninteracting approximations to study photon assisted transport.
In particular, systems where charge accumulation does occur, like
the weakly coupled superlattices we shall describe in Section
\ref{Superlattices}, always need selfconsistency to some extent in
order to achieve charge and current conservation.

Without ac driving, Eq.~(\ref{tiengordon}) becomes
\begin{eqnarray}
I_1^{dc} &=& - \frac{e}{h} 
\int dE
\sum_{m}T_m(E) [ f_1(E) - f_2(E)]. \label{Landauer1}
\end{eqnarray}                       
which in the linear response limit gives the so-called Landauer formula for the linear conductance:
\begin{eqnarray}
\mathcal{G}&=& \frac{e^2}{h}\sum_{m}T_m(E_F)\label{Landauer2}  
\end{eqnarray}   
\section{Non equilibrium Green's functions formulation of transport \label{Keldysh}}
We have seen in the previous section the needfulness to go beyond a noninteracting picture when discussing
photon assisted tunneling. Here, we briefly review
the nonequilibrium Green's functions formalism, which allows the study of photon-assisted transport and
the inclusion of other effects, like impurity scattering
or electron-electron interactions, within a common scheme.

Early studies of
nonequilibrium tunneling problems
were presented already
in the 70s by Caroli and
co-workers
\cite{Caroli1,Caroli2,Caroli3,Combescot}. 
Among the first studies of photon assisted transport using non-equilibrium Green's functions techniques
we can mention that of Chen and Ting (ac conductance of resonant tunneling diodes \cite{ChenPRL(90)}),
Levy-Yeyati and Flores (photocurrent effects in scanning tunneling microscopes \cite{YeyaPRB(91)})
and the work of Datta and Anatram \cite{DattaPRB(92)}
where the possibility of obtaining a Landauer-type expression for the dc current by using
nonequilibrium Green's functions was first analyzed.

\subsection{General formulation for tunneling systems}
The complete theory of time-dependent transport in interacting resonant tunneling systems
was put forward by Jauho et al in Ref.~\cite{JauPRB(94)}\footnote{For further details, we refer the reader
to the texbook of Haug and Jauho \cite{Jauhobook} where a complete
description of the nonequilibrium Green's functions formalism and
its application to transport in mesoscopic physics can be found.}, here we closely follow their derivation.

The basic idea when applying the nonequilibrium formalism to
tunneling problems is to assume that the leads and the "central
region", see below, are decoupled in the remote past. Also, it is
assumed that each region is in thermal equilibrium (each
equilibrium distribution function being characterized by its
respective chemical potential). The couplings between the
different regions are then established and treated as
perturbations via nonequilibrium perturbation theory
\cite{nonequilibrium1,nonequilibrium2}.

The contacts are assumed to be noninteracting, and the
single-particle energy in lead $\alpha$ is given by
$\varepsilon_{k,\alpha}(t)=\epsilon_{k,\alpha} +
\Delta_\alpha(t)$, where $\Delta_\alpha(t)$ is the external time
modulation. The leads are connected to the central region via a
hopping term with matrix element $V_{k\alpha;n}(t)$, where $n$
labels the eigenstates of the central region.  The total
Hamiltonian describing the coupled system is $H=H_L+H_R+H_T+H_{\rm
cen}$ with:
\begin{eqnarray}
H_\alpha &=& \sum_{k,\alpha\in L,R} \epsilon_{k,\alpha}(t) c^\dagger_{k,\alpha} c_{k,\alpha}\nonumber\\
H_T&=&\sum_{k,\alpha\in L,R;n}
\left[V_{k\alpha;n}(t) c^\dagger_{k,\alpha} d_n + {\rm h.c.}\right]\nonumber\\
H_{\rm cen}&=&H_{\rm cen}\left[\{d_n\},\{d^\dagger_n\},t\right]\;,
\end{eqnarray}
where the central part Hamiltonian depends on the particular
system under consideration.  The operators
$\{d_n\},\{d^\dagger_n\}$ refer to a complete set of
single-particle states of the central region. For instance, for
the Anderson impurity model (see Section \ref{Kondo}):
\begin{eqnarray}
H_{\rm cen}=\sum_{\sigma =\uparrow,\downarrow}
\epsilon_{\sigma}(t) d^\dagger_{\sigma} d_{\sigma} + U
d_{\uparrow}^\dagger d_{\uparrow} d_{\downarrow}^\dagger
d_{\downarrow} \label{HAnd}\;.
\end{eqnarray}
The derivation starts by considering the current from the, e.g.,
left contact to the central part:
\begin{equation}
J_L(t) = \langle I_L(t) \rangle = -e \langle {\dot N}_L (t)
\rangle = -ie \langle \left[ H,N_L \right]\rangle \;.
\end{equation}
evaluating the commutator $[H,N_L]$ one finds
\begin{equation}
J_L(t) = {2e\over\hbar} {\rm Re} \Big\{ \sum_{k,\alpha,n}
V_{k\alpha,n}(t) G^<_{n,k\alpha}(t,t)\Big \}\;,
\label{time-dependent-Glesser}
\end{equation}
which involves the time-diagonal part of the correlation function
\begin{equation}
G^<_{n,k\alpha}(t,t') = i \langle c^\dagger_{k\alpha}(t') d_n (t)
\rangle\;.
\end{equation}
$G^<_{n,k\alpha}(t,t')$ can be obtained by applying analytic
continuation with the Langreth rules \cite{LangrethRules} to the
equation-of-motion for the time-ordered (along a complex contour)
function $G^t_{n,k\alpha}(t,t')=-i\langle T_\tau
c^\dagger_{k\alpha}(t')d_n (t) \rangle$ which lead to
\begin{equation}
G^<_{n,k\alpha}(t,t') = \sum_m \int dt_1 V^*_{k\alpha,m } \Big [
G^r_{nm}(t,t_1) g^<_{k\alpha}(t_1,t') + G^<_{nm}(t,t_1)
g^a_{k\alpha}(t_1,t') \Big ]. \label{Glesser}
\end{equation}
As mentioned above, the occupations of the leads are determined by
equilibrium distribution functions
\begin{eqnarray}
g_{k\alpha}^<(t,t') &\equiv&
i\langle{\bf{c}}_{k\alpha}^{\dagger}(t'){\bf{c}}_{k\alpha}(t)\rangle
=i f(\varepsilon^0_{k\alpha}) \exp\big [ -i \int_{t'}^t d
t_1\varepsilon_{k\alpha}(t_1)\big ]\;,
\label{gcontactless} \\
g_{k\alpha}^{r,a}(t,t') &\equiv& \mp i \theta(\pm t \mp t')
\langle\lbrace{\bf{c}}_{k\alpha}(t),
{\bf{c}}_{k\alpha}^{\dagger}(t')\rbrace\rangle
\nonumber\\
&=& \mp i\theta(\pm t \mp t') \exp\big [ -i \int_{t'}^t
dt_1\varepsilon_{k\alpha}(t_1)\big ] \;. \label{gcontactr}
\end{eqnarray}
Substituting Eqs.~(\ref{gcontactless})--(\ref{gcontactr}) into
Eq.~(\ref{Glesser}), one gets
\begin{eqnarray}
J_L(t) &=& - {2e\over \hbar} \int_{-\infty}^t dt_1 \int
{d\epsilon\over 2\pi} {\rm ImTr} \Big \{ e^{-i\epsilon(t_1-t)}
{\bf \Gamma}^L(\epsilon,t_1,t)
\nonumber\\
&\quad&\quad\times \left[ {\bf G}^<(t,t_1) + f_L(\epsilon) {\bf
G}^r(t,t_1)\right] \Big \}\;. \label{jfinal}
\end{eqnarray}
Here the Green functions ${\bf G}^{<,r}$ are {\it matrices} in the
indices $(m,n)$, and the functions ${\bf \Gamma}$ are defined as
\begin{equation}
\left[ {\bf \Gamma}^L(\epsilon,t_1,t)\right]_{mn} =
2\pi\sum_{\alpha\in L} \rho_\alpha(\epsilon)
V_{\alpha,n}(\epsilon,t) V^*_{\alpha,m}(\epsilon,t_1) e^{\big[-i
\int_t^{t_1} dt_2 \Delta_\alpha (\epsilon,t_2)\big]}\;,
\end{equation}
where $\rho_\alpha(\epsilon)$ is the density of states. It is
important to note that the current formula only involves the Green
function of the central region.  However, ${\bf G}^<(t,t_1)$ must
be calculated in the presence of the coupling to the leads, which
is a highly nontrivial task for an interacting system.

In the absence of time dependent fields, Eq.~(\ref{jfinal}) can be further
simplified and one gets \cite{MeiPRL(92)}:
\begin{eqnarray}
J &=& {ie\over 2\hbar} \int {d\epsilon\over 2\pi} {\rm Tr}\Big \{
\left[ {\bf \Gamma}^L(\epsilon) - {\bf \Gamma}^R(\epsilon)\right]
{\bf G}^<(\epsilon)\nonumber\\
&\quad&\quad + \left[ f_L(\epsilon){\bf \Gamma}^L(\epsilon) -
f_R(\epsilon) {\bf \Gamma}^R(\epsilon)\right] \left[ {\bf
G}^r(\epsilon) - {\bf G}^a(\epsilon)\right]\Big\}.
\end{eqnarray}
Often it is a good approximation to assume that the couplings
${\bf \Gamma}^{L(R)}(\epsilon)$ are proportional. In this case,
\begin{eqnarray}
J =  {i e\over\hbar}\int{d\varepsilon\over 2\pi}
\left[f_L(\varepsilon)-f_R(\varepsilon)\right] {\mathcal
T}(\varepsilon)\;, \label{jprop}
\end{eqnarray}
where
\begin{equation}
{\mathcal T}(\varepsilon)  = {\rm {Tr}} \left\lbrace {{\bf
\Gamma}^L(\varepsilon){\bf \Gamma}^R(\varepsilon)\over {\bf
\Gamma}^L(\varepsilon)+{\bf\Gamma}^R(\varepsilon)} \bigl[{\bf
G}^r(\varepsilon)-{\bf  G}^a(\varepsilon)\bigr ]\right\rbrace \;.
\label{calT}
\end{equation}
The difference between the retarded and advanced Green functions
is essentially the density of states.
Eq.~(\ref{jprop}) is very similar to the Landauer formula (see Eq.~(\ref{Landauer1}): 
it expresses the current as the 
integral of a weighted density of states, ${\mathcal T}(\varepsilon)$, 
times the difference of occupation factors in the contacts.
Note, however, that ${\mathcal T}(\varepsilon)$ is not just the transmission coefficient
but rather the fully interacting density of states of the central region 
(including the electron-electron interaction
or spin-flip processes\footnote{A discussion about this issue can be found,
for instance, in Ref.~\cite{KonPRB(02)}.}, for instance). 

Assuming proportional couplings, the time average of
Eq.~(\ref{jfinal}) becomes:
\begin{equation}
\langle J_L(t) \rangle = - {2e\over\hbar} \int {d\epsilon\over
2\pi}\left[ f_L(\epsilon) - f_R(\epsilon) \right] {\rm Im Tr} \{
{\Gamma_L(\epsilon) \Gamma_R(\epsilon)\over \Gamma_L(\epsilon) +
\Gamma_R(\epsilon)} \langle {\bf A}(\epsilon,t) \rangle\},
\label{Jauho-jfinal-average}
\end{equation}
where
\begin{equation}
{\bf A}(\epsilon,t)=\int dt_1 {\bf
G}^r(t,t_1)e^{[i\epsilon(t-t_1)+i\int_{t_1}^t dt_2\Delta(t_2)]}
\end{equation}

This expression for the current in terms of a time-averaged density of states 
has been extensively used in the literature for studying the average current of ac-driven 
{\it interacting} systems. We shall describe various examples 
of the use of Eq.~(\ref{Jauho-jfinal-average}) along the review.

If one considers a single, noninteracting state with
energy $\epsilon_0$ in the central
region under the influence of a harmonically varying field
with amplitude $V_{ac}$, the function
$\langle {\rm Im}
A(\epsilon,t)\rangle=-\frac{\Gamma}{2} \langle
|A(\epsilon,t)|^2\rangle$ is given by
\begin{equation}
\langle {\rm Im} A(\epsilon,t)\rangle = - {\Gamma\over 2}
\sum_{n=-\infty}^{\infty}J_n^2(\frac{eV_{ac}}{\hbar\omega})
{1 \over
(\epsilon-\epsilon_0-n\hbar\omega)^2 + (\Gamma/2)^2} \;
\end{equation}
and the
current can be written as
\begin{equation}\label{cur}
\langle J_L(t) \rangle =
{e\over h}\sum_{n=-\infty}^{\infty}J^2_n\left(
{eV_{ac}\over\hbar\omega}\right) \int d\epsilon
[f_L(\epsilon)-f_R(\epsilon)]
T(\epsilon-n\hbar\omega)
\end{equation}
where $T(\epsilon)$ is the elastic transmission coefficient
through the resonant system.        
In the linear response (with respect to the {\it dc bias voltage}) regime the linear conductance is then
\begin{equation}\label{Lac}
\mathcal{G}_{\rm ac} = {e^2\over h}\sum_{n=-\infty}^\infty 
J^2_n\left({eV_{ac}\over\hbar\omega}\right)
T(E_F-n\hbar\omega)\;,
\end{equation}
which is a generalization of the Landauer formula in Eq.~(\ref{Landauer2})
to the ac driven situation. Note that Eq.~(\ref{cur}) despite being similar 
to the Tien-Gordon formula, compare with Eq.~(\ref{Tien-Gordon-rectif}),
is {\it not} of this form.
The main difference between both expressions 
is that Eq.~(\ref{cur}) is obtained by considering the coupling of the driven resonant level to
{\it equilibrium} contacts to all orders, while Eq.~(\ref{Tien-Gordon-rectif}) is obtained by considering
only the lowest order coupling between the different parts of a system {\it biased} with a voltage
$V(t)=V_{dc} + V_{ac}\cos\omega t$.
\begin{figure}
\begin{center}
\includegraphics[width=0.75\columnwidth,angle=0]
{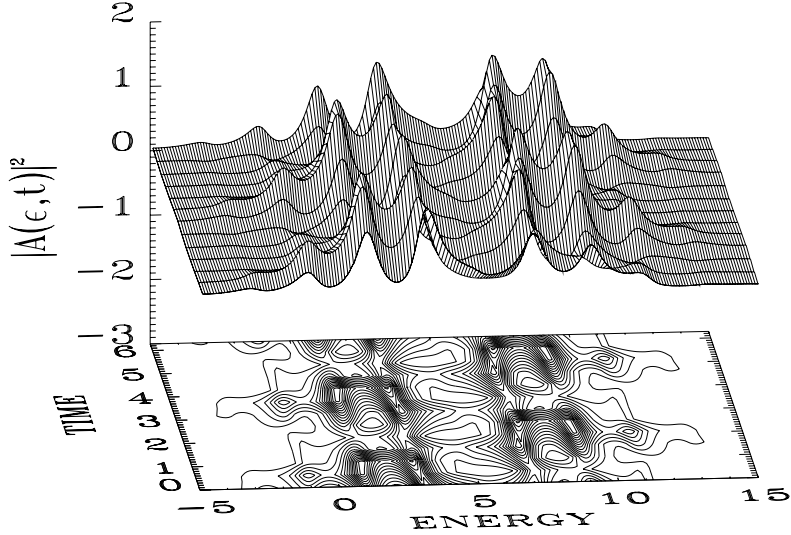}
\end{center}
\caption{$|A(\epsilon,t)|^2$ as a function of time for harmonic modulation
for a symmetric structure, $\Gamma_L=\Gamma_R=\Gamma/2$.
The unit for the time-axis is $\hbar/\Gamma$, and all energies
are measured in units of $\Gamma$, with the values
$\mu_L=10$, $\mu_R=0$, $\epsilon_0=5$, $\Delta=5$, $\Delta_L=10$,
and $\Delta_R=5$.  The modulation frequency is
$\omega=2\Gamma/\hbar$. Reprinted with permission from \cite{JauPRB(94)}.
\copyright 1994 American Physical Society.}
\label{Jauho_fig1}
\end{figure}          

In the time dependent case, the occupation of the central level, $N(t)=-iG^<(t,t)$, reads:
\begin{eqnarray}
N(t) &=& \sum_{L/R} \int {d\epsilon\over 2\pi} f_{L/R}(\epsilon)
|A_{L/R}(\epsilon,t)|^2.
\end{eqnarray}
Using this result, the time dependent current can be written as a sum of currents flowing out from 
the central region into the left (right) contact $J_{L/R}^{out}(t)={e\over \hbar} \Gamma_{L/R}N(t)$
and currents flowing into the central region from the left (right) contact 
$J_{L/R}^{in}(t)=- {e\over \hbar} \Gamma_{L/R}\int \frac{d\epsilon}{\pi} f_{L/R}(\epsilon)
{\rm Im}\left\{ A_{L/R}(\epsilon,t)\right\}$, such that the total time dependent 
current is\footnote{The effects of nonorthogonality of the electronic states in the 
leads on the time-dependent current have been
analyzed by Fransson et al in Ref.~\cite{FranPRB(02)} using the same Green's functions approach.}:
\begin{eqnarray}
J_{L/R}(t) &=& - {e\over \hbar} \Gamma_{L/R} \Big [
N(t) + {1\over \pi} \int d\epsilon f_{L/R}(\epsilon)
{\rm Im}\left\{ A_{L/R}(\epsilon,t)\right\} \Big ].
\label{time-dependent-Jauho}
\end{eqnarray}
In the case of harmonic modulation $\Delta_{L/R,0}(t)= \Delta_{L/R,0}\cos\left(\omega t\right)$ one has:
\begin{eqnarray}
A_{L/R}(\epsilon,t) &=& \int dt_1 e^{i\epsilon(t-t_1)}
e^{-i\int_t^{t_1}dt_2 \Delta_{L/R}(t_2)}
G^r(t,t_1)\nonumber\\
&=&e^{-i{\Delta_0\over\hbar\omega}\sin(\omega t)}
\sum_{n=-\infty}^\infty
{J_n\left(\Delta_0-\Delta_{L/R}\over \hbar\omega\right)
e^{in\omega t}\over \epsilon-\epsilon_0 -n\hbar\omega + i \Gamma/2}
\;.
\end{eqnarray}
Fig.~\ref{Jauho_fig1} shows an example of the time dependence of $|A(\epsilon,t)|^2$ \cite{JauPRB(94)}.
The maxima in the plot are related to
photonic side-bands occurring at $\epsilon=\epsilon_0 \pm n\hbar\omega$.

\begin{figure}
\begin{center}
\includegraphics[width=0.75\columnwidth,angle=0]
{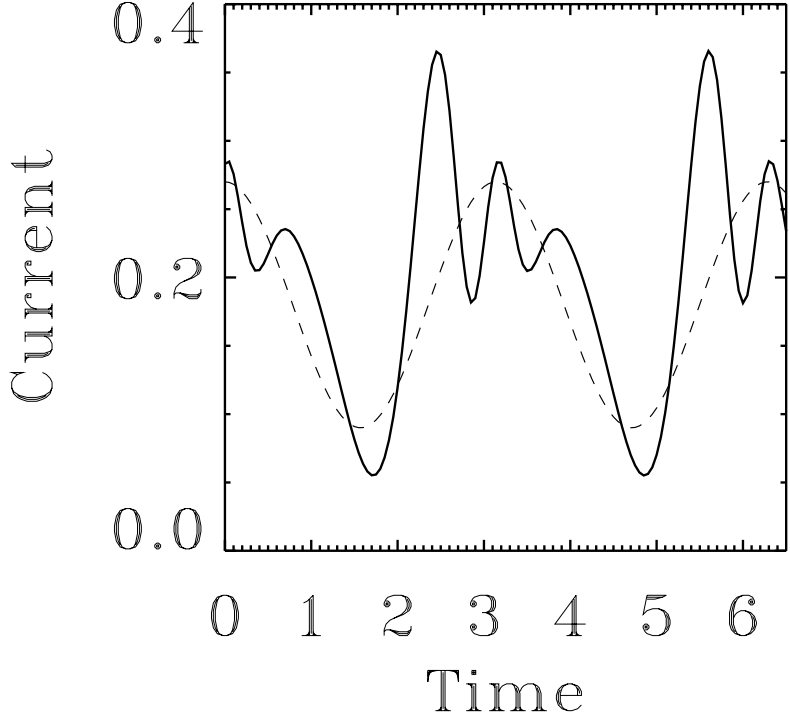}
\end{center} 
\caption{The time-dependent current $J(t)$ for harmonic modulation.
The dc bias is defined via $\mu_L=10$ and $\mu_R=0$, respectively.
The dotted line
shows (not drawn to scale) the time dependence of the drive signal.
The temperature is $k_BT=0.1\Gamma$. Reprinted with permission from \cite{JauPRB(94)}.
\copyright 1994 American Physical Society.}
\label{Jauho_fig2}
\end{figure}

The time-dependent current is shown in Fig.~\ref{Jauho_fig2}. 
$J(t)$ displays a non-adiabatic time-dependence which reflects
the complex structure of $|A(\epsilon,t)|^2$ and $ImA(\epsilon,t)$, which determine the out and in currents, 
respectively, see Eq.~(\ref{time-dependent-Jauho}).
The basic physical mechanism underlying the secondary maxima and
minima in the current is the line-up of a photon-assisted
resonant tunneling peak with the contact chemical potentials. The rapid time variations are due to $J_{in}$ 
(or, equivalently, to $ImA(\epsilon,t)$). The out-current $J_{out}$ is determined by the occupation $N(t)$ and 
hence varies only on a time scale $\Gamma/\hbar$.

The above derivation exemplifies how nonequilibrium Green's functions are powerful tools 
to study high frequencies and far-from equilibrium situations. 
Of particular importance is the
analysis of current partition
and displacement currents in multiprobe samples; 
issues which are essential in order to have theories which 
are current-conserving and Gauge invariant 
(see the discussion about Gauge invariance in Section \ref{Scattering approach}).
A multiterminal conductance formula describing resonant tunneling through an 
interacting mesoscopic system was derived by Stafford in Ref.~\cite{Stafford96}. In this work, 
an explicit gauge-invariant formulation for the nonlinear dc case is obtained.
The generalization to ac situations has been put forward by
Anantram and Datta \cite{Ana95} and Wang et al \cite{Wang99}.

In a multiprobe system,
the dynamic conductance $G_{\alpha\beta}$
due to the tunneling current is defined
as
\begin{equation}
J_\alpha(\omega)=\sum_\beta G_{\alpha\beta}(\omega)V_\beta(\omega).
\end{equation}
In the time-dependent case the tunneling currents do not add up
to zero, due to charge accumulation/depletion.  The total current,
however, is conserved:
\begin{equation}
\sum_\alpha J_\alpha(\omega) = i\omega Q(\omega)\;,
\label{total-sum}
\end{equation}
where 
\begin{equation}
Q(\omega) = -\sum_\beta ie \int {d\epsilon\over 2\pi}
{\rm Tr}[g^<_\beta(\epsilon+\omega,\epsilon)]V_\beta,
\label{charge-frequency}
\end{equation}
is the accumulated charge.
The function $g^<_\beta (\epsilon,\epsilon ')$ is the double Fourier transform of
the {\it small signal} component of the full Green function
$G^<$, defined as $\sum_\beta g^<_\beta V_\beta=G^<-G_{eq}^<$ in the linear regime \cite{Ana95,Wang99} 
(here, $G_{eq}^<$ is the equilibrium lesser function).  
The total current in probe $\alpha$
is $J_\alpha^{\rm tot}=J_\alpha+J_\alpha^d$, where
$\sum_\alpha J_\alpha^d=J^d=dQ/dt$, is the {\it total} displacement current, and current conservation means
$\sum_\alpha J_\alpha^{\rm tot}=0$.  Additional information is
required to partition $J^d$, because only the sum of
the various displacement currents is
known via Eq.(\ref{total-sum}), namely:
\begin{equation}
\sum_\alpha J_\alpha^d(\omega) = -i\omega Q(\omega).
\label{total-displacement}
\end{equation}  
This can be easily done
in a model where coupling constants
between the central region and the contacts are independent of
energy where the partitioning is 
\begin{equation}
J^d_\alpha=
(\Gamma_\alpha/\sum_\beta\Gamma_\beta)J^d.
\end{equation}
By requiring charge conservation and gauge invariance
Wang et al \cite{Wang99}
have outlined a procedure which allows to partition the current in a more general situation:
Since $J_\alpha^{\rm tot}=\sum_\beta G_{\alpha\beta}^{\rm tot}V_\beta$, current conservation means 
$\sum_\alpha G_{\alpha\beta}^{\rm tot}=0$, while Gauge invariance means 
$\sum_\beta G_{\alpha\beta}^{\rm tot}=0$.
If now the total displacement current is partitioned into the contributions from individual probes, 
$J_\alpha^{\rm tot}=J_\alpha+A_\alpha J^d$, with partition 
coefficients that satisfy $\sum_\alpha A_\alpha=1$,
the result for
the dynamical conductance is
\begin{equation}
G_{\alpha\beta}^{\rm tot}(\omega)=G_{\alpha\beta}
- G^d_\beta{\sum_\gamma G_{\alpha\gamma}\over \sum_\gamma G^d_\gamma}\;,
\label{dyncond}
\end{equation}
where
\begin{equation}
G_\beta^d = -e\omega \int{d\epsilon\over 2\pi} {\rm Tr}
[g^<_\beta(\epsilon+\omega,\epsilon)]\;,
\end{equation}
is obtained from Eq.~(\ref{charge-frequency}).
The result (\ref{dyncond}) formally agrees with the scattering
matrix results of B{\"u}ttiker et al \cite{PedPRB(98),ButtPRL(93)} 
in the zero-capacitance limit, namely if the Coulomb interaction insures 
local charge neutrality, but now the various terms
are expressed in terms of nonequilibrium Green functions which, in principle,
allow for inclusion of different effects like impurity scattering
or electron-electron interactions within the same scheme.   
\subsection{Truncation method for Green's functions in 
time-dependent fields \label{truncation-brandes}}
Another application
of the nonequilibrium Green's functions technique was proposed by
Brandes in Ref.~\cite{BranPRB(97)}.
Interestingly, this method is not restricted to tunneling systems and
can be applied in two-dimensional and three dimensional problems
and for arbitrary static potentials. This is of importance for
investigations related to transport experiments as,
e.g., in the quantum Hall regime \cite{Meietal92,Sunetal94,Meietal96}, where
microwave irradiation gives rise to peculiar changes of the DC 
conductivities\footnote{See Section \ref{MicrowaveHall}.}. 
This formalism allows to calculate the Green's
function of non--interacting electrons moving in an arbitrary static
potential under the influence of a time--dependent electric field with
frequency $\omega_0$.
By using the Dyson equation, the Keldysh Green's function can be obtained in two different
calculation schemes.

The starting point of both approaches is the Dyson equation for the
Keldysh Green's function matrices. Starting from a basis of eigenstates
labeled with $\alpha $ and defining Green's functions \cite{Jauhobook}
$iG_{\alpha \beta }^{T}(t_{1},t_{2})\equiv\langle Tc_{\alpha }(t_{1})c_{\beta
}^{\dagger}(t_{2})\rangle$,
$iG_{\alpha \beta }^{\tilde{T}}(t_{1},t_{2})\equiv\langle \tilde{T}c_{\alpha
}(t_{1})c_{\beta }^{\dagger}(t_{2})\rangle$,
$iG_{\alpha \beta }^>(t_{1},t_{2})\equiv\langle c_{\alpha
}(t_{1})c_{\beta }^{\dagger}(t_{2})\rangle$,
$iG_{\alpha \beta }^<(t_{1},t_{2})\equiv-\langle c_{\beta
}^{\dagger}(t_{2})c_{\alpha }(t_{1})\rangle$,
where $T$ ($\tilde{T}$) denote (anti)-chronological time ordering.
The Green's functions are written in a matrix block
\begin{eqnarray}
\label{matrix}
G(t_{1},t_{2})=
\left( \begin{array}{cc}
G^{T}(t_{1},t_{2}) & -G^{<}(t_{1},t_{2})  \\
G^{>}(t_{1},t_{2})   &
-G^{\tilde{T}}(t_{1},t_{2})
\end{array}\right),
\end{eqnarray}
for which the Dyson-integral equation has the same form as in equilibrium
theory,
\begin{eqnarray}
\label{Brandes-dyson}
G(t_{1},t_{2})=G^{0}(t_{1},t_{2})+\int\int dtdt'G^{0}(t_{1},t)\Sigma (t,t')
G(t',t_{2}),
\end{eqnarray}
where the matrix $\Sigma(t,t')$ is composed of $\Sigma^{T}(t,t')$,
$-\Sigma^{\tilde{T}}(t,t')$, $\Sigma^{>}(t,t')$, $-\Sigma^{<}(t,t')$, in analogy to
Eq.~(\ref{matrix}). Here, $G^{0}$ denotes the unperturbed Greens' function,
e.g. $iG_{\alpha \beta }^{T,0}(t_{1},t_{2})\equiv\langle T \hat{c}_{\alpha
}(t_{1})\hat{c}_{\beta }^{\dagger}(t_{2})\rangle$, where the electron
creation (annihilation) operators $\hat{c}^{(\dagger)}_{\alpha }(t)$ are
given in the interaction picture which is defined according to the
splitting of the total time dependent Hamiltonian $H(t)$. This splitting
defines the way that the perturbation theory is performed.

Since one is interested in a Hamiltonian a part of which oscillates with
frequency $\omega_0 $, it is useful to perform a Fourier analysis according to
time 'center of mass' and relative coordinates
(''Wigner--coordinates''), namely $T=(t_{1}+t_{2})/2$
and $t=t_{1}-t_{2}$. This decomposition is defined according to
\begin{eqnarray}
\label{fourier}
G(t_{1}=T+t/2,t_{2}=T-t/2)=\frac{1}{2\pi}\sum_{n}\int_{-\infty}^{\infty}
d\omega
e^{-i\omega t}e^{i\omega_0 n T} G(\omega ,n)
\end{eqnarray}
and correspondingly for $G^{0}$ and $\Sigma $. The inverse transformation is
\begin{eqnarray}
G(\omega ,n)=\int_{-\infty}^{\infty}dt e^{i\omega t}
\int_{0}^{2\pi} \frac{d(\omega_0 T)}{2\pi}
e^{-i\omega_0 nT}
G(t_{1}=T+t/2,t_{2}=T-t/2).
\end{eqnarray}
Of special interest is the component $n=0$ which determines the
average over the 'center-of-mass' coordinate $T$.
In particular, in the case of an equilibrium situation
(no electric field), all components $G(\omega ,n)$ with $n\ne 0$ vanish because
the Green's function depends on the relative coordinate $t=t_{1}-t_{2}$ only.

Inserting Eq.~(\ref{fourier}) into the Dyson equation Eq.~(\ref{Brandes-dyson}), a
straightforward calculation yields
\begin{eqnarray}
\label{Brandes-dyson2}
G(\omega ,n)&=&G^{0}(\omega ,n)+\sum_{n_{1} n_{2}}G^{0}\left(\omega
+\frac{n_{1}-n}{2}\omega_0 ,n_{1}\right)\nonumber\\
&\times&\Sigma \left(\omega +
\frac{n_{1}-n_{2}}{2}\omega_0 ,n-n_{1}-n_{2}\right)
G\left(\omega
+\frac{n-n_{2}}{2}\omega_0 ,n_{2}\right).
\end{eqnarray}

Brandes concentrates on the non-
interacting case where the perturbation is a one-particle operator and
the self energy is (Keldysh) block-diagonal
\begin{eqnarray}
\label{self}
\Sigma (t,t')=
\left( \begin{array}{cc}
V(t) & 0  \\
0    & V(t)
\end{array}\right)\delta (t-t').
\end{eqnarray}
An impurity average effectively introduces interactions among the electrons
and the self energy becomes different from Eq.~(\ref{self}). However, as long
as no impurity average is performed, one merely has to deal with a one-
particle problem and the integral equation Eq.~(\ref{Brandes-dyson}) together with
Eq.~(\ref{self}) exactly determines the Green's function $G(t_{1},t_{2})$.

Because of the linear relation $G^{T}+G^{\tilde{T}}=G^{>}+G^{<}$, a rotation
to tridiagonal form can be performed in Eq.~(\ref{Brandes-dyson}). One of the
resulting equations is the one for the retarded
Green's function\footnote{This Dyson equation for the retarded Green's function can be obtained as well with the
Langreth's rules for analytical continuation. See Refs.~\cite{Jauhobook,LangrethRules}.}
$iG_{\alpha \beta }^{R}(t_{1},t_{2})\equiv
\theta(t_{1}-t_{2})\langle c_{\alpha }(t_{1})c_{\beta
}^{\dagger}(t_{2})+c_{\beta}^{\dagger}(t_{2})c_{\alpha }(t_{1})\rangle$,
which reads, using $\Sigma ^{R}(t,t')=V(t)\delta (t-t')$,
\begin{eqnarray}
\label{Brandes-dysonretarded}
G^{R}(t_{1},t_{2})=G^{0,R}(t_{1},t_{2})+\int dtG^{0,R}
(t_{1},t)V(t)G^{R}(t,t_{2}).
\end{eqnarray}

Let us assume now that
the system is subject to a spatially
homogeneous electric field which oscillates in time with a
frequency $\omega_0 $ and is polarized in direction ${\bf e}$,
${\bf E}(t)\equiv{\bf e}E_{0}\cos(\omega_0 t)$
The associated vector potential  $ {\bf A}^{e}(t) = -({\bf e}E_{0}/\omega_0 )\sin(\omega_0 t)$
couples to the momentum ${\bf p}_{i}$ of the $i$-th
electron via
\footnote{An external magnetic field would give an extra contribution to the vector potential.}
$
{\bf p}_{i}\to {\bf p}_{i}-{\bf A}^{e}(t),
$
(we take $\hbar=e=c=1$ throughout).
The {\em additional} energy through the electric field is
\begin{eqnarray}
\label{firstq}
H_{e}(t)= \frac{E_{0}}{\omega_0 }\sin(\omega_0 t)\sum_{i=1}^{N_{e}}
{\bf e}{\bf v}_{i}
\end{eqnarray}
where $m^{*}$ is the effective mass, $m^{*}\bf{v}_{i}=  {\bf p}_{i}-{\bf A}({\bf x_{i}})$, and
$N_{e}$ the number of electrons in the
system\footnote{The term $N_{e}\frac{1}{2m^{*}\omega_0^{2}}E_{0}^{2}\sin^{2}(\omega_0 t)$ is neglected
in this derivation. This term gives rise to a global phase factor in the Green's functions.}.
Note that although the electric field ${\bf E}(t)$ is
homogeneous, in an alternative gauge a corresponding {\em scalar potential}
would be linear in the space coordinate, namely $\sim {\bf e x}\cos(\omega_0 t)$.

As we mentioned above, the Keldysh Green's function can be obtained in two different
calculation schemes by using the Dyson equation in Eq.~(\ref{Brandes-dyson2}).
First, if one regards the static potential as
a perturbation and includes the electric field in the unperturbed Hamiltonian,
a formally exact solution can be obtained which can be evaluated approximately by
inverting a truncated matrix containing a finite number of 'photo-blocks'.
The advantage of this method is its exactness in the electric field; it
furthermore sums up the static potential to infinite order and is
perturbative in the higher {\em Fourier components} of the Green's function
which correspond to the 'center--of--mass' time coordinate.
Second, starting from the exact eigenstates of the static potential, again an
exact formal solution is derived in which the Green's function is represented
as the inverse of an infinite tridiagonal matrix, which is the Green's
function analogue of the Floquet state Hamiltonian. This approach in
particular is useful in situations where the static scattering problem
is already solved and one is interested in
the effect of an additional, time-dependent electric field. Here, we briefly review the first method.

Starting from
an eigenstate basis of plane waves, i.e. $|\alpha \rangle=|{\bf k}\rangle$
where $\langle{\bf x}|{\bf k}\rangle\equiv \phi_{{\bf k}}({\bf x})=
(1/L^{d/2})\exp(-i{\bf kx})$ and $L^{d}$
is the system volume ($L\to \infty$ in the thermodynamic limit),
the velocity matrix element is diagonal\footnote{Note that Eq.~(\ref{velocitydia}) is exact only for plane waves.
The results by Brandes are also valid for a general
eigenstate basis with the approximation that non-diagonal elements of the
velocity $\langle \alpha |v|\beta \rangle$ are zero.},
\begin{equation}
\label{velocitydia}
\langle \alpha |v|\beta \rangle=\delta _{\alpha ,\beta }\langle \alpha
|v|\alpha \rangle\equiv \delta _{\alpha \beta }v_{\alpha },
\end{equation}
namely $v_{\alpha }=v_{{\bf k}}={\bf e}{\bf k}/m^{*}$.
The bare Hamiltonian (system + time dependent field) is thus
\begin{equation}
\label{hamepstime}
H_{0}(t)=\sum_{\alpha }\varepsilon _{\alpha }(t)c^{\dagger}_{\alpha }c
^{\phantom{\dagger}}_{\alpha };
\quad \varepsilon _{\alpha }(t)\equiv \varepsilon _{\alpha }+\frac{E_{0}}
{\omega_0 }v_{\alpha }\sin(\omega_0 t),
\end{equation}
where $\varepsilon _{\alpha }$ is the energy of state $\alpha $.

The static potential $V({\bf x})$ now gives rise to an
additional part to the total Hamiltonian which reads
\begin{equation}
\label{fullh}
H(t)=H_{0}(t)+V,\quad V=\sum_{\alpha \beta }V_{\alpha \beta }
c_{\alpha}^{\dagger}c_{\beta }^{\phantom{\dagger}}.
\end{equation}
At this stage, $V$ is not yet specified further. Depending on the physical
situation, it describes a single impurity, a distribution of random
scatterers, a double barrier etc.

It is straightforward to define an interaction picture with respect to the unperturbed
part $H_{0}$ of the Hamiltonian which relates bare Green's functions
with and without electric field as:
\begin{equation}
\label{gfree}
G^{T,0}_{\alpha \beta }(t_{1},t_{2})=e^{ig_{\alpha }[\cos(\omega_0 t_{1})
-\cos(\omega_0 t_{2})]} G^{T,free}_{\alpha}(t_{1}-t_{2})\delta _{\alpha \beta },
\end{equation}
with
$g_{\alpha }\equiv\frac{E_{0}v_{\alpha }}{\omega_0 ^{2}}$.
In Fourier space, the equation for the Keldysh Green's function matrix
can be written as:
\begin{equation}
\label{gbessel}
G^{0}_{\alpha \beta }(\omega ,N)=i^{N}\sum_{n}J_{n}(g_{\alpha })J_{N-n}(-g_{\alpha })
G^{free}_{\alpha }(\omega +[n-N/2]\omega_0 )\delta _{\alpha \beta }.
\end{equation}

Since the perturbation potential $V$ is time-independent, the selfenergy
in Eq.~(\ref{Brandes-dyson2}) is
\begin{equation}
\Sigma (\omega ,N)=S \cdot\delta _{N,0},\quad
S=
\left( \begin{array}{cc}
V & 0  \\
0    & V
\end{array}\right).
\end{equation}
Thus,
\begin{eqnarray}
\label{dyson3}
G(\omega ,N)&=&G^{0}(\omega ,N)+\sum_{N_{1}}G^{0}\left(\omega
+\frac{N_{1}-N}{2}\omega_0 ,N_{1}\right)\nonumber\\
&\times&SG\left(\omega
+\frac{N_{1}}{2}\omega_0 ,N-N_{1}\right).
\end{eqnarray}
One immediately sees the fundamental difficulty in Eq.~(\ref{dyson3}): Even for
obtaining only the Fourier component $G(\omega ,N=0)$, corresponding to an
average of the 'center-of-mass' time coordinate, the $N_{1}$-sum couples the
different $N$--components in Eq.~(\ref{dyson3}).

Brandes demonstrates that Eq.~(\ref{dyson3}) can be solved
formally to {\em all orders} in the self energy and for all components $N$.
This can be done by rewriting
Eq.~(\ref{dyson3}) as
\begin{equation}
g_{ln} = g^{0}_{ln}+\sum_{n'}g^{0}_{l+n',n'+n}Sg_{l+n+n',-n'},
\end{equation}
where the notation $g_{l,n}\equiv G\left(\omega +\frac{l}{2}\omega_0 ,n\right)$ has been introduced and
$n'$-summation index is shifted. Next, if one introduces
a matrix $\gamma $ with elements $\gamma
_{rs}=g_{ln}\equiv g_{r+s,s-r}$, together with a matrix
$\gamma ^{0}$ with the corresponding elements of $g^{0}$, the Dyson equation can be rewritten as:
\begin{equation}
\gamma _{rs}=\gamma ^{0}_{rs}+\sum_{n'}\gamma ^{0}_{rn'}S\gamma _{n's}.
\end{equation}
Therefore, in the space of the $(r,s)$-indices, the solution of the matrix
equation becomes
\begin{equation}
\gamma =(1-\gamma ^{0}S)^{-1}\gamma ^{0}.
\end{equation}
Explicitely, the matrix $\gamma $ has the form
\begin{eqnarray}
\label{exactpot}
\gamma =\left( \begin{array}{ccccc}
... & & & &   \\
    &1-g_{-20}^{0}S &-g^{0}_{-11}S  &-g_{02}^{0}S &...   \\
... &-g^{0}_{-1-1}S  &1-g_{00}^{0}S &-g^{0}_{11}S   &...   \\
    &-g^{0}_{0-2}S  &-g^{0}_{1-1}S &1-g_{20}^{0}S  &...  \\
... & & & &.... \\
\end{array}\right)^{-1}\times \gamma^{0}.
\end{eqnarray}

The effective dimension of the matrix
$\gamma$ is $D\times n_{ph}$, where where $D$ is the dimension of the eigenstate basis (which is finite
in any numerical calculation) and $n_{ph}$ is the number of
'photoblocks' in $(1-\gamma ^{0}S)$, e.g. $n_{ph}=3$ for the three blocks
per row in Eq.~(\ref{exactpot}). Since
there are infinitely many Fourier components of the Green's function, in the
exact solution Eq.~(\ref{exactpot}), $n_{ph}=\infty$, and $\gamma $ is of
infinite dimension.  In practical calculations, however,
numerical convergence is reached quickly by truncating
$\gamma $ at relatively small $n_{ph}$. Moreover, Eq.~(\ref{exactpot}) offers the possibility
of a systematic investigation of
the 'truncation method', in particular of known approximations like
the so-called 'fast approximation'
which is obtained by cutting the matrix $(1-\gamma^{0}S)$ such
that only the $(r=0,s=0)$, i.e. the 'central' element is retained.
This fast approximation, which neglects the center-of-mass time coordinate,
is used typically in the 'non-adiabatic regime' of resonant
tunneling where the frequency $\omega_0$ of the time dependent
field is much larger than the inverse tunneling time.
\section{Photon assisted tunneling in double barrier systems \label{Double barriers}}
During the last decades, resonant tunneling
through semiconducting double barrier (DB) structures
\cite{TsuAPL(73),GolPRL(87),GolPRB(87),AlvEL(88),SheAPL(88)}
 has been one of the most
active research fields
in solid state physics, both from
the theoretical and experimental standpoints, the main reason being
their great applicability in electronics.
In particular, high-frequency device applications have been one of the major motivations for studying
resonant tunneling devices: the experimental demonstration of detectors up to THz \cite{SolAPL(83)},
quantum well oscillators up to hundreds of GHz \cite{BrownAPL(87)}, or
the fabrication of resonant tunneling
transistors \cite{CapIEEE(86),ReedAPL(89)}
as well as equivalent circuits\cite{YoungAPL(89)}, are just some of the relevant examples.
From the theoretical point of view, the first studies focused on the frequency limitations
of these kind of devices \cite{LurAPL(85),FrensPRL(86),JacoSolidSC(90)}.
After these early works,
a great deal of papers studying photon assisted tunneling in resonant tunneling devices
have appeared in the literature
\cite{ChittaJPC(94),CooJAP(85),LiuPRB(91),WagnerPRB(94),WagnerPRA(95),WagnerPRL(96),KuttPhd,IñaPRB(94),IñaEL(97),SokPRB(88),JonPRB(89),CaiPRL(90),JohPRB(90),WinAPL(90),ApePRB(92),JohPRB(92),ZouPRB(94),IñaPRB(95),AguPRB(96),YakPRB(96),KeayPRB(96),LiPRB(96),LyoPRB(01)}.

\begin{figure}
\begin{center}
\includegraphics[width=0.75\columnwidth,angle=0]
{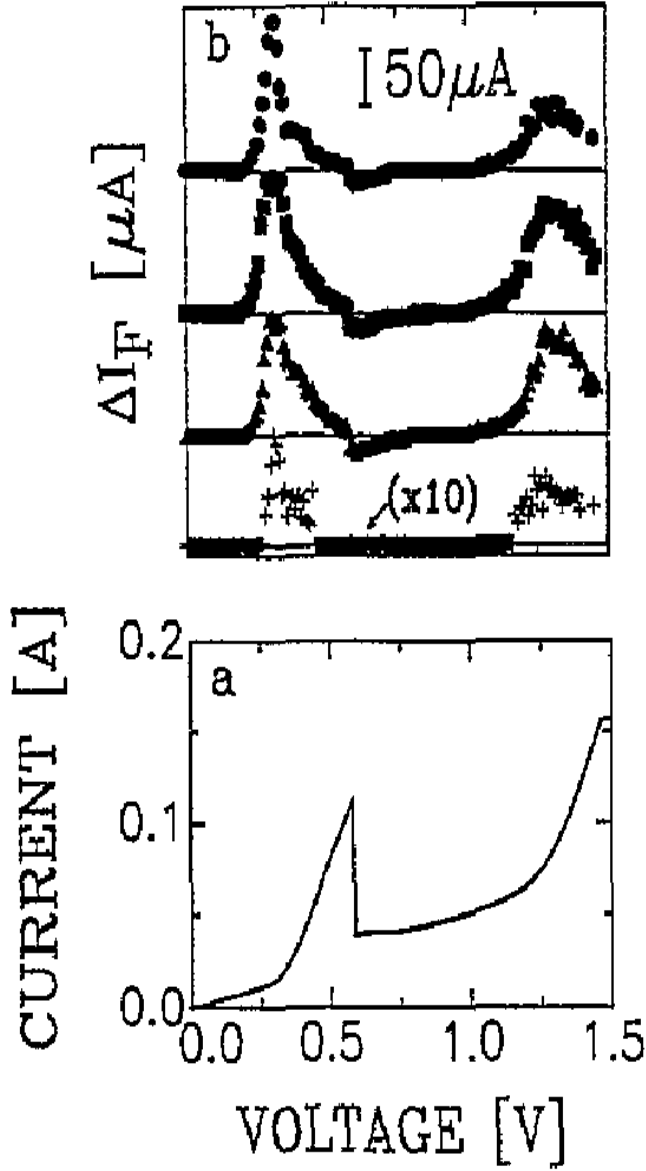}
\end{center}
\caption{Current-voltage characteristics (a) and FIR response (b) measured at
4.2 K of a double barrier device consisting
of two 5.6 nm thick
$Ga_0.6Al_0.4As$ barriers and one undoped quantum well of 5nm thickness.
The frequencies of the FIR radiation are
$\hbar\omega$=13.6 meV (closed circles),
$\hbar\omega$= 8.2 meV (closed squares), $\hbar\omega$
 =4.2 meV (closed triangles) and $\hbar\omega$=2.5 meV (crosses).
Reprinted with permission from \cite{ChittaJPC(94)}.
\copyright 1994 Institute of Physics}
\label {chitta_94_fig3}
\end{figure}

The work of Sollner et al \cite{SolAPL(83)}, is the experimental
starting point for studies on the effect of time-dependent
potentials in resonant tunneling through semiconductor
nanostructures: they studied the influence of electromagnetic
radiation on the resonant tunneling current flowing through
semiconductor diodes. The experiments of Chitta et
al \cite{ChittaJPC(94)} analyzed the far infrared response (FIR) of
double barrier resonant tunneling structures.
Their results are shown in Fig.~\ref{chitta_94_fig3} where the current-voltage characteristics
and FIR response measured at 4.2 K are plotted for different FIR
wavelengths. The most intriguing effect in the experimental curve
is that the current difference between the irradiated and the non
irradiated cases present a main peak at the same dc bias voltage
and {\em the threshold of the current is frequency independent}.

These experimental features
could not be reproduced with Tien-Gordon-like models and were
related to sample heating. However,
careful experiments \cite{KuttPhd}
excluded heating effects in the measured FIR response.

I\~narrea et al \cite{IñaPRB(94),Iñatesis} showed theoretically that
the experiments of Chitta et al can indeed be explained in terms of quantum
photon-assisted transport. The crucial point missed in previous theoretical analysis,
as we shall describe in the next subsection, is to
include the mixing, induced by the photon field, of electronic states with different wave numbers
{\em within each spatial region} and not only
between different regions.
\subsection{Transfer matrix description of photoassisted
coherent tunneling \label{Iñarrea1}}
As we have mentioned in section \ref{Tien-Gordon}, theoretical work on
tunneling devices under the influence of a time-dependent
potential has a long history. After the work of Tien and
Gordon\cite{tien} on superconducting tunneling devices in
microwave cavities, several authors have investigated the effect
that external ac potentials have on different sample
configurations. Among them, Jonson\cite{JonPRB(89)}, Johansson \cite{JohPRB(90)}
Apell and Penn \cite{ApePRB(92)}
have studied the sequential contribution to the tunneling through
a resonant tunneling diode with an applied electromagnetic field, using models based on
the TH formalism\cite{BardPRL(61)} (see section \ref{THPAT}).
In the spirit of Tien and Gordon, see Eq.~(\ref{eq-TG1}), most of these works consider that
the coupling between electrons and the electromagnetic field
takes place just in one region of the structure, the well or the reservoirs,
but not in the whole structure.

I\~narrea et al \cite{IñaPRB(94)} analyzed
the coherent transmission coefficient
and tunneling current for non-interacting electrons considering
that the radiation interacts with barriers, well and contacts.
As mentioned in the previous section, the photon field mixes electronic states
with different wave numbers
{\em within each spatial region}. The inclusion of this mixing
is crucial in order to explain the
experimental evidence.
The most popular model
for photon-assisted tunneling developed by Tien and Gordon
\cite{tien} assumes that the interaction with the
electromagnetic field
can be described as an effective
time-dependent potential $eV_{i}cos\omega t$ within each spatial region $i$ of the system.
Importantly, this model assumes that the potential drop due to the external time-dependent field occurs
only between two regions, namely $e(V_{i+1}-V_{i})cos\omega t$, but {\it not} inside each region.
This has to be compared with the exact coupling term in the Hamiltonian which, depending on the chosen gauge,
is of the form
$\vec{A}.\vec{p}$ ($\vec{A}$ is the vector potential of the electromagnetic field
and $\vec{p}$ is the electronic momentum operator) or $\vec{E}.\vec{r}$ ($\vec{E}$ is the electric field
and $\vec{r}$ is the position operator).
These kind of terms produce transitions between
different electronic states within each region of the structure (leads and well).
In DB's the coupling of the well states with the continuum states of the contacts
produces a quasi-continuum of states in the well such that
the transmission coefficient and the tunneling current in the presence of radiation
can be strongly modified due to the aforementioned mixing.
Thus, a description in terms of the TG model
is not suitable for describing photon-assisted tunneling
through a DB. However, in systems like multiple quantum wells
(where the interwell sequential tunneling through quasidiscrete
states determines the tunneling current) the TG model can be
applied in most cases.

Following I\~narrea et al,
the one-electron Hamiltonian for a double barrier system in the presence
of an electromagnetic field (EMF) represented by a plane
electromagnetic wave of wave vector $\vec{k}$, parallel to the $x$
direction and polarized in the $z$ direction $\vec{E}=(0,0,F)$,
can be written as:
\begin{eqnarray}
H_{tot}=(1/2m^{*})(\vec{p}+e\vec{A}(\vec{R},t))^{2}+V(\vec{R}).
\label{EqIñarrea1}
\end{eqnarray}
Where the Hamiltonian is written within the effective mass approximation and
$V(\vec{R})\equiv V(z)$ is the double barrier
potential accross the growth direction, $z$, of the heterostructure
(an applied dc bias voltage is included
in the definition of $V(z)$).
In the Coulomb gauge $\vec{\bigtriangledown} . \vec{A}    =0 $,
Eq.~(\ref{EqIñarrea1}) becomes:
\begin{eqnarray}
H_{tot} =  p^{2}/2m^{*} + (e/m^{*}) \vec{p}.\vec{A}(\vec{R},t)+
(e^{2}/2m^{*})A^{2} (\vec{R},t)+V(z)
\end{eqnarray}
where the vector potential operator is
$\vec{A}(\vec{R},t)=A_{z}(x,t)$.

In general, $A^{2}(R,t)$ is negligible compared to the
$(e/m^{*})\vec{p}.\vec{A}(\vec{R},t)$ term, therefore  in second
quantization the total Hamiltonian becomes:
\begin{eqnarray}
H_{tot}&=&H_{e}^{0}+H_{ph}^{0}+W_{D}(t)+W_{OD}(t)\nonumber\\
&=&\sum_{k} \epsilon_{k} c_{k}^{+} c_{k}+
\hbar \omega a^{+}a+W_{D}(t)+W_{OD}(t)
\label{Htot}
\end{eqnarray}
where
\begin{eqnarray}
W_{D}(t)&=&\sum_{k} [(e/m^{*})<k|p_{z}|k> c_{k}^{+} c_{k}
  (\hbar/2\epsilon V\omega)^{1/2} (a e^{-i\omega t} +a^{+} e^{i\omega t})]
\label{Htot1}
\end{eqnarray}
\begin{eqnarray}
W_{OD}(t)&=& \sum_{k}\sum_{k^{'}\neq k} [(e/m^{*})<k^{'}|p_{z}|k>
  c_{k^{'}}^{+} c_{k} (\hbar/2\epsilon V\omega)^{1/2}
   (a e^{-i\omega t} + a^{+} e^{i\omega t})],\nonumber\\
\label{Htot2}
\end{eqnarray}
with $A_{z}(x,t)=(\hbar/2\epsilon V\omega)^{1/2} \vec{\varepsilon_{z}}
(a e^{-i\omega t} + a^{+} e^{i\omega t}) $ ($\omega$ is the photon frequency and the
wave vector of the EMF has been neglected).
$H_{e}^{0}$ is the independent, electronic Hamiltonian and
includes the double barrier potential and the external applied
bias voltage. 
The operators $c_{k}^{+}$ create the eigenstates of $H_{e}^{0}$, $\Psi_{0}(k)$ which
describe tunneling states for bare electrons (the factor
$e^{-iEt}$ is already included in the state vector $|k>$).
$H_{ph}^{0}$, is the
photon field Hamiltonian whithout coupling with electrons and
$W_{D}$ and $W_{OD}$, describe the coupling between electrons and
photons in the total Hamiltonian.

The coupling term can be divided into the
"diagonal" $W_D$ and the "off-diagonal" $W_{OD}$ contributions.
In these systems a quasi-localized state
is connected by the EMF with a continuum of extended states.
Therefore $W_{OD}$, can be treated in first order time dependent
perturbation theory. For problems in which two o more
quasi-localized states should be connected by the light, the
method could not be applied in the same way, requiring some
generalization.
Therefore, the total Hamiltonian can be written as:
\begin{eqnarray}
H_{tot}=H_{D}(t)+W_{OD}(t)
\label {Ham}
\end{eqnarray}
where $H_{D}(t)=H_{e}^{0}+H_{ph}^{0}+W_{D}(t)$.
The Hamiltonian $H_{D}$, can be solved exactly considering a
canonical transformation\cite{IñaPRB(94),JonPRB(89)}. It allows the
exact electronic wave function dressed by photons to be obtained:
$\Psi_{D}(k)=U^{+}\Psi_{0}(k)$,  where $\Psi_{0}(k)$ is the
electronic double barrier eigenstate with no photon field present
in the sample. Once the eigenstates for $H_{D}$ are obtained, time
dependent perturbation theory is applied in order to treat the
$W_{OD}$ term.
The operator $U$ for the canonical transformation
is given by $U=e^{s}$, where $s$ can be written as:
\begin{eqnarray}
s&=&\frac{e}{m^{*}\hbar\omega}(\frac{\hbar}{2\epsilon V\omega})^{1/2}
<p_{z}>c_{k}^{+}c_{k} (a^{+} e^{i\omega t} - a e^{-i\omega t})\nonumber\\
&=&\frac{M}{\hbar\omega} c_{k}^{+}c_{k} (a^{+} e^{i\omega t} - a e^{-i\omega t})
 \label{canonical}
\end{eqnarray}
The Hamiltonian under this transformation becomes:
\begin{eqnarray}
\tilde{H}_{D}&=&\tilde{c}_{k}^{+}\tilde{c}_{k}
(\epsilon_{k}-\frac{M^{2}}{\hbar\omega})+ \hbar\omega \tilde{a}^{+}
\tilde{a}
\end{eqnarray}
where $\tilde{a}^{+}=a^{+}-\frac{M}{\hbar\omega} c_{k}^{+}c_{k}$ and
$\tilde{a}=a-\frac{M}{\hbar\omega} c_{k}^{+}c_{k}$. In the transformed
Hamiltonian $\tilde{H}_{D}$ the electrons and photons are not
coupled anymore and the electronic eigenvalues are shifted in
$\Delta = \frac{M^{2}}{\hbar\omega}$ which is negligible with respect
to the free electron eigenvalues. Assuming a semiclassical EMF,
$(a+a^{+})\rightarrow 2\sqrt N cos(\omega t)$ where N is the number
of photons\footnote{\label{semiclassical}A classical treatment of the electromagnetic
field is justified
for the range of field parameters in typical experimental setups.
To estimate the number of photons $N$,
one can use the relation between intensity, $I$, and field strength, $E$:
$I={\frac{1}{2}\sqrt{\frac{\epsilon}{\mu}}}|E|^2$ (for GaAs, $\mu=1$ and $\epsilon =10.9$)
together with $I/c=\frac{\hbar\omega}{V}(N+1/2)$, namely the
relation between the energy content per unit volume, $I/c$,
the photon energy and the number of photons.
Solving for $N$, one gets for the experiments of Sollner et al \cite{SolAPL(83)}
(typical frequencies of 2.5 THz and intensities
$I\approx 10^8 W/m^2$)
$N\approx 2\times 10^6$ photons in a volume of $100 \AA \times 1 mm^2$.}.
From the above expressions $s$ can be written as:
$s=\frac{-ieF}{m^{*}\hbar\omega^{2}}sin\omega t <k|p_{z}|k> c_{k}^{+}c_{k}$.
Finally the exact eigenstate for $H_{D}$ can be expressed in terms
of the electric field intensity $F$:
\begin{eqnarray}
\Psi_{D}(k)&=&exp[\frac{-ieF}{m^{*}\hbar\omega^{2}}<p_{z}>sin(\omega t)]\Psi_{0}(k)\nonumber\\
&=&\Psi_{0}(k)\sum_{n=-\infty}^{\infty}J_{n}(\beta_{k}) e^{-in\omega t}
\label{wf}
\end{eqnarray}
where
\begin{eqnarray}
\beta_{k}=\frac{eF<p_{z}>}{m^{*}\hbar\omega^{2}}.
\label{beta_k}
\end{eqnarray}
Note that the scaling parameter of the Bessel functions characterizing the photonic sidebands
depends on the frequency as $\omega^{-2}$
as compared with the TG model where the dependence goes as $\omega^{-1}$.
This difference in the scaling parameter has been
studied in detail by Wagner and Zwerger in Ref.~\cite{WagPRB(97)} and indicates that the TG model
is not applicable in situations involving ac fields rather than ac potentials.

Once the eigenstate for $H_{D}$ is obtained, time-dependent
perturbation theory is applied in order to treat the $W_{OD}(t)$
term. By doing this, the expression for the total wave function of
the tunneling electron under the influence of the EMF becomes:
\begin{eqnarray}
\Psi(t)=\alpha[\Psi_{D}(k_{0})+\sum_{m}b_{m}^{(1)}(t)\Psi_{D}(k_{m})].
\end{eqnarray}
$k_{0}$ is the wave vector of the initial electron,
$k_{m}$ is the wave vector of the corresponding electronic coupled
states and $\alpha$ is a normalization constant.
The time dependent coefficients are given by:
\begin{eqnarray}
b_{m}^{(1)}(t)=\lim_{\alpha \rightarrow 0} \int_{-\infty}^{t}
  (1/i \hbar) <\Psi_{D}(k^{'})|W(k)|\Psi_{D}(k)> e^{\alpha t} dt
\end{eqnarray}
where
\begin{eqnarray}
W(k)=(eF/m^{*} \omega) \sum_{k^{'}}<k^{'}|p_{z}|k> c_{k^{'}}^{+} c_{k}
cos(\omega t),
\end{eqnarray}
such that
\begin{eqnarray}
b_{m}^{(1)}=\frac{-ieFL}{4 \hbar^{2}\omega} \sum_{n^{'},n} \left[
J_{n^{'}}(\beta_{k_{m}})J_{n}(\beta_{k_{0}})
\frac{<k_{m}|p_{z}|k_{0}>}{k_{m}} \right].
\end{eqnarray}
$n^{'}$ and $n$ run from $-\infty$ to $\infty$ and $m=n^{'}-n\pm1
=\pm1, \pm2,\pm3,....$. The normalization constant
$\alpha=\frac{1}{\sqrt{1+\sum_{m}|b_{m}^{(1)}(t)|^{2}}}$,
guarantees current conservation.
$\Psi_{D}(k_{0})$ is the
"dressed" or diagonal reference state and
 $\Psi_{D}(k_{m})$, represents the coupled "dressed"
states due to photon absorption and emission. The spectral density
consists in a central peak
 (weighted
by $J_{0}^{2}$) and infinite n-sidebands separated in $n \hbar
\omega $ from the central peak and weighted by $ J_{n} $. If the
argument of the Bessel functions is very small, the sidebands
intensities are negligible and it is enough to consider
transitions between the main side bands (the ones weighted by
$J_{0}$) of different electronic states separated in energy by
$\hbar \omega$.
For higher values of $ \beta_{k} $, the spectral density weight is
shared between the satellite peaks and their contribution cannot
be neglected.
Applying the
current operator to the transmitted and incident wave function,
the time-averaged coherent transmission coefficient following the
Transfer Matrix technique \cite{IñaPRB(94)} becomes:
\begin{eqnarray}
T&=&\sum_{i=-\infty}^{\infty}\frac{T_{i}|b_{i}|^{2}}{\sum_{l=0}^{\infty}{\frac{k_{l}
|b_{l}|^{2}}{k_{i}} +\frac{k_{-l}|b_{-l}|^{2}}{k_{i}}}}
\label{transmission}
\end{eqnarray}
Where $|b_{0}|$=1. $T_{i}$ is the coherent transmission
coefficient for a
double barrier at the energy corresponding to a wavevector $k_{i}$.
In the case of very small $\beta$ only
one photon absorption and emission processes are considered and:
\begin{eqnarray}
b_{1,(-1)}^{(1)}=(-ieFL/4
\hbar^{2}\omega)J_{0}(\beta_{k_{1,(-1)}})J_{0} (\beta_{k_{0}})
<k_{1,(-1)}|p_{z}|k_{0}>/k_{1,(-1)},
\end{eqnarray}
such that
\begin{eqnarray}
T&=&T_{0}/(1+k_{1}/k_{0}|b_{1}^{(1)}|^{2}+k_{-1}/k_{0}
|b_{-1}^{(1)}|^{2})+\nonumber\\
 & &T_{1}|b_{1}^{(1)}|^{2}/(k_{0}/k_{1}+|b_{1}^{(1)}|^{2}+k_{-1}/k_{1}
 |b_{-1}^{(1)}|^{2})+\nonumber\\
 & &T_{-1}|b_{-1}^{(1)}|^{2}/(k_{0}/k_{-1}+k_{1}/k_{-1}
|b_{1}^{(1)}|^{2}+ |b_{-1}^{(1)}|^{2})
\label{transmission0}
\end{eqnarray}
A similar expression can be obtained for the reflection
coefficient replacing $T_{0}$, $T_{1}$ and $T_{-1}$ by $R_{0}$,
$R_{1}$ and $R_{-1}$ respectively. $R_{0}$, $R_{1}$ and $R_{-1}$,
($T_{0}$, $T_{1}$ and $T_{-1}$),
are the standard coherent double barrier
reflection (transmission) coefficients, evaluated at the reference
energy, at one photon above and at one photon below the reference
energy, respectively.
\begin{figure}
\begin{center}
\includegraphics[width=0.75\columnwidth,angle=0]
{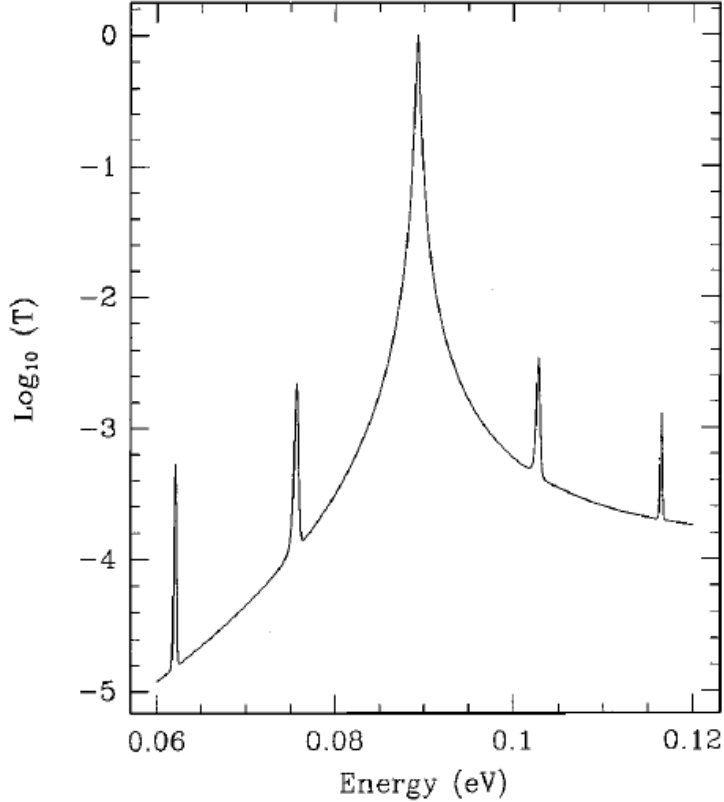}
\end{center}
\caption{$ Log_{10}$ of coherent T(E) through a DB (well and
barriers $ 50 \AA $ wide).
 ($F=4 \times 10^{5} V/m$, $\hbar\omega=13.6 meV$).}
\label{InarreaEPL_97_fig1}
\end{figure}
This expression for the reflection coefficient verifies current
conservation:  $|T|^{2}+|R|^{2}=1$, meaning that the probability
for an electron to tunnel with no photon absorption or emission is
smaller than the corresponding with no light present in the
sample which is a  consequence of unitarity.
As the EMF intensity increases, multiphoton
processes play a role in the transmission probability and formula
Eq.~(\ref{transmission})
has to be considered.

Using the above formalism, I\~narrea et al \cite{IñaPRB(94),IñaEL(97)}
studied the experiments of Chitta et al \cite{ChittaJPC(94)}, see
Fig.~\ref{chitta_94_fig3}.
Their results are shown in Fig.~\ref{InarreaEPL_97_fig1} where
the multiphoton transmission
coefficient for coherent tunneling is plotted (field
intensity $F=4\times 10^{5} V/m$, $\hbar\omega=13.6 meV$ and
zero bias voltage). The
main features observed in the transmission coefficient T(E), are
multiple satellite peaks at both sides of the central one, coming
from photon absorption and emission. The two closest peaks to the
central one correspond to one photon processes, mainly to the
transitions between the zero-side bands of electronic states
differing in one photon energy.
The other two peaks separated 2$ \hbar \omega$ from the main one
correspond to processes involving two photons. Higher
multiphoton transitions have much weaker intensities.

\begin{figure}
\begin{center}
\includegraphics[width=0.75\columnwidth,angle=0]
{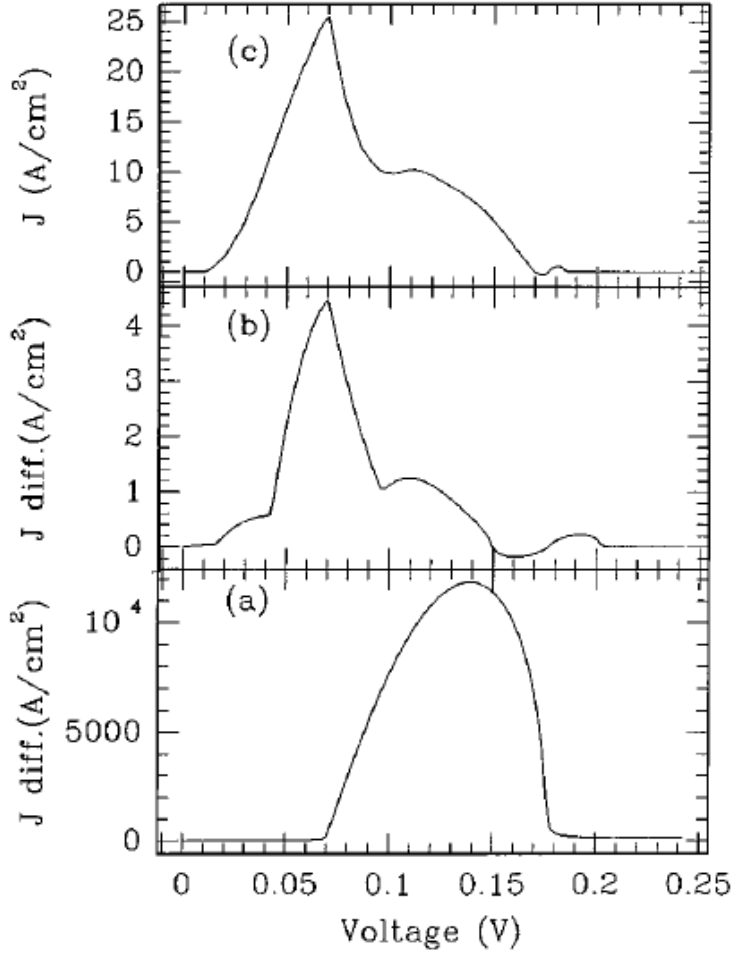}
\end{center}
\caption{ a) Coherent current as a function of dc bias voltage for a $
Ga_{.7}Al_{.3}As$ DB (well and barriers $ 50 \AA$ wide). b) $
\Delta J$ /V
 (F=$4\times 10^{5} V/m $, $ \hbar \omega=13.6 meV$).
c) Same as in b) for $\hbar \omega=4.2 meV$).}
\label{InarreaEPL_97_fig2}
\end{figure}

In Fig.~\ref{InarreaEPL_97_fig2}(a) the coherent tunneling current
density as a function of the dc voltage in the presence of the FIR
laser is plotted.
The effect of the light on J can be observed in Fig.~\ref{InarreaEPL_97_fig2}(b) and
Fig.~\ref{InarreaEPL_97_fig2}(c)
corresponding to $ \hbar\omega$=13.6 meV and $ \hbar\omega
$=4.2 meV respectively.
In these figures,
the current difference, with and without EMF, $\Delta J$,
is plotted versus the applied dc bias voltage.
One observes firstly that the current threshold
takes place at the same voltage for both frequencies. Also, a main peak appears for
both cases
at the same voltage. This central peak corresponds
to the current threshold without light.
A shoulder appears for voltages close to
the center of the current peak at the higher voltage side, a weak
negative contribution shows up for higher bias voltage and, finally,
a small structure, associated with the current cutoff, appears.
The fact that the current threshold takes place at a lower
bias voltage than in the case without radiation is easy to understand: the
electrons in the emitter can absorb photons and the current flows when
$ E_{r}=n\hbar\omega+E_F$ ($E_{r}$ is the energy of the resonant state in the well and
$E_F$ the Fermi energy at the emitter).
On the contrary, the fact that the threshold is frequency-independent is not expected.
This counterintuitive behavior
(at first sight, one would expect a linear shift with respect to $\omega$
of both the current threshold and the main peak of $\Delta J$)
is in qualitative agreement with the experimental results of Fig.~\ref{chitta_94_fig3} \cite{ChittaJPC(94)}.
The first attemps made in order
to explain these results, first in terms of classical response and
secondly relating the experiments to sample heating due to the
laser, failed. The dependence of J with the temperature
\cite{KuttPhd} shows a qualitative different behavior
than the obtained in the presence of the laser, and, therefore,
heating does not explain the experiments (see Fig. 2 and Fig. 3 of
Ref.~\cite{ChittaJPC(94)}.
Regarding to the classical response,
the photon energies considered are much larger than the energy
broadening of the DB resonant state and a quantum behavior is
expected.

The qualitative agreement between the experimental photon-assisted tunneling current and
the transfer matrix calculations by I\~narrea et al \cite{IñaPRB(94),IñaEL(97)}
allows to understand the features discussed above. The fact that
the current threshold for different $\omega$ takes place at
practically the same bias voltage is related to the multiphotonic
processes. If one compares the two cases corresponding to
$\omega$=13.6 meV and $\omega$= 4.2meV (fixed F) in  Fig.~\ref{InarreaEPL_97_fig2}, it
is obvious that $ \beta_{k}$ is larger for the smallest frequency, $\omega$=4.2 meV,
such that more side bands contribute
efficiently to the current. Therefore, the current threshold moves to
a lower bias voltage as compared with cases 
where only single photon processes do contribute. These
multiphoton contributions wash out the linear dependence that the
threshold voltage should follow as a function of $\hbar \omega$ if
only one photon process took place.
Concerning the main peak in $\Delta J$, its frequency-independent position is related with the number
of parallel states available to tunnel resonantly with the
absorption of one or more photons. The main peak corresponds to
the voltage where the resonance energy, $ E_{r}$, is just above the
Fermi energy, $ E_{F}$, at the emitter. In this situation the
number of parallel states which can tunnel resonantly via
absorption of
one or more photons is maximum no matter the $\omega$ considered.\\
\subsection{Sequential tunneling current \label{Iñarrea2}}
In the sequential tunneling regime the electrons suffer
scattering processes and the electronic wave function looses
memory in the quantum well.

The sequential tunneling contribution
to the current can be obtained evaluating separately the current
for the first and the second barriers, $J_{1}$, and $J_{2}$. These
currents are related to the Fermi level in the well $E_{w}$
(which defines the amount of electronic charge stored into the
well). The Fermi Level is obtained selfconsistently by imposing current
conservation, namely $J_{1}=J_{2}$.
This procedure takes into account the possible
scattering processes within the well in a phenomenological way.

In order to calculate $J_{1}$, and $J_{2}$ without light, the TH
method \cite{BardPRL(61)} should be considered. The probability
$P_{1}$ for the electron to cross from the emitter to the well
reads:
\begin{eqnarray}
P_{1}&=&(2\pi/\hbar)(2\pi/L^{2})^{2}[\frac{\hbar^{4}k_{e}k_{w}}
{2m^{*2}L(w_{2}+(1/\alpha_{b})+ (1/\alpha_{d}))}\nonumber\\
&\times&T_{s}\delta
(k_{p}^{w}-k_{p}^{e}) \delta(E_{z}-E_{tn})]
\end{eqnarray}
where $T_{s}$ is the transmission coefficient for a single
barrier; $k_{e}$ ($k_{p}^{e}$) and $k_{w}$ ($k_{p}^{w}$) are the
perpendicular (parallel) component for the electronic wave vector
in the emitter and well respectively. For a given bias voltage $V_{dc}$,
$E_{tn}$ is the well state
energy referred to conduction band bottom:
$E_{tn}=E_{R}-V_{dc}(w_{1}+w_{2}/2)/w_{t}$ (where $E_{R}$, is the
well state energy referred to well bottom and $w_{1}$, $w_{2}$ and $w_{t}$
are the left barrier, well and total widths respectively).
Finally, $\alpha_{b}
=\sqrt{\frac{2m^{*}(V_{0}-E_{R}+V_{dc}(w_{1}+w_{2})/
2w_{t})}{\hbar^{2}}}$, $\alpha_{d}=\sqrt{\frac{2m^{*}(V_{0}-E_{R}+
V_{dc}(w_{2}+w_{3})/2w_{t})}{\hbar^{2}}}$,
where $w_{3}$ is the width of the right barrier.

For small values of $\beta_{k}$, it is enough to consider one-photon processes only
and the sequential current through the left barrier $J_1$ reads:
\begin{eqnarray}
J_{1}&=&(e/2\pi\hbar)\int_{0}^{E_{F}} dE_{z} \frac{k_{w}}
{w_{2}+(1/\alpha_{b})+(1/\alpha_{d})}\times \nonumber\\
     & &\Big[\delta[E_{z}-E_{tn}]
     \frac{T_{s,0}}{1+k_{1}/k_{0}|b_{1}^{(1)}|^{2}+k_{-1}/k_{0}
       |b_{-1}^{(1)}|^{2}}+\nonumber\\
     & &\delta[E_{z}-(E_{tn}-\hbar w)]
     \frac{T_{s,1}|b_{1}^{(1)}|^{2}}{k_{0}/k_{1}+|b_{1}^{(1)}|^{2}+
       k_{-1}/k_{1}|b_{-1}^{(1)}|^{2}}+\nonumber\\
     & &\delta[E_{z}-(E_{tn}+\hbar w)]
     \frac{T_{s,-1}|b_{-1}^{(1)}|^{2}}{k_{0}/k_{-1}+k_{1}/k_{-1}
      |b_{1}^{(1)}|^{2}+ |b_{-1}^{(1)}|^{2}}
     \Big] \int_{E_{w}}^{E_{F}-E_{z}}dE_{p}
\label{jseq}
\end{eqnarray}
where the subscript "0" means the reference  state energy that in
our case is the resonant well state energy. The subscript "1" and
"-1" mean one photon energy above and below respectively, and
$T_{s,0}$, $T_{s,1}$ and $T_{s,-1}$ are the transmission
coefficients through a single barrier, in this case, the emitter
one.
For the second barrier a similar expression is obtained for $J_2$.

By using the method above to calculate the sequential contribution to the current,
I\~narrea et al conclude in Ref.~\cite{IñaPRB(94)} that the current
in the experiments of Chitta et al \cite{ChittaJPC(94)} is mainly given by a coherent contribution.
The calculated sequential current is of the same order as the coherent one but the current
difference (with and without irradiation) for sequential tunneling
is one order of magnitud smaller than the corresponding to the
coherent tunneling.
\subsection{Magnetotunneling current in the presence of
radiation \label{Iñarrea3}}
The analysis of the magnetotunneling current for both configurations
of the magnetic field,
parallel and perpendicular to the growth direction of an
heterostructure,
gives a great deal of information on the density of states in the well
corresponding to the Landau level (LL) ladder, in the first case
and on the edge states in the second configuration \cite{MenPRB(86),SchPRB(90),PlaPRB(89)}.
As described in Ref.~\cite{IñaPRB(95)},
an external radiation field can be used to widen
the information obtained upon the application of a magnetic field.

The coherent magnetotunneling current
through a DB when a magnetic field is applied parallel to the
current in the presence of light can be analyzed
following a similar scheme than the one described in section \ref{Iñarrea1}.
If the FIR radiation is
linearly polarized in
the same direction as the static magnetic field,
the electronic motion is modified
by the light only in the transport direction
and the electronic lateral states remain unaffected by light. With
no magnetic field present, the parallel component for the
electronic wave vector is conserved during the coherent tunneling
process. In the presence of a magnetic field, it is the LL index
what is conserved: the current presents a peak as a function of
the external bias voltage when a LL in the emitter aligns with
the corresponding one in the well. As the magnetic field increases, the degeneracy of each LL increases,
less number of LL contribute to the current and, as a result,
the current presents less, but more intense, peaks.

In the presence of FIR radiation, the electronic part of the Hamiltonian in Eq.~(\ref{Htot})
is modified as:
\begin{eqnarray}
H_{e}^{0}&=&\sum_{k} \epsilon_{z} c_{z}^{+} c_{z}+
     \hbar \omega_{c}(a_{B}^{+} a_{B} + 1/2)
\end{eqnarray}
where $B$ is the magnetic field intensity, $\omega_{c}=eB/m^{*}$ is the
cyclotron frequency, $a_{B}^{+}$ and $a_{B}$ are
the creation and destruction operators for the Landau states, and
$\epsilon_{z}$ is the electronic energy perpendicular part.
$H_{e}^{0}$ is the independent
electronic Hamiltonian and includes the double barrier potential
and the external applied bias voltage, therefore the eigenstates of
$H_{e}^{0}$, $\Psi_{0}(k)$, are the tunneling states for bare
electrons in the presence of a magnetic field. The rest of the
terms in the Hamiltonian are as in Eqs.~(\ref{Htot1}-\ref{Htot2}).
Following the derivation in section \ref{Iñarrea1}, the coupling term is divided in the
"diagonal"  and the "off-diagonal" contributions:
$H_{tot}=H_{D}(t)+W_{OD}(t)$
where $H_{D}(t)=H_{e}^{0}+H_{ph}^{0}+W_{D}(t)$. Again, the Hamiltonian
$H_{D}$, can be solved exactly by a canonical transformation and
the off-diagonal term is treated in time dependent perturbation
theory.\\
The expression for the
coherent magnetotunneling current is:
\begin{eqnarray}
J=(2/2\pi^{2})(e/\hbar)^{2} B \sum_{n=0}^{N}\int_{(n+1/2)\hbar w}
 ^{E_{F}} dE [f(E)-f(E+V_{dc})] T(E,n).
\end{eqnarray}
$n$ the LL index, $N$ is the maximum LL index occupied, and
$T(E,n)$ the  coherent transmission coefficient through a double
barrier structure
when the photon field is present in the sample.

\begin{figure}
\includegraphics[width=0.75\columnwidth,angle=0]
{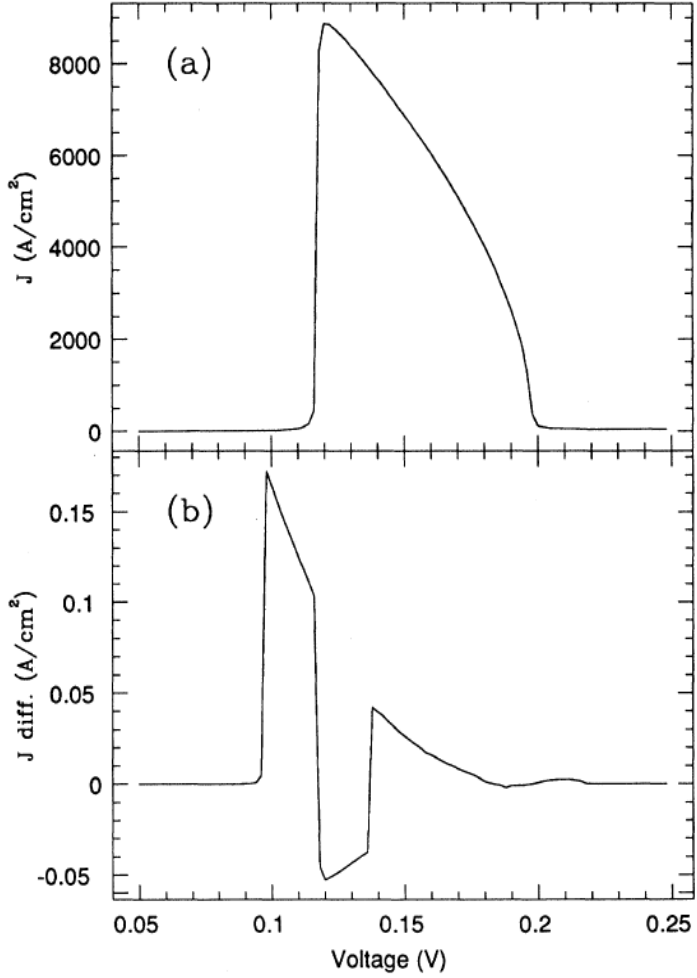} \caption{a)  Coherent magnetocurrent, magnetic field B=24T,
versus dc bias voltage in the
presence of radiation ($F=5\times 10^{4} V/m$, $\hbar \omega=10.3 meV$.
b) Coherent magnetocurrent density difference, with and without light, versus dc bias voltage.}
\label {InarreaPRB_95_fig2}
\end{figure}

An example of the effect of FIR radiation on the coherent magnetocurrent
through a DB is plotted
in Fig.~\ref{InarreaPRB_95_fig2}(a), where the coherent magnetocurrent
density is represented as a function of the
external bias voltage.
In this case, only one LL contributes to the tunneling current and
the analysis
of the effect of the light on the current can be done in a simpler
way than in the case where more LL participate in the current.

\begin{figure}
\includegraphics[width=0.75\columnwidth,angle=0]
{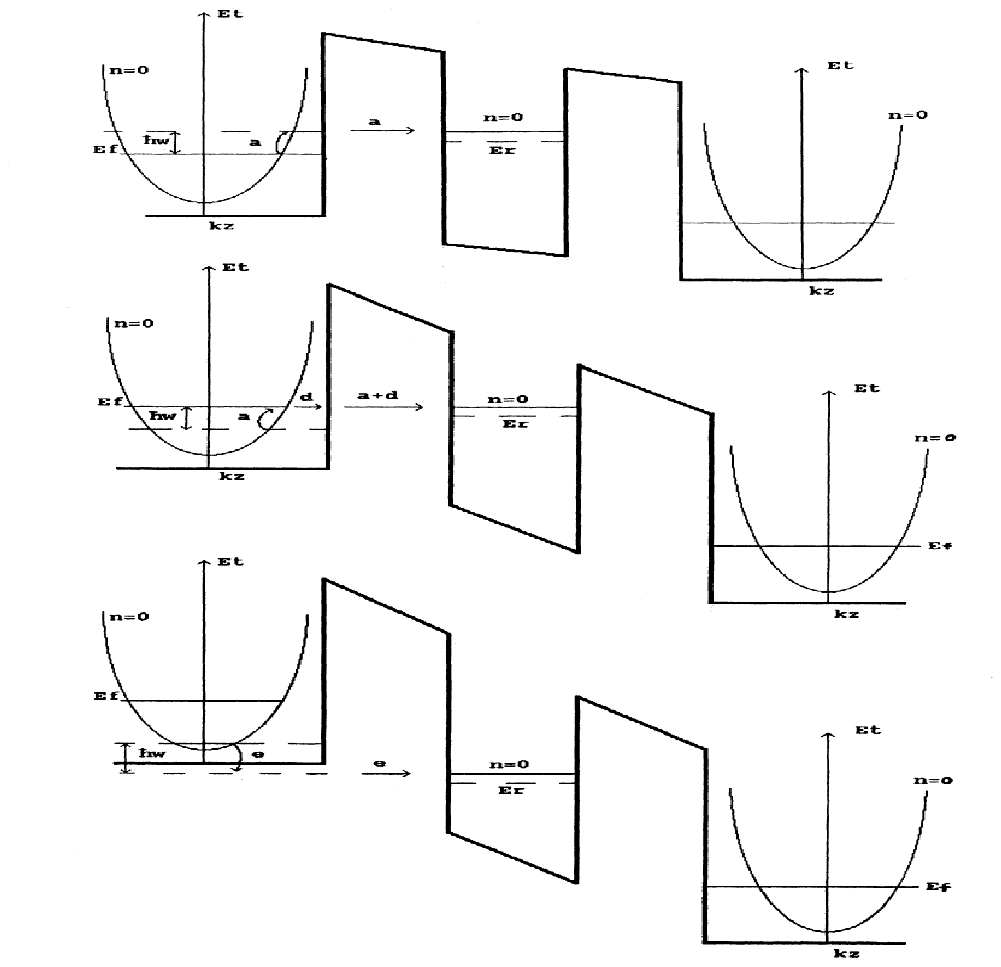} \caption{Schematic picture of the photon-assisted tunneling
processes} \label {InarreaPRB_95_fig3}
\end{figure}

The
current difference, with and without light,
is plotted in Fig.~\ref{InarreaPRB_95_fig2}(b). In this case there is a main peak which
appears for smaller bias voltage than the corresponding to the threshold
bias voltage for the magnetotunneling current with no light present. As
$V_{dc}$ increases the current difference decreases and becomes
negative. There is also a small positive and a negative structure
for higher voltages and as the voltage corresponding to the cut off of
the current is reached
 there is an additional
positive contribution to the current difference.
These features are schematically  explained
in Fig.~\ref{InarreaPRB_95_fig3}: for small voltages the resonant
state in the well with energy corresponding to the first LL is
higher in energy than the Fermi energy $ E_F$ in the emitter.

As the voltage increases there are electrons close to $E_F$ which are
able to absorb a photon and tunnel resonantly from the first LL in
the emitter with LL index conservation, therefore the threshold
voltage for the current is smaller than the corresponding one for no
light present (it moves twice the photon energy) and there is a
positive peak in the current difference. For higher voltages the first
LL in the well crosses the $E_F$ in the emitter and the current
difference becomes negative abruptly due to the fact that the
electrons in the emitter have the possibility to absorb a photon
and it reduces the number of electrons efficient to tunnel
resonantly. For higher voltages there are absorption, emission and
direct tunneling processes whose combinations give the positive
structure observed.
\begin{figure}
\includegraphics[width=0.75\columnwidth,angle=0]
{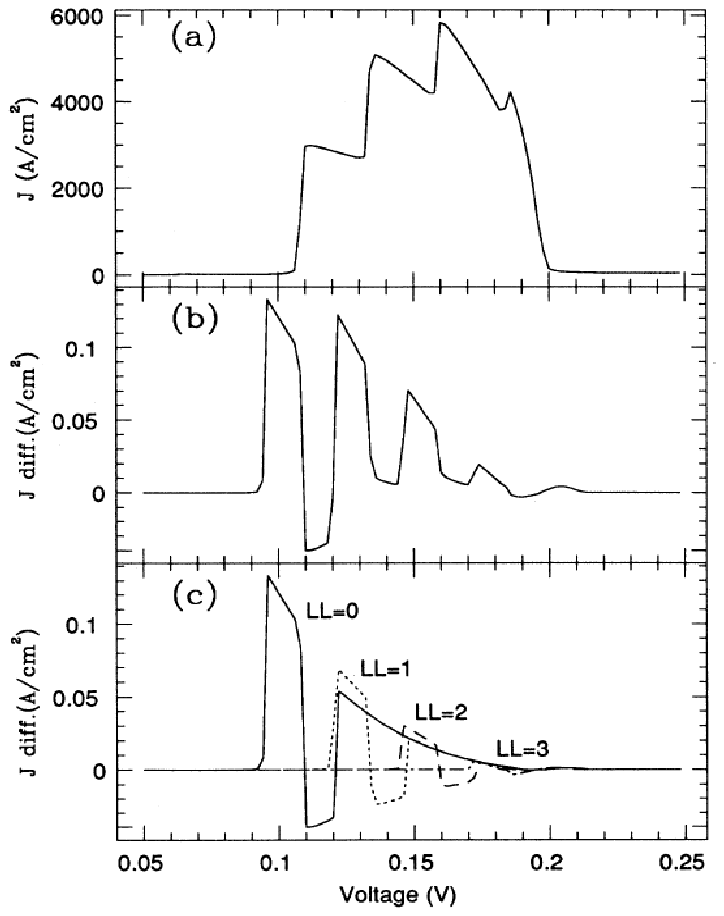}
\caption{a) Coherent contribution $J/V_{dc}$ for $F= 5\times 10^{4}V/m$,
$\hbar\omega= 6.9 meV$, B= 8T. b) $J_{diff.}$ for
 $F= 5\times 10^{4} V/m$, $\hbar\omega= 6.9 meV$,B= 8T.
c) $J_{diff.}$/$V_{dc}$ for
 each LL
separately.}
\label {InarreaPRB_95_fig4}
\end{figure}
As $V_{dc}$ increases
and the energy of the resonant state in the well for the first LL
lies one photon higher than the conduction band bottom of the
emitter the electrons have a probability to emit a photon below
the bottom of the conduction band and the resonant current is
reduced (it corresponds to the small negative contribution to the
current difference for large voltages).
Once the resonant state crosses the bottom of the
conduction band there are electrons in the emitter which can emit
a photon and tunnel resonantly, therefore there is a positive peak
in the current difference and the current cut off moves to larger
voltages.
\begin{figure}
\includegraphics[width=0.75\columnwidth,angle=0]
{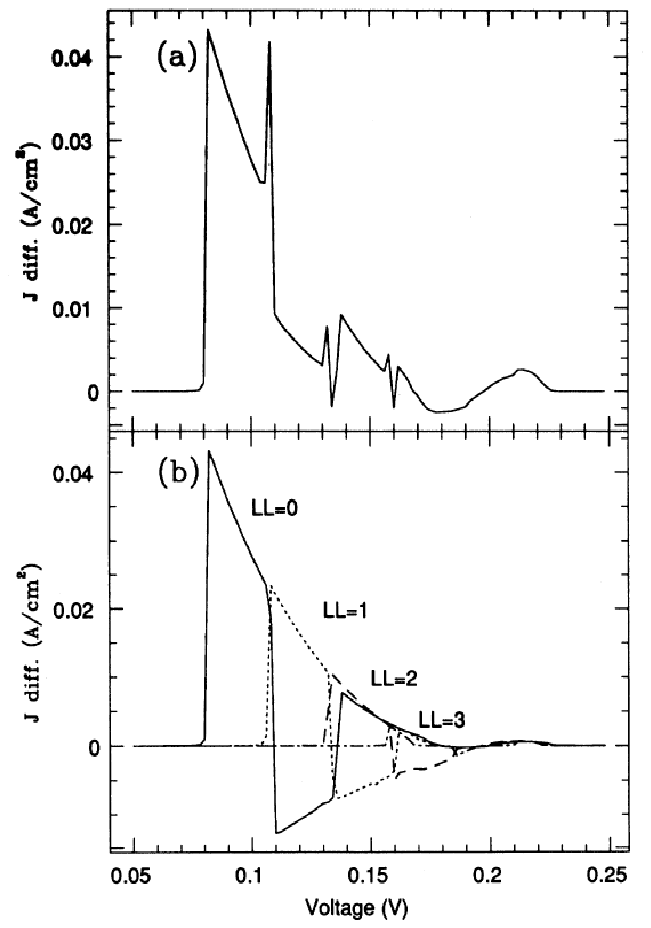}
\caption{a) Coherent magnetocurrent
difference for B= 8T and $\hbar\omega=\hbar\omega_{cyclotron} $ . b)
Same as in a) for each Landau level separately. }
\label
{InarreaPRB_95_fig5}
\end{figure}
As the magnetic field decreases there are more LL´s which
contribute coherently to J. In Fig.~\ref{InarreaPRB_95_fig4}(a)
the current as a function of the external dc bias voltage is plotted for
a field of 8 Tesla. We observe four LL´s which contribute to J.
The photon frequency $\omega_{0}$ is
6.9 meV (one half of the cyclotron frequency $\omega_{c}$). The current difference is
shown in Fig.~\ref{InarreaPRB_95_fig4}(b). In this case the
main peak in the current difference due to the effect of the light
appears at different bias voltage for the different LL´s and the
contribution at the cut off is added up for the four levels. In
Fig.~\ref{InarreaPRB_95_fig4}(b) the current difference has been
plotted separately for each LL. If the photon frequency is tuned to
the same value as the cyclotron one for the same magnetic field
(8T) the current difference changes dramatically and the main
contribution comes from the peak at the threshold bias voltage and an
additional narrow structure in this region of voltages
(Fig.~\ref{InarreaPRB_95_fig5}). For higher voltages the additional
 features to the current difference
are much  smaller  in intensity than in the previous case.
 The reason for this difference between both cases is not
only the change of the threshold voltage and cut off of J due to the
difference of photon frequencies (the threshold voltage is lower for
higher photon frequencies and the cut off voltage is larger for
higher photon frequencies) but also is due to the fact that when
the ratio $\omega_{c}/\omega_{0}$ is one, there are absorption and
emission processes taking place for electrons coming from
different LL´s which compensate each other. This feature can be
observed in Fig.~\ref{InarreaPRB_95_fig5}(b),
 where the
contribution to the current density coming from each LL is
represented. Due to this compensation it is possible to control
the the effect of the light on the magnetocurrent by tuning the
ratio between the cyclotron
and the photon frequency.

The sequential magnetotunneling current for the same
cases as the coherent one was evaluated in Ref.~\cite{IñaPRB(95)}.
\begin{figure}
\includegraphics[width=100mm,angle=0,keepaspectratio,clip=true]
{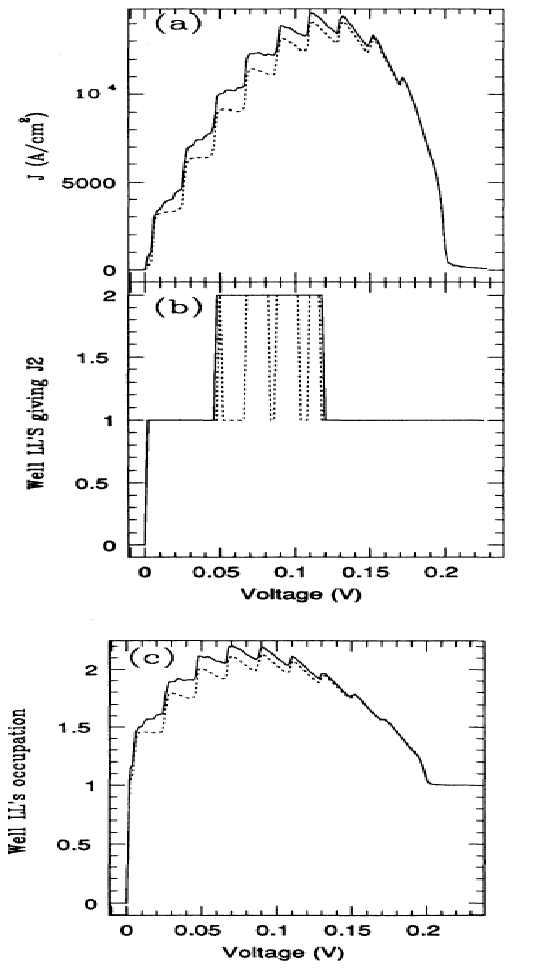} \caption{a) Sequential magnetotunneling
current assisted by light as a function of $V_{dc}$.
( $F=5\times 10^{6} V/m$, $\hbar\omega=7 meV$ , B=6 Tesla). b)
 Total number of LL´s into the well contributing to J as a function of
 $V_{dc}$ with (continuous line) and without (dotted line) light.
 c) LL´s occupation into de well as a function of $V_{dc}$.
 Continuous line, light present; dotted line, no light present. }
\label {InarreaPRB_95_fig9}
\end{figure}
Again, the calculated sequential current is of the same order as the coherent one but the current
difference (with and without irradiation) for sequential tunneling
is much smaller than the corresponding to
coherent magnetotunneling.
When the frequency of the applied laser
($\hbar\omega= 13.8 meV$) equals the cyclotron frequency,
there are compensations in the current difference coming
from different LL´s and the light affects mainly the current
density at
the threshold and the cut off voltage.

The LL index, which is conserved if tunneling occurs coherently,
is not conserved for sequential tunneling.
When an electron tunnels through the first barrier,
inmediately relaxes inside the quantum well and, thus,
looses information about the LL index carried during the tunneling event.
Therefore, the
number of LL´s at the emitter contributing to the current can be
different than the ones
participating in the tunneling current from the well to the collector.
Interestingly, these numbers can be modified by the external radiation.
An example is shown in Fig.~\ref{InarreaPRB_95_fig9}, where at low
field (6T) there are many Landau levels giving current. The dotted
line represents the magnetocurrent when no light is present
and the continuous line corresponds
to the case where the sample is irradiated ($\hbar\omega=7 meV$ and $F=5\times 10^{6} V/m$).
Without radiation, the current presents a sawtooth profile coming from the
participation of additional LL's as the bias voltage increases.
 When the light is switched on, the
current threshold moves to lower voltages and there is a three step
like structure between each jump.

The total number of LL´s partially occupied in the well (as a function of the bias voltage) are
represented for both cases in Fig.~\ref{InarreaPRB_95_fig9}(b):
with (continuous line) and without
(dotted line) light. For values of $V_{dc}\sim 0.04 V$
the second LL begins to be occupied in the irradiated case.
Without light, the second LL level
begins to contribute to the current through the second barrier at a slightly larger voltage.
As $V_{dc}$ increases, the well is discharged and the second LL
becomes empty for the case in which there is no light present.
Finally, at large voltages, the bias voltage at which the second
LL becomes discharged
is smaller in
the case whitout light than in the irradiated case.
\subsection{The Transfer Hamiltonian method for photon assisted tunneling: some examples
\label{THPAT}}
The TH formalism is suitable to study systems
with localized states (for instance to analyze the effect of an
external magnetic field parallel to the interfaces of the
heterostructure, where the edge states are the electronic
eigenstates) and situations where external fields, scattering
proccesses or other perturbations affect in different way the
different spatial regions of the system.

Jonson investigated inelastic resonant
tunneling in the presence of a boson field by means of the TH method \cite{JonPRB(89)}.
Within this approach, the Hamiltonian describing tunneling from the left reservoir to the
well is\footnote{Along this discussion we consider non interacting electrons.}:
\begin{equation}
H_{1}=H_L+H_{W}+H_{T}
\end{equation}
where
\begin{eqnarray}
H_L&=&\sum_{\bf k_L} \epsilon({\bf k_L}) a_{\bf k_L}^{\dagger}a_{\bf k_L},\nonumber\\
H_{T}&=&\sum_{\bf k_L,k_W}(T_{\bf k_L,k_W}a_{\bf k_L}^{\dagger}a_{\bf k_W}+H.c),\nonumber\\
H_{W}&=&\sum_{\bf k_W} \left[\epsilon({\bf k_W})+\sum_{{\bf q}}M_{{\bf q}}(b_{\bf q}+b_{\bf q}^{\dagger})\right]
a_{\bf k_W}^{\dagger}a_{\bf k_W}
+\sum_{\bf q} \hbar \omega (b_{\bf q}^{\dagger}b_{\bf q}+1/2).
\end{eqnarray}
It is assumed that the boson energy has no dispersion and that
the electron-boson interaction takes place only in the quantum well (with only one resonant level).
The wave vector is decomposed in a component perpendicular to the heterostructure $k_z$ and a
parallel component ${\bf K_{||}}$, namely ${\bf k_W}=(k_z,{\bf K_{||}})$, such that:
$\epsilon({\bf k_W})=E_z(k_z)+E_{||}({\bf K_{||}})=\epsilon_r+E_{||}({\bf K_{||}})$.

Within the TH formalism the tunneling current through the barrier is obtained from the rate of change of
the number operator for electrons in the left reservoir, namely
$N_L=\sum_{\bf k_L} a_{\bf k_L}^{\dagger}a_{\bf k_L}$, as
$J_{L,W}=-e\langle \dot{N}_L\rangle$.
By using $\dot{N}_L=\frac{i}{\hbar}[N_L,H_T]=\frac{i}{\hbar}\sum_{\bf k_L,k_W}
(T_{\bf k_L,k_W}a_{\bf k_L}^{\dagger}a_{\bf k_W}-H.c)$ it can be
shown that the current reads (see also Section \ref{Keldysh}):
\begin{eqnarray}
J_{L,W}&=&\frac{2e}{\hbar}\sum_{\bf k_L,k_W}|T_{\bf k_L,k_W}|^2
\int_{-\infty}^{\infty}\frac{d\epsilon}{2\pi}A_W({\bf k_W},\epsilon)
A_L({\bf k_L},\epsilon+\mu_L-\mu_W)\nonumber\\
&\times&[f_L(\epsilon)-f_W(\epsilon)].
\label{current-TH}
\end{eqnarray}
A similar description holds for the current $J_{W,R}$.
In Eq.~(\ref{current-TH}), $f_L(\epsilon)$ is the equilibrium (Fermi-Dirac) distribution
function of electrons the left reservoir
and $f_W(\epsilon)$ is the distribution function of electrons inside the well.
If some scattering mechanism inside the well is implicitly assumed in order to maintain
thermal equilibrium, $f_W(\epsilon)$ is a Fermi-Dirac
distribution function and $\mu_W$ has to be calculated by imposing current conservation
\begin{equation}
J_{L,W}(\mu_W)=J_{W,R}(\mu_W).
\end{equation}
If, on the other hand, resonant tunneling occurs without any scattering, $f_W(\epsilon)$ itself has
to be calculated by imposing current conservation
\begin{equation}
J_{L,W}(f_W)=J_{W,R}(f_W),
\end{equation}
and a {\it nonequilibrium} distribution function is obtained.

Moreover, the current depends on the product of the two spectral functions $A_L$ and $A_W$.
Due to the coupling to the bosons, the spectral function in the well changes from the delta-function
form $A_W^0({\bf k_W},\epsilon)=2\pi\delta(\epsilon-\epsilon({\bf k_W}))$ to a more complicated
function involving
a distribution of free-electron
spectral functions\footnote{This can be shown by considering the canonical tranformation
$\bar{H}=e^{s}He^{-s}$ with
$s=\sum_{{\bf q}} \frac{M_{{\bf q}}}{\hbar\omega}(b_{\bf q}-b_{\bf q}^{\dagger})
\sum_{\bf k_W}a_{\bf k_W}^{\dagger}a_{\bf k_W}$, see also Eq.~(\ref{canonical})
in section \ref{Double barriers}.}:
\begin{equation}
A_W({\bf k_W},\epsilon)=\sum_{n=-\infty}^{\infty}S_n A_W^0({\bf k_W},\epsilon+n\hbar\omega-\Delta),
\end{equation}
where the strenght $S_n$ is given by:
\begin{equation}
S_n=e^{-g(2n_{BE}+1)}I_n(2g\sqrt{n_{BE}(n_{BE}+1)})e^{\frac{n\hbar\omega}{2k_BT}}.
\label{Jonson-strength}
\end{equation}
In Eq.~(\ref{Jonson-strength}) $I_n$ is a modified Bessel function, $g$ is the adimensional coupling
\begin{equation}
g=\sum_{\bf q}\frac{M_{{\bf q}}^2}{(\hbar\omega)^2}\equiv\frac{\Delta}{\hbar\omega},
\end{equation}
and $n_{BE}$ is the Bose-Einstein distribution function $n_{BE}=1/(e^{\frac{\hbar\omega}{k_BT}}+1)$.

Jonson applied the above model
to study photoassisted tunneling through resonant devices.
By treating the photon field semiclassically (see footnote \ref{semiclassical}),
one recovers a model similar to the Tien and Gordon model \cite{JonPRB(89)}.

Foden and Whitaker presented in Ref.~\cite{FodPRB(98)} a
quantum electrodynamic treatment of photon-assisted sidebands that appear when an electron tunnels
through a resonant state and interacts with a {\it coherent field}, namely a Hamiltonian of the form:
\begin{equation}
H_{tot} = \sum_i \epsilon_i a_i^{\dagger}a_i +\sum_q \hbar \omega_q (b_q^{\dagger}b_q+1/2)
+\sum_{q,i,j}M_{q,i,j}(b_q+b_q^{\dagger})a_j^{\dagger}a_i
\end{equation}
where $M_{q,i,j}=-(e/m)(2\pi\hbar/V\omega_q\epsilon)^{1/2}\langle i|e^{i{\bf qr}}\hat{p}|j\rangle$.
Again, the electron-photon coupling is assumed to take place only in the quantum well.
The photon field is described by a single-mode coherent state
$|\alpha\rangle=e^{-|\alpha|^2/2}\sum_n(\alpha^n/\sqrt{n!})|n\rangle$, namely a Poisson distribution of
number states with mean value $\bar{n}=|\alpha|^2$ (the expectation value of the electric field in this state is
$\sqrt{2\bar{n}\hbar\omega/\epsilon V}$).
Considering just one resonant state,
they obtained the quantum well spectral density:
\begin{equation}
\chi(\epsilon)=2\pi e^{-g^2}\sum_{n=-\infty}^{\infty} \delta (\epsilon
-E_0-n\hbar\omega+\Delta) \sum_{m=-\infty}^n {J_m}^2(2g\alpha)
\frac{g^{2(n-m)}}{(n-m)!}
\end{equation}
with $g=M/\hbar\omega$ and $\Delta=\lambda^2\hbar\omega$.
Under the influence of radiation, the $\delta$ spectral
function is shifted by $\Delta$ due to the renormalization
of the electron energy as a result of its interaction with
the electromagnetic field, and splits in a series of side bands.
This result differs from the one obtained from a semiclassical
model (see Eq.~(\ref{tien-gordon-dos})):
\begin{equation}
 \chi_{class}(\omega)=2\pi \sum_{n=-\infty}^{\infty}{{J_n}^2}(eV/\hbar\omega)
\delta (\epsilon-E_0-n\hbar\omega)
\end{equation}
In the quantum case the relative intensities of the spectral lines
are different from those obtained classically and the spectrum becomes
asymmetric with respect to the $n=0$ line \cite{FodPRB(98)}.
This can be understood in terms of spontaneous emission
of photons by the electron due to the interaction with the vacuum fluctuations of the electromagnetic field.
This asymmetry is, of course, not present in the clasical limit.

Johansson analyzed in Ref.~\cite{JohPRB(90)}
the effect of an ac component in the voltage across a double barrier.
The ac voltage is included
only in the reservoirs as:
\begin{eqnarray}
H_L&=&\sum_{\bf k_L} \left[\epsilon({\bf k_L})+V_0cos(\omega t)\right] a_{\bf k_L}^{\dagger}a_{\bf k_L}
\nonumber\\
H_R&=&\sum_{\bf k_R} \left[\epsilon({\bf k_R})-V_0cos(\omega t)\right] a_{\bf k_R}^{\dagger}a_{\bf k_R}
\label{ac-voltage-leads}
\end{eqnarray}
and the quantum well (with a single resonant state)
is assumed to remain unafected by the time dependent field.
By using the TH method to calculate the transmission probability for incoming electrons,
Johansson concludes that when the period of the ac voltage is short compared with the lifetime of an
electron inside the
quantum well, photon-assisted tunneling occurs.
If, on the other hand, the period is long the transmission probability is governed by the
instantaneous value of the voltage.
The effect of interlevel transitions due to the coupling with radiation
was considered by Johansson and Wendin in Ref.~\cite{JohPRB(92)}.
They reported shifts of the positions
of the transmission resonances due to the Dynamic Stark
effect.

All the treatments above consider the coupling with the external field just in one part of the structure.
In particular, neglecting the effects of the field outside the quantum well
is only a good approximation as long as the frequency of the field is higher
than the plasma frequency of the electron gas in the doped contacts.
When this is not true, the electrons in the contacts screen the external field
strongly and most of the interaction between the field and
tunneling electrons will take place outside the well. Above the plasma
frequency, on the other hand, most of the absorption and emmission of
radiation by the tunneling electrons occur when they are in the
quantum well. For an electron gas in GaAs with Fermi Energy of 50 meV the
plasma frequency is of the order of 40 meV. The plasma frequency increases
with increasing Fermi energy roughly as $(E_F)^{3/4}$ \cite{ApePRB(92)}.
Usually, in experimental setups for resonant tunneling diodes
the Fermi energy at the contacts is around
50 meV and the photon frequency is of the order of 10 meV.
In order to compare with available experiments in these systems, it is thus crucial to include
the coupling of the radiation with the contacts,
as we have described in section \ref{Iñarrea1}.
\subsection{Generalized Transfer Hamiltonian for coherent photon assisted tunneling
\label{Aguado1}}
The Transfer Matrix technique described in Sections \ref{TM} and \ref{Double barriers}
is a very powerful tool to analyze coherent tunneling.
Here, we review an alternative method.
This method, which generalizes the
TH approach we have described in Section \ref{THPAT}, accounts for high order tunneling events
(beyond the first order described by the TH).

The extension of the TH method was put forward by Brey et al in Ref.~\cite{BreyPRB(88)}
to the study of tunneling through resonant states in heterostructures.
The method, dubbed generalized Transfer Hamiltonian (GTH),
allows to describe not only the
sequential tunneling (which considers the electrons tunneling
through each single barrier in a sequential way)
but the coherent one
which includes the virtual transitions
through the resonant states for electrons crossing coherently the
system.
The GTH was later extended
by Aguado et al in Refs.~\cite{AguPRB(96),Aguadotesis} to include the effect of ac
potentials using the two configurations discussed along this section: constant
potentials (no mixing of electronic states, case a) and a coupling of the form 
$\vec{A}.\vec{p}$ (mixing of electronic states, case b).

\begin{figure}
\begin{center}
\includegraphics[width=0.75\columnwidth,angle=0]{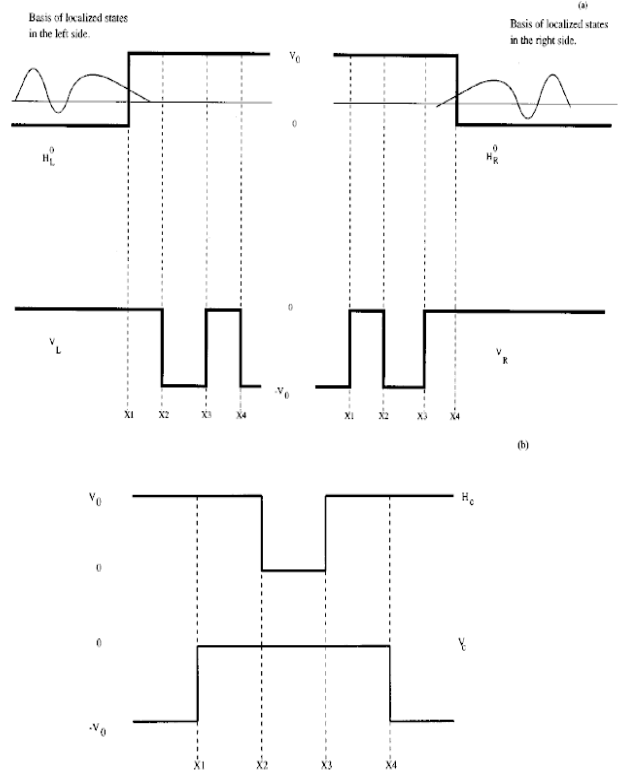}
\end{center}
\caption[]{Sketch of the different Hamiltonians used to study
tunneling with the GTH method:a) left and right Hamiltonian; b)
center Hamiltonian} \label{AguadoPRB_96_fig1}
\end{figure}

The TH considers a localized basis representation, namely
approximated Hamiltonians whose eigenstates are spatially
localized (see $H^{0}_{L}$ and $H^{0}_{R}$ in Fig.~\ref{AguadoPRB_96_fig1}(a).
By means of the interaction picture the
required perturbations are switched on adiabatically:
\begin{eqnarray}
  H=H_{L}+V_{L}(t)e^{\eta t} = H_{R}+V_{R}(t) e^{\eta t}
\end{eqnarray}
Where $ V_{L} $ and $ V_{R} $ are sketched in Fig.~\ref{AguadoPRB_96_fig1}(b).
Following Aguado et al in Ref.~\cite{AguPRB(96)},
the time evolution of the wave function, including ac potentials (case a) of the form
$H_{2}(t)=\pm V_{ac}cos\omega_{0}t$, see Eq.~(\ref{ac-voltage-leads}),
on both sides (left and right),
can be written as:
\begin{eqnarray}
|\Psi(t)\rangle&=&f(t)\sum_{m=-\infty}^{\infty}J_{m}
(\frac{V_{ac}}{\hbar\omega_{0}})e^{-im\omega_{0}t}e^{-i\omega_{k_{L}}t}|k_{L}
\rangle\nonumber\\
&+&\sum_{n=-\infty}^{\infty}J_{n}(\frac{V_{ac}}{\hbar\omega_{0}})e^{in\omega_{0}t}
\sum_{p_{R}}U_{R}(t,-\infty)e^{-i\omega_{p_{R}}t}|p_{R}\rangle
\end{eqnarray}
This wave function must describe a particle initially on the left
side. This is satisfied by taking $f(-\infty)=1$ and
$U_{R}(-\infty,-\infty)=0$. The electrons in a particular state
$|k_{L}\rangle$ can in principle evolve to any state
$|p_{R}\rangle$ in the right side so that a summation over right
states is required in the expression of the wave function. The
time evolution operator $ U_{R}(t,-\infty)$ gives the evolution of
an electron to a right state and is determined at every order from
the Schr\"odinger equation by an expansion in a perturbation
series.
$U_{R}^{(j)}$ becomes in terms of the retarded Green's function:
\begin{eqnarray}
&&U_{R}^{(j)}(t,-\infty)=\sum_{n=-\infty}^{\infty}
\sum_{m=-\infty}^{\infty}J_{n}J_{m}
\frac{e^{\frac{i}{\hbar}(\epsilon_{p_{R}}-\epsilon_{k_{L}}-n\hbar\omega_{0}
-m\hbar\omega_{0}-ij\eta)t}}{(\epsilon_{p_{R}}-\epsilon_{k_{L}}
-n\hbar\omega_{0}-m\hbar\omega_{0}
-ij\eta)}\nonumber\\
&&\langle
p_{R}|V_{R}G_{R}^{r}(\epsilon_{k_{L}}+m\hbar\omega_{0}+i(j-1)\eta)V_{R}
....V_{R}G_{R}^{r}(\epsilon_{k_{L}}+m\hbar\omega_{0})V_{L}|k_{L}\rangle,
\end{eqnarray}
where
\begin{eqnarray}
G_{R}^{r}(\epsilon)=\sum_{n=-\infty}^{\infty}\sum_{p_{R}} J_{n}^{2}
\frac{ |p_{R}\rangle
\langle p_{R}| }
{\epsilon-\epsilon_{p_{R}}+n\hbar\omega_{0}+i\eta}.
\end{eqnarray}  
The transition probability from left to right per unit time
can be expressed as
$P_{RL}=\lim_{\eta\rightarrow
0}2Re[U_{R}^{*}(t,-\infty)\frac{dU_{R}(t,-\infty)}{dt}]$,
where $U_{R}$ includes the sum over all orders in perturbation
theory.
\begin{figure}
\begin{center}
\includegraphics[width=0.75\columnwidth,angle=0]{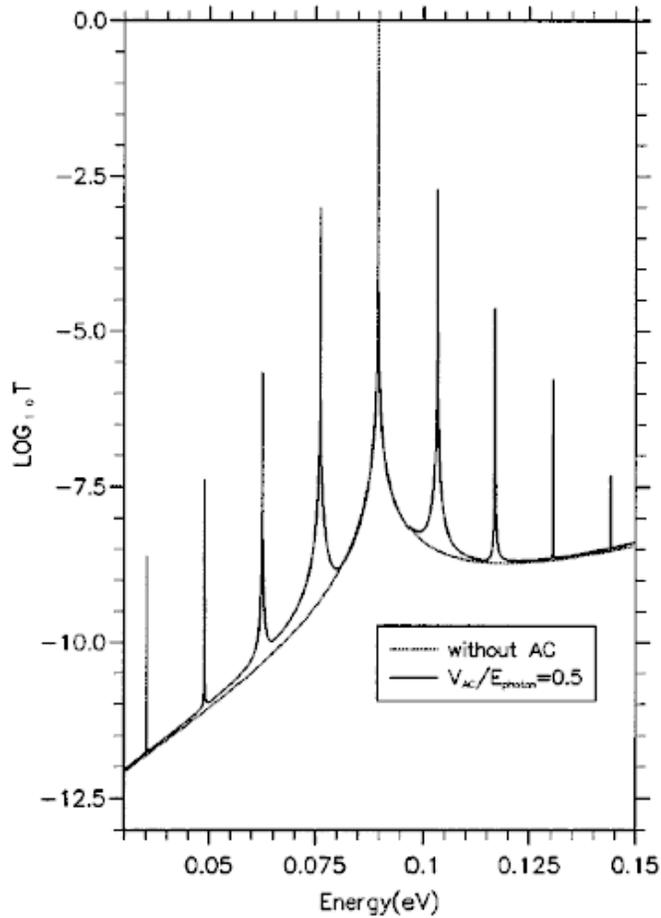}
\end{center}
\caption[]{$Log_{10}$ of the coherent transmission coefficient as
a function of energy for a
GaAs/AlGaAs double barrier of 100-50-100$\AA$ with and without an ac potential
($V_{ac}/\hbar\omega_{0}=0.5,\hbar\omega_{0}=$13.6 meV).} 
\label{AguadoPRB_96_fig2}
\end{figure}
Performing the summation and taking the stationary limit,
the expression for the transition
probability reads:
\begin{eqnarray}
P_{RL}&=&\frac{2\pi}{\hbar}\sum_{n,m}J_{n}^{2}J_{m}^{2}\delta(\epsilon_{p_{R}}
-\epsilon_{k_{L}}-n\hbar\omega_{0}-m\hbar\omega_{0})\nonumber\\
&\times&|\langle
p_{R}|V_{L}+V_{R}G^{r}(\epsilon_{k_{L}}+m\hbar\omega_{0})V_{L}|{k_{
L}} \rangle|^{2}, \label{PRL1}
\end{eqnarray}
Where $G^{r}$ is the total
retarded Green's function of the system 
\begin{equation}
G^{r}(\epsilon)=G^{r}_{R}(\epsilon)+G^{r}_{R}(\epsilon)V_{R}G^{r}_{R}(\epsilon)
+...
\end{equation}    

This formula for the transition probability is a natural extension
of the Fermi Golden Rule formula \cite{JonPRB(89),JohPRB(90),ApePRB(92)}.
The first term
corresponds to first order
perturbation theory, 
$P_{RL}=\frac{2\pi}{\hbar}\sum_{n,m}J_{n}^{2}J_{m}^{2}
|\langle p_{R}|V_{L} |{k_{L}} \rangle|^{2} 
\delta(\epsilon_{p_{R}}-\epsilon_{k_{L}}-(n+m)\hbar\omega_{0})$
and is the only one appearing in the TH
method \cite{BardPRL(61)}. 
The term containing the
retarded Green's function is the one which includes processes which involve
intermediate states
and therefore describes correctly the coherent resonant tunneling.
\begin{figure}
\begin{center}
\includegraphics[width=0.75\columnwidth,angle=0] {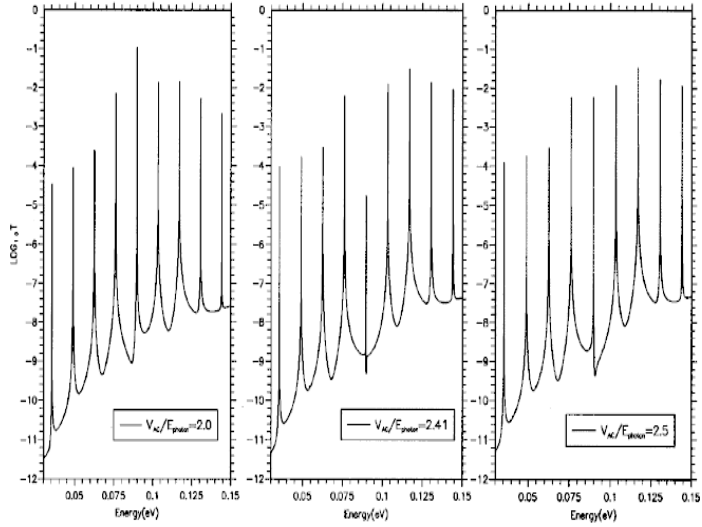}
\end{center}
\caption[]{$Log_{10}$ of the coherent transmission coefficient as
a function of energy for a
double barrier with an applied ac potential 
close to the first zero of $J_0$: $V_{ac}/\hbar\omega_{0}=2, 2.4, 2.5$, with $\hbar\omega_{0}$=13.6meV).
Same sample as in
Fig.~\ref{AguadoPRB_96_fig2}}
\label{AguadoPRB_96_fig3}
\end{figure}
\begin{figure}
\begin{center}
\includegraphics[width=0.75\columnwidth,angle=0]{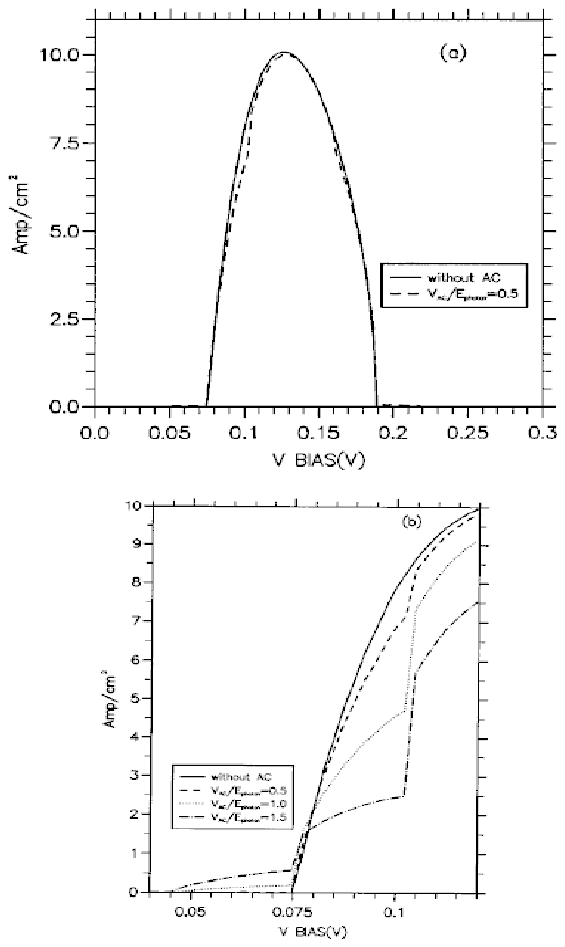}
\end{center}
\caption[]{Coherent tunneling current as a function of $V_{dc}$
for the sample of Fig.~\ref{AguadoPRB_96_fig2}: a) with and without
ac potential for the parameters: $V_{ac}/\hbar\omega_{0}=0.5,
\hbar\omega_{0}$=13.6 meV;b) For different ratios between the
intensity and frequency of the ac potential
($\hbar\omega_{0}$=13.6 meV)}
\label{AguadoPRB_96_fig4}
\end{figure}

The transmission coefficient through the structure
can be calculated from Eq.~(\ref{PRL1}).
An example is shown in Fig.~\ref{AguadoPRB_96_fig2}.
In this
case, processes up to fourth order contribute to the transmission
as reflected in the four satellites which appear at both sides of
the main central peak.
Note that the broadening of the resonant states
in the well
is not a constant but depends on the side band index
$m$ \cite{AguPRB(96)}. In the present case the contribution of the
photoside band of index  $m>$0 to the transmission coefficient
is smaller than the one coming from the main peak (m=0).
In Fig.~\ref{AguadoPRB_96_fig3} an example of dynamical localization is shown.

The current through a double barrier versus the applied bias voltage is plotted in
Fig.~\ref{AguadoPRB_96_fig4} for different parameters of the ac potential. The main
features are: a shift of the threshold current to lower
voltages and the reduction of the current for larger voltages in the presence of the
ac potential. This effect increases as the ratio
$\frac{V_{ac}}{\hbar\omega_{0}}$ increases. Also a step-like
behavior is observed (see Fig.~\ref{AguadoPRB_96_fig4}(b)).
These features can be explained in terms of
the photoside bands.
The contribution to the current at low dc voltages comes from photoside bands
associated to electronic states close to the Fermi energy in the
emitter. They contribute to the resonant tunneling
when the resonant state $E_{r}$ is above $E_{F}$, $E_{r}=m\hbar\omega_0+E_{F}$, via the absorption
of $m$ photons. This process has a low
spectral weight and its contribution to the current is small.
By increasing the applied dc voltage,
$E_r$ approaches $E_F$ such that another photo-sideband corresponding
to the process $E_{r}=n\hbar\omega_0+E_{F}$ ($n<m$) starts to contribute.
The spectral weight corresponding to this process is higher and the current increases.
The reduction of current (with respect to the case without ac) can be explained similarly in terms of
the photo-sidebands.

\begin{figure}
\begin{center}
\includegraphics[width=0.75\columnwidth,angle=0]{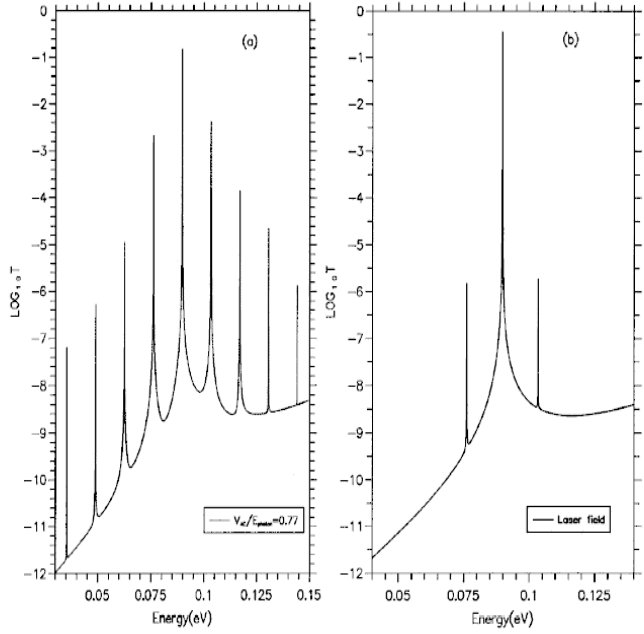}
\end{center}
\caption[]{Comparison of $Log_{10}$ of coherent transmission
coefficient as a function of energy for constant ac potentials and for
electromagnetic radiation: a) $V_{ac}/\hbar\omega$=0.77; b)
$F=4.10^{5}$V/m, $\hbar\omega$=13.6 meV. Same sample as in
Fig.~\ref{AguadoPRB_96_fig2}}
\label{AguadoPRB_96_fig5}
\end{figure}

As discussed in section \ref{Iñarrea1}, the effect of an homogeneous EMF
affecting the whole sample is quite different from the one obtained from
spatially constant ac potentials. Including the EMF within the GTH scheme in a similar fashion
as the one described by Eqs.~(\ref{Htot1}-\ref{Htot2}),
the transition
probability (only one photon processes are considered) for this second 
configuration (case b) can be written as \cite{AguPRB(96)}:
\begin{eqnarray}
P_{RL}&=&\frac{2\pi}{\hbar}\{|A_{RL}|^{2}\delta(\epsilon_{p_{R}}-\epsilon_{k_{L}})\nonumber\\
&+&|B_{RL}|^{2}\delta(\epsilon_{p_{R}}-\epsilon_{k_{L}}+\hbar\omega_{0})
+(\omega_{0} \rightarrow -\omega_{0})\}
\end{eqnarray}
where $A_{RL}$ and $B_{RL}$ contain the matrix elements:
\begin{eqnarray}
A_{RL}&=&J_{0}(\beta_{p_{R}})J_{0}(\beta_{k_{L}})
\langle p_{R}|V_{L} + V_{R}G^{r}(\epsilon_{k_{L}})V_{L}|k_{L}\rangle\nonumber\\
B_{RL}&=&\frac{eF J_{0}(\beta_{p_{R}})J_{0}(\beta_{k_{L}})
}{2m^{*}\omega_{0}}
\{\langle p_{R}|V_{R}G^{a}_{L}(\epsilon_{p_{R}}) P_{z}|
k_{L}\rangle+
\langle p_{R}|P_{z}G^{r}_{R}(\epsilon_{k_{L}})V_{L}|
k_{L}\rangle+\nonumber\\
&&\langle p_{R}|V_{R}G^{a}(\epsilon_{p_{R}})V_{L}G^{a}_{L}(\epsilon_{p_{R}})
P_{z}|
k_{L}\rangle+\langle p_{R}|P_{z}G^{r}_{R}(\epsilon_{k_{L}})V_{R}G^{r}
(\epsilon_{k_{L}}) V_{L}|k_{L}\rangle\nonumber\\
&&\langle p_{R}|V_{R}G^{a}(\epsilon_{p_{R}})P_{z}G^{r}(\epsilon_{k_{L}})
V_{L}|
k_{L}\rangle\},
\end{eqnarray} 
where $G^{r}$ is the total Green's function and
\begin{equation}
G_{L(R)}^{r(a)}(\epsilon)=\sum_{k_{L(R)}} J_{0}^{2}(\beta_{k_{L(R)}})
\frac{ |k_{L(R)}\rangle
\langle k_{L(R)}| }
{\epsilon-\epsilon_{k_{L(R)}}\pm i\eta}.
\end{equation}
In these expressions, the argument of the Bessel functions is now governed by 
matrix elements of the momentum operator, 
cf. Eq.~(\ref{beta_k}). 
An example of the comparison between cases (a) and (b) is shown in 
Fig.~\ref{AguadoPRB_96_fig5} where the transmission coefficient is plotted
for both configurations.
Typically, the argument of the Bessel functions of high order $m>0$
are negligible in the second case (the momentum matrix elements are very small)
such that only two satellites show up in the transmission coefficient
Fig.~\ref{AguadoPRB_96_fig5}(b). Furthermore,
these side-peaks have another origin than
the photoside bands of Fig.~\ref{AguadoPRB_96_fig5}(a): they come from the mixing of electronic
states due to the homogeneous light, namely they appear from
the off-diagonal matrix elements of the electronic
momentum coupled by the light.
In other words, the tunneling channels for the two configurations are
different. In the case of an ac potential the off-diagonal terms
cancel if the time dependent field is considered constant within
each region (left, center and right). In this case, the main
tunneling channels (the only ones within this approximation)
are the photoside bands: those in the emitter
align in energy with the photoside bands in the well producing
additional contributions to the transmission probability and the
resonant current. Their contribution can be important even for
high order photoside bands if the ratio
$\frac{V_{ac}}{\hbar\omega_{0}}\geq 1$.
In the case of homogeneous light, the situation is different. The
off-diagonal electron-photon coupling terms in the Hamiltonian are
the ones which modify the transmission and, thus, the current.
These channels, involving different
electronic states, Fig.~\ref{AguadoPRB_96_fig5}, contribute in
principle also with all
their photoside bands. However, since the argument of the Bessel
functions, which is
controlled by the momentum matrix elements (see Eq.~\ref{beta_k}), remains very small,
just the zero index photoside band (the main one) is
non-negligible and gives a contribution to the transmission. Therefore
the three peaks in the transmission coefficient come from the main
bands (index zero) corresponding to three electronic states which
differ in one photon energy and which are mixed by the EMF field.
\subsection{Beyond the single electron picture: charge accumulation
effects and current bistability \label {bistability}}
So far, we have discussed photon-assisted tunneling of independent electrons.
The effect of the electrostatic fields induced by the
electronic charge are not considered in the model calculations of sections
\ref{Iñarrea1}-\ref{Aguado1}.
The space charge
alters the voltage distribution over the whole structure such
that the actual electrostatic potential profile does
not correspond to the simple description in terms of
abrupt interfaces. One consequence of that is the appearance of
intrinsic current bistability, two values of the current for a given dc bias voltage,
which is caused by the electrostatic feedback of the
space charge dynamically stored in the well.

The first experimental evidence of intrinsic current bistability in the electronic transport through DB's
was observed by Goldman et al in Ref.~\cite{GolPRL(87),GolPRB(87)}. Although these experimental results
were first questioned in Refs.~\cite{SolPRB(87),FosPRB(89)},
intrinsic bistability in the current through resonant tunneling diodes was unambiguosly confirmed
in subsequent experiments \cite{ZasAPL(88),ZasPRB(90),JimPRB(95)}.
On the theory side, many papers have been devoted to the
study of current bistability in resonant tunneling diodes
\cite{GolPRB(87),SheAPL(88),KluPRB(89),SofPRB(90),JenPRL(91),FiiSS(92),PerPRB(93),WagnerJJAP(93)}.
\subsubsection{Selfconsistent model \label{Selfconsmodel}}
As we mentioned, the Coulomb interaction between electrons
induces electrostatic
fields which modify the distribution of the electrostatic
potential through the heterostructure. Therefore, the energy of the
resonant states in the quantum well is modified and so the
current density for a given bias voltage changes.
In a resonant tunneling diode there are three regions spatially separated: emitter, well
and collector, where the charge is accumulated. The potential
profile through the whole heterostructure is not abrupt and
accumulation and depletion layers in the emitter and collector
are built up. By solving simultaneously the Poisson
and Schr\"odinger equations these potential
profiles, together with the current, can be obtained.
This procedure can be simplified by assuming
that the accumulated charge in each region
is distributed as a two dimensional sheet of charge \cite{GolPRB(87),IñaEL(96)} as we shall
describe below.

Assuming that the electrons in the well are in
local equilibrium with Fermi energy $\epsilon_{\omega}$ which
define the electronic density $n_{w}$, current conservation
can be used to obtain the Fermi energy in the well as:
\begin{equation}
J_{1}(\epsilon_{w},\Phi)=J_{2}(\epsilon_{w},\Phi),
\label{ratedb}
\end{equation}
with the (zero-temperature) currents:
\begin{eqnarray}
J_{1}&=&(e/2\pi\hbar)\int_{0}^{E_{F}} \frac{T_1 k_{w}}
{w_{2}+(1/\alpha_{b})+(1/\alpha_{d})}\frac{\Gamma}{(E_{z}-E_{tn})^2+\Gamma^2}\nonumber\\
&\times&(E_F-\epsilon_{w}-E_{z})dE_{z}\nonumber\\
J_{2}&=&(e/2\pi\hbar)\int_{0}^{E_{F}} \frac{T_2 k_{w}}
{w_{2}+(1/\alpha_{b})+(1/\alpha_{d})}\frac{\Gamma}{(E_{z}-E_{tn})^2+\Gamma^2}
\epsilon_{w}dE_{z}.
\end{eqnarray}          
$\Gamma$ is the half-width of the resonant level
and the rest of parameters are defined as in subsection \ref{Iñarrea2}.
In Eqs.~(\ref{ratedb}), 
$\Phi$ denotes the set
of voltage drops through the structure.
The Poisson equation yields the potential drops in the
barriers, $V_{1}$ and $V_{2}$, and in the well, $V_{w}$ :
\begin{eqnarray}
\frac{V_{w}}{w}&=&\frac{V_{1}}{d}+\frac{
n_{w}(\epsilon_{\omega})-eN_{D}^{w}}{2\varepsilon}
\label{field.inside1}\\
\frac{V_{2}}{d}&=&\frac{V_{1}}{d}+\frac{n_{w}(
\epsilon_{\omega})-eN_{D}^{w}}{\varepsilon}\,,
\label{field.inside2}
\end{eqnarray}
 where $\varepsilon$ is the GaAs static permittivity,
$n(\epsilon_{\omega})$ is the 2D (areal) charge density at the
 well (to be determined), $w$ and $d$ are the well and
barrier thickness respectively,
and $N_{D}^{w}$ is the 2D intentional doping at the
wells. The emitter and collector layers can be described by the
following equations \cite{GolPRL(87),IñaEL(96)}:
\begin{equation}
\frac{\Delta_{1}}{\delta_{1}} = \frac{eV_{1}}{d}\,,
\quad\quad\quad\quad \sigma = 2 \varepsilon\,\frac{V_{1}}{d}\simeq
eN(E_{F})\Delta_{1}\delta_{1}\
\label{emitter}
\end{equation}
\begin{equation}
\frac{\Delta_{2}}{e}=\frac{V_{2}\delta_{2}}{d}-\frac{1}{2
\varepsilon} eN_{D}\delta_{2}^{2}\,,\quad\quad\quad\quad
\delta_{3} =
\frac{\delta_{2}E_{F}}{\Delta_{2}}\
\label{collector}
\end{equation}
To write the emitter equations (\ref{emitter}), we assume that
there are no charges in the emitter barrier. Then the electric
field across $\delta_{1}$ (see Fig.\ref{AguadoRC_97_fig1} in section \ref{Superlattices})
is equal to that
in the emitter barrier. Furthermore, the areal charge density
required to create this electric field is provided by the emitter.
$N(E_{F})$ is the density of states at the emitter $ E_{F}$. To
write the collector equations (\ref{collector}), it is assumed
that the region of length $\delta_2$ in the collector is
completely depleted of electrons and local charge
neutrality in the region of length $\delta_3$ between the end of
the depletion layer $\delta_2$ and the collector holds.

In order
to close the set of equations, two extra equations are employed.
The first one imposes global charge conservation:
\begin{equation}
\sigma+n_{w}(\varepsilon_{\omega})-eN_{D}^{w}
=eN_{D}(\delta_{2}+{1\over 2}\delta_{3}) \,.
\label{charge.conserv}
\end{equation}
Finally, all voltage drops across the different regions
must add up to the applied bias voltage:
\begin{equation}
V_{dc} = V_{1}+V_{2}+V_{w} +
\frac{\Delta_{1}+\Delta_{2}+E_{F}}{e} .\label{bias}
\end{equation}
Note that the right hand side of Eq.(\ref{charge.conserv}) is the
positive 2D charge density depleted in the collector region.

This system of equations, together with appropriate initial
conditions, determine completely and self-consistently the
current.
The generalization of the selfconsistent method to treat multiple-quantum well structures was done
in Ref.~\cite{AguPRB(97)} and will be
described in more detail in section \ref{Superlattices}.

\begin{figure}
\begin{center}
\includegraphics[width=0.75\columnwidth,angle=0]
{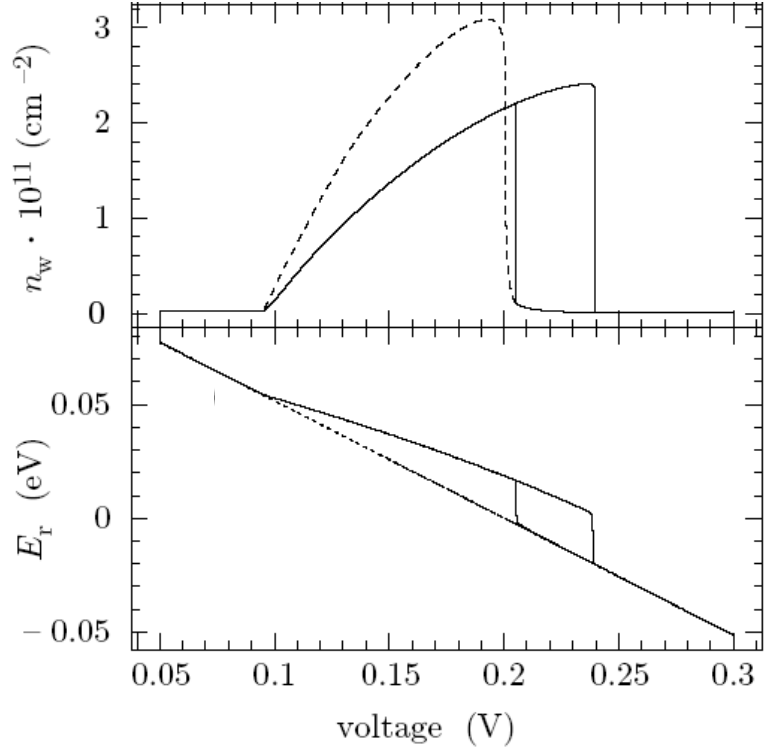}
\end{center}
\caption{Top: Electronic charge density accumulated in the
quantum well as a function of the bias voltage.
Bottom: Resonant level, $E_r$, as a function of the bias voltage.
The continuous (dotted) line corresponds to the case with (without) selfconsistency.
The results correspond to a DB consisting of 5nm-5nm-5nm GaAs-GaAlAs and n=$10^{18} cm^{-3}$.}
\label{InarreaEPL_96_fig1}
\end{figure}

As we mentioned at the beginning of the section, charge accumulation may lead to current bistability
in resonant tunneling diodes. An example is shown in Fig.~\ref{InarreaEPL_96_fig1} (top), where
the charge accumulated (continuous line) in a
50\AA-50\AA-50\AA
 $GaAs-Al_{.3}GaAs$ DB is plotted versus the applied bias voltage.
Both directions of the bias voltage, forward and backward, are shown.
For comparison, the charge without Coulomb interaction
(dotted line) is also plotted.
The selfconsistent curve (solid line) presents a clear bistable region around $V_{dc}\sim 0.2-0.25 V$
in contrast with the non
interacting electrons case (dotted line).
The accumulated charge into the well
is of the order of $10^{11}$ $cm^{2}$. This large value
is responsible for the highly non linear
distribution of the electrostatic potentials through the structure
which eventually produces current bistability.
These features can be
easily understood by analysing the voltage dependence of the resonant state
(Fig.~\ref{InarreaEPL_96_fig1}, bottom).
As expected, the resonant state in the non interacting case (dotted line)
drops linearly with $V_{dc}$.
The selfconsistent solution strongly deviates from a linear behavior
as soon as charge accumulates in the well.
This charge accumulation produces the electrostatic fields
which are responsible of the non-linear behavior of $E_r$.
The linear dependence is recovered as soon as the well is discharged.
This happens for different voltages in the forward and backward directions
leading to bistability.

\subsubsection{Current bistability in the presence of light}
I\~narrea et al demonstrated in Ref.~\cite{IñaEL(96)} that the application of an external EMF field
modifies the intrinsic bistability properties of a resonant tunneling diode.
In particular, the external time-dependent field induces new bistable regions.
Furthermore, the bistability region of the unirradiated
sample becomes reduced in the presence of radiation. This latter effect has been recently confirmed by
Orellana et al \cite{OrePRB(00)}.

\begin{figure}
\begin{center}
\includegraphics[width=0.75\columnwidth,angle=0]
{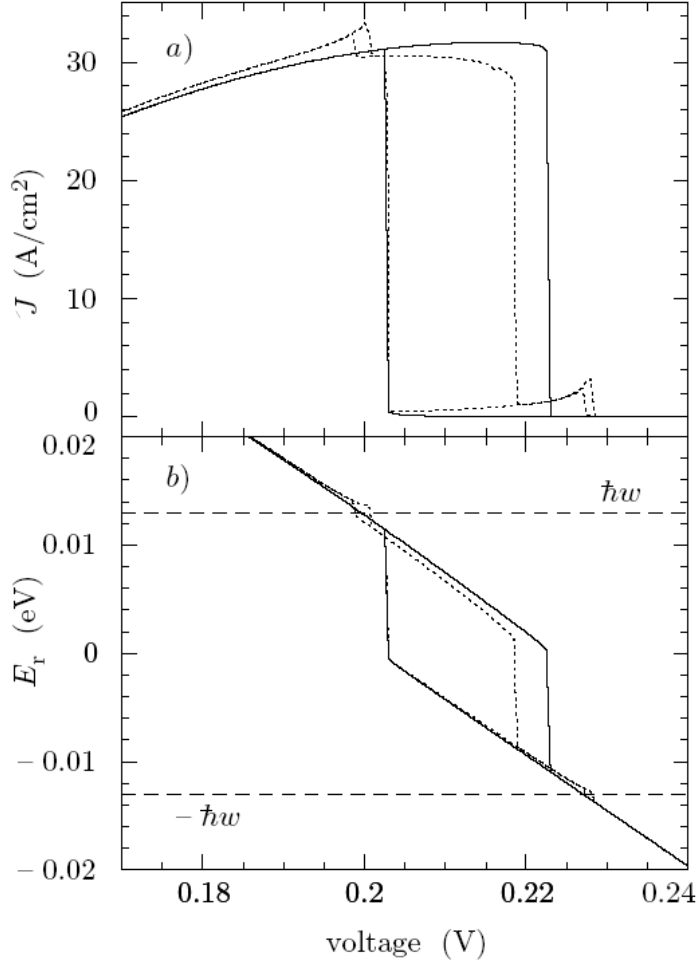}
\end{center}
\caption{a) Current versus applied voltage with (dotted line) and without (solid line) radiation.
b) Position of the resonant state versus applied voltage.
$F= 8\times 10^{6} V/m$ and $\hbar\omega=13 meV$.
Sample: 10nm-5nm-10nm GaAs-GaAlAs DB with n=$10^{18} cm^{-3}$.}
\label{InarreaEPL_96_fig2}
\end{figure}

An example of the effects of an external EMF on bistability is shown
in Fig.~\ref{InarreaEPL_96_fig2}(a) where the current with (dotted line) and without (solid line)
radiation is plotted versus the applied voltage. The results are obtained along the lines presented in section
\ref{Iñarrea2} together with the selfconsistent method defined by Eqs.~(\ref{field.inside1})-(\ref{bias}).
\begin{figure}
\begin{center}
\includegraphics[width=0.75\columnwidth,angle=0]
{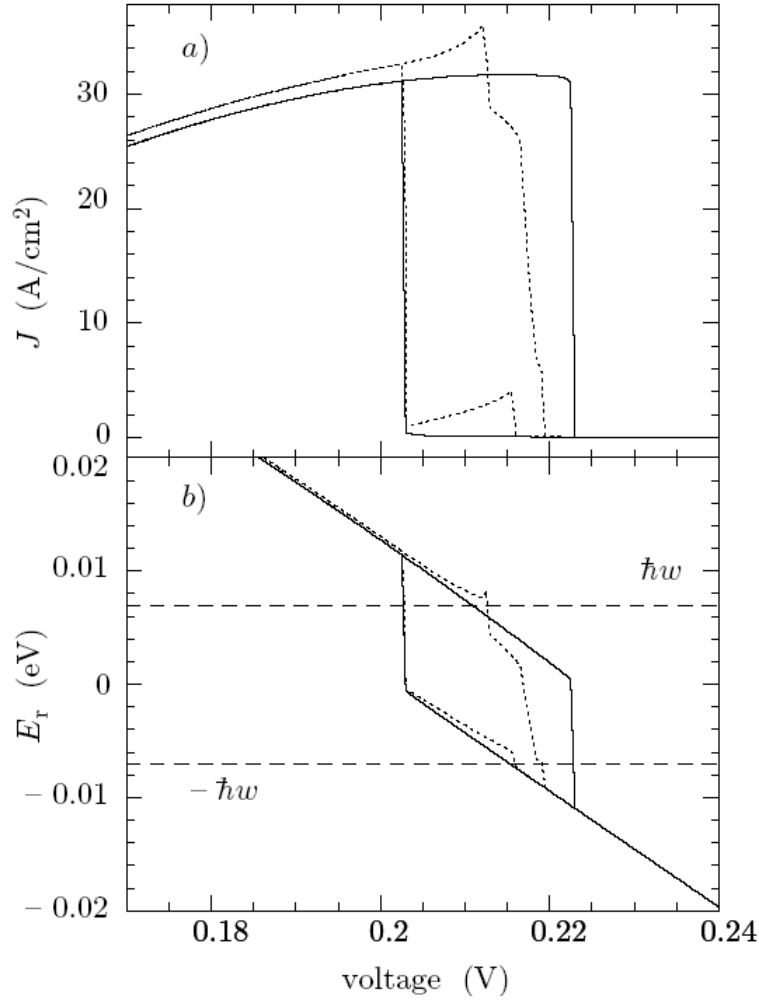}
\end{center}
\caption{a)
Current versus applied voltage with (dotted line) and without (solid line) radiation
($F= 8\times 10^{6} V/m$ and $\hbar\omega=7 meV$).                                              b) Position of the resonant state versus applied voltage. 
Sample: 10nm-5nm-10nm GaAs-GaAlAs DB with n=$10^{18} cm^{-3}$.}
\label{InarreaEPL_96_fig3}
\end{figure}

The new bistable regions can be understood by
analyzing the voltage dependence of the resonant state (Fig.~\ref{InarreaEPL_96_fig2}(b)):
The possibility of exchanging photons with the EMF field changes the voltage positions at which the
the quantum well becomes charged and discharged. The two new bistable regions thus correspond to absorption
and emission of one photon.
For instance, if
$E_r$ is one photon above the bottom of the emitter conduction
band, the electrons
have a finite probability of emitting a photon below the
conduction band and then the light acts discharging the well and so
$E_r$ drops abruptly as $V_{dc}$ increases. This effect only
occurs in the presence of light and explains the first peak in the
current and the first bistability region.
Further application of (forward) voltage in the region which
corresponds to the resonant state above the emitter conduction band
and below one photon energy gives a contribution to the current smaller than
the values obtained without light. The reason
being again the decrease of the charge density due to
the emission of one photon for $ 0 < E_{r} < \hbar \omega$.
As a consequence, the strong non linear effect of the
charge on the electrostatic potentials moves $E_r$ below the
bottom of the emitter conduction band and the current decreases abruptly
at a voltage which is smaller than the one corresponding to the case without EMF.
Then, in the presence of light, $E_r$ drops abruptly to a value in the range
$- \hbar \omega  < E_{r} < 0 $.
Without EMF, the current
should drop to zero by energy and momentum conservation if $E_{r} < 0$.
Increasing $V_{dc}$ further,
$E_r$ becomes smaller than $-\hbar\omega$ and, thus, the current drops to zero.
In the backward direction, the current begins to flow at a
different voltage than the one corresponding to the current cut off in
forward bias voltage which results in a new bistability region. The other bistable regions
can be explained similarly.

At lower frequencies, the main bistable region can be also modified as shown in
Fig.~\ref{InarreaEPL_96_fig3}.
\subsection{Dynamical selfconsistency for ac-driven
resonant tunneling diodes \label{Dynamical-self}}
Noninteracting models
assume that the driving field is known and equals the external
field.
\begin{figure}
\begin{center}
\includegraphics[width=0.75\columnwidth,angle=0]
{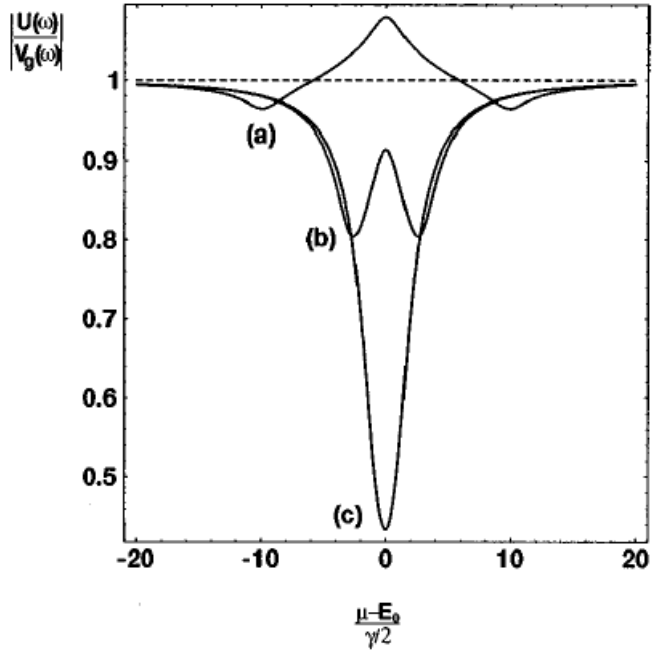}
\end{center}
\caption{Ratio of the internal potential to the gate voltage as function of the
Fermi energy, for $C=\frac{e^2}{\pi\gamma}$, $V_{dc}=0$ and for the frequencies
a) $\hbar\omega/(\gamma/2)=10$,
b) $\hbar\omega/(\gamma/2)=3$, and
c) $\hbar\omega/(\gamma/2)=1$. $\mu$, $E_0$ and $\gamma$, correspond, respectively,
to $E_F$, $E_r$ and $\Gamma$ in the text.
Reprinted with permission from \cite{PedPRB(98)}.
\copyright 1998 American Physical Society.}
\label{Pedersen}
\end{figure}
However, the long-range Coulomb interaction will screen the
external field and generates an internal potential that
deviates considerably from the applied one. We have discussed in
Section \ref{Selfconsmodel} how to calculate the internal field selfconsistently
by solving rate equations together with Poisson equations.
Importantly, the selfconsistent model of Section \ref{Selfconsmodel} assumes that only
{\em time-averaged} quantities do play a role in order to study stationary transport.
This is a simplification: the selfconsistent effect of displacement currents on the internal
field can be relevant for some ranges of frequencies. This is of most relevance when treating
time-dependent problems.

The theory of frequency-dependent transport in which all frequency components of the current are treated
selfconsistently has been put forward by B\"uttiker and coworkers in a series of papers
\cite{ButtPRL(93),PedPRB(98),PretPRB96,BuJPCM93,BlaEPL(98),BuNato97}.
Based on
the scattering formalism, the
analysis of linear ac-conduction in response to oscillating
potentials and considering the long-range Coulomb interaction
has been discussed in Refs.~\cite{ButtPRL(93),PretPRB96} for zero-dimensional systems and in
Refs.~\cite{BuJPCM93,BlaEPL(98)} for extended systems.
The generalization of the scattering formalism to include the nonlinear dependence on oscillating
potentials was put forward by Pedersen and B\"uttiker in Ref.~\cite{PedPRB(98)}.

Here, we briefly describe their selfconsistent calculation for a resonant tunneling diode
capacitively coupled to a gate with capacitance $C$ \cite{PedPRB(98)}. The effect of screening
is taken into account to second order in the oscillating potentials by using
a random phase approximation (RPA) treatment.
Assuming a sample subject to a dc bias $V_{dc}$
and an oscillating voltage $V_{g}(\omega)$ applied only to the
gate,
Pedersen and B\"uttiker find that the ratio of the applied to the
external potential is determined by the ac-conductances
$g_{\alpha\beta}^{(0)}(\omega;V_{dc})$:
\begin{equation}
    \frac{U(\omega)}{V_{g}(\omega)} = \left[ 1+ \frac{i}{\omega C} \sum_{\alpha\beta}
        g_{\alpha\beta}^{(0)}(\omega;V_{dc}) \right]^{-1}.
\label{uratio}
\end{equation}
The ac-conductances
read at zero temperature \cite{BuNato97,FuPRL(93)}:
\begin{eqnarray}
    g_{11}^{(0)}(\omega) &=& g_{21}^{(0)}(\omega) \left[ \frac{\Gamma_L}{\Gamma_R}-\frac{\Gamma}
        {\Gamma_R} \left( 1-i\frac{\hbar\omega}{\Gamma} \right)\right],
        \nonumber\\
    g_{22}^{(0)}(\omega) &=& g_{12}^{(0)}(\omega) \left[ \frac{\Gamma_R}{\Gamma_L}-\frac{\Gamma}
        {\Gamma_L} \left( 1-i\frac{\hbar\omega}{\Gamma} \right)\right],\\
    g_{12}^{(0)}(\omega) &=& \frac{e^2}{h} \frac{\Gamma_L\Gamma_R}{\Gamma\hbar\omega}
        \frac{1}{1-i\frac{\hbar\omega}{\Gamma}}
         \left[
        \frac{i}{2} \ln \frac{[(E_F-\hbar\omega-E_r-W-eV_{dc}/2)^2+(\Gamma/2)^2]}
        {[(E_F-E_r-W-eV_{dc}/2)^2+(\Gamma/2)^2]}\right. \nonumber \\
        && \hspace*{-0.8cm}
        + \frac{i}{2} \ln \frac{[(E_F+\hbar\omega-E_r-W-eV_{dc}/2)^2+(\Gamma/2)^2]}
                {[(E_F-E_r-W-eV_{dc}/2)^2+(\Gamma/2)^2]} \nonumber \\
        && \hspace*{-0.8cm}
        + \arctan\left(\frac{E_F+\hbar\omega-E_r-W-eV_{dc}/2}{\Gamma/2}\right) \nonumber \\
        && \hspace*{-0.8cm}
        - \left.
        \arctan\left(\frac{E_F-\hbar\omega-E_r-W-eV_{dc}/2}{\Gamma/2}\right) \right] \nonumber\\
        g_{21}^{(0)}(\omega; V_{dc}) &=& g_{12}^{(0)}(\omega; -V_{dc}).
\end{eqnarray}
The ratio in Eq.~(\ref{uratio}) has two simple limits.
In the non-interacting limit $C\rightarrow\infty$,
the internal potential directly
follows the applied potential.
In the limit $C\rightarrow0$,
the sample is charge neutral and $U(\omega)=0$.

An example of the behavior of
$\frac{U(\omega)}{V_{g}(\omega)}$ as a function of the Fermi energy is shown in Fig.~\ref{Pedersen}
for different frequencies. When the Fermi energy is close to resonance, screening induces a
large renormalization of the internal potential. This is expected because at resonance there are
more screening electrons. For a given Fermi energy, the ratio changes considerably as a function of frequency.
Away from resonance, the ratio converges to the nonscreened case.

Since screening depends on the position of the resonant level compared to the Fermi level, the central peak
and the sidebands will experience a different degree of screening and, thus, their
intensity will no longer be given by a Bessel function behavior (Tien-Gordon model).
\begin{figure}
\begin{center}
\includegraphics[width=0.75\columnwidth,angle=0]
{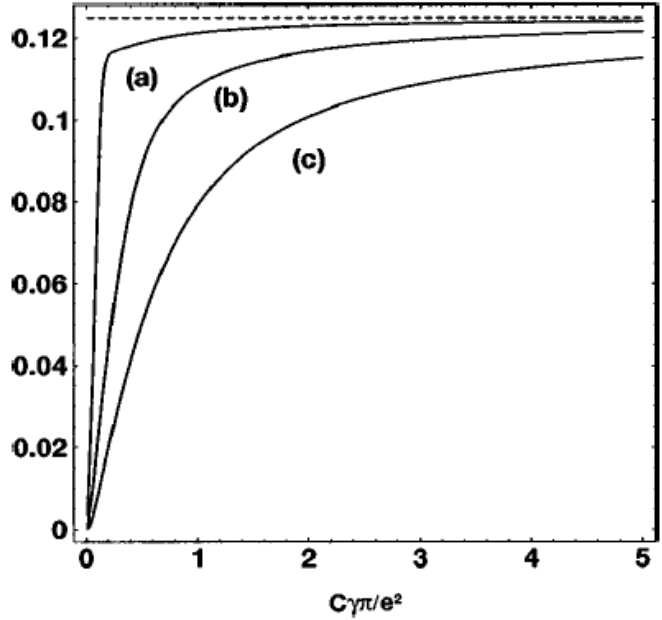}
\end{center}
\caption{Ratio of the sideband weight to central peak weight as function of capacitance
in the current versus gate voltage characteristic $I(V_g)$
for frequencies
$a)\hspace{0.15cm} \hbar\omega/(\gamma/2)=3$,
$b)\hspace{0.15cm} \hbar\omega/(\gamma/2)=5$, and
$c)\hspace{0.15cm} \hbar\omega/(\gamma/2)=10$.
The dashed line shows the result when no screening is present.
Reprinted with permission from \cite{PedPRB(98)}.
\copyright 1998 American Physical Society.}
\label{Pedersenb}
\end{figure}
This is similar to the effect described in Section \ref{Selfconsmodel} where the selfconsistent field
changes considerably as the quantum well becomes charged. Here, the frequency dependence is also
included so one can expect that the ratio of the sideband weight to the central resonance should
be frequency dependent.
This is demonstrated in Fig.~\ref{Pedersenb} \cite{PedPRB(98)}.
where the ratio of the sideband peak to the central peak is plotted.
The non-interacting
theory predicts a ratio of 0.125 for the parameters chosen (dashed line).
Depending on capacitance and frequency, the ratio changes completely. Again, as $C\rightarrow\infty$
the noninteracting limit is reached and the Tien and Gordon answer is recovered.

The asymmetry between the $\pm$ sidebands is another manifestation of the effects of screening on photon-assisted tunneling.
\begin{figure}
\begin{center}
\includegraphics[width=0.75\columnwidth,angle=0]
{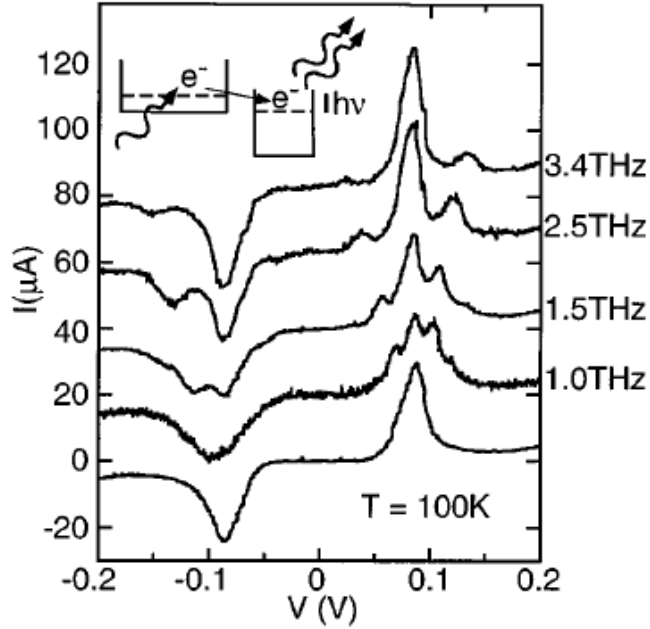}
\end{center}
\caption{Current-voltage characteristics of a triple barrier resonant tunnelin diode for different frequencies
of the ac field. Inset: schematics of the one-photon stimulated processes.
Reprinted with permission from \cite{DrexAPL(95)}.
\copyright 1995 American Institute of Physics}
\label{DrexlerAPL_95_fig1}
\end{figure}
When the Fermi energy if off resonance, emission and absorption of photons occur at different
potentials and, therefore, screening will occur asymmetrically for the two peaks. This effect has been observed
in photon-assisted tunneling experiments by Drexler et al \cite{DrexAPL(95)} performed on triple barrier resonant tunneling diodes.
An example is shown in Fig~\ref{DrexlerAPL_95_fig1}
where the asymmetry between emission and absorption sidebands is clearly
seen. A detailed selfconsistent analysis of these experiments can be found in Ref.~\cite{AguPRBbr(97)}.
\section{ac-driven superlattices \label{Superlattices}}
The first unambiguous evidence of discrete photon exchange coming
from photon assisted tunneling in a semiconductor structure was
obtained by Guimar\~aes et al \cite{GuiPRL(93)} who studied the
current-voltage characteristics of a GaAs/AlGaAs superlattice
under intense THz irradiation from a free-electron laser. These
pioneer experiments performed at the University of Santa Barbara
were followed up by a series of studies
\cite{KeayPRL(95)a,KeayPRL(95)b,ZeuPRB(96)} were different
interesting phenomena like absolute negative conductance or photon
assisted electric field domains were observed in the transport
through THz irradiated superlattices.
In this section we describe some of these phenomena and explain
the basic physics behind them.
\subsection{THz irradiated superlattices in the linear regime \label{SLlinear}}
\begin{figure}
\begin{center}
\includegraphics[width=0.75\columnwidth,angle=0]
{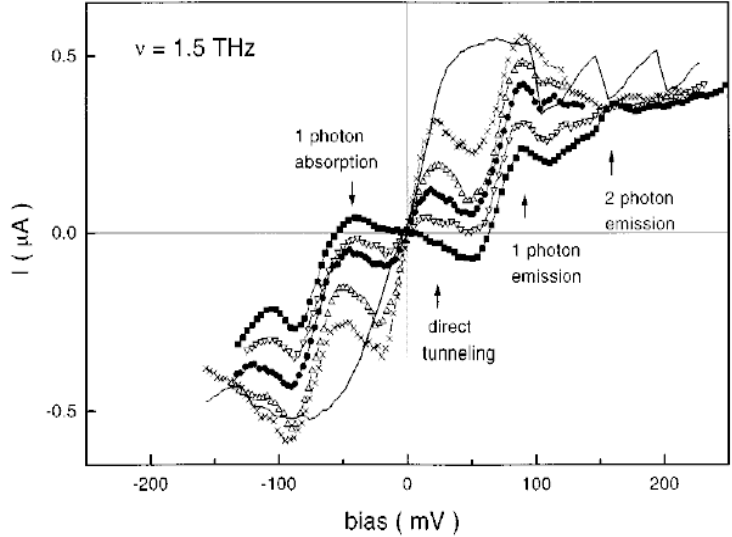}
\end{center}
\caption{Current-voltage characteristics measured in intense 1.5
THz fields. The laser power increases from the top to the bottom
trace by a factor of four. Reprinted with permission from
\cite{ZeuPRB(96)}. \copyright 1996 American Physical Society.}
\label{ZeunerPRB_96__fig2}
\end{figure}

\begin{figure}
\begin{center}
\includegraphics[width=0.75\columnwidth,angle=0]
{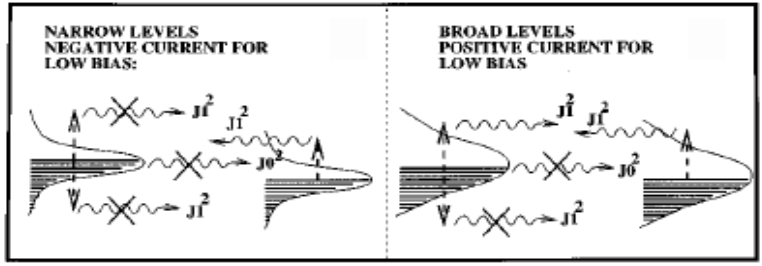}
\end{center}
\caption{Left: Resonant levels in neighbor wells at low dc bias
voltage. For $ed E_{dc}<\hbar\omega$ the emission channels are
inhibited while the absorption ones are enhanced; Right: At higher
dc bias voltages, $ed E_{dc}>\hbar\omega$, the absorption channels
are inhibited while the emission channels are not. The vertical
arrows represent fixed photon energy and the horizontal arrows
indicate the resonantly enhanced process (thick arrow) and the
impeded process (thin arrow), respectively.}
\label{PlateroAPL_97__fig2}
\end{figure}

\begin{figure}
\begin{center}
\includegraphics[width=0.75\columnwidth,angle=0]{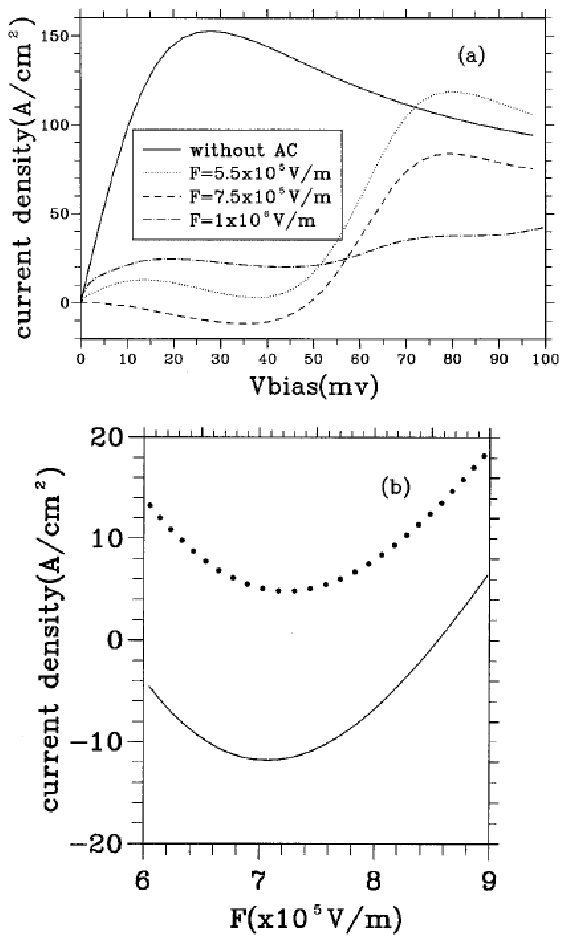}
\end{center}
\caption{a) I-V characteristics (low voltages region) of an
irradiated superlattice (fixed frequency $f_{ac}=1.5 THz$ and
different intensities) consisting of 10 undoped wells of GaAs $150
\AA$ wide and 11 barriers of AlGaAs of $50 \AA$ thickness. The
emitter and collector are n doped, n=$2 \times 10^{18}cm^{-3}$,
the temperature is $T=100 K$ and $\gamma=1meV$. b)Current versus
intensity of the ac field for fixed $V_{dc}=30mV$, $\gamma=1meV$
(continuos line) and $\gamma=2meV$ (dotted line).}
\label{PlateroAPL_97__fig1}
\end{figure}

One of the most spectacular manifestations of photon assisted
tunneling in a semiconductor superlattice is the possibility of
obtaining absolute negative conductance (ANC), namely a net {\it
negative} dc current in a sample biased by a {\it positive} dc
voltage. An example is shown in Fig.~\ref{ZeunerPRB_96__fig2}
\cite{ZeuPRB(96)} where the current at low dc bias voltages is
plotted for different intensities of the external THz source
($f_{ac}=1.5THz$). As the intensity increases, the conductance
near zero bias voltage is progressively reduced to zero and, at
the highest intensity of the external THz radiation, to absolute
negative values. At the same time, several features, not present
in the unirradiated curves, start to develop at large voltages.
The position of these new peaks are intensity independent and
shift linearly with the applied frequency (not shown),
unambiguously demonstrating photon-assisted tunneling.
The explanation for this effect is schematically described in
Fig.~\ref{PlateroAPL_97__fig2}: Absolute negative conductance
occurs near the condition for dynamical localization. As we
described in Section \ref{Floquet}, this phenomenon consists in
the complete quenching of the direct tunneling channel when the
frequency and intensity of the ac field are such that the first
zero of $J_0$ is reached. In this case, transport can only occur
via absorption or emission of photons. If there is a mechanism
that breaks the symmetry between absorption and emission, like
broadening of the resonant levels due to, e.g., disorder, it is
possible for an electron to absorb a photon an flow opposite to
the applied dc bias voltage. This mechanism for ANC is confirmed
in more elaborated calculations like the ones presented in
Refs.~\cite{Aguadotesis,FriPRB93,DakPRB95,PlatAPL(97),WackPRB(97)a}. An
example from Ref.~\cite{PlatAPL(97)} is shown in Fig.~\ref{PlateroAPL_97__fig1} where a
calculation of the current through a superlattice consisting of 10
undoped wells of GaAs $150 \AA$ wide and 11 barriers of AlGaAs of
$50 \AA$ thickness is presented. Scattering effects are included
phenomenologically by asuming that the spectral functions in the
wells are Lorentzians whose half width, $\gamma$, is a parameter
of the model. It is assumed that the electrons in each well are in
local equilibrium with Fermi energies $\epsilon_{\omega_{i}}$
which define the electronic densities $n_i$. For a given set
$\{\epsilon_{\omega_{i}}\}$ the densities evolve according to rate
equations (for $N$ wells):
\begin{equation}
\frac{dn_{i}}{dt} = J_{i-1,i}-J_{i,i+1}, \hspace{2cm}
i=1,\ldots,N.\label{SLrate1}
\end{equation}
The interwell currents $J_{i,i+1}$ are calculated within the
Transfer Hamiltonian framework (in the sequential tunneling
regime). In the presence of the external field, the levels in each
well change as $E_{ki}(t)=E_{k_{i}}^{0}+eFz_{i}cos\omega t $
($z_{i}$ is approximated for the mean position in the $i$-th well,
$f_{ac}=\omega/2\pi$ is the field frequency and F its intensity)
such that the local spectral functions in each well are modified
by the ac field as in the Tien-Gordon model
\cite{PlatAPL(97),WackPRB(97)a}. The dc current and the final set
of densities and Fermi levels $\epsilon_{\omega_{i}}$ is obtained
by taking the stationary limit $\frac{dn_{i}}{dt}\rightarrow 0$ of
Eqs.~(\ref{SLrate1}). Fig.~\ref{PlateroAPL_97__fig1}(a) shows the
current-voltage characteristics in the region of low bias voltages
(without and with external irradiation) for fixed external
frequency $f_{ac}=1.5 THz$ and different field intensities $F$.
The low bias voltage peak in the curve
without irradiation (solid line) corresponds to the ground to
ground state tunnel between the wells. Once the ac field is
applied, the current at low bias voltages is strongly reduced and
becomes negative for $F= 7.5\times 10^5 V/m$ (dashed line). As the
voltage increases further it becomes positive again. For higher
intensities ($F=10^6 V/m$, dotted line), the current is always
positive. At a fixed voltage, the negative current occurs as a
result of the intriguing interplay between the final local Fermi
level in each well, the scattering induced broadening of the
density of states and dynamical localization.
For this structure the dynamical localization condition is reached
for $F\sim 7.5\times 10^{5} V/m$ such that the current is
inhibited through the central channel. This channel is open again
increasing F as one can see for F=$10^{6}V/m$.
This mechanism for ANC is further substantiated by studying the
current at fixed dc voltage for different scattering times. This is
shown in Fig.~\ref{PlateroAPL_97__fig1}(b) \cite{PlatAPL(97)} where the current at a
fixed voltage $V_{dc}=30 mV$ is plotted versus the intensity of
the external field $F$ for $f_{ac}=1.5THz$ and two different
$\gamma$. For $\gamma=1meV$, ANC can occur for some values of $F$,
the electrons are able to overcome the static voltage ($eV\leq
m_{max}\hbar\omega$) and electronic pumping in the opposite
direction occurs. The current presents a minimum exactly at the
first zero of $J_0$, i.e. $F\sim 7.5\times 10^{5}V/m$. As $\gamma$
increases ($\gamma=$2meV in Fig.~\ref{PlateroAPL_97__fig1}(b)) the
emission channel is opened i.e., there are empty available states
in the next well to tunnel and the flow of current occurs in the
direction of the applied dc voltage.

So far, we have restricted ourselves to describe the effect of an
external high frequency field in the linear transport properties
of semiconductor superlattices. By increasing the dc voltage,
charge accumulation in the wells typically occurs and, as a
consequence, new phenomena arising from the strong nonlinearity of
the problem do start to play a role. As an example, the
unirradiated current-voltage characteristics of the experiments in
Fig.~\ref{ZeunerPRB_96__fig2} develops a sawtooth structure
reflecting the formation of electric field domains (EFD). The
electric field domains develop a complicated structure in the
presence of radiation. A description of transport beyond the
single-particle picture is thus called for in order to explain
these, and others, nonlinear phenomena.
\subsection{Weakly-coupled Superlattices as a paradigm of a nonlinear dynamical system I: statics in the undriven case
\label{SLnonlinear-undriven1}} As we just mentioned, the large dc
voltage transport regime in semiconductor superlattices is
typically accompanied by strong nonlinear effects. This is
expected, because solid state electronic devices presenting
negative differential conductance (NDC), such as resonant
tunneling diodes, Gunn diodes or Josephson
junctions~\cite{Schollbook}, are nonlinear dynamical systems with
many degrees of freedom. Semiconductor superlattices display typical nonlinear phenomena
such as multistability, oscillations, pattern formation or
bifurcation to chaos, all these nonlinear phenomena have their origin in the interplay between
Coulomb interaction and NDC.
In this section we briefly describe the static and dynamical
transport properties of biased heterostructures whose main
mechanism is sequential tunneling. This is a topic which has
attracted a great deal of attention in recent times. In n-doped
weakly coupled superlattices, multistability due to domain
formation has been much studied both theoretically and
experimentally,
\cite{ChoPRB(87),GraPRL(91),BonPRB(94),KasAPL(94),PrePRB(94),WacPRB(97)}.
When the doping in the wells is reduced, self-sustained current
oscillations
\cite{MerProc(95),KasPRB(97),KanPSS(97),BonSIAM(97),BonJPC(02)}
and chaos
\cite{BulPRB(95),ZhaPRL(96),LuoPRL(98),LuoPRB(98),ZwoPRB(03)} due
to domain dynamics are possible.
\begin{figure}[!htp]
\begin{center}
\includegraphics[width=0.65\columnwidth,angle=0]
{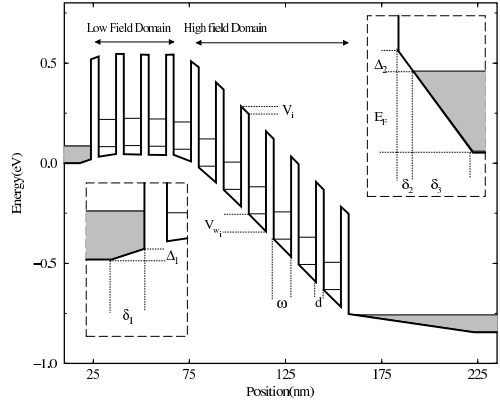}
\end{center}
\caption{Electrostatic potential profile of a superlattice in the
nonlinear regime: $\Delta_1$ and $\Delta_2$ represent the
potential drops at the contacts and $\delta_1, \delta_2$ and
$\delta_3$ are the accumulation and depletion lengths.}
\label{AguadoRC_97_fig1}
\end{figure}
As we have discussed previously, Coulomb interaction in
heterostructures with large area wells is a small effect compared
with the energy difference between non-interacting eigenstates of
the structure. Therefore a mean field description is, for many
purposes, a good approximation. Concerning transport, the most
successful modeling of these nonlinear phenomena use discrete rate
equations for the electron density and electric field in each
well, plus constitutive laws for the current, bias voltage, boundary and
initial conditions \cite{GraPRL(91),BonPRB(94),BonSIAM(97)}. The
laws may be phenomenological \cite{BonPRB(94)} or obtained from
microscopic considerations,
\cite{PrePRB(94),LaiPRB(93),WacPS(97)}. In all cases cited, the
boundary conditions for electrostatics were selected in a more or
less ad hoc manner by using the available information from
experiments. This is particularly annoying because the boundary
conditions select the relevant dynamics of EFD in the oscillatory
regime \cite{HigPD(92)}.

In order to include in a natural way boundary conditions due to
the emitter and collector regions, Aguado et al \cite{Aguadotesis,AguPRB(97)}
extended the model described in Section \ref{Selfconsmodel}
\cite{GolPRL(87),IñaEL(96)} 
to consider transport in multiwell
structures\footnote{It is possible to obtain a discrete drift difussion model 
from the microscopic model that we present in this subsection \cite{BonPRB(00)}.
In this discrete drift difussion model realistic transport coefficients and contact current-field
characteristic curves are calculated from microscopic expressions,
knowing the design parameters of the superlattice. 
The detailed boundary conditions obtained from the discrete model clarify the
analysis of the electron dynamics and when possible self-sustained
oscillations of the current are due to monopole or dipole
recycling (see section \ref{SLnonlinear-undriven2} below).}. 
The main ingredients of the sequential
tunneling model are as follows: it is assumed
that the characteristic time of intersubband relaxation due to
scattering is much smaller than the tunneling time, which is in
turn much smaller than the dielectric relaxation times responsible
for reaching a steady state. This separation of time scales, as
well as the configuration of a typical sample allows one to
consider that: {\it i}) only the ground state of each well is
populated, {\it ii}) the tunneling processes are stationary and
{\it iii}) the local density in each well can be calculated from
an equilibrium distribution function (Fermi-Dirac). These
assumptions justify the use of rate equations for the electron
densities at each well with relations for the currents calculated
by means of the Transfer Hamiltonian approach \cite{BardPRL(61)}
which gives the following expressions for
the interwell tunneling currents:
\begin{eqnarray}
J_{i,i+1}& = &\frac{2e\hbar k_{B}T}{\pi^{2}m^{*}}\sum_{j=1}^{n}
\int \frac{\gamma}{(\epsilon-\epsilon_{C1})^{2}+\gamma^{2}}
\frac{\gamma}{(\epsilon-\epsilon_{Cj})^{2}+\gamma^{2}}\,\nonumber\\
&\times&T_{i+1}(\epsilon)\, \ln
\left[\frac{1+e^{\frac{\epsilon_{\omega_{i}}-\epsilon}{k_{B}T}}}
{1+e^{\frac{\epsilon_{\omega_{i+1}}-\epsilon}{k_{B}T}}} \right]\,
d\epsilon.
\end{eqnarray}
$n$ is the number of subbands in the well with energies
$\epsilon_{Cj}$ (referred with respect to the origin of
potential drops), and $T_{i+1}(\epsilon)$ is the transmission coefficient
through the $i+1$-th barrier.
The spectral function of each well is a
Lorentzian, its half-width $\gamma$ is a phenomenological parameter and roughly
corresponds to the LO phonon lifetimes
($\simeq$ 1-10 meV) in a typical quantum well: $A_{Cj}(\epsilon) = \gamma/[(\epsilon -
\epsilon_{Cj})^2 +\gamma^2]$. Of course this model can be improved
by calculating microscopically the self-energies, which could
include other scattering mechanisms (e.g.\ interface roughness,
impurity effects) or even exchange-correlation effects (which
affect the electron lifetime in a self-consistent way
\cite{ZouPRB(94)}). 
The tunneling current from the emitter to the nearest neighbor
well and the current to the collector coming from its neighbor
well are
\begin{eqnarray}
J_{e,1} = \frac{2ek_{B}T}{\pi^{2}\hbar}\sum_{j=1}^{n} \int
A_{Cj}(\epsilon)\,
T_{1}(\epsilon)\, \ln
\left[\frac{1+e^{\frac{\epsilon_{F}-\epsilon}{k_{B}T}}}
{1+e^{\frac{\epsilon_{\omega_1}-\epsilon}{k_{B}T}}}\right]\,
d\epsilon\nonumber\\
J_{N,c} = \frac{2ek_{B}T}{\pi^{2}\hbar} \int A_{C1}(\epsilon)\
T_{N+1}(\epsilon)\,
\ln
\left[\frac{1+e^{\frac{\epsilon_{\omega_N}-\epsilon}{k_{B}T}}}
{1+e^{\frac{\epsilon_{F}-eV-\epsilon}{k_{B}T}}} \right]\,
d\epsilon. \label{THM}
\end{eqnarray}
Again, for a given set of Fermi energies
$\{\epsilon_{\omega_{i}}\}$ the densities evolve according to the
following rate equations:
\begin{equation}
\frac{dn_{i}}{dt} =
J_{i-1,i}(\epsilon_{\omega_{i-1}},\epsilon_{\omega_{i}},\Phi)
-J_{i,i+1}(\epsilon_{\omega_{i}},\epsilon_{\omega_{i+1}},\Phi)
\hspace{1cm} i=1,\ldots,N. \label{SLrate2}
\end{equation}
The rate equations for the electron densities imply that the
interwell currents and the currents from the emitter and to the
collector are all equal to the total current in the stationary
case.
In these equations $J_{e,1}\equiv
J_{e,1}(\epsilon_{\omega_{1}},\Phi)$ is the current from the
emitter to the first well and $J_{N,c}\equiv
J_{N,c}(\epsilon_{\omega_{N}},\Phi)$ the current from the $N$-th
well to the collector. $\Phi$ denotes the set of voltage drops
through the structure which are calculated from the electrostatics
of the problem. Eqs.~(\ref{field.inside1}-\ref{field.inside2}) are
now generalized to describe the 2N+1 potential drops corresponding
to N wells $V_{wi}$ and N+1 barriers, $V_{i}$:
\begin{eqnarray}
\frac{V_{w_{i}}}{w}&=&\frac{V_{i}}{d}+\frac{
n_{i}(\epsilon_{\omega_{i}})-eN_{D}^{w}}{2\varepsilon}
\label{field.inside1b}\\
\frac{V_{i+1}}{d}&=&\frac{V_{i}}{d}+\frac{n_{i}(
\epsilon_{\omega_{i}})-eN_{D}^{w}}{\varepsilon}\,,
\label{field.inside2b}
\end{eqnarray}
where $n_{i}(\epsilon_{\omega_{i}})$ is the 2D (areal) charge
density at the $i$th well (to be determined), $w$ and $d$ are the
well and barrier thickness respectively.
The emitter and collector layers are described by
Eqs.~(\ref{emitter}) and (\ref{collector}).

\begin{figure}
\begin{center}
\includegraphics[width=0.75\columnwidth,angle=0]
{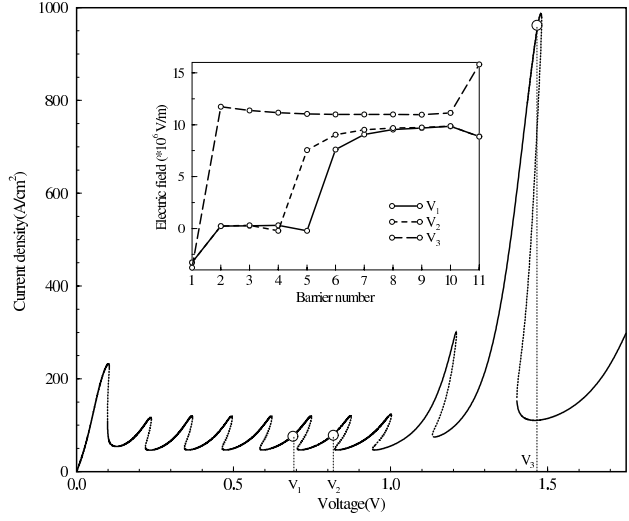}
\end{center}
\caption{ Current--voltage characteristic curve of a superlattice
($90\AA$GaAs/$40\AA$Ga$_{.5}$Al$_{.5}$As superlattice with 11
barriers and 10 wells, $N_{D} = 2\times 10^{18}$cm$^{-3}$ and
$N^{w}_{D}=1.5 \times 10^{11}$cm$^{-2}$ are the contact and well
dopings, respectively). The inset shows the electric field
distribution through the superlattice for three voltages: $V_1 =
0.69$ V; $V_2 = 0.81$V; $V_3 = 1.48$V.} \label{AguadoRC_97_fig5}
\end{figure}

In order to close the set of equations, global charge conservation
(see Eq.~(\ref{charge.conserv})) and applied voltage conservation
(see Eq.~(\ref{bias})) are imposed. Instead of the rate equations
(\ref{SLrate2}), we can derive a form of Amp\`ere's law which
explicitly contains the total current density $J(t)$.
Differentiating (\ref{field.inside2b}) with respect to time and
eliminating $n_i$ by using (\ref{SLrate2}) one gets
\begin{equation}
{\varepsilon\over d}\frac{dV_{i}}{dt} + J_{i-1,i} = J(t),
\quad\quad\quad i=1,\ldots,N+1 ,\label{ampereb}
\end{equation}
where $J(t)$ is the sum of displacement and tunneling currents.

The time-dependent model consists now of the $3N+8$ equations
which contain the $3N+8$ unknowns $\epsilon_{\omega i}$,
$V_{w_{i}}$, ($i=1,\ldots, N$), $V_j$ ($j=1,\ldots,N+1$),
$\Delta_1$, $\Delta_2$, $\delta_k$ ($k=1,2,3$), $\sigma$, and $J$.
This system of equations , together with appropriate initial
conditions, determine
completely and self-consistently the problem.\\
One way to analyse the statics of the model and the stability of
the stationary solutions is to numerically solve the
algebraic-differential system (plus appropriate initial
conditions) for each voltage until a stationary profile is reached.
This is rather costly, so a good strategy is to follow this
procedure for a given value of the bias voltage and then use a
numerical continuation method to obtain all stationary solution
branches in the current--voltage characteristic diagram. This
yields both unstable and stable solution branches. Direct
integration of the stationary equations [dropping the displacement
current in (\ref{ampereb})] usually presents important problems of
numerical convergence to the appropriate solutions in regions of
multistability.
\begin{figure}[!htp]
\begin{center}
\includegraphics[width=0.75\columnwidth,angle=0]{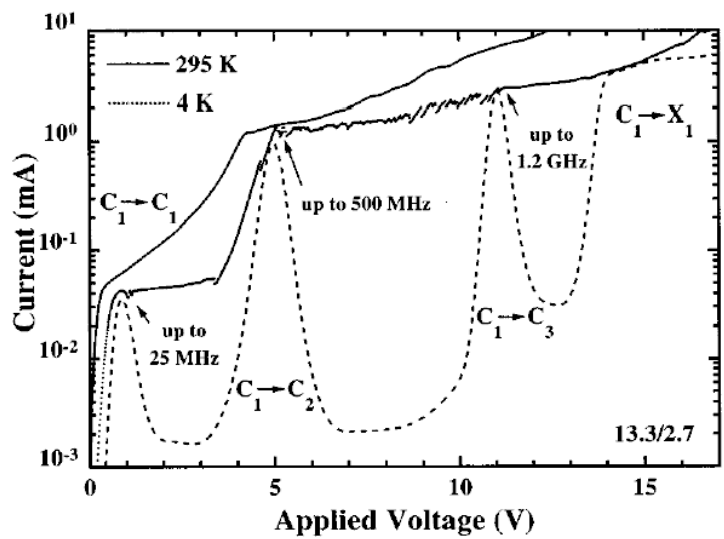}
\end{center}
\caption{Current--voltage characteristics for a 50 period
superlattice (13.3 nm GaAs/2.7nm AlAs). The dashed line shows
schematically the expected drift velocity versus field
characteristic of this sample at homogeneous field. C1, C2 and C3
are the ground, first excited and second excited quantum well
states. Reprinted with permission from \cite{KasPRB(97)}.
\copyright 1997 American Physical Society.}
\label{KastrupPRB_97__fig3}
\end{figure}

The formation of sharp discontinuities in the current--voltage
characteristics (the stable branches are connected by unstable
ones) can be explained by the formation of a charge accumulation
layer in one of the wells (domain wall) that splits the
superlattice in two regions with low and high electric field
respectively. Once the ground state of the quantum well closer to
the collector becomes disaligned with that of the neighbor quantum
well the charge is accumulated there, producing a high electric
field towards the collector and the current drops abruptly
(negative differencial conductance) Increasing the voltage, this
charge cannot move continuously through the superlattice. This
motion can only occur for voltages allowing resonant interwell
tunneling, it happens as the first excited state of the right most
quantum well (the one closest to the collector) is aligned with
the ground state of its neighbor well. Increasing further the voltage a
new region of negative differential conductance appears and the
domain wall moves from the $i$-th well to the $i-1$-th well. It
produces a sawtooh like profile for the current. At large bias
voltage, the high electric field region contains all the quantum
wells, the field is homogeneous and the tunneling current take
place from the ground state of a quantum well to the first excited
level of the well located in the current direction.

An example of this system configuration is shown in
Fig.~\ref{AguadoRC_97_fig5} \cite{AguPRB(97)} where the
current--voltage characteristics of a superlattice presenting EFD
formation is plotted. The stable (unstable) branches 
are shown as continuous (dotted)
lines. The inset shows three electric field profiles corresponding
to three different voltages. They show the presence of domains in
the superlattice with a domain wall which moves one well as the
bias voltages changes from one branch to the next one. Domain
formation is also shown in the superlattice electrostatic
potential profile; see Fig.~\ref{AguadoRC_97_fig1} for a fixed
voltage $V_{2}=0.81$V.

The first branch in the characteristics I/V
Fig.~\ref{AguadoRC_97_fig5} corresponds to $C1\rightarrow C1$
tunneling ($Ci$ are the conduction subbands ordered starting from
that with lowest energy). As $V$ increases, $C1\rightarrow C2$
tunneling becomes possible in part of the structure and domain
formation does take place. The last branch (large current) appears
when transport is regulated by $C1\rightarrow C2$ tunneling across the whole
structure.
The voltage region between the $C1\rightarrow C1$ peak to
the $C1\rightarrow C2$ peak is dubbed first plateau. At higher voltages, 
more plateaus can appear due to tunneling to other excited states.
An interesting feature in
Fig.~\ref{AguadoRC_97_fig5} is that neighbor peaks have a smaller
current than the $C1\rightarrow C1$ peak. 
Another interesting feature due
to the voltage drop at the contacts is that the number of branches
in the current--voltage characteristics is less than the number of
wells, in agreement with the experiments. 
This behavior can be understood by looking at the branch at
1.21 V where the low field domain occupies the two wells closer to
the emitter. $C1\rightarrow C2$ tunneling occurs between all the
wells in the branch with $V_{3} = 1.48$V corresponding to an
intense peak of the current, at this voltage all the quantum wells
have dropped in the high field domain.

We finish this part by mentioning in passing the non-linear
transport properties of weakly coupled diluted magnetic
semiconductor superlattices which have been recently studied in
Refs.~\cite{SanPRB(02),BejNATO(03),BejPRB(03)}. The main interest
of these systems is that the exchange interaction of the local
moments of the magnetic impurities with the spin of the carriers,
electrons or holes, produce very interesting spin-dependent
properties as for instance, a very large Zeeman splitting in the
presence of a small external magnetic field. The transport
properties were studied for II-VI n-doped semiconductor
superlattices doped with Mn. There, the interplay of the negative
differential resistance regions associated with resonant
tunneling, the Coulomb and the exchange interaction add a new
dimension to the problem which depends crucially on the spin.
Multistability of the spin polarized current and of the spin
polarization of the magnetically doped quantum wells as a function
of the dc voltage (i.e., multistability driven by electric fields)
are an example of the interesting non-linear transport properties
of these devices with potential applications in Spintronics.

\begin{figure}[!htp]
\begin{center}
\includegraphics[width=0.75\columnwidth,angle=0]{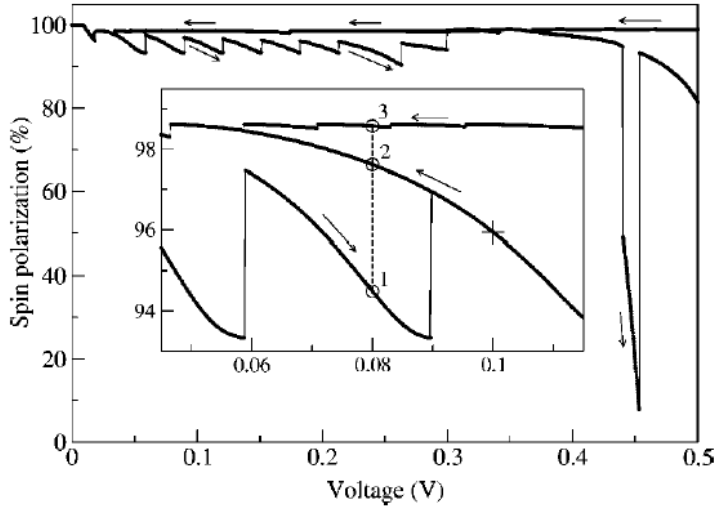}
\end{center}
\caption{(a) Multistability of distinct spin polarization steady
states within the magnetic well (doped with Mn) located in the
center of an n-doped superlattice with 9 wells of ZnSe/(Zn,Cd)Se
(electron doping at the contacts is: $N_c=3 \times 10^{11} cm^{-2}$). The
inset shows a blow up of three different steady states reached at
V=0.08 mV. The state labeled 1 (3) is achieved by sweeping voltage
up (down) from a high (low) initial bias voltage. The state
labeled 2 is obtained by sweeping voltage up to V=0.1 V (marked
with a cross) and then reversing the sweep direction.}
\label{SanPRB01_fig10}
\end{figure}
\subsection{Weakly Coupled Superlattices as a paradigm of a nonlinear dynamical system II:
 dynamics in the undriven case
\label{SLnonlinear-undriven2}} Stationary electric  field domains
appear in voltage biased superlattices if the doping is large
enough. When the carrier density is below a critical value,
self-sustained oscillations of the current may appear. They are
due to the dynamics of the domain wall (which is a charge monopole
accumulation layer or, briefly, a {\em monopole}) separating the
electric field domains. This domain wall moves through the
structure and is periodically recycled. The frequencies of the
corresponding oscillation depend on the applied bias voltage and
range from the  kHz to the GHz regime. Self-oscillations persist
even at room temperature, which makes these devices promising
candidates for microwave generation~\cite{KasPRB(97),KanPSS(97)}.
An experimental example from Kastrup et al in
Ref.~\cite{KasPRB(97)} is shown in Fig.~\ref{KastrupPRB_97__fig3}.
\begin{figure}[!htp]
\begin{center}
\includegraphics[width=0.75\columnwidth,angle=0]{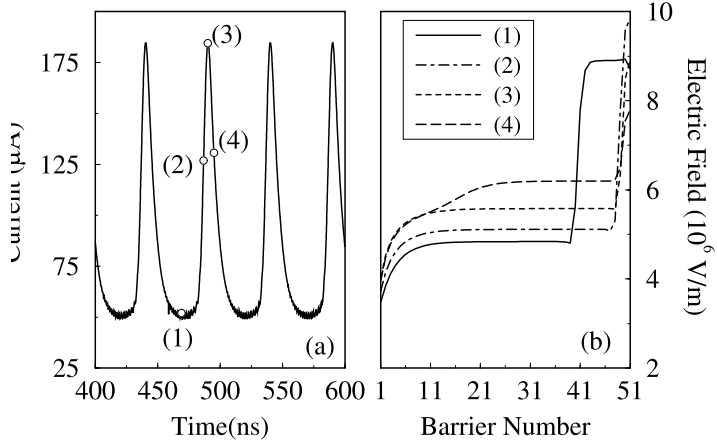}
\end{center}
\caption{(a) Self-sustained oscillations of the total current
through the superlattice due to monopole recycling and motion.
$V_{dc}$=$5.5 V$ and emitter doping, $N_{c}=2\times
10^{16}$~cm$^{-3}$. (b) Electric field profiles at the times
marked in (a) during one period of the current oscillation.}
\label{monosc}
\end{figure}
Theoretical and experimental work on these systems have gone hand
in hand. Thus the paramount role of monopole dynamics has been
demonstrated by theory and experiments. Monopole motion and
recycling can be experimentally shown by counting the spikes
--high frequency modulation--  superimposed on one period of the
current self-oscillations: current spikes correspond to
well-to-well hopping of a domain wall through the superlattice. In
typical experiments the number of spikes per oscillation period is
clearly less than the number of superlattice
wells~\cite{KasPRB(97),KanPSS(97)}. It is known that monopoles are
nucleated well inside the
superlattice~\cite{KasPRB(97),BonSIAM(97)} so that the number of
spikes tells over which part of the superlattice they move.\\
Self-sustained oscillations are not the only type of superlattice
oscillations, there are two other type that appear in the miniband
regime (in the miniband regime the quantum wells are strongly
coupled and the electronic spectrum becomes coherent minibands
extended all through the superlattice): first, oscillations occur
when the carriers within the miniband are accelerated beyond the
Brillouin zone boundary, where the drift velocity becomes
negative. In absence of scattering the electron wave packet
performs Bloch oscillations with frequency: $\omega_{B}=eFd/h$
where $d$ and $h$ are the superlattice period and the Planck
constant respectively. These Bloch oscillations were predicted by
Esaki and Tsu\cite{TsuAPL(73)} and many papers have followed,
inspired in the perspective of producing a superlattice Bloch
oscillator. Secondly, a different type of oscillations occur in
the miniband regime, when scattering times are shorter than the
tunneling time. In this case a transient charge accumulation
traveling through the superlattice may lead to current
oscillations.

S\'anchez et al \cite{SanPRB(99),Santesis} used the model proposed above to
investigate electron dynamics in superlattices, i.e., situations
where the displacement currents are non-zero and the electronic
current is time-dependent. An example is shown in
Fig.~\ref{monosc}(a) \cite{SanPRB(99)} which depicts the current
as a function of time for a dc bias voltage of 5.5~V on the second
plateau of the $I-V$ characteristic curve of a 13.3~nm GaAs/2.7~nm
AlAs superlattice consisting of 50 wells and 51 barriers, as
described in \cite{KasPRB(97)}. Doping in the wells and in the
contacts are $N_{w} = 2\times 10^{10}$~cm$^{-2}$ and $N_{c}= 2
\times 10^{16}$~cm$^{-3}$ respectively.
\begin{figure}[!htp]
\begin{center}
\includegraphics[width=0.75\columnwidth,angle=0]{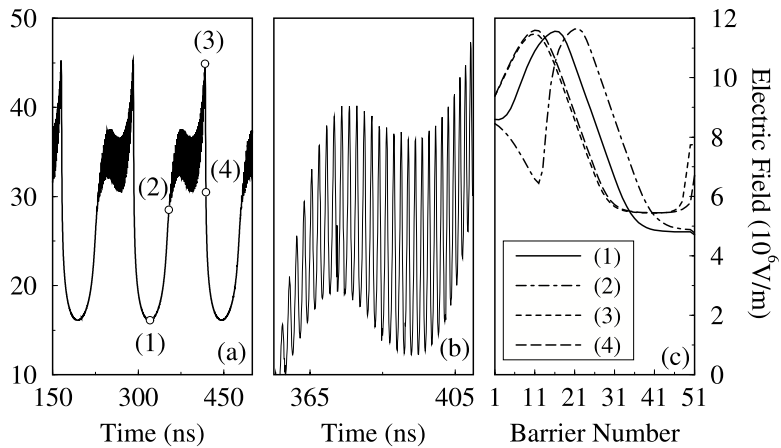}
\end{center}
\caption{(a) Dipole-mediated self-oscillations of the current at
$5.5 V$ for $N_{c}=2\times 10^{16}$cm$^{-3}$. (b) Detail of the
current spikes. (c) Electric field profiles at the times marked in
(a).} \label{diposc}
\end{figure}
$J(t)$ oscillates periodically at 20~MHz. Between each two peaks
of $J(t)$, 18 additional spikes can be observed. The electric
field profile is plotted in Fig.~\ref{monosc}(b) at the four
different times of one oscillation period marked in
Fig.~\ref{monosc}(a). What is remarkable in Fig.~\ref{monosc}(a)
are the spikes superimposed near the minima of the current
oscillations. Such spikes have been observed experimentally and
attributed to well-to-well hopping of the domain
wall~\cite{KanPSS(97),KasPRB(96)}. They are a cornerstone to
interpret the experimental results and in fact support the
theoretical picture of monopole recycling in part (about $40\%$)
of the superlattice during self-oscillations. The identification
between number of spikes and of wells traversed by the monopole
rests on voltage turn-on measurements supported by numerical
simulations of simple models during early stages of stationary
domain formation~\cite{KasPRB(96)}. These simplified models do not
predict spikes superimposed on current self-oscillations due to
monopole motion~\cite{BonPRB(94),KasPRB(97),WacPRB(97)}. To
predict large spikes, an artificial time delay in the tunneling
current~\cite{KanPSS(97)} or random doping in the
wells~\cite{PreProc(96)} have to be added.

When contact doping is reduced below a certain value, there appear
dipole-mediated self-oscillations, where the domain wall consists
on a charge accumulation and a charge depletion layers. There is a
range of voltages for which dipole and monopole oscillations
coexist as stable solutions. This range changes for different
plateaus. When the emitter doping is further lowered, only the
dipole self-oscillations remain. Fig.~\ref{diposc}
\cite{SanPRB(99)} presents data in the crossover range (below $N_c
= 4.1\times 10^{16}$cm$^{-3}$ and above $N_c = 1.7\times
10^{16}$cm$^{-3}$), for the same electron doping in the quantum wells
and bias voltage as in Fig.\ \ref{monosc}. Except for the presence
of spikes of the current, dipole recycling and motion in
superlattices are similar to those observed in models of the Gunn
effect in bulk GaAs~\cite{HigPD(92)}. These self-oscillations have
not been observed so far in experiments due to the high values of
the contact doping adopted in all the present experimental
settings. However, the contact doping is not the only parameter
which can be modified in order to have dipole oscillations in the
current. It also can be reached by modifying the sample
configuration, as for instance the quantum well widths
\cite{BonPRB(00)}.
\\
Bejar et al. \cite{BejPRB(03)} have explored interesting features
that occur in weakly doped superlattices that support
self-sustained oscillations when they are doped with magnetic
impurities. In this case, the interplay between strongly nonlinear
interwell charge transport and the large tunable spin splitting
induced by exchange interactions with spin-polarized Mn ions
produces interesting spin features. Time dependent periodic
oscillations of the spin polarized current and of the spin
polarization in both magnetic and nonmagnetic quantum wells were
predicted \cite{BejPRB(03)}. These spin-dependent features can
potentially be exploited for device applications, as spin
polarized current injection oscillators.

\begin{figure}[!htp]
\begin{center}
\includegraphics[width=0.75\columnwidth,angle=0]{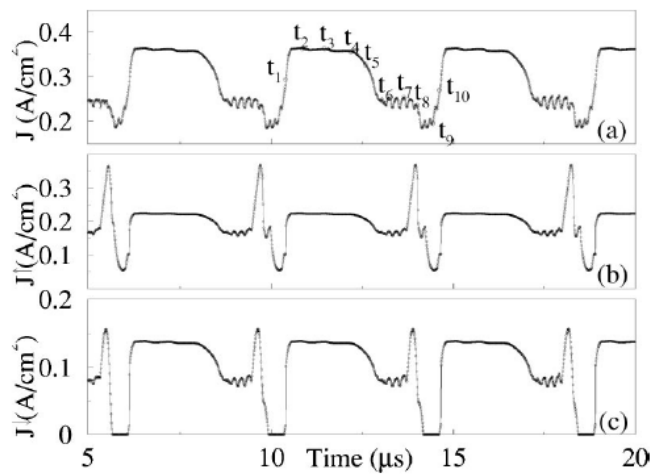}
\end{center}
\caption{(a)  Total time-dependent current (tunneling plus
displacement), b) spin-up, c) spin-down time dependent current at
fixed dc voltage $V_{dc}$ = 0.5 V for a 50-well n-doped
ZnSe/ZnCdSe system doped with fractional MnSe monolayers at the
1st and 50th quantum wells. Contact doping $N_c = 9.9 \times
10^{10} cm^{-2}$ (intermediate n-doped sample). The current
oscillations present a flat region and overimposed spikes.
Comparision of (b) and (c) indicates that the current towards the
collector is partially spin-up polarized.} \label{BejarPRB03_fig2}
\end{figure}

\begin{figure}[!htp]
\begin{center}
\includegraphics[width=0.75\columnwidth,angle=0]{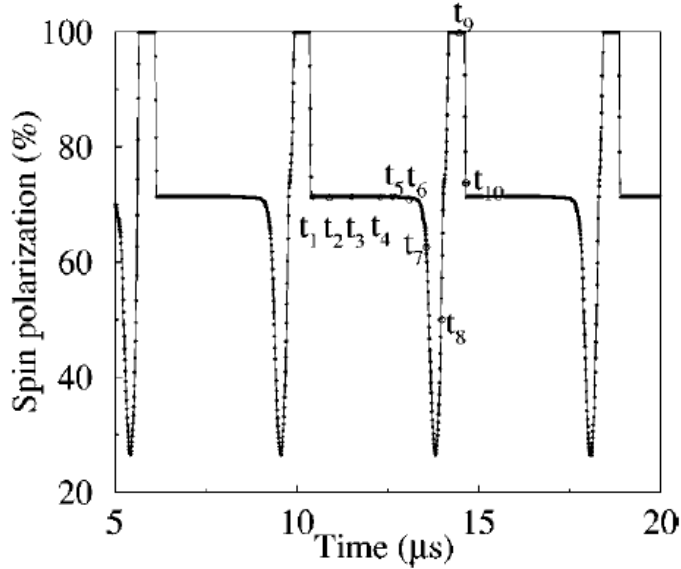}
\end{center}
\caption{Spin polarization P in the quantum well closest to the
collector as a function of t for $V_{dc}$=0.5 V for the sample of
fig. \ref{BejarPRB03_fig2}. The fractional polarization of the
isolated quantum well is $0.75$. Within the superlattice, in the
strong nonlinear regime, the polarization oscillates and reaches,
for a small time window of the period, full spin-up polarization.}
\label{BejarPRB03_fig4}
\end{figure}
\subsection{Weakly-coupled Superlattices as a paradigm of a nonlinear dynamical system III:
stationary transport in the ac-driven case
\label{SLnonlinear-driven1}} The aplication of an external ac
signal, superimposed to the applied dc bias voltage, drastically
changes some of the nonlinear phenomena discussed above and brings
about new physics not present in undriven samples \footnote{The
calculations presented in this Section, and in Section
\ref{SLnonlinear-driven2}, were performed using the selfconsistent
model of Section \ref{SLnonlinear-undriven1} (only {\em
time-averaged} quantities are included in the selfconsistent
equations). It is thus assumed that the separation of time scales
is such that it is a good approximation to neglect the
selfconsistent effect of displacement currents. The modelling of
nonlinear transport through superlattices in a fully dynamical and
selfconsistent way constitutes an extremely dificult problem
which, to our knowledge, remains open.}. As an example, the
current--voltage characteristics of an ac-driven superlattice
displaying electric-field domain formation develop new
multistability regions: Depending on the parameters of the
external high-frequency field, many stable operating points,
giving different dc currents, do appear at a fixed dc voltage as a
result of the interplay between the strong nonlinearity and the
ac-induced photoassisted tunneling channels.
\begin{figure}
\begin{center}
\includegraphics[width=0.75\columnwidth,angle=0]{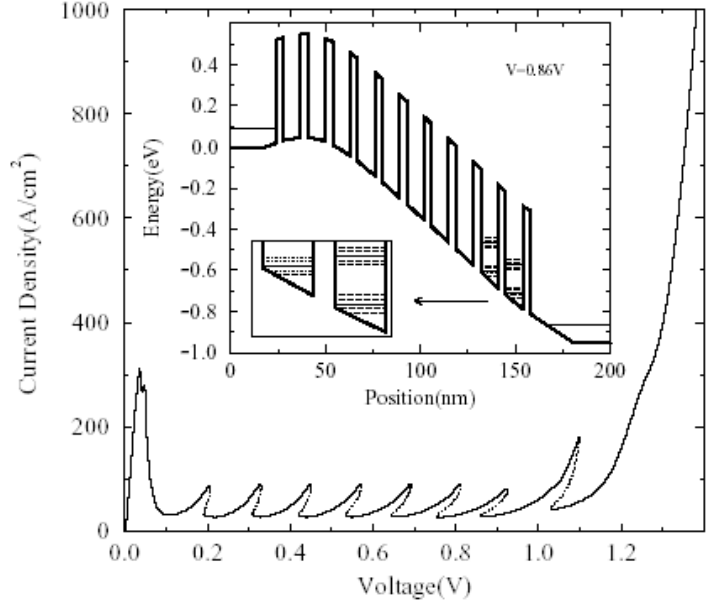}
\end{center}
\caption{I--V characteristic for a superlattice (consisting of 10
wells with $90 \AA  GaAs$ wells and $40 \AA  Ga_{0.5}Al_{0.5}As$
barriers. $N_{D}=2\times 10^{18} cm^{-3}$ and $N_{D}^{w}=1.5\times
10^{11}cm^{-2}$ are the contact and well dopings, respectively).
The solid (dotted) lines are the stable (unstable) solutions.
Parameters of the ac field:  $ F=0.47 \times 10^{6} $ and
$f_{ac}=3 THz$ The inset shows the calculated potential profile at
$V_{dc}=0.86 V$.} \label{AguadoPRL_98__fig1}
\end{figure}
Furthermore, electric field domains supported by absorption and
emission sidebands corresponding to resonant states in neighbor
wells are possible: As we explained in the previous section,
transport in the high (low) electric field domain is only possible
by $C1\rightarrow C2$ ($C1\rightarrow C1$) resonances. New
tunneling channels open up in the presence of an ac field thus
allowing more electric field domain configurations.
Experimentally, this was studied in
Refs.~\cite{KeayPRL(95)b,ZeuPRB(96)} (see
Fig.~\ref{ZeunerPRB_96__fig2}).
\begin{figure}
\begin{center}
\includegraphics[width=0.75\columnwidth,angle=0]{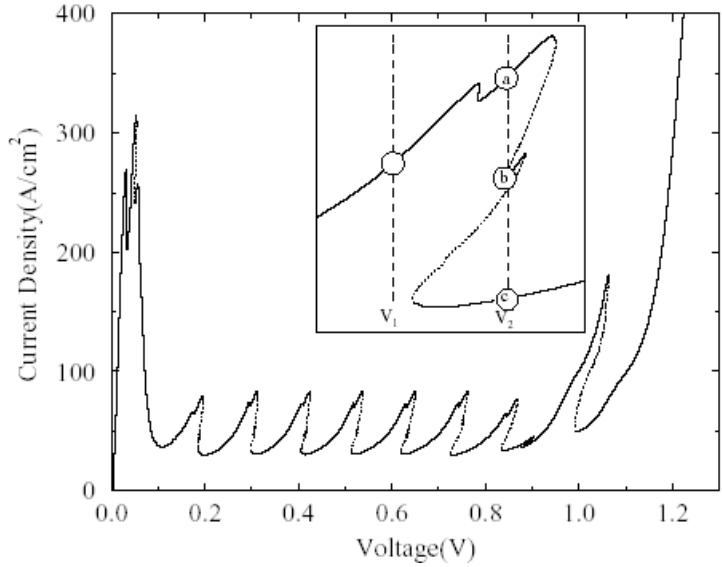}
\end{center}
\caption{I--V characteristic for an irradiated superlattice (same
parameters as in Fig.~(\ref{AguadoPRL_98__fig1})).
$F=0.95\times10^{6} V/m$ and $f_{ac}=3THz$.The inset shows a blow
up of the first branch.} \label{AguadoPRL_98__fig2}
\end{figure}
The extension of the selfconsistent method presented in subsection \ref{SLnonlinear-undriven1}
to include high
frequency fields was put forward in Refs.~\cite{Aguadotesis,AguPRL(98)a}. The
results are presented in Fig.~(\ref{AguadoPRL_98__fig1}) and
Fig.~(\ref{AguadoPRL_98__fig2}). The first plot demonstrates that
high field domains can be supported by photon-assisted tunneling
(for this particular case $C1\rightarrow C2$ tunneling involving
absorption of two photons (see inset).
\begin{figure}
\begin{center}
\includegraphics[width=0.75\columnwidth,angle=0]{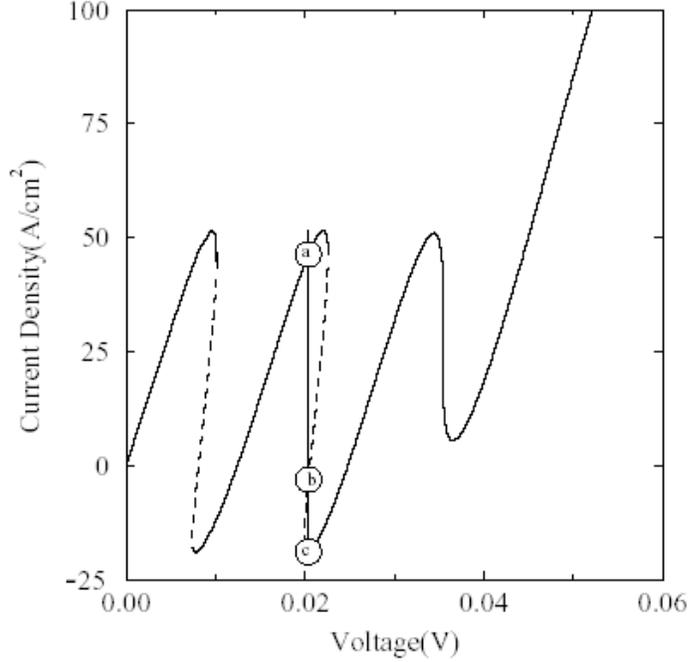}
\end{center}
\caption{ I--V characteristic for an irradiated superlattice (same
parameters as in Fig.~(\ref{AguadoPRL_98__fig1})) at low bias
voltage. $F=1.14\times10^{6} V/m$ and $f_{ac}=1.5THz$.}
\label{AguadoPRL_98__fig3}
\end{figure}
The second plot focuses on the effects of a very intense ac field
when there is a high probability of having multiphotonic effects,
leading to multistability of the branches. The inset shows a
magnification of the first branch, the circles mark the stable
operating points for a fixed voltage. At $V_{1}=0.16V$ transport
in the high field domain occurs via tunneling between the
two-photon absorption virtual state associated with C1 and the
two-photon emission virtual state associated with C2. At
$V_{2}=0.19V$ the branch develops a multistable solution (five
solution coexist, three stable, two unstable). These solutions
correspond to a different number of photons emitted in C2: one
photon in the highest current stable solution (circle a), two
photons in the lowest current stable solution (circle c); the
process from the highest current to the lowest one involves the
motion of the domain wall. The situation repeats periodically as
the domain wall moves, giving the sawtooth structure in the
current.

We described in Section \ref{SLlinear} how it is possible to
obtain absolute negative conductance in superlattices near the
dynamical localization condition. In doped samples the effect can
be even more spectacular {\it for the current develops regions of
bistability between positive and negative current near Dynamical
Localization}. This is shown in Fig.~\ref{AguadoPRL_98__fig3}
\cite{AguPRL(98)a} where the current--voltage characteristics for
an irradiated sample ($ F=1.14 \times 10^{6} V/m $ and $f_{ac}=1.5
THz$) is plotted, for low voltages.

\subsection{Weakly Coupled Superlattices as a paradigm of a nonlinear dynamical system IV:
dynamics in the ac-driven case \label{SLnonlinear-driven2}}
\subsubsection{High frequency driving: photon-assisted tunneling \label{SLnonlinear-driven2-osc}}
If the carrier density of a doped superlattice is below a critical
value, self-sustained oscillations of the current may appear as we
have described in Section \ref{SLnonlinear-undriven2}. The way an
external high frequency field affects this complex dynamics has
been analyzed in Ref.~\cite{LopPRB(03)} by using the
non-equilibrium Green's function formalism described Section
\ref{Keldysh}. It is straightforward to generalize the derivation of the time dependent
current in Section \ref{Keldysh}, see Eqs.~(\ref{time-dependent-Glesser}-\ref{Glesser}),  
and calculate the tunneling current traversing the $i$-th quantum well 
from the time evolution of the particle density $n_i$:
\begin{eqnarray}
I_{i,i+1}(t)&=&\frac{2e}{\hbar}\mbox{Re} \sum_{k_{i} k_{i+1}}
|T_{k_{i} k_{i+1}}|^2 \int d\tau [G^r_{k_{i+1}}(t,\tau)
g^<_{k_{i}}(\tau,t)\nonumber\\
&+&G^<_{k_{i+1}}(t,\tau)g^a_{k_{i}}(\tau,t)] \label{currentii+1}
\end{eqnarray}
Here, $g^{a (<)}_{k_{i}}$ is the advanced (lesser) Green's
function which includes the effect of the ac signal and scattering
processes for an isolated quantum well. The scattering processes
allow a non-equilibrium quasiparticle to relax its excess energy
(e.g., due to interactions with ionized impurities or LO phonons).
As in the model of previous sections, a phenomenological
relaxation time approximation is made by introducing a self-energy
as an energy independent constant (which is denoted by
$\gamma=\mbox{Im}\Sigma_{\rm sc}$). Of course, this model might be
improved by means of a microscopic calculation of $\Sigma_{\rm
sc}$  due to the aforementioned scattering processes. $G^{r
(<)}_{k_{i}}$ in Eq.~(\ref{currentii+1}) corresponds to the
retarded (lesser) Green's function which includes tunneling
events. If the same separation of time scales used in the selfconsistent model 
presented in subsection \ref{SLnonlinear-undriven1}
holds, one can assume
an \emph{equilibrium} distribution function for each quantum well,
since the electrons that tunnel relax their energy excess almost
instantaneously. Taking into account these considerations, the
effect of the ac potential consists of introducing a global phase
in the expression for these Green's functions:
$G^{r(<)}_{k_{i}}(t,t')=\exp [(ieV_i^{\rm ac}/\hbar\omega)
\left(\sin\omega t-\sin\omega t'\right) ]
\bar{G}^{r(<)}_{k_{i}}(t-t')$, where $\bar{G}^{r,<}_{k_{i}}(t-t')$
are the static retarded and lesser quantum well Green's functions.
They have the following expressions:
\begin{eqnarray}
{\bar G}^{r}_{k_{i}}(t-t')=-i\theta(t-t')\exp\left
[-i(E_{k_{i}}+\gamma)(t-t')\right] \,, \label{chap5equa18}
\end{eqnarray}
and
\begin{eqnarray}
 {\bar G}_{k_{i}}^{<}(t-t')\approx
\int\frac{d\epsilon}{2\pi }e^{i\epsilon(t-t')}\frac{2\gamma}
{(\epsilon - E_{k_{i}})^2+\gamma^2}f_{i}(\epsilon),
\label{chap5equa25}
\end{eqnarray}
where $f_i(\epsilon)$ is the Fermi-Dirac distribution function for
the $i$-th quantum well,
$f_i(\epsilon)=1/[1+\exp{(\epsilon-\epsilon_{\omega_i})/k_B T}]$.
A similar transformation applies for $g^{a(<)}_{k_{i}}(t,t')$.
Eventually, by inserting the obtained expressions for the
nonequilibrium Green's functions [$G^{r,<}_{k_{i+1}}(t,t')$ and
$g^{a,<}_{k_{i}}(t,t')$] into Eq.~(\ref{currentii+1}), one arrives
at the expression for the tunneling current between two quantum
wells irradiated with a THz-field in the sequential tunneling
regime:
\begin{eqnarray}
I_{i,i+1}(t)&=& \frac{2e}{\hbar}\sum_{k_{i} k_{i+1}} T_{k_{i}
k_{i+1}} \sum_{m=-\infty}^{m=\infty}J_m(\beta) \bigg\{
\cos\left(\beta \sin\omega t-m\omega t\right)\nonumber\\
&\times&\int d\epsilon\left[
A_{k_{i+1}}(\epsilon+m\hbar\omega)A_{k_{i}}(\epsilon)
\left(f_i(\epsilon)-f_{i+1}(\epsilon+m\hbar\omega)\right)\right]
\nonumber\\
&+&\sin\left(\beta\sin \omega t-m\omega t\right) \int d\epsilon
[A_{k_{i+1}}(\epsilon+m\hbar\omega)\mbox{Re} \bar{g}^a_{k_{i}
k_{i}}(\epsilon)f_{i+1}(\epsilon+m\hbar\omega)\nonumber\\
&+& \mbox{Re} \bar{G}^r_{k_{i+1} k_{i+1}}(\epsilon+m\hbar\omega)
A_{k_{i}}(\epsilon)f_i(\epsilon)] \bigg\} \label{eq2}
\end{eqnarray}
where $A_{k_{i}}$ is the spectral function for the $i$-th isolated
quantum well including scattering. The arguments of the Bessel
functions are given by $\beta=e(V_{i}^{\rm ac}-V_{i+1}^{\rm
ac})/\hbar\omega$. Notice that it is assumed that the ac potential
is spatially uniform along a quantum well (but different from that
of its neighbors) and $\beta$ is independent of the quantum well
index.

The current~(\ref{eq2}) may be written as
\begin{equation}
I(t)= I_0+\sum_{l>0} [I^{\cos}_l \,\cos\left(l\omega t\right)
+I^{\sin}_l \,\sin\left(l\omega t\right)],
\end{equation}
where $I_0$ is the time-averaged current. $I^{\cos}_l$ and
$I^{\sin}_l$ contain higher harmonics for $l>0$. In the
photoassisted tunneling regime $\hbar\omega>\gamma$. This means
that the electrons experience at least one cycle of the ac
potential between two successive scattering events. In addition,
the scattering lifetime represents the lowest temporal cutoff
above which the assumption of local equilibrium within each
quantum well holds. Therefore, the \emph{explicit} time variation
of $I(t)$ vanishes and one is left with the \emph{implicit} change
of $I_0$ with respect to time. This variation (in time scales
larger than $\hbar/\gamma$) results from the evaluation of the
continuity equation for $i=1,\ldots,N$, where $N$ is the number of
wells, supplemented with Poisson equations, constitutive
relations, and boundary conditions, similarly to the method
explained in Section \ref{SLnonlinear-undriven1}, such that the
current~(\ref{eq2}) is a functional of the Fermi energies and the
set of voltage drops in the superlattice (denoted by $\Phi$):
$I_{i,i+1}= I_{i,i+1}(\epsilon_{\omega_{i}},
\epsilon_{\omega_{i+1}},\Phi)$. The total current density
traversing the sample is the sum of the tunneling current plus the
displacement current, i.e.,
\begin{equation}
{\mathcal I}(t)=I_{i,i+1}+(\epsilon/d)(dV_i/dt),
\end{equation}
 where $\epsilon$ is the
static permittivity, $d$ the barrier width, and $V_i$ the voltage
drop in the $i$th barrier.
\begin{figure}[t]
\begin{center}
\includegraphics[width=0.75\columnwidth,angle=0]{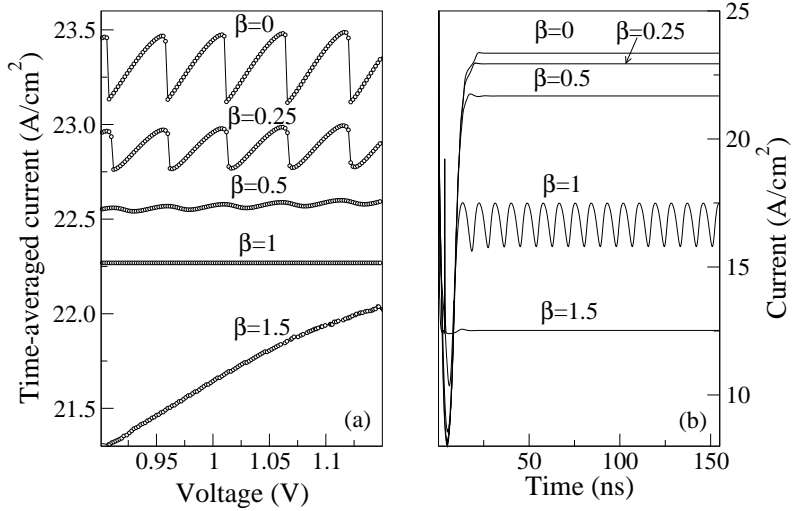}
\end{center}
\caption{(a) Current--voltage characteristics for a superlattice
consisting of 50 wells with 13.3-nm GaAs (wells) and 2.7-nm AlAs
(barriers). Well doping is $N_w=2\times10^{-10}$cm$^{-2}$ and
$\gamma=7$~meV. The frequency of the ac field is fixed to $f_{\rm
ac}=3$~THz and different intensities are taken. Lines are used to
guide the eye. Curves for $\beta=0$, $\beta=0.5$, $\beta=1$, and
$\beta=1.5$ have been shifted 0.05, 0.91, 5.46, and
9.35~A/cm$^{2}$, respectively, for clarity. At $\beta=0$ and
$\beta=0.25$ the electric field domain formation is stable, the
total current is stationary and it results in discontinuous
branches. With increasing $\beta$, branches coalesce, causing the
development an oscillatory pattern at $\beta=0.5$, followed by a
flat plateau that is formed at $\beta=1$. Larger values of $\beta$
involve a smooth, increasing curve of current with voltage (see
$\beta=1.5$) (b) Time-resolved electric current for a dc bias voltage
$V_{dc}=1.1$~V. The variation with $\beta$ shows the dependence of
the state character (static or dynamic) on the ac potential.
Schematically, the transition (static electric field
domains)$\longrightarrow$(moving electric field domains) takes
place at around $\beta=1$ whereas the process (moving electric
field domains)$\longrightarrow$(homogeneous electric field) occurs
at around $\beta=1.5$.} \label{LopezPRB_03__fig1}
\end{figure}
Solving selfconsistently at each time step the above set of
equations L\'opez et al \cite{LopPRB(03)} demonstrate that the
photon field {\it with frequency in the range of THz} is able to
induce low-frequency self-sustained current oscillations in
superlattices {\it with frequencies in the range of MHz} as
in the undriven case.\\
This is shown in Fig.~\ref{LopezPRB_03__fig1}a, where the time
average of ${\mathcal I}(t)$ is plotted as a function of the
applied dc bias voltage, $V_{dc}$. Without ac, the ${\mathcal I}$--$V$
curve shows branches after the first peak. As we have discussed in
Section \ref{SLnonlinear-undriven1} this is characteristic of
static electric field domain formation. In the presence of an ac
signal, the branches become smoother ($\beta=0.25$), and finally
they coalesce and a plateau clearly forms ($\beta=1$). This is the
key signature of current self-oscillations. We described in
Section \ref{SLnonlinear-undriven1} how the electric field domain
configuration becomes unstable with decreasing the doping density
such that self-sustained current oscillations occur due to the
periodic recycling of the domain wall. Here the same effect is
induced by the ac signal. By increasing $\beta$ further, the
plateau starts to be replaced by a positive differential
resistance region. There is a similar well-known phenomenon in
weakly coupled superlattices driven only by dc voltages: under a
critical value of the carrier density neither static nor moving
domain walls exist and the electric field drops homogeneously
across the \emph{whole} sample. The transition from static to
time-dependent current maybe also effectively achieved at constant
carrier doping by either applying a transverse magnetic field
\cite{SunPRB(99)} or raising the temperature \cite{WanAPL(99)}.
Here, the doping density is constant and it is the ac potential
that tunes this transition. This is illustrated in
Fig.~\ref{LopezPRB_03__fig1}b where ${\mathcal I}(t)$ for a fixed
bias voltage $V_{dc}=1.1~V$ is plotted. For $\beta=0$, the current
achieves a constant value after a transient time. As $\beta$
increases ($\beta=1$), the current oscillates with a frequency in
the range of MHz, much smaller than $f_{\rm ac}$. This is a result
of the motion of the accumulating layer of electrons, and its
recycling in the highly-doped contacts (see below). Then the ac
potential induces a transition from a stationary configuration
toward a dynamic state likely via a supercritical Hopf
bifurcation. Below, it is shown that the existence of
photosidebands and their influence on the non-linear behavior of
the system drives the superlattice toward oscillations. For
$\beta=1.5$ the current is damped and ${\mathcal I}(t)$ reaches a
uniform value. This is a striking feature--- {\it an oscillation
disappearance induced by an ac potential}.

\begin{figure*}[t]
\begin{center}
\includegraphics[width=0.75\columnwidth,angle=0]{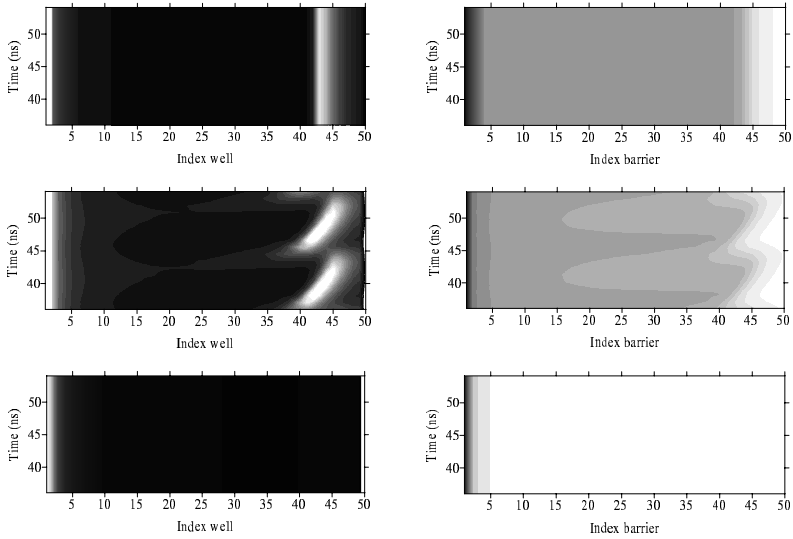}
\end{center}
\caption{Left panels: Time evolution of electron densities as a
function of the index well. Lighter areas mean larger densities.
Right panels: Time evolution of the voltage drop at the barriers
(last barrier has been omitted for simplicity). Lighter areas
indicate larger values of the electric field. Top: $\beta=0$ (no
ac potential is present). Electrons are accumulated mainly in well
43, forming a domain wall which separates high and low electric
field regions. Middle: $\beta=1$ (self-sustained oscillations).
The domain wall drifts along part of the superlattice. The
monopole is clearly visible at well 39, moves toward well 47 and
dissolves at the collector. Notice the oscillatory behavior of the
electric fields, which is correlated with the monopole motion.
Bottom: $\beta=1.5$ (homogeneous case). Voltage drops almost
linearly across the sample and consequently no accumulation layer
is formed.} \label{LopezPRB_03__fig3}
\end{figure*}

The ac induced transition from static electric field domains
toward homogeneous field distributions through self-sustained
current oscillations is illustrated in
Fig.~\ref{LopezPRB_03__fig3}. We observe how the charge density
through the structure, at fixed dc bias voltage, undergoes a transition
from being accumulated in the 43th quantum well, independently of
time (stationary electric field domains) at zero ac potential, to
presenting periodic oscillations ($\beta$=1). Increasing $\beta$
further ($\beta$=1.5) a homogeneous charge distribution is reached
and the electric field and charge are uniformly distributed
through the sample (with small inhomogeneities at the emitter
contact). A qualitative explanation of this transition is as
follows:\\
Let $v(F)$ denote the average drift velocity due to tunneling
between two QW's with local electric field $F$. Within a
semiclassical approximation, the current~(\ref{eq2}) can be
approximated by  $I_{i,i+1}=e n_i v(F_i)/{\mathcal L}$, where the
electronic drift velocity is given by $v(F)=I(N_w,N_w,F) {\mathcal
L}/e N_w$. Here, the current $I(N_w,N_w,F)$ is evaluated by using
Eq.~(\ref{eq2}) after imposing $n_i=n_{i+1}=N_w$ and setting an
average interwell electric field $F$ along the superlattice period
${\mathcal L}=d+w$ \cite{BonPRB(00)}. The contribution from
diffusivity, which can be important at very low electric fields
\cite{BonPRB(00)}, is neglected. As shown in
Ref.~\cite{WacBook(97)}, the sufficient condition for stationary
electric field domains to form reads:
\begin{equation}
N_w\gtrsim N_{w}^{\rm eff} \equiv \varepsilon v_m \frac{F_m- F_M}{
e\, (v_M - v_m)} \,,
\end{equation}
where $v_{M}$ ($v_m$) is the maximum (minimum) electron drift
velocity attained at an electric field given by $F_{M}$ ($F_{m}$).
Unlike the minimum velocity, the maximum drift velocity is very
sensitive to the external ac potential. We see from the time
average of Eq.~(\ref{eq2}) that first current peak (i.e., $v_{M}$)
is weighted by $J_0^2(\beta)$ at low values of $\beta$ (the
zero-photon peak). As $\beta$ increases, the THz potential
produces photoassisted tunneling with absorption and emission of
photons. As a result, the zero-photon peak is quenched as the
contribution of terms with $J_{p\neq 0}^2(\beta)$ begins to grow.
As we have seen in previous Sections, this is a consequence of the
photoassisted formation of sidebands. The overall effect is that
$N_{w}^{\rm eff}$ decreases as $\beta$ increases. For a certain
critical value of $\beta$ [$\beta_{\rm crit}\sim 1$; see
Fig.~\ref{LopezPRB_03__fig1}(b)], one finds $N_w\lesssim
N_{w}^{\rm eff}$ and the steady electric field domain
configuration is no longer stable. The system evolves
spontaneously toward self-sustained current oscillations. On the
other hand, once the dynamical configuration is stable, increasing
$\beta$ will tend to drive the superlattice to a trivially
homogenous electric field profile (see
Fig.~\ref{LopezPRB_03__fig3}, lower panel). The reason for that is
the complicated shape of the time-averaged drift velocity induced
by the ac potential. The ac potential opens up new tunneling
channels due to photon absorption and emission and their relative
weight and their contribution to $v(F)$ depend in a non trivial
way on the ac frequency and intensity, the sample characteristics
and the scattering processes involved. This can lead to a
${\mathcal I}$--$V$ curve exhibiting positive differential
resistance with a Z shape unlike the electric field domain case,
which exhibits a ${\mathcal I}$--$V$ curve with a N shape
\cite{WacBook(97)} Of course, this qualitative argument does not
provide with an estimate of the different transition points but
still shows conclusively that an ac field may induce a dynamical
transition from stable stationary domains to traveling field
domains and a homogeneous electrostatic configuration by modifying
the effective electronic drift velocity with the dimensionless ac
parameter $\beta$.

Finally, we finish this part by mentioning a recent work by
Batista et al \cite{BatisPRB(03)} where the intersubband
transitions in n-$\delta$-doped quantum wells strongly driven by far
infrarred radiation is studied. They demonstrate that a suitably
taylored quantum well can exhibit superharmonic generation and
nonlinear phenomena in their absorption lineshapes. In their study
they show that intersubband transitions can produce strong
subharmonic (period doubling) or a strong inconmensurate (Hopf)
frequency response by varying the density and the intensity.
\subsubsection{Adiabatic driving: routes to chaos \label{chaos}}
The rich dynamical behavior an ac signal induces in a
superlattice is not restricted to the high frequency regime.
Driving a superlattice with a low-frequency signal, in particular
for frequencies of the ac field inconmensurated with the natural
frequency of the system, produces quasiperiodicity, frequency
locking or chaotic current as a function of the intensity of the
driving field. Intriguing routes to chaos,  reflected in complex
bifurcation diagramms, have been experimentally observed in
semiconductor superlattices driven by an ac field.
\begin{figure}
\begin{center}
\includegraphics[width=0.75\columnwidth,angle=0]{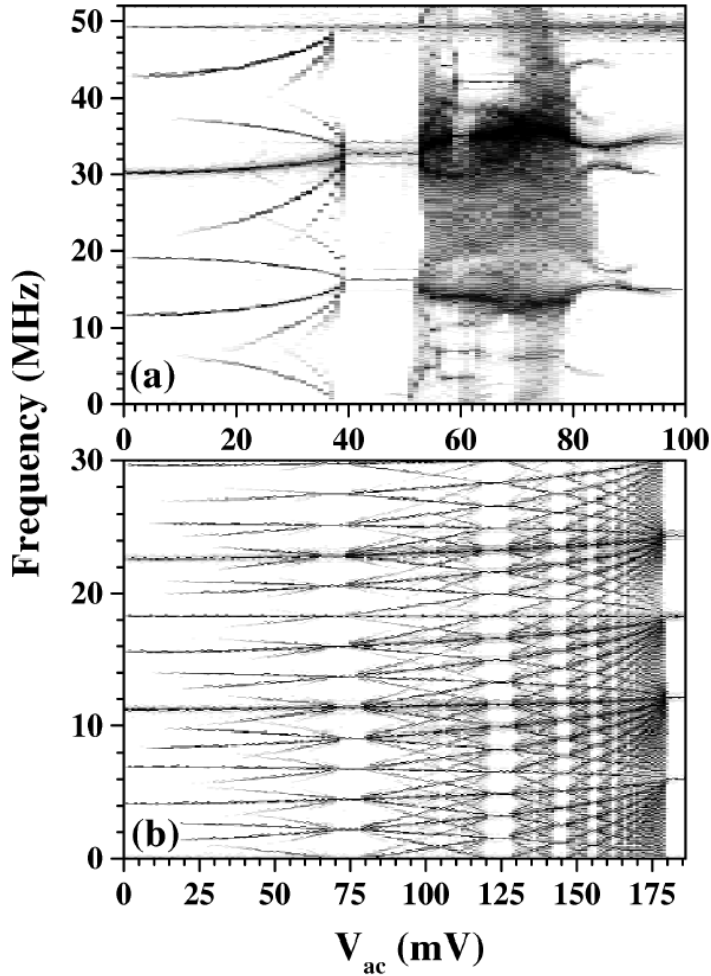}
\end{center}
\caption{Frequency bifurcation diagram for a) $V_{dc}$=6.574 V and
$f_{ac}$=49.4 MHz . b) $V_{dc}$=7.08 V and $f_ac$=18.4 MHz  at 5K,
for a 40-period weakly coupled superlattice with 9 nm GaAs wells
and 4 nm AlAs barriers. The current power spectra are shown as
density plots vs the amplitude of the driving voltage $V_{ac}$,
where dark areas correspond to large amplitudes. Reprinted with
permission from \cite{LuoPRL(98)}. \copyright 1998 American
Physical Society.} \label{LuoPRL_98__fig1}
\end{figure}
\cite{BulPRB(95),ZhaPRL(96),LuoPRL(98),LuoPRB(98),BulPRB(99)}.
Many studies fix the frequency of the ac drive as the golden mean
number $(1+\sqrt 5)/2\approx 1.618$ times the frequency of the
natural oscillations (i.e., the frequency ratio is an irrational
number hard to approximate by rational numbers), which is
convenient to obtain complex dynamical behavior. In this case, the
system presents a rich power spectrum, a complex bifurcation
diagram and different routes to chaos including quasiperiodicity,
frequency locking, etc.
\cite{BulPRB(95),ZhaPRL(96),LuoPRL(98),LuoPRB(98),BulPRB(99)}.
\begin{figure}[!htp]
\begin{center}
\includegraphics[width=0.75\columnwidth,angle=0]{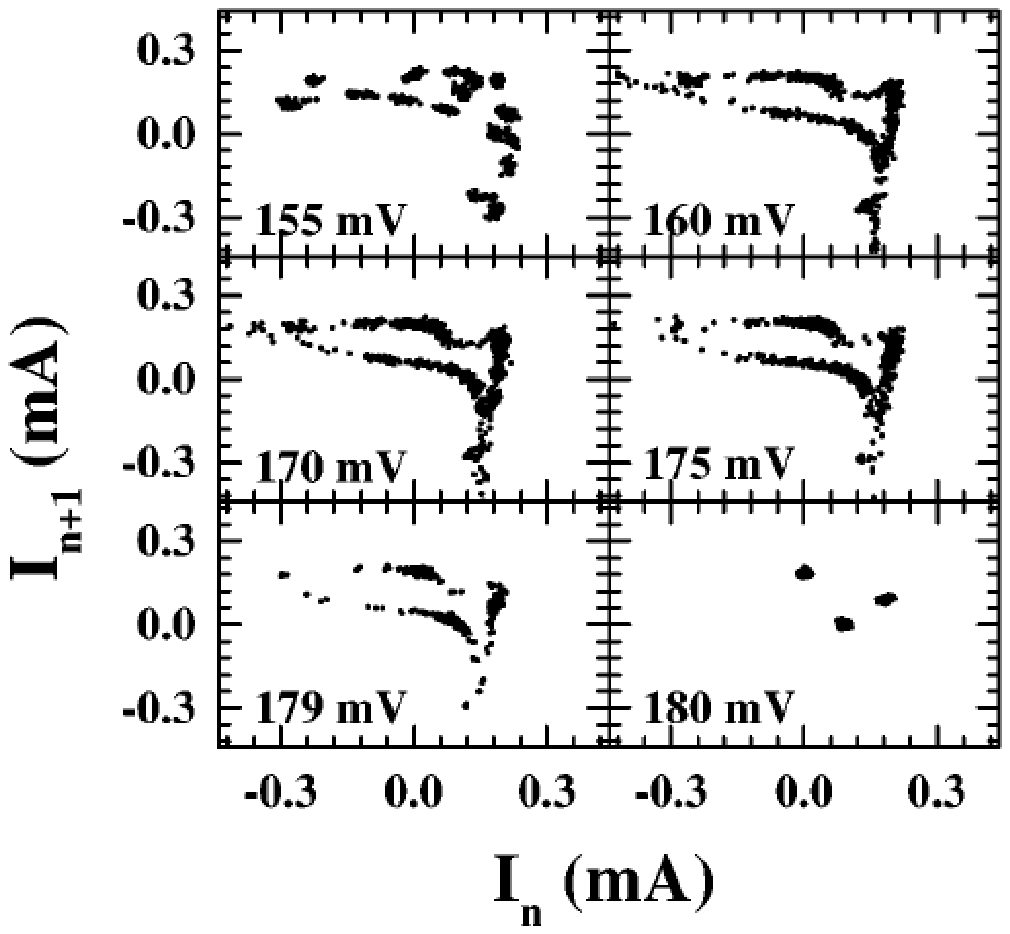}
\end{center}
\caption{Poincar\'e maps for several driving amplitudes $V_{ac}$
between 155 and 180 mV from current oscillation traces for the
same conditions as in Fig.~\ref{LuoPRL_98__fig1} b). Reprinted
with permission from \cite{LuoPRL(98)}. \copyright 1998 American
Physical Society.} \label{LuoPRL_98__fig3}
\end{figure}
First return or Poincar\'e maps are used to analyze unambiguously
the underlying attractors \cite{LuoPRB(98)}. In the quasiperiodic
case, Poincar\'e maps usually consist of smooth loops, whereas
they are a set of discrete points in the case of frequency
locking. More exotic Poincar\'e maps resembling distorted double
loops in the quasiperiodic case have been experimentally observed
in middle of the second plateau of the current--voltage
characteristic of a superlattice \cite{LuoPRL(98),LuoPRB(98)}. At
the onset of this plateau, Poincar\'e maps are smooth and not
distorted. The origin of distorted maps was not understood at the
time of their observation, although disorder and sample
imperfections were invoked \cite{LuoPRL(98)}. Luo et al
\cite{LuoPRB(98)} showed that a combination of signals with
different frequency was needed in order to reproduce
experimentally observed distorted double layer Poincar\'e maps.
The origin of this combination was not ascertained in that work.
By using the model of section \ref{SLnonlinear-undriven1},
S\'anchez et al have shown in Ref.~\cite{SanPRB(01)} that
high-frequency current spikes of the self-oscillations give rise
to these exotic Poincar\'e maps. In turn, current spikes are due
to the well-to-well motion of the domain wall during each period
of the self-oscillations. Thus distorted Poincar\'e maps reflect
the domain wall motion in ac driven superlattices. They analyze
the sequential tunneling current with an applied voltage, $V(t) =
V_{ac}(t) + V_{dc}$, where $V_{ac}(t)= V_{ac}\sin (2\pi f_{ac}t)$
where $f_{ac}$ is set to the golden mean times the natural
frequency of the system.
\begin{figure}[!htp]
\begin{center}
\includegraphics[width=0.75\columnwidth,angle=0]{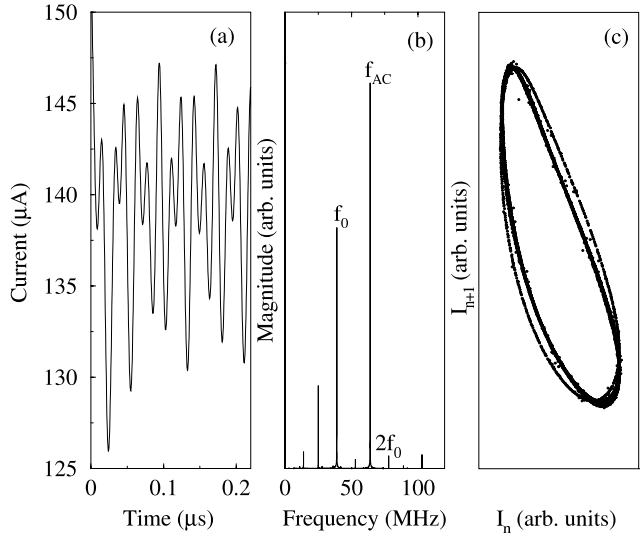}
\end{center}
\caption{(a) I(t) for $V_{dc}=4.2$~V, $f_{0}=39$~MHz,
$V_{ac}=19$~mV for a 50-period superlattice consisting of 13.3 nm
GaAs wells and 2.7 nm AlAs barriers. Doping in the wells and in
the  contacts are $N_{w} = 2\times 10^{10}$~cm$^{-2}$ and $N_{c}=
2\times 10^{18}$~cm$^{-3}$ respectively. With these doping values,
self-oscillations are due to recycling of monopole domain walls;
Spikes are not resolved. (b) Power spectrum. Notice that higher
harmonics of the fundamental frequency are barely formed. (c)
Poincar\'e map, constructed by plotting the current at the (n+1)st
ac period versus the current in the preceding period.}
\label{notrans}
\end{figure}
Namely, $hf_{ac}$ is very small compared with typical energy
scales of the system such that $V_{ac}(t)$ modifies adiabatically
the potential profile of the superlattice. Thus, the condition
that all voltage drops across the different regions of the
nanostructure must add up to the applied bias voltage, c.f.
Eq.~(\ref{bias}), is in this case:
\begin{eqnarray}
V(t) = \sum_{i=1}^{N+1}V_{i}(t)+\sum_{i=1}^{N}V_{wi}(t) +
\frac{\Delta_{1}+\Delta_{2}+E_{F}}{e} .\label{bias-time}
\end{eqnarray}
where $ V_{i}(t)$ and $V_{wi}(t)$ are the potential drops in the
i-barrier and well respectively and $\Delta_{1}$ and $\Delta_{2}$
correspond to the potential drops at the contacts, 
c.f. Eqs.~(\ref{field.inside1}--\ref{collector}).\\
The results are shown in Fig.~\ref{notrans} where the evolution of
the current through the superlattice, its Fourier spectrum and its
Poincar\'e map for $V_{dc}=4.2$~V and $V_{ac}= 19$~mV are plotted.
These values correspond to the onset of the second plateau of the
$I$--$V$ characteristic curve. For the doping values taken (see
caption) self-oscillations are due to recycling of monopole domain
walls.
\begin{figure}
\begin{center}
\includegraphics[width=0.75\columnwidth,angle=0]{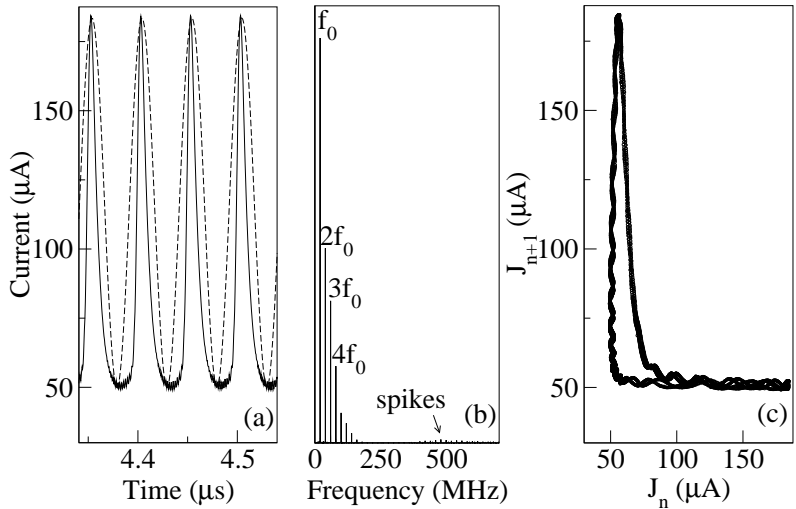}
\end{center}
\caption{a) I(t) versus time (soild line) with $V_{dc}=5.5V$ and
$V_{ac}=2mV$ for a 50-period superlattice consisting of 13.3 nm
GaAs wells and 2.7 nm AlAs barriers. Fitting to a sine function
(dotted line) is shown for comparison. Doping in the wells and in
the  contacts are $N_{w} = 2\times 10^{10}$~cm$^{-2}$ and $N_{c}=
2\times 10^{16}$~cm$^{-3}$ respectively; b) power spectrum; c)
Poincar\'e map.} \label{mon}
\end{figure}
The current trace of Fig.~\ref{notrans}(a) is quasiperiodic and
does not present observable superimposed high-frequency
oscillations (spikes).  The natural oscillation near the onset of
the plateau is caused by monopole recycling very close to the
collector contact. Thus the DW does not move over many wells and
the current trace does not present an appreciable number of
spikes. In the power spectrum of Fig.~\ref{notrans}(b) there are
contributions coming from the  fundamental frequency $
f_{0}\approx39$ MHz, the frequency of the applied ac field
$f_{ac}$, the combination of both and their higher harmonics. The
Poincar\'e map depicted in Fig.~\ref{notrans}(c) is a smooth loop
with a nontrivial double layer structure indicating quasiperiodic
oscillations. By reducing the doping of the contacts, $I(t)$
(solid line in Fig.~\ref{mon}(a)) deviates from a sine (dotted
line Fig.~\ref{mon}(a)) due to the presence of spikes at low
currents values which results in higher harmonics in the frequency
spectrum (Fig.~\ref{mon}(b)). The first return map gets a strong
distortion (see Fig.~\ref{mon}(c)) such that it can be concluded
that the presence of spikes give rise to the wiggles that
ultimately cause the twist of the loop (note that the twisted arm
ranges from about 49~$\mu$A to 55~$\mu$A, exactly the region
covered by the spikes in Fig.~\ref{mon}(a)).

\begin{figure}
\begin{center}
\includegraphics[width=0.75\columnwidth,angle=0]{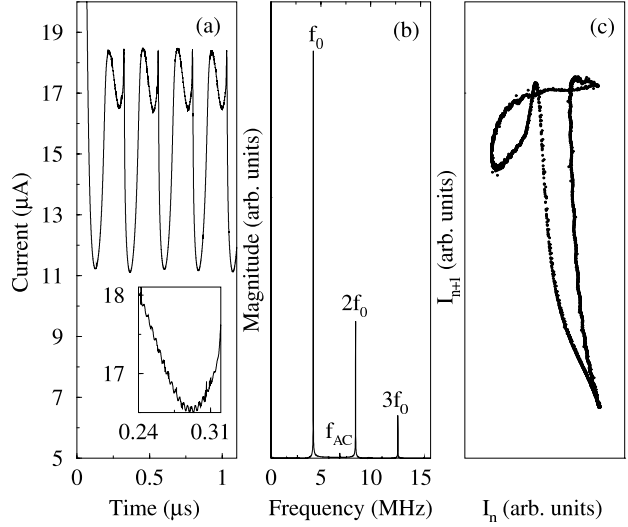}
\end{center}
\caption{(a) I(t) for $V_{dc}=1.5$~V, $f_{0}=4$~MHz,
$V_{ac}=2$~mV. Spikes are superimposed on the current throughout
the signal (see inset). (b) Power spectrum. Higher harmonics of
$f_{0}$ contribute with a finite amplitude to the power spectrum.
(c) Poincar\'e map. The distortion is greater than in
Fig.~\ref{mon}(c) (see text).} \label{dip}
\end{figure}
The previous conclusion may be reinforced by changing the dc
voltage to $V_{dc}$=1.5 V (middle of the first plateau) such that
the frequency of the natural oscillation is now reduced to 4 MHz,
$I(t)$ presents dipole-like oscillations and superimposed finite
amplitude spikes (Fig.~\ref{dip}(a)). The Poincar\'e map,
Fig.~\ref{dip}(c), is much more complicated than in the previous
case, showing three well defined distorted loops. Since loops in
the Poincar\'e map are due to combination of strong enough signals
of different frequencies \cite{LuoPRB(98)}, the greater strength
of the high-frequency spikes gives rise to the additional loop
structure and higher harmonic content (Fig.~\ref{dip}(b)).
Therefore, the high frequency selfsustained oscillations or spikes
are essential to explain the observed electronic trajectories and
the complicated electronic dynamics in a low frequency ac-driven
multiquantum well structure \cite{SanPRB(01)}.

Finally, we finish this part by mentioning that a detailed analysis of the routes
to chaos in low frequency ac-driven weakly coupled superlattices with magnetic impurities
has been recently performed \cite{Bej-preprint}.
\subsection{Strongly Coupled Superlattices in ac potentials
 \label{Bloch}}
In a pioneering paper, Esaki and Tsu proposed high frequency
oscillators by tayloring the non linear electronic transport
properties of semiconductor superlattices \cite{Esa70}. Strongly
coupled superlattices are characterized by energy bands, the
so-called minibands, instead of discrete levels. The corresponding
extended states are Bloch states. In the presence of a constant
electric field $E$, the Bloch states are no longer eigenstates of
the Hamiltonian: the electrons are accelerated by the electric
field and perform a repetitive motion of acceleration and Bragg
reflexion called Bloch oscillation, characterized by the Bloch
frequency: $\omega_{Bloch}=eEd/\hbar$ where $d$ is the
superlattice period\footnote{For a review on transport in
superlattices see Ref.~\cite{WackPR(02)}.}.

The effect of radiation on the transport properties of
superlattices in the miniband regime has been experimentally
analyzed by different groups during the last two decades. These
experiments demonstrate that the application of an external ac
field to a superlattice in the miniband regime induces a great
deal of interesting phenomena.
\begin{figure}
\begin{center}
\includegraphics[width=0.7\columnwidth,angle=0]{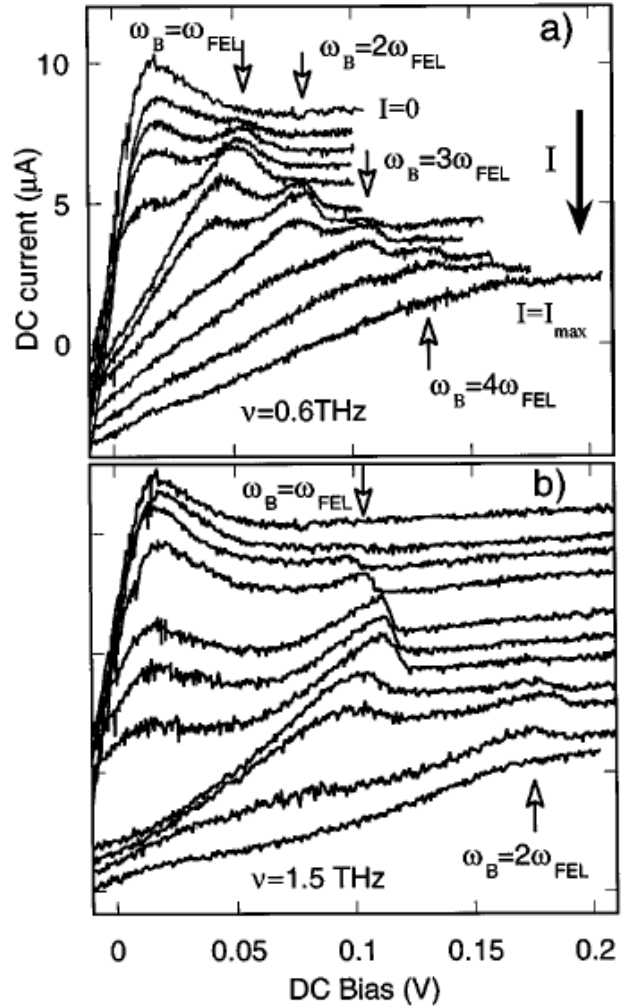}
\end{center}
\caption{Current-voltage characteristics for miniband
superlattices irradiated by 0.6 THz (a) and 1.5 THz (b) radiation.
The curves are shifted downwards for increasing intensities. In
the negative differential conductance region additional features
occur attributed to resonances at the Bloch frequency and its
subharmonics. Reprinted with permission from \cite{UntPRL96}.
\copyright 1996 American Physical Society.} \label{UntPRL_96}
\end{figure}
As we described in subsection \ref{miniband-collapse} an external
ac field can produce the collapse of the miniband width at the
dynamical localization condition as demonstrated by Holthaus in
Ref.~\cite{HoltPRL(92)}. A similar effect was observed by Ignatov
and coworkers \cite{IgnADP94,SchomburAPL(96)} who reported
ac-field-induced reduction of the dc current. They attributed such
a reduction to a frequency modulation of the Bloch oscillations of
electrons at the frequency of the external ac field. The group of
Santa Barbara \cite{UntPRL96} observed what they called {\it the
inverse Bloch oscillator effect}, which consists of resonant
changes in the current-voltage characteristics when the Bloch
frequency is resonant with a THz field and its harmonics. An
example from these experiments is shown in Fig.~\ref{UntPRL_96}.

The nonlinear dynamics of miniband superlattices under irradiation
have been analyzed by Alekseev and coworkers in a series of papers
\cite{Alek1,Alek2,Alek3}. Interestingly, they show in Ref.~\cite
{Alek3} that a purely ac external field applied to an unbiased
superlattice can create a dc bias voltage and thus spontaneously
generate a dc current. In technical terms, this can be explained
as a spontaneous breaking of spatial symmetry induced by the
external field.

Finally, we mention another interesting effect induced by
dynamical localization: Meier et al \cite{MeierPRL(95)} analyzed
how and external ac field can alter the effective dimensionality
of the excitons in a superlattice. Based on a full three
dimensional description of both coherent and incoherent phenomena
in anisotropic structures, they found that appropiated applied
oscillating fields change the exciton wave function from
anisotropic three dimensional to basically two dimensional. This
effective dimension change is caused by dynamical localization
which leads to an increase of the exciton binding energy and of
the corresponding oscillator strength.
\section{Microwave-induced zero resistance in two-dimensional electron gases 
\label{MicrowaveHall}}
\begin{figure}
\begin{center}
\includegraphics[width=0.75\columnwidth,angle=0]
{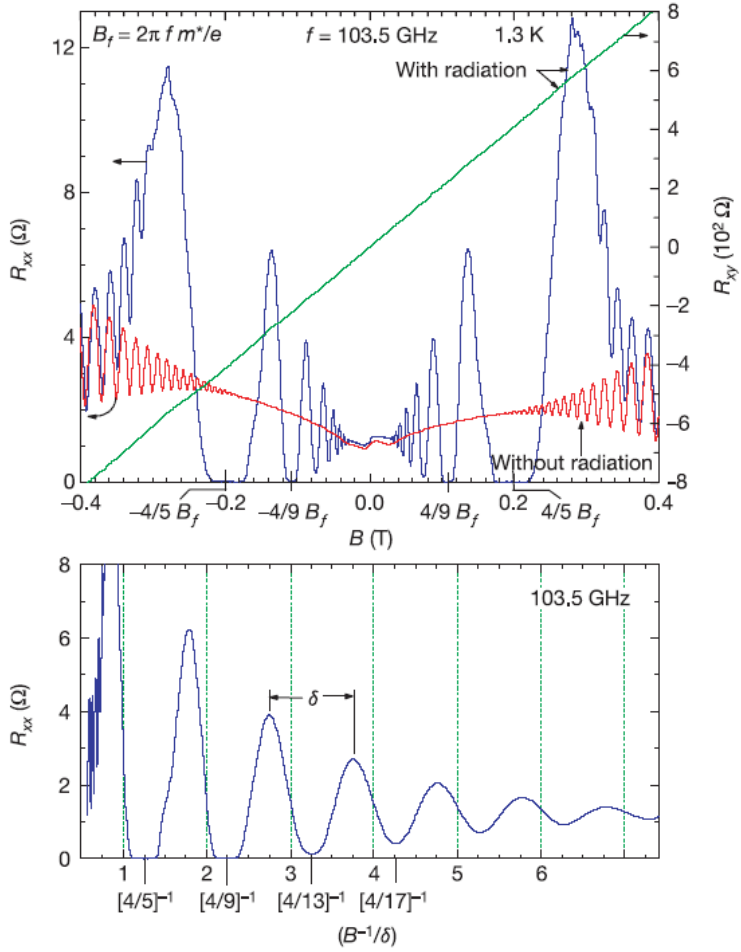}
\end{center}
\caption{Top: Resistance of a high mobility two-dimensional electron system under 
microwave irradiation. 
Shubnikov-de Haas oscillations are seen in the longitudinal resistance $R_{xx}$ for 
magnetic fields above 0.2 T. Below that, $R_{xx}$ without microwaves is featureless; 
with microwaves $R_{xx}$ presents strong oscillations although the transverse Hall 
resistance $R_{xy}$ remains unafected. Bottom: The oscillations in $R_{xx}$ are periodic in 
$1/B$ with a period given by $B_f^{-1}=(\omega m^*/e)^{-1}$. 
The maxima are found at $B_f/B=n+1/4$ for $n$ an integer; the minima at  $B_f/B=n-1/4$.
Reprinted with permission from \cite{ManiNat(02)}.
\copyright 2002 Nature Publishing Group.}
\label {ManiNature_fig1}
\end{figure}

Recently,
the study of two-dimensional electron gases irradiated by microwaves has received a great
deal of attention due to the observation of microwave-induced zero longitudinal resistance
in two-dimensional electron systems at low magnetic fields
(just below the onset of Shubnikov-de Haas oscillations) \cite{ManiNat(02),ZudovPRL(03)}.
In the presence of microwaves,
the longitudinal resistance oscillates (with minima reaching zero-resistance and even
negative values) as a function of the inverse of the applied magnetic field with a
period given by $(\omega m^*/e)^{-1}$, where $\omega$ is the microwave
frequency and $m^*$ the effective electron mass, see Fig.~\ref{ManiNature_fig1}. 
An explanation for this intriguing
phenomenon was given by Durst et al in
Ref.~\cite{DurstPRL(03)}. Similar arguments were presented by 
Anderson and Brinkman in Ref.~\cite{Andcondmat(03)} and by
Shi and Xie in Ref.~\cite{ShiPRL(03)}:
If an electron absorbs a microwave photon $\omega=n\omega_c$,
where $\omega_c$ is the cyclotron energy which defines the spacing of the
Landau-level ladder (which is tilted by the applied dc voltage),
no transport is possible.
If, on the other hand,  $\omega\gtrsim n\omega_c$, energy can be conserved 
if impurities scatter the electron laterally.
The upstream or downstream motion will reduce or increase the conductivity
of the sample. If the final density of states to the left exceeds that to the right
the current is enhanced. If vice versa, the current is diminished such that
the scattering events can drive the conductivity to zero or even
negative values if the electrons tend to flow uphill.

The above physical picture can be understood
by the following simple model \cite{ShiPRL(03)}:
Let us consider a junction biased by an ac voltage $V_{\mathrm{ac}}=\Delta \cos \omega t$
and assume that the left and right regions have the same density of states 
$\rho _{L}(\epsilon )=\rho _{R}(\epsilon )=\rho (\epsilon )$ 
such that the current and the conductance can be written as (see Section \ref{Tien-Gordon}):
\begin{eqnarray}
I & = & eT_{LR}\int d\epsilon \sum _{n}J_{n}^{2}\left(\frac{\Delta }{\hbar \omega }\right)
\left[f(\epsilon )-f(\epsilon +n\hbar \omega +eV)\right]\nonumber \\
 &  & \times \rho (\epsilon )\rho(\epsilon +n\hbar \omega +eV)\nonumber\\
\sigma  & = & e^{2}T_{LR}\int d\epsilon \sum _{n}J_{n}^{2}\left(\frac{\Delta }{\hbar \omega }
\right)\left\{ \left[-f^{\prime }(\epsilon )\right]\rho (\epsilon )
\rho (\epsilon +n\hbar \omega )\right.\nonumber \\
 &  & \left.+[f(\epsilon )-f(\epsilon +n\hbar \omega )]\rho (\epsilon )
\rho ^{\prime }(\epsilon +n\hbar \omega )\right\}.
\label{eq:conductivity-hall1}
\end{eqnarray}      
The second
term in the expression for the conductance depends on the derivative of the density of states, and can be
either positive or negative. The contribution from the second term
is purely due to the photon-assisted tunneling process, and vanishes
when there is no ac field. If one now assumes a density of states
which is a periodic function of energy near the Fermi surface, with period $\hbar \omega _{c}$,
the simple toy model in Eq.~(\ref{eq:conductivity-hall1}) 
captures most of the important features of the experiments~\cite{ManiNat(02),ZudovPRL(03)}.
In particular, if one takes
\begin{equation}
\rho (\epsilon )=\left(1+\lambda \cos \frac{2\pi \epsilon }{\hbar \omega _{c}}\right)
\rho _{0}\, ,\label{eq:dos}
\end{equation}
where $\lambda $ is a dimensionless constant representing the fluctuation
amplitude of the density of states, a straightforward calculation
yields the conductance of the system,
\begin{eqnarray}
\sigma (T)/\sigma _{0} & = & \sum _{n=-\infty }^{\infty }
J_{n}^{2}\left(\frac{\Delta }{\hbar \omega }\right)
\left[1+\frac{\lambda ^{2}}{2}\cos \left(2\pi n\frac{\omega }{\omega _{c}}\right)\right.
\nonumber \\
&  & \left.-n\pi \lambda ^{2}\frac{\omega }{\omega _{c}}
\sin \left(2\pi n\frac{\omega }{\omega _{c}}\right)\right]
+g\left(\frac{\mu }{\hbar \omega _{c}},T\right),\label{eq:sigmaT}
\end{eqnarray}
where $\sigma _{0}=e^{2}D\rho _{0}^{2}$, and $g(\mu /\hbar \omega _{c},T)$
is the contribution from the Shubnikov-de Hass oscillation which diminishes
rapidly at finite temperatures. The conductance oscillation minima
can be easily determined from Eq.~\ref{eq:sigmaT}: for the $k$th
harmonics of the oscillation, the positions of the conductance minima
are given by the equation $\tan x=-x/2$, where $x=2\pi k\omega /\omega _{c}$.
For $k=1$, it yields the conductance minimum positions very close
to $\omega /\omega _{c}=n+1/4$, although not exactly. When the higher
orders of harmonics become important, one can expect that the conductance
minima deviate from the $n+1/4$ rule. The amplitude of oscillation
is independent on the temperature, namely any temperature dependence
observed in the experiments should come from the temperature dependence
of the density of states, \emph{i.e.,} $\lambda $.  
By using a more realistic density of states taking into account
that in a 2DEG under a perpendicular weak magnetic field
$\lambda $ is a function of $\omega _{c}$
\begin{equation}
\lambda =2\exp \left(-\frac{\pi }{\omega _{c}\tau _{f}}\right),
\label{eq:lambda}
\end{equation}
where $\tau _{f}$
is the relaxation time of electron which depends on the scattering
mechanisms of the system and the temperature, 
the agreement with the experimental observation is quite good:
The conductance
minima are found at the positions near $\omega /\omega _{c}=n+1/4$
for the low and intermediate intensities of the microwaves. 
As in the experiments~\cite{ManiNat(02),ZudovPRL(03)}, one 
gets two sets of crossing points at $\omega /\omega _{c}=n$ and $\omega /\omega _{c}=n+1/2$,
where the conductances equal their dark field values. 
\begin{figure}
\begin{center}
\includegraphics[width=0.75\columnwidth,angle=0]
{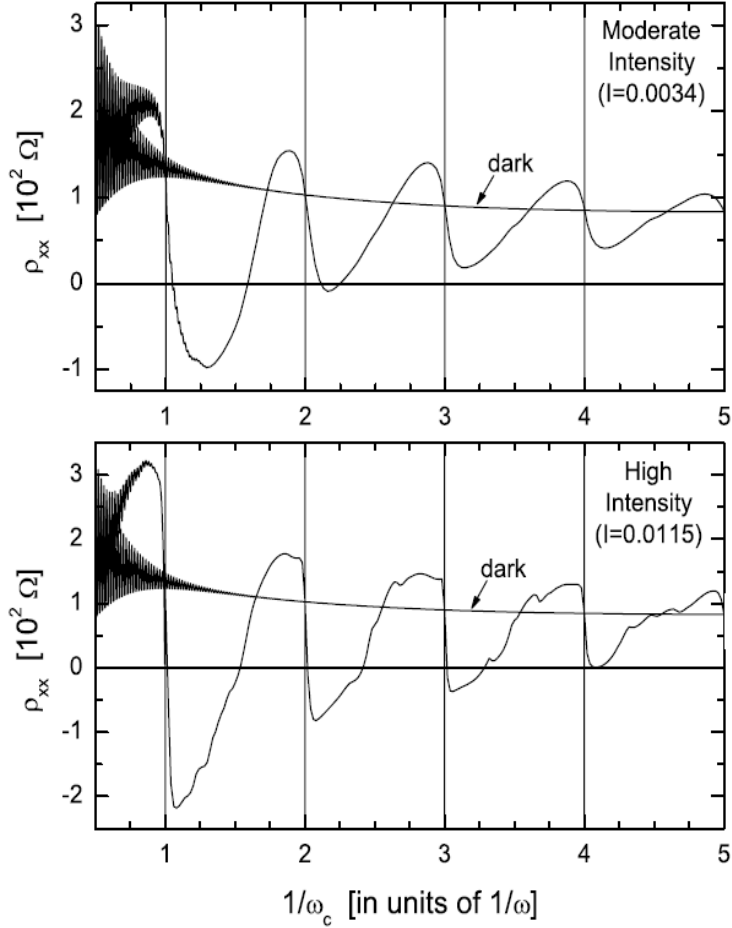}
\end{center}
\caption{$\rho_{xx}$ vs $1/\omega_c$ at fixed $\omega$ and three values of the intensity
(in units of $m^*\omega^3$): I=0 (dark), I=0.0034 (upper pannel) and I=0.0115 (lower pannel).
Reprinted with permission from \cite{DurstPRL(03)}.
\copyright 2003 American Physical Society.}
\label{DurstPRL_03_fig1}
\end{figure}      
Note that we have discussed here the behavior of the conductance whereas the
experiments measure resistance.
The connection can be made by rebembering that in a Hall bar the relation between
conductance and resistance is:
\begin{eqnarray}
\rho_{xx}=\frac{\sigma_{xx}}{\sigma^2_{xx}+\sigma^2_{xy}}\nonumber\\
\rho_{xy}=\frac{\sigma_{xy}}{\sigma^2_{xx}+\sigma^2_{xy}}.
\end{eqnarray}
As pointed out by the authors of Ref.~\cite{DurstPRL(03)},
the experiments are in the regime where $\rho_{xx}\approx \rho^2_{xy}\sigma_{xx}$ and
$\rho_{xy}\approx \sigma_{xy}$ such that $\rho_{xx}$ and $\sigma_{xx}$
have the same microwave-induced period and phase ($\rho_{xy}$ remains unafected by the
external radiation).                      

The simple picture given by Eqs.~(\ref{eq:conductivity-hall1}-\ref{eq:lambda}) 
is confirmed by a more ellaborated calculation by Durst et 
al \cite{DurstPRL(03)} who evaluate the conductante  
using a diagramatic selfconsistent Born approximation including 
radiation and disorder. An example of their calculation in shown in 
Fig.~\ref{DurstPRL_03_fig1} where $\rho_{xx}$ vs $1/\omega_c$ at fixed $\omega$ 
and different values of the radiation intensity is plotted. In agreement with the experiment, 
the period of the oscillations is $1/\omega$ and minima are found near 
$\omega/\omega_c=n+1/4$. 
The authors note that the 1/4 phase shift is not universal, varying between 0 and 1/2
depending upon disorder and intensity.

Andreev et al \cite{AndrePRL(03)} noted that a 
negative conductivity makes the two-dimensional electron gas unstable. 
Due to this instability the systems develops a domain structure 
with an inhomogeneous current pattern, for which the measured resistance would be zero.
Other explanations have also been proposed in the 
literature \cite{Hallrest1,Hallrest2,Hallrest3}.

Finally we mention that the phenomenon of ac-induced negative conductance in 
two-dimensional electron gases bears close resemblance with the ac-induced negative resistance
observed in THz-irradiated superlattices we have discussed in 
subsection \ref{SLlinear}.
\section{Electron pumps \label{pumps}}
We have seen in previous sections that the application of an ac
signal to a semiconductor heterostructure brings about a good deal
of new phenomena like, for instance, coherent destruction of
tunneling or absolute negative conductance in semiconductor
samples which are biased with a {\it positive} voltage. This
latter phenomenon is just an example of a general class of systems
dubbed {\it electron pumps} where current rectification is
achieved by combining nonlinear ac driving with either absence of
inversion symmetry in the device, or lack of time-reversal
symmetry in the ac signal. The range of possible electron pumps
includes turnstiles \cite{GeePRL(90),KouPRL(91)}, photon-assisted
pumps \cite{HN91,StaPRL(96),WagPRL(99),SolsAnnPhys(00),WWG02} or
adiabatic pumps
\cite{ThouPRB(83),NiuPRL(90),Brouwer98,SMCG99,AltSci(99),ZSA99,SAA00,Simon00,WWG00,Brouwer01,PB01,MB01,AEGS01,VAA01,MM01,CB02,EWAL02,blau,MoskPRB(02),MoskPRB(02)b}.
Ratchets are another example of systems where the combined action
of nonlinearity, noise and asymmetry also produces current
rectification \cite{magn93,bart94,zapa96}. Here we focus on
coherent quantum pumping.

In general, coherent quantum pumping appears when traversal paths
of different energy interfere in the presence of an oscillating
scatterer. A complete description of quantum pumps, both adiabatic
and nonadiabatic, in terms of Floquet scattering theory
\cite{WagnerPRB(94),WagnerPRA(95),PedPRB(98),ShirPR(65),LR99} has
been put forward by Moskalets and B\"uttiker in
Ref.~\cite{MoskPRB(02)b}. As we have described in subsection
\ref{Scattering approach}, the Floquet scattering theory deals
with the scattering matrix dependent on two energies (incident and
outgoing). The matrix element $S_{\alpha\beta}(E_n,E)$, with $E_n
= E + n\hbar\omega$, is the quantum mechanical amplitude for an
electron with energy $E$ entering the scatterer through lead
$\beta$ to leave the scatterer through lead $\alpha$ having
absorbed ($n > 0$) or emitted ($n < 0$) energy quanta. $\alpha$,
$\beta$,  number the leads connecting the sample to $N_{r}$
reservoirs. Thus, all the quantities of interest are expressed in
terms of the side bands \cite{BLPRL(82)} corresponding to
particles which have gained or lost one or several modulation
quanta $\hbar \omega$.

In particular, by expressing the annihilation operator for
outgoing particles in the lead $\alpha$ in terms of annihilation
operators for incoming particles in leads $\beta$, $ \hat
b_{\alpha}(E) = \sum_\beta \sum_{E_n>0} S_{\alpha\beta}(E,E_{n})
\hat a_{\beta}(E_n)$ \footnote{$\sum_{E_n>0}$ means a sum over
those $n$ (positive and negative) for which $E_n = E +
n\hbar\omega > 0$ (the negative values $E_n < 0$ correspond to
bound states near the oscillating scatterer which do not directly
contribute to the current) \cite{MoskPRB(02)b}.}, the distribution
function for electrons leaving the scatterer through the lead
$\alpha$, $f^{(out)}_{\alpha}(E)$, can be related with the
distribution function for electrons entering the scatterer through
lead $\beta$, $f^{(in)}_{\beta}(E_n)$, as:
\begin{equation}
f^{(out)}_{\alpha}(E) = \sum\limits_\beta \sum\limits_{E_n>0}
|S_{\alpha\beta}(E,E_{n})|^2 f^{(in)}_{\beta}(E_n).
\label{distri-funct}
\end{equation}
Using Eq.~(\ref{distri-funct}), the current directed from the
scatterer towards the reservoir
\begin{equation}
 I_{\alpha} = \frac{e}{h} \int\limits_{0}^{\infty} dE
[f^{(out)}_{\alpha}(E) - f^{(in)}_{\alpha}(E)]
\label{current-moskalets1}
\end{equation}
can be rewritten as:
\begin{equation}
 I_{\alpha} = \frac{e}{h} \int\limits_{0}^{\infty} dE
 \sum\limits_\beta \sum\limits_{E_n>0} |S_{\alpha\beta}(E_{n},E)|^2
[f^{(in)}_{\beta}(E) - f^{(in)}_{\alpha}(E_{n})].
\label{current-moskalets2}
\end{equation}
By using current conservation,
$\sum_\alpha \sum_{E_n>0} |S_{\alpha\beta}(E_n,E)|^2 = 1$,
Eq.~(\ref{current-moskalets2}) can be expressed in a very useful
representation:
\begin{eqnarray}
 I_{\alpha} &=& \frac{e}{h} \int\limits_{0}^{\infty} dE
 \sum\limits_{\beta\neq\alpha} \sum\limits_{E_n>0}
[|S_{\alpha\beta}(E_{n},E)|^2 f^{(in)}_{\beta}(E)\nonumber\\
&-& |S_{\beta\alpha}(E_{n},E)|^2 f^{(in)}_{\alpha}(E)].
\label{current-moskalets3}
\end{eqnarray}

In a typical pump setup, incoming electrons in all the channels
can be described by the same Fermi distribution function $f(E)$
(the electrochemical potential $\mu$ of the incoming electrons is
the same throughout the whole structure) such that the pumped
current is:
\begin{eqnarray}
 I_{\alpha} &=& \frac{e}{h} \int\limits_{0}^{\infty} dE
 \sum\limits_{\beta\neq\alpha} \sum\limits_{E_n>0}
f(E)[|S_{\alpha\beta}(E_{n},E)|^2 - |S_{\beta\alpha}(E_{n},E)|^2
]. \label{current-moskalets4}
\end{eqnarray}
The existence of the pump effect is thus directly related to the
symmetry of the scattering problem. By considering all possible
symmetries, Moskalets and Buttiker conclude that in the adiabatic
case ($\omega\to 0$) only a scatterer without spatial and time
reversal symmetry can produce a directed current. On the other
hand in the nonadiabatic case (at large pumping frequency) to
achieve pumping it is necessary to have a scatterer with either
broken spatial or time reversal symmetry.

\begin{figure}
\begin{center}
\includegraphics[width=0.75\columnwidth,angle=0]
{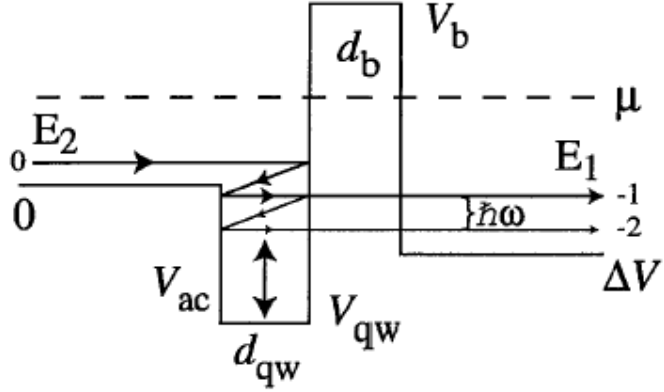}
\end{center}
\caption{Schematic potential profile for a Fermi-sea pump. The
chemical potential $\mu$ is the same in both contacts. Reprinted
with permission from \cite{WagPRL(99)}. \copyright 1999 American
Physical Society.} \label{Wagner-Sols}
\end{figure}

As an example, let us now consider the effect of time reversal
symmetry (TRS) on the pumped current \footnote{the time reversal
$t\to -t$ interchanges incoming and outgoing channels $\left[
S_{\alpha\beta}(E_{n},E)\right]^{(TR)}
=S_{\beta\alpha}(E,E_{n})$.}: in the presence of TRS,
Eq.~(\ref{current-moskalets4}) reads:
\begin{equation}
 I_{\alpha}^{(TRS)} = \frac{e}{h} \int\limits_{0}^{\infty} dE
f(E)
 \sum\limits_{\beta\neq\alpha} \sum\limits_{E_n>0}
\left( |S_{\alpha\beta}(E_{n},E)|^2 - |S_{\alpha\beta}(E,E_{n})|^2
\right). \label{current-moskalets5}
\end{equation}
If $S_{\alpha\beta}(E_{n},E) \neq S_{\alpha\beta}(E,E_{n})$ the
pump generates a current. Based on this principle, Wagner and Sols
proposed in Ref.~\cite{WagPRL(99)} a pump in which the current is
carried deep within the Fermi sea. An schematic potential profile
of this pumping device is shown in Fig.~\ref{Wagner-Sols}.
In this device, a quantum well is driven harmonically by an
external ac potential $V_{\rm ac}\cos\omega t$. Adjacent to the
well is a static barrier, and the overall potential profile
features a band offset $\Delta V$ between the left and right
leads. The chemical potential is the same throughout the whole
structure such that any dc current flowing is thus due to the
combined effect of the driving ac force and the spatial asymmetry
of the model. The current can thus be written as:
\begin{eqnarray}
I &=& \int_{\Delta V}^\infty dE f(E-\mu) J(E)
\label{eq:Current_Wagner-Sols}\\
J(E) &=& \frac{2e}{h} \int_{\Delta V}^{E} dE_z D_{\bot}(E-E_z)
T_{\rm net}(E_z). \label{Spectral_Wagner-Sols}
\end{eqnarray}
where $D_{\bot}$ is the density of states in the dimensions
perpendicular to the direction of transport and $T_{\rm net}
\equiv T_{\rm \rightarrow} - T_{\rm \leftarrow}$ where
\begin{eqnarray}
T_{\rm \rightarrow}(E_z) = \int dE'_z T_{\rm LR}(E_z,E'_z)
\quad,\quad T_{\rm \leftarrow}(E_z)  = \int dE'_z T_{\rm
RL}(E_z,E'_z).
        \label{eq:Tnet_Wagner-Sols}
\end{eqnarray}
The pumped current in Eq.~(\ref{eq:Current_Wagner-Sols}) can be
numerically obtained by employing the transfer-matrix technique
\cite{WagPRL(99),SolsAnnPhys(00)}. Interestingly, it is shown that
the pumped current is carried by electrons which, for sufficiently
high Fermi energies, may stay well below the Fermi surface,
thereby rendering the total current insensitive to temperature.
This remarkable effect can be explained in terms of the {\it
pipelines} displayed by the total transmission probability of the
device. Pipelines are pairs of left and right scattering channels,
of energies $E_2= E_1+ \hbar\omega$, that are strongly coupled.

Indeed, the minimal model, that consisting of a single pipeline
\cite{WagPRL(99),SolsAnnPhys(00)}, reproduces analytically the
main features obtained from the full numerical calculation of the
transfer matrix equations. Let us consider  a single pipeline of
strength $T_p$ connecting the energies $E_1$ on the right and
$E_2$ to the left. Assuming incident electrons approaching the
device outside these two channels to be reflected with unit
probability, the transmission probabilities read:
\begin{eqnarray}
T_{\rm LR}(E_z,E'_z) &=&
           T_p \delta(E_z-E_2) \delta(E'_z-E_1) \label{TLR} \\
T_{\rm RL}(E_z,E'_z) &=&
           T_p \delta(E_z-E_1) \delta(E'_z-E_2), \label{TRL}
\end{eqnarray}
such that
\begin{eqnarray}
        T_{\rm net}(E_z) =
           T_p \left[\delta(E_z-E_2) - \delta(E_z-E_1)\right] .
        \label{eq:PipelineModel}
\end{eqnarray}
In the single-pipeline model, the pumped current in
Eq.~(\ref{eq:Current_Wagner-Sols}) reads:
\begin{equation}
I=\frac{2e}{h}T_p \int_{0}^{\infty} dE_{\bot} D_{\bot}(E_{\bot})
[f(E_{\bot}+E_2-\mu)-f(E_{\bot}+E_1-\mu)] .
\label{current-pipeline}
\end{equation}

For one spatial dimension, Eq.~(\ref{current-pipeline}) translates
into
\begin{eqnarray}
        I_{\rm 1D} = \frac{2e}{h} T_p
              \left[f(E_2-\mu) - f(E_1-\mu)\right] ,
        \label{eq:Current-1D}
\end{eqnarray}
which has a peak at $\mu$ = $(E_1+E_2)/2$, and an {\it
exponential} decay for $\mu$ $\gg$ $k_BT$. In 2D,
\begin{eqnarray}
        I_{\rm 2D} \approx {2 e \over h^2} T_p \sqrt{\frac{2 \pi m}{k_BT}}
             \Bigl[
                 {\rm Li}_{-{1\over 2}}(-{\rm e}^{\frac{\mu}{k_BT}})
                        (E_2-E_1)
          -{\frac{1}{2k_BT}} {\rm Li}_{-{3\over 2}}(-{\rm e}^{\frac{\mu}{k_BT}})
                        (E_2^2-E_1^2)
              \Bigr],\nonumber\\
        \label{eq:Current-2D}
\end{eqnarray}
where Li is the polylogarithm function. Expanding for $\mu$ $\gg$
$k_BT$, one finds that in 2D the pump current decays only {\it
algebraically} as $1/\sqrt{\mu}$. Finally, in 3D
\begin{eqnarray}
        I_{\rm 3D} \approx {4\pi m e \over h^3} T_p
              \left[f(-\mu)(E_1-E_2)
          + {f'(-\mu)\over 2} (E_1^2-E_2^2)\right] .
        \label{eq:Current-3D}
\end{eqnarray}
For $\mu$ $\gg$ $k_BT$ one has $f(-\mu)$ $\approx$ 1 and
$f'(-\mu)$ $\approx$ 0, i.e., the current in 3D becomes {\it
independent} of $\mu$ in this limit, $I_{\rm 3D}$ = $-(4\pi m
e/h^3) T_p (E_2-E_1)$. It is interesting to analyze the spectral
function (\ref{Spectral_Wagner-Sols}) that leads to the pump
current. Within the single--pipeline model, one obtains
\begin{equation}
J(E)=\frac{2e}{h} T_p [D_{\bot}(E-E_2)\theta(E-E_2)-
D_{\bot}(E-E_1)\theta(E- E_1)]. \label{spectral-pipeline}
\end{equation}
In the particular case of three dimensions,
$D_{\bot}(E_{\bot})=2\pi m/h^2 \equiv D_0$ and
Eq.~(\ref{spectral-pipeline}) yields a square function localized
between $E_2$ and $E_1$. The total current is a convolution of
$J(E)$ with a thermal population of incoming electrons, see
Eq.~(\ref{eq:Current_Wagner-Sols}), such that for $\mu \gg E_2$
the pump current is sustained by scattering states with incident
energy well below the Fermi surface. As a consequence the current
in this regime {\it is insensitive to temperature}, even for $k_BT
\sim \hbar\omega$.
\section{Photon assisted tunneling in quantum dots I: Coulomb blockade regime \label{quantumdotsI}}
In this part we discuss electron transport on semiconductor
quantum dots that are driven by microwaves. In particular, we
shall focus on lateral quantum dots \footnote{Other
configurations, also termed quantum dots in the literature,
include nanocristals, self-assembled quantum dots and vertical
quantum dots. For a review, see
Refs.~\cite{AliviSci(96),review-dots}.}. The starting point for
these devices is a two-dimensional electron gas (2DEG) at the
interface of a semiconductor heterostructure (typically
GaAs/AlGaAs). To define the quantum dot, metallic gates are
patterned on the surface of the wafer by means of electron-beam
lithography. Negative voltages applied to the metallic surface
gates deplete the 2DEG underneath, defining a small confined
region (the quantum dot) with a typical size of $\sim 100nm$. The
resulting dot contains a few electrons and is coupled to the large
2DEG regions (electron reservoirs) by tunnel barriers. These kind
of systems are very suitable for quantum transport studies because
of their tunability (level spacing, charging energy, barriers,
etc). One can estimate the charging energy $E_C$ (energy needed to
add an extra electron to the system, see below) and the level
spacing $\Delta\varepsilon$ from the dimensions of the dot.
Typical values are $E_C\sim 1meV$ and $\Delta\varepsilon\sim
0.1-0.01meV$. Although the typical number of electrons in lateral quantum
dots is of the order of hundred, the experimental challenge of realizing
few-electrons quantum dots in lateral geometries has
been recently achieved by the groups of Ottawa \cite{CiorgaPRB(00)} and Delft \cite{ElzerPRB(03)}.

Transport through a quantum dot occurs when the Fermi energy of
the leads is aligned with one of the discrete energy levels of the
confined region. This resonant current, due to elastic tunneling
of electrons between the leads and the dot, is strongly modified
in the presence of microwaves: when an additional time-dependent
potential $eV_{ac}cos(\omega t)$ is applied to the central gate,
the electrons can exchange photons of energy $\hbar\omega$ with
the external field with typical experimental frequencies
$f=\omega/2\pi$ which range from 1-75GHz. These inelastic
tunneling processes, namely photon assisted tunneling, lead to
drastic changes in the dc transport through these devices
\cite{KouPRB(94),KouPRL(94),BliAPL(95),FujSM(97),OosPRL(97)}.

At zero temperature (and neglecting cotunneling) transport occurs
if the electrochemical potential of the quantum dot $\mu_{dot}(N)$
lies between the electrochemical potentials of the reservoirs
$\mu_{left}$ and $\mu_{right}$, where $\mu_{right}-\mu_{left}$ is
the applied bias voltage $V_{sd}$.
\begin{figure}
\begin{center}
\includegraphics[width=0.75\columnwidth,angle=0]{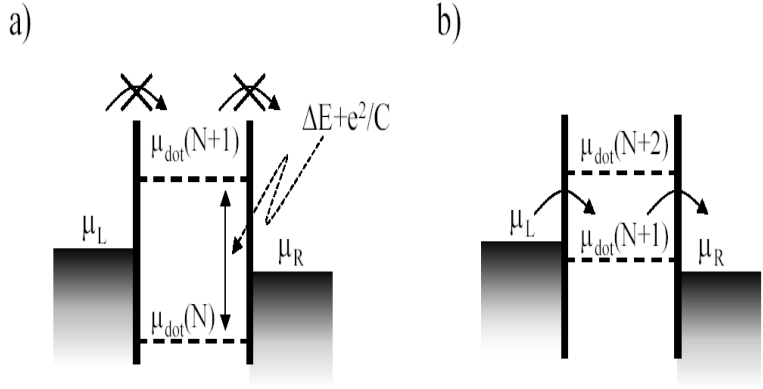}
\end{center}
\caption[]{Schematic diagram of the energy profile of a quantum
dot} \label{scheme1}
\end{figure}
The electrochemical potential of the dot is, by definition, the
minimum energy needed to add the Nth electron to the dot:
$\mu_{dot}(N)=U(N)-U(N-1)$, where $U(N)$ is the total ground state
energy for N electrons on the dot at zero temperature. Apart from
the quantization of the energy levels, the confinement leads to
charge quantization if $R_t>>h/e^2$, where $R_t$ is the tunnel
resistance of the barriers and $h/ e^2=25.813k\Omega$ is the
resistance quantum. This charge quantization makes it essential to
take into account Coulomb interactions when calculating the ground
state energy of a quantum dot.

The simplest model taking into account charge quantization for
describing transport is the Coulomb blockade model. This model
parametrizes the Coulomb interaction by means of a capacitance
$C=C_L+C_R+C_g$, i.e the sum of the capacitances of the barriers
and the capacitance of between the dot and the gate, such that
\begin{equation}
\label{electrochemical1}
\mu_{dot}(N)=E_N+\frac{(N-N_0+1/2)e^2}{C}-e\frac{C_g}{C}V_g.
\end{equation}
$E_N$ is the total energy of N independent electrons. When at
fixed gate voltage $V_g$ the number of electrons changes by one,
the change in electrochemical potential is:
\begin{equation}
\label{electrochemical2}
\mu_{dot}(N+1)-\mu_{dot}(N)=E_{N+1}-E_N+\frac{e^2}{C}\equiv\Delta
E+ E_C.
\end{equation}
The {\it addition energy} $\mu_{dot}(N+1)-\mu_{dot}(N)$ consists
of two terms: a purely electrostatic part $E_C$, which is large
for a small capacitance, and the energy spacing between two
discrete quantum levels. Note that if two electrons are added to
the same spin degenerate level $\Delta E=0$. At low temperatures,
$E_C>>k_BT$, the charging energy dominates transport. When
$\mu_{dot}(N)<\mu_{left},\mu_{right}<\mu_{dot}(N+1)$ the electron
transport is blocked, namely the quantum dot is in the {\it
Coulomb Blockade regime} (see Fig.~\ref{scheme1}a).
\begin{figure}
\begin{center}
\includegraphics[width=0.75\columnwidth,angle=0]{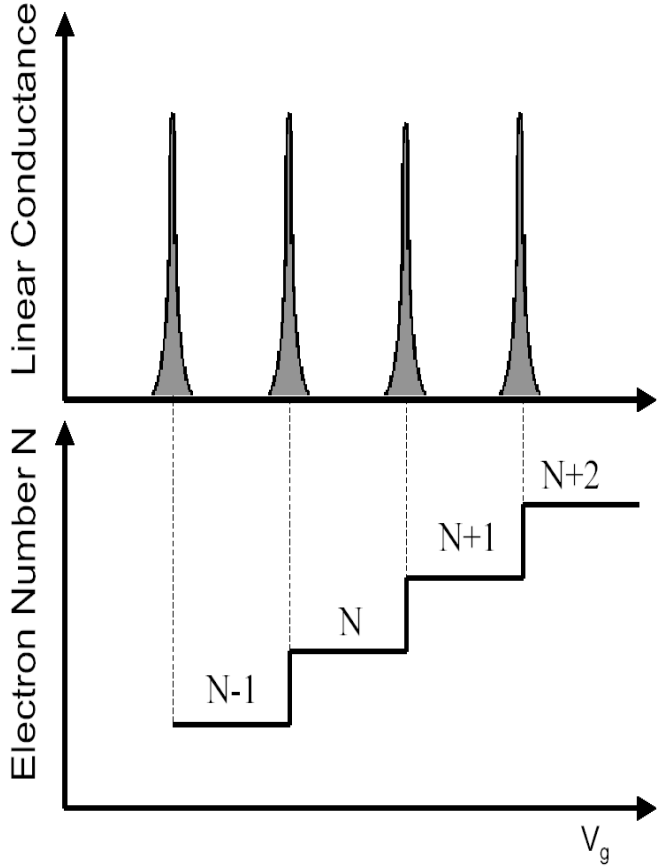}
\end{center}
\caption[]{Schematic diagram of the Coulomb oscillations of the
conductance as a function of the gate voltage (top figure), and
the number of electrons inside the dot (bottom figure). Note that
in the regions of Coulomb blockade (zero conductance) the electron
number is fixed} \label{QD_fig0}
\end{figure}
The Coulomb blockade can be removed by changing the gate voltage,
to align $\mu_{dot}(N+1)$ between the chemical potentials of the
reservoirs (Fig.~\ref{scheme1}b) such that an electron can tunnel
from the left reservoir to the dot and from the dot to the right
reservoir, which causes the electrochemical potential to drop back
to $\mu_{dot}(N)$. A new electron can enter now the dot such that
the cycle $N\rightarrow N+1\rightarrow N$ is repeated. This
process is called {\it single electron tunneling}. By changing the
gate voltage, the linear conductance oscillates between zero
(Coulomb blockade) and non-zero. In the regions of zero
conductance, the number of electrons inside the quantum dot is
fixed (see Fig.~\ref{QD_fig0}).

Assuming sequential tunneling of single electrons, the current can
be calculated with a master equation approach \cite{master} or by
means of nonequilibrium Green´s function techniques
\cite{MeiPRL(92)} (see section \ref{Keldysh}). Here, we describe
the master equation approach which is probably the simplest method
that allows for a qualitative explanation of the Coulomb
oscillations of the conductance. The master equation method
generalizes the "orthodox theory" \cite{SET} for SET in metal
systems to include 0D-states.
\begin{figure}
\begin{center}
\includegraphics[width=0.75\columnwidth,angle=0]{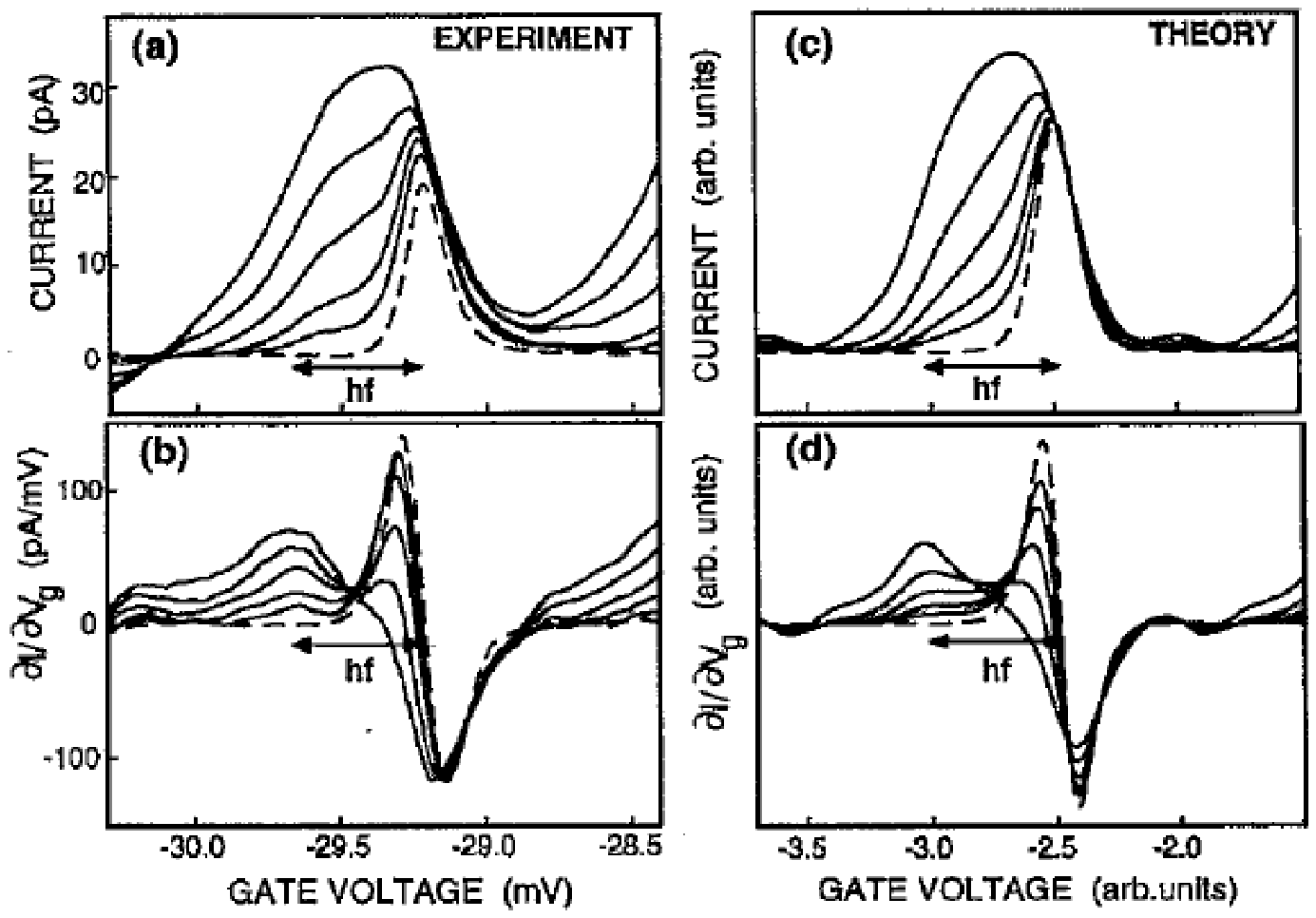}
\end{center}
\caption[]{Comparison between experimental curves and the Tien and
Gordon theory for photon-assisted tunneling through a single
quantum dot with $f=27GHz$. The only ajusted parameters in the
theoretical curves are the ac amplitudes. Reprinted with
permission from \cite{KouPRL(94)}. \copyright 1994 American
Physical Society.} \label{QD_fig1}
\end{figure}
\begin{figure}
\begin{center}
\includegraphics[width=0.75\columnwidth,angle=0]{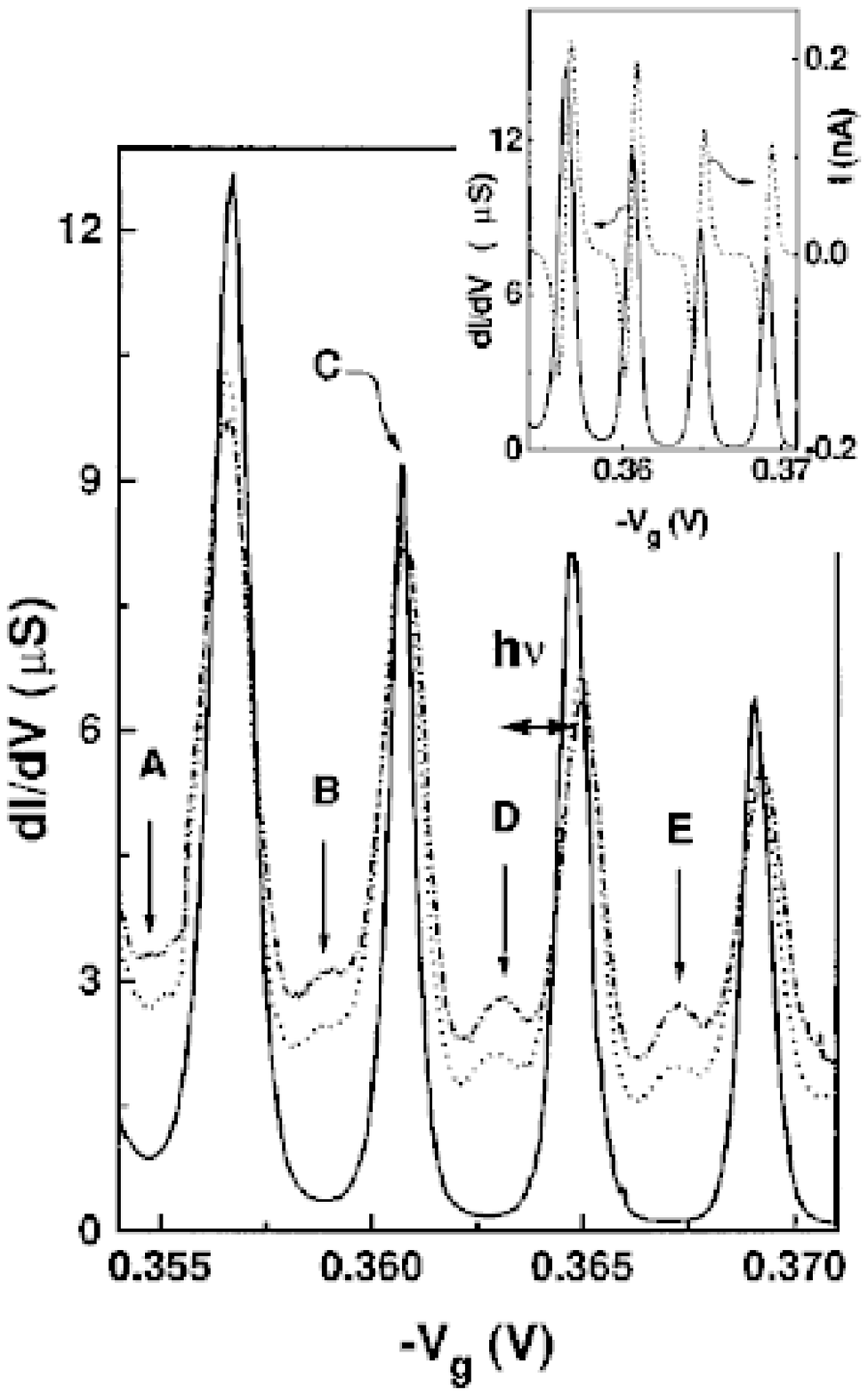}
\end{center}
\caption[]{Coulomb oscillations in the presence of microwaves
$f=155GHz$ with increasing power. The only ajusted parameters in
the theoretical curves are the ac amplitudes. Reprinted with
permission from \cite{BliAPL(95)}. \copyright 1995 American
Institute of Physics.} \label{blick}
\end{figure}
The Coulomb oscillations are modified by the application of
microwaves to the gate voltage. This effect can be included in a
master equation that takes into account Coulomb blockade and
photon-assisted tunneling by writing the tunnel rate through each
barrier in the presence of microwaves $\tilde{\Gamma}(E)$ in terms
of the rates without microwaves $\Gamma(E)$ in the Tien and Gordon
spirit \cite{LikPhysB(94),HadPhysB(94),BruPRL(94)}:
\begin{equation}
\label{rate} \tilde{\Gamma}(E)=\sum_{n=-\infty}^{\infty}
J_n^2(\beta)\Gamma(E+n\hbar\omega).
\end{equation}
The modified rate $\tilde{\Gamma}(E)$, reflects the appearance of
new channels for transport, the so-called photon sidebands, which
correspond to emission and absorption processes.
\begin{figure}
\begin{center}
\includegraphics[width=0.75\columnwidth,angle=0]{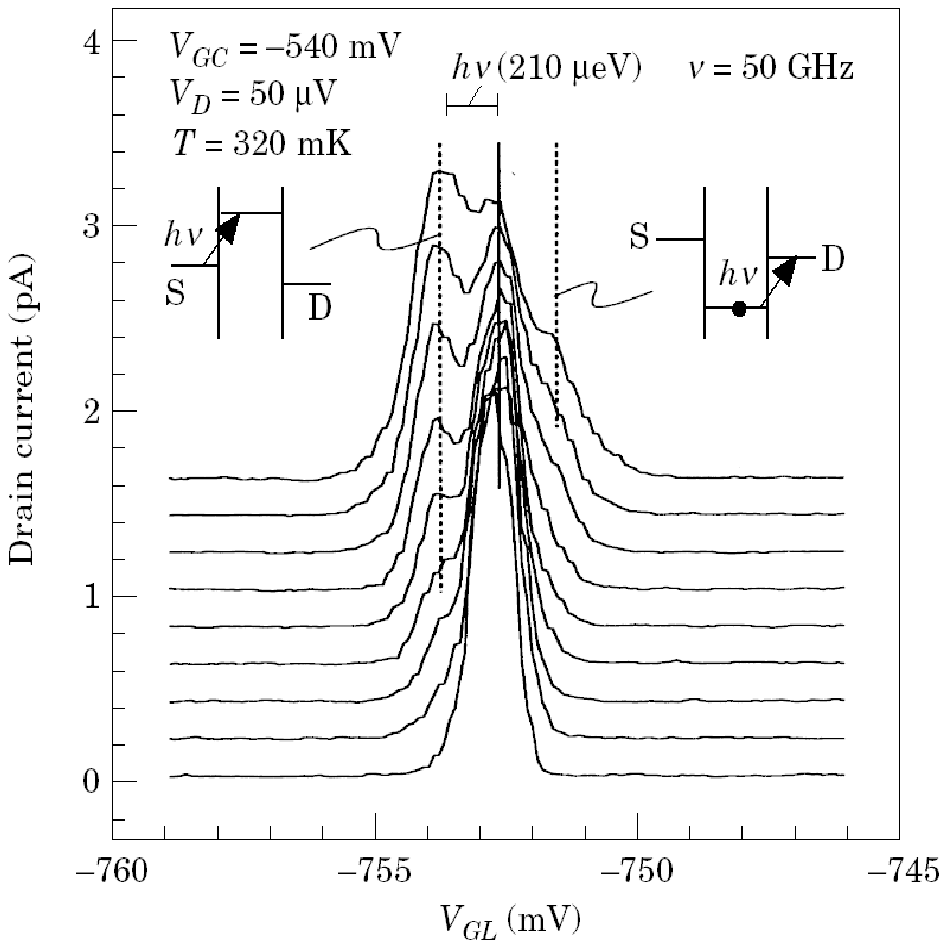}
\end{center}
\caption[]{Photon-assisted tunneling current through a quantum dot
as a function of the gate voltage for different microwave powers.
Each curve is offset by 0.2 pA for clarity. Reprinted with
permission from \cite{FujSM(97)}. \copyright 1997 Academic Press
Limited.} \label{toshiPAT1}
\end{figure}
In the master equation, one has to keep track of the particular
occupation of the electrons in the single particle levels
${\varepsilon_l}$ for a given number of electrons $N$. If the $N$
electrons are distributed over $n$ levels, the number of different
configurations $M$ is given by a binomial factor ${n \choose N}$.
Assuming $E_C>>\Delta\varepsilon,k_{BT},V_{sd}$, only two charge
states, $N,N+1$ have to be taken into account such that the
probability $P_{N,M}$ for state $(N,M)$ is calculated from the set
of master equations:
\begin{eqnarray}
\label{master} \dot{P}_{N,M}&=&\sum_{\alpha\in
L,R}\Big\{\sum_{M'}P_{N+1,M'}\tilde{\Gamma}_{\alpha,i_{M'}}^{out}-
P_{N,M}\sum_{l=empty}\tilde{\Gamma}_{\alpha,l}^{in}\Big\}
\nonumber\\&+&\sum_{M''\neq M}P_{N,M''}\Gamma_{M''\rightarrow
M}-P_{N,M}\sum_{M'''\neq M}\Gamma_{M\rightarrow M'''},
\end{eqnarray}
and a similar equation for $\dot{P}_{N+1,M'}$. In the stationary
limit $\dot{P}\rightarrow 0$, which together with the boundary
condition $\sum_M P_{N,M}+\sum_{M'} P_{N+1,M'}=1$, close the set
of equations to be solved.
\begin{figure}
\begin{center}
\includegraphics[width=0.75\columnwidth,angle=0]{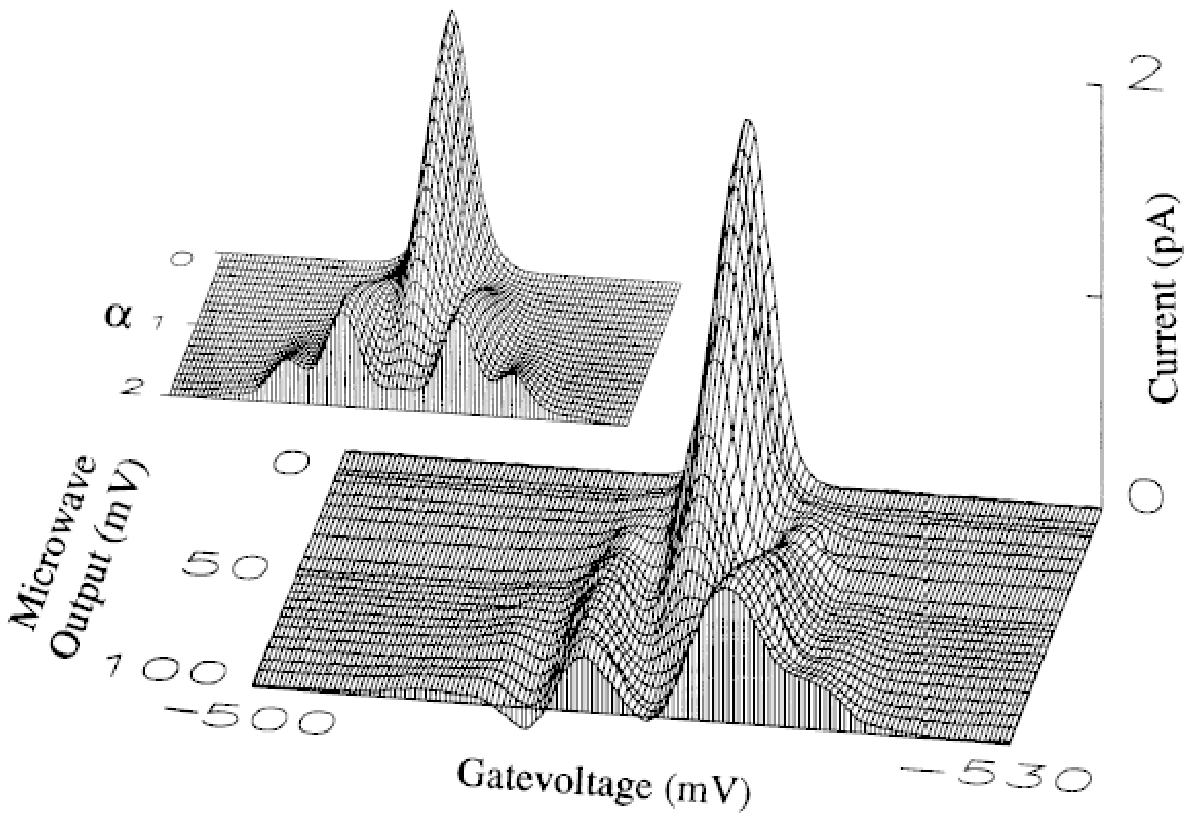}
\end{center}
\caption[]{Current through a quantum dot as a function of the gate
voltage and the microwave output. Parameters:
$\Delta\varepsilon=165\mu eV$, $hf=110\mu eV$ corresponding to
$f=27GHz$. The data are taken at $V_{sd}=13\mu V$ and a magnetic
field of $B=0.84T$. The inset shows a calculation for the same set
of parameters. Reprinted with permission from
\cite{OosThesis(99)}.} \label{Oost_thesis}
\end{figure}
The first term of Eq. (\ref{master}) describes tunneling out of
states $i_{M'}$ that leave the dot in $(N,M)$. The second term
describes tunneling onto the dot and all the empty states have to
be taken into account. These processes are described by the rates:
\begin{eqnarray}
\tilde{\Gamma}_{\alpha,l}^{in}(\varepsilon_l)&=&\Gamma_{\alpha,l}\sum_{n=-\infty}^{\infty}J_n^2(\beta)
f(\varepsilon_l-\frac{C_g}{C}eV_g+n\hbar\omega+e\eta_\alpha V_{sd})\nonumber\\
\tilde{\Gamma}_{\alpha,i_{M'}}^{out}(\varepsilon_l)&=&\Gamma_{\alpha,l}\sum_{n=-\infty}^{\infty}J_n^2(\beta)
[1-f(\varepsilon_l-\frac{C_g}{C}eV_g+n\hbar\omega+e\eta_\alpha
V_{sd})],
\end{eqnarray}
where $\Gamma_{\alpha,l}$ are the tunneling rates through the left
and right barrier evaluated at energy $\varepsilon_l$, $f(E)$ are
the Fermi functions of the reservoirs and the coefficients
$\eta_\alpha$ describe the asymmetry of the dc voltage drop across
each barrier.
\begin{figure}
\begin{center}
\includegraphics[width=0.75\columnwidth,angle=0]{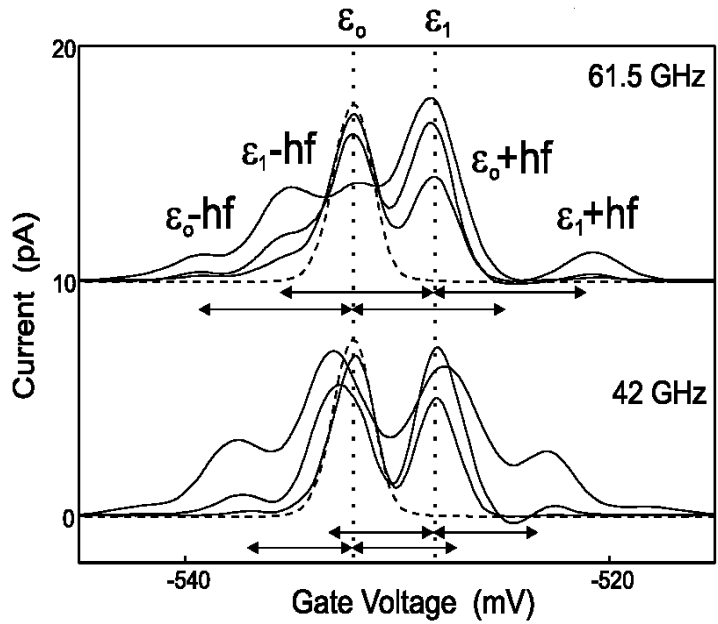}
\end{center}
\caption[]{Current through a quantum dot as a function of the gate
voltage for different microwave powers (dashed curve without
microwaves). Top figure $f=61.5GHz$, bottom figure $f=42GHz$. As
the microwave power increases a new resonance corresponding to the
excited state $\varepsilon_1$ emerges. Reprinted with permission
from \cite{OosPRL(97)}. \copyright 1997 American Physical
Society.} \label{QD_fig4}
\end{figure}
\begin{figure}
\begin{center}
\includegraphics[width=0.75\columnwidth,angle=0]{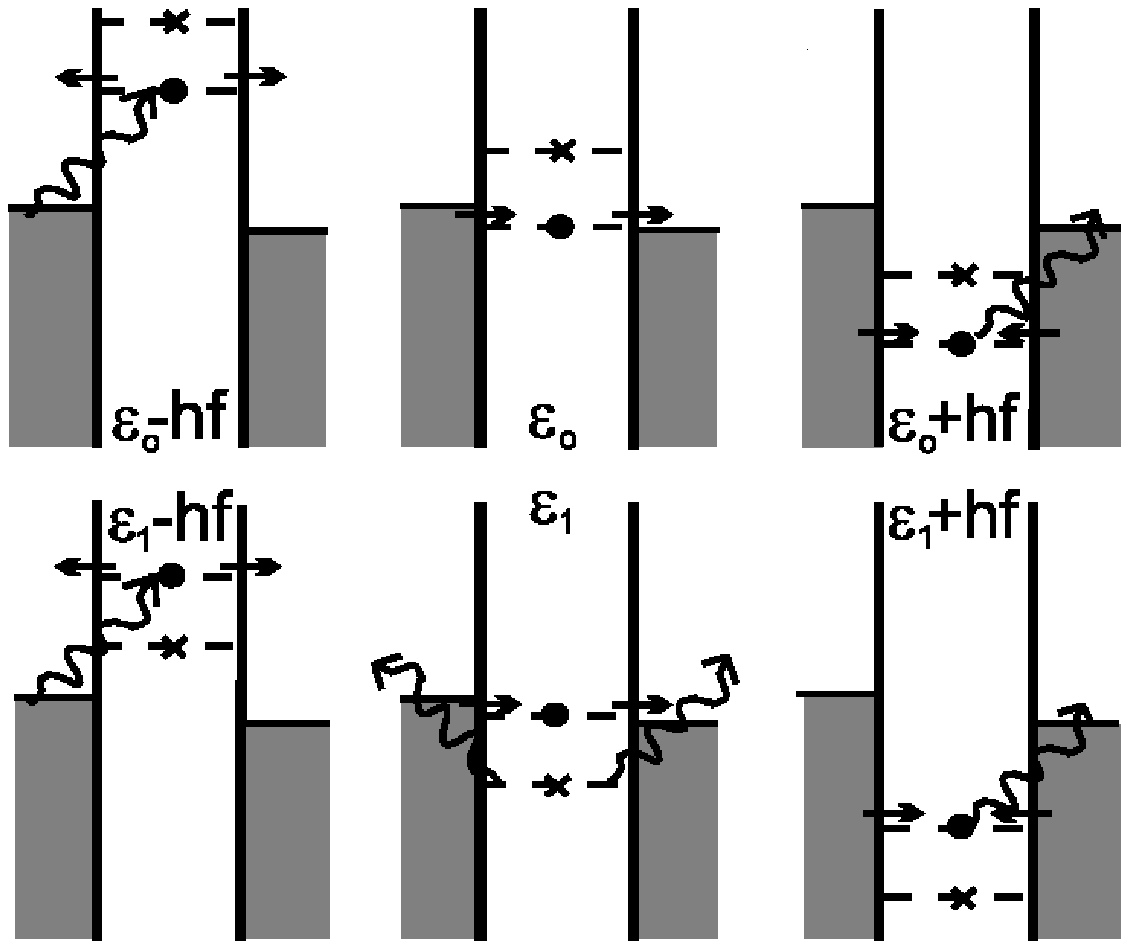}
\end{center}
\caption[]{Tunneling events that contribute to sequential
transport when $\Delta\varepsilon <\hbar\omega$. Reprinted with
permission from \cite{OosPRL(97)}. \copyright 1997 American
Physical Society.} \label{QD_fig3}
\end{figure}
The last two terms of Eq. (\ref{master}) describe relaxation and
excitation processes where the number of electrons inside the dot
remains fixed. The dc current can be calculated, for instance,
from the net tunneling rate through the left barrier:
\begin{eqnarray}
I=e\Big [\sum_M \sum_{l=empty} P_{N,M} \tilde{\Gamma}_{L,l}^{in}
-\sum_{M'} \sum_{l=full} P_{N+1,M'} \tilde{\Gamma}_{L,l}^{out}\Big
]
\end{eqnarray}

We mention in passing that Eq. (\ref{rate}) is a particular
example of a general description of the effects of an external
fluctuating environment on single-electron tunneling. In general,
\begin{equation}
\label{rate2}
\tilde{\Gamma}(E)=\int_{-\infty}^{\infty}d(\hbar\omega)P(\hbar\omega)
\Gamma(E+\hbar\omega),
\end{equation}
where $P(\hbar\omega)$ is the espectral density characterizing the
environment. Eq. (\ref{rate2}) describes a modified rate in the
presence of an environment with fluctuations which are broad band
in frequency: finite impedance of the leads \cite{Ing(92)},
phonons \cite{FujSci(98)}, quantum noise \cite{AguPRL(98)}, etc.
Eq. (\ref{rate}) is thus a particular example describing a
monochromatic environment.

Many of the experiments on photon-assisted tunneling in quantum
dots can be explained in terms of the Tien-Gordon theory. An
example is shown in Fig.~\ref{QD_fig1} where we plot
 a comparison between experimental and theoretical data by
Kouwenhoven et al \cite{KouPRL(94)}. In this experiment, the
Coulomb blockade peaks develop shoulder structures in the presence
of the microwave signal. Later, Sun and Lin anayzed in detail
these experiments \cite{SunPRB(97)} by using the nonequilibrium
Green's functions technique (see section \ref{Keldysh}).  They
concluded that the shoulder structure can be explained if one
assumes a strong asymmetry of the applied ac signal (their results
are in good agreement with the experiments of
Ref.~\cite{KouPRL(94)} when the ac potential is applied only to
one lead). Similar experiments at higher frequencies were
performed by Blick et al in Ref.~\cite{BliAPL(95)} (see
Fig.~\ref{blick}). The position of the shoulder in the
photon-assisted tunneling curves of Fig.~\ref{QD_fig1} is
independent of power and shifts linearly with frequency, which
unambigously indicates photon-assisted tunneling. In this
experiment the effective density of states of the dot is
continuous and there is no evidence of the 0D-states.

Later experiments \cite{FujSM(97),OosPRL(97),OosThesis(99)}
demonstrated that it is possible to perform transport spectroscopy
through 0D states by studying photon-assisted tunneling on smaller
dots.
If $\Delta\varepsilon >>\hbar\omega$, microwave frequencies
smaller than the average level spacing, the transport occurs
through only a single level. In this case, at both sides of the
main peak, sidebands do develop at multiples of $\hbar\omega$
corresponding to the emission and absorption of photons. An
example from the experiment by Fujisawa and Tarucha
\cite{FujSM(97)} is shown in Fig.~\ref{toshiPAT1} where the
sidebands due to photon-assisted tunneling can be clearly resolved
as one increases the microwave power at fixed frequency. The
amplitude of the {\it nth} sideband, namely the probability of
absorption and emission of the {\it nth} photon, is given by
$J_n^2(\beta\equiv\frac{eV_{ac}}{\hbar\omega})$, i.e. the
probability changes nonlinearly when the microwave power is
increased, as predicted by the rates of Eq.(\ref{rate}). A
systematic study of this power dependence was performed by
Oosterkamp et al \cite{OosThesis(99)}. We show in
Fig.~\ref{Oost_thesis} one of their experimental curves where the
current as a function of the gate voltage, for different microwave
powers and fixed frequency ($\hbar\omega=110\mu eV$, $f=27GHz$ and
$\Delta\varepsilon =165\mu eV$), is plotted. For comparison, a
calculation of the current using Eq.~(\ref{master}) (and assuming
equal ac voltages across each barrier) is also shown. The
agreement between experiment and theory is very good, the small
difference bewteen both curves can be explained by taking into
account the asymmetry in the ac voltage drop across each barrier.

When $\Delta\varepsilon <\hbar\omega$, photon-assisted tunneling
can induce current through excited states such that new peaks at
gate voltages given by $(m\Delta\varepsilon+n\hbar\omega)$ do
appear. These kind of experiments have been performed by
Oosterkamp et al \cite{OosPRL(97)} where they show that photon
assisted tunneling can lead to transport through excited states in
a process similar to the photo-ionization of atoms. In the
presence of microwaves a new resonance appears on the right side
of the main resonance (Fig.~\ref{QD_fig4}). This can be explained
as photo ionization of the quantum dot followed by tunneling
through the first excited state as illustrated in
Fig.~\ref{QD_fig3}. These experiments were analyzed theoretically
in Refs.~\cite{BrunePRB(97),SunPRB(98)}.
\section{Photon assisted tunneling in double quantum dots \label{doubledots}}
In the previous section we have described how photon-assisted
tunneling can be used as a powerful spectroscopic tool to extract
information about internal energy scales of a quantum dot. The
natural extension of this idea, namely performing photon-assisted
tunneling in systems consisting of two or more coupled dots has
proven to be extremely fruitful during the last years. By coupling
two quantum dots in series one defines a double quantum dot, which
can be regarded as an 'artificial molecule'.
\begin{figure}
\begin{center}
\includegraphics[width=0.75\columnwidth,angle=0]{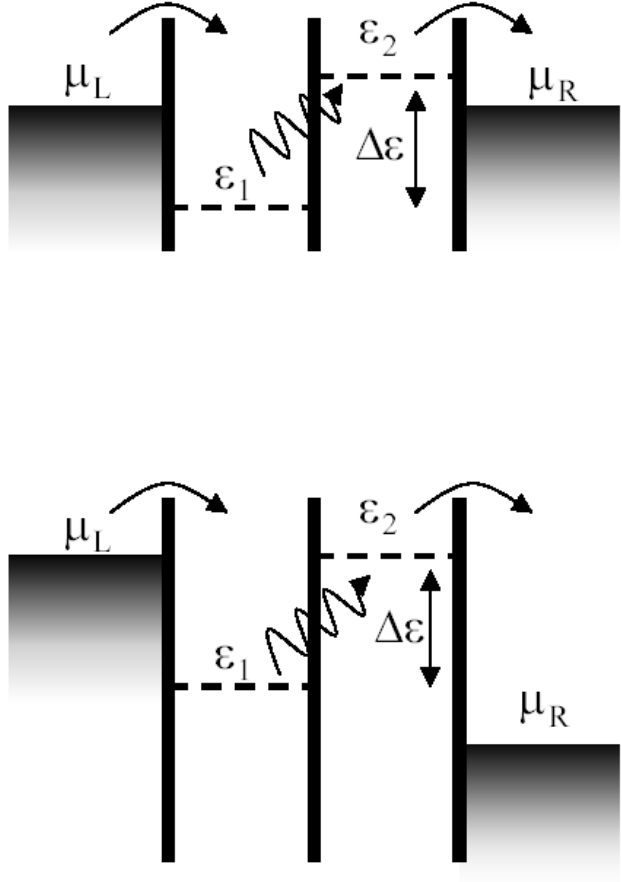}
\end{center}
\caption[]{Schematic energy diagrams of photon assisted tunneling
in double quantum dots (only one discrete level per dot is
considered, see main text). The upper diagram corresponds to the
pumping configuration where only absortion of photons contributes
to a dc current $I>0$ for $\Delta\varepsilon <0$ ($I<0$ for
$\Delta\varepsilon >0$). The lower diagram shows the large voltage
bias regime. In this case the dc current is always positive $I>0$
corresponding to absorption ($\Delta\varepsilon <0$) or emission
($\Delta\varepsilon >0$).} \label{configurations}
\end{figure}
\begin{figure}
\begin{center}
\includegraphics[width=0.75\columnwidth,angle=0]{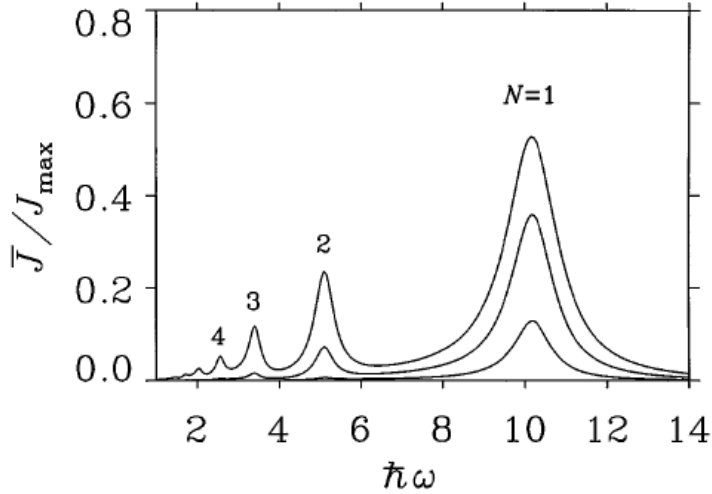}
\end{center}
\caption[]{Pumped dc current $\bar{J}$ (in units of
$J_{max}=e\Gamma/2\hbar$) through a weakly coupled double quantum
dot with $\Delta\varepsilon=-10$, $\Gamma=0.5$ and ac amplitude
$eV_{ac}=2,4,6$ (increasing currents). All energies in units of
the interdot tunneling coupling $t_C$. Reprinted with permission
from \cite{StaPRL(96)}. \copyright 1996 American Physical
Society.} \label{Staff-Wingr}
\end{figure}
\begin{figure}
\begin{center}
\includegraphics[width=0.75\columnwidth,angle=0]{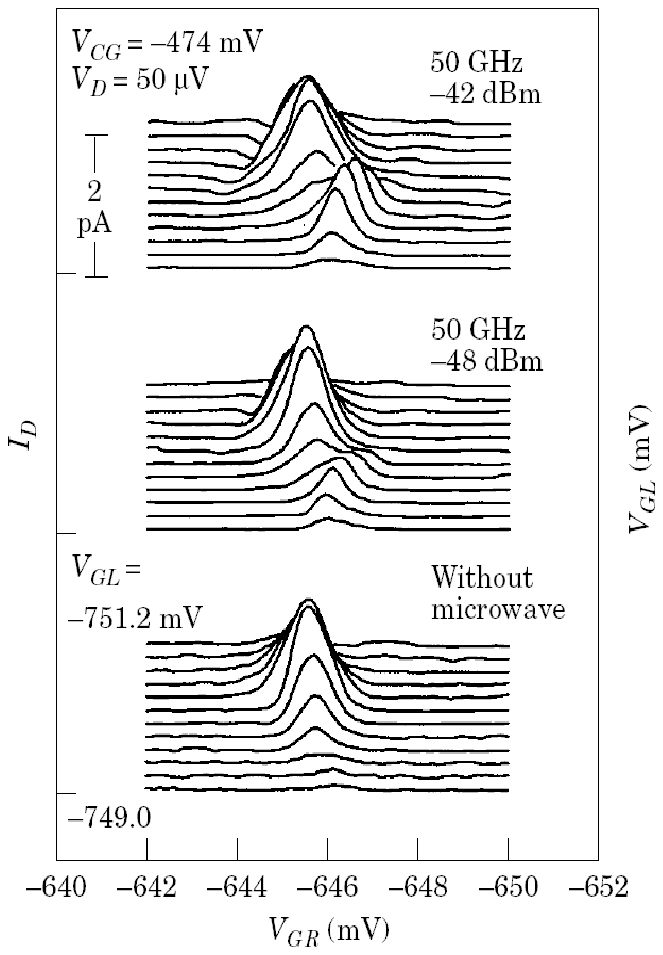}
\end{center}
\caption[]{Current vs. gate voltage. By increasing the microwave
power (from bottom to top) the photon sideband becomes apparent.
Reprinted with permission from \cite{FujSM(97)}. \copyright 1997
Academic Press Limited.} \label{toshiPAT2}
\end{figure}
Depending on how strong is the inter-dot tunneling coupling (which
can be tuned by a gate voltage), the two dots can form ionic-like
molecules (weak inter-dot tunneling coupling) or covalent-like
molecules (strong inter-dot tunneling coupling). If the double
quantum dot is tuned such that only the topmost occupied level on
each dot is taken into account, this device is an artificial
realization of a quantum two level system. This description of the
double quantum dot as an effective two level system is correct as
long as transport occurs due to resonant tunneling between the
ground states of both dots: namely, starting from the ground state
$(N,M)$ with N electrons in the left dot and M electrons in the
right dot, the transport occurs between the states $(N+1,M)$ and
$(N,M+1)$ \footnote{A detailed description of transport in double
quantum dots can be found in the review of Van der Wiel et al in
Ref.~\cite{VanRMP(03)}.}.
\begin{figure}
\begin{center}
\includegraphics[width=0.75\columnwidth,angle=0]{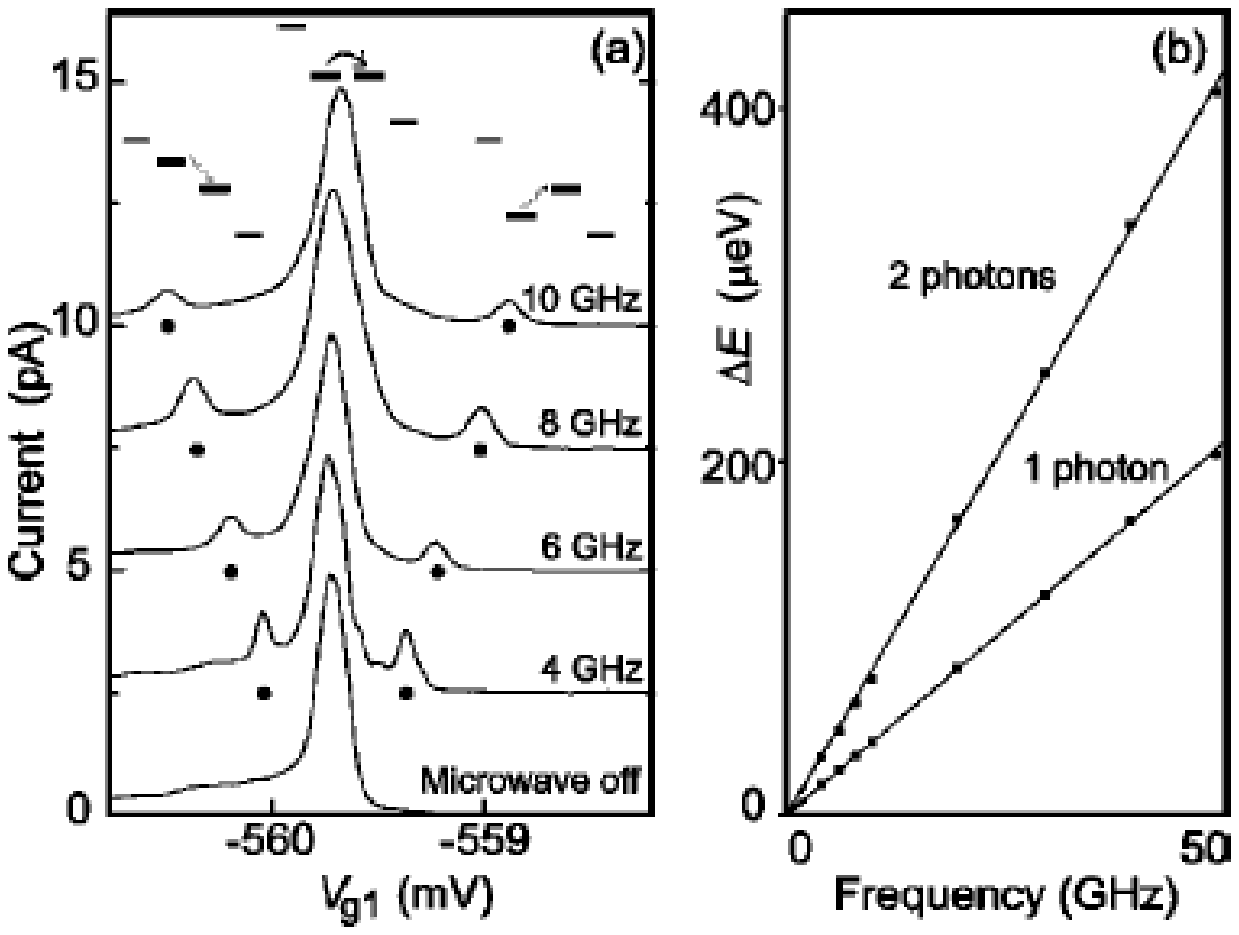}
\end{center}
\caption[]{Photon-assisted tunneling current through weakly
coupled quantum dots in the large bias voltage ($V_{sd}=500\mu V$)
regime. a) dc current as a function of the level separation
$\Delta\varepsilon$ (which corresponds to the gate voltage
$V_{g1}$ in the experiment). Different curves (offset for clarity)
show the current for increasing frequencies (0-10GHz). The central
resonance corresponds to elastic tunneling while the satellites
correspond to the absorption (right satellite) and emission (left
satellite) of one photon when $\Delta\varepsilon=\hbar\omega$. b)
Distance between the central resonance and the satellites as a
function of the microwave frequency. Reprinted with permission
from \cite{VanRMP(03)}. \copyright 2003 American Physical
Society.} \label{QD_fig5}
\end{figure}
\begin{figure}
\begin{center}
\includegraphics[width=0.75\columnwidth,angle=0]{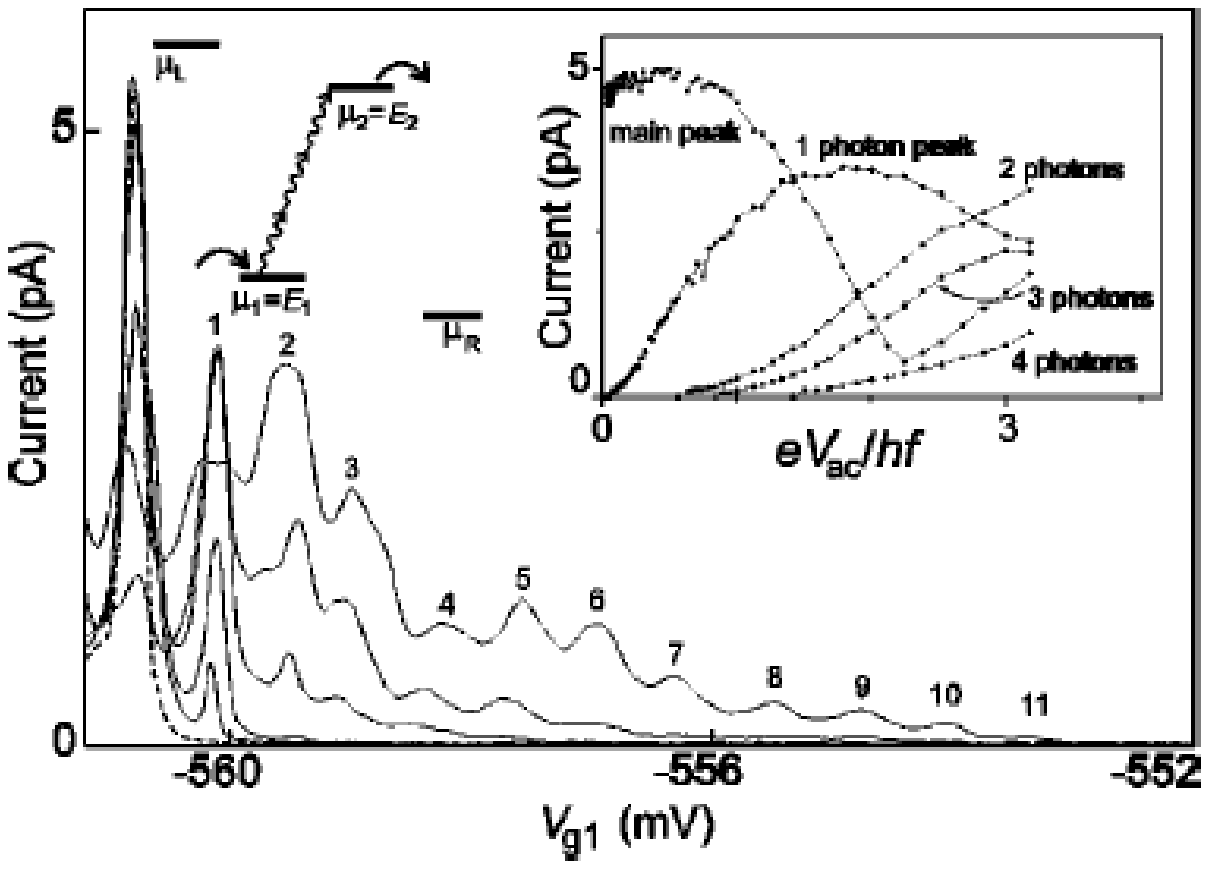}
\end{center}
\caption[]{Photon-assisted tunneling current through weakly
coupled quantum dots in the high microwave power and large bias
voltage ($V_{sd}=700\mu V$) regimes. Dashed curve is without
microwaves. The solid lines correspond to data taken at fixed
frequency and increasing microwave powers. At the highest power,
absorption of 11 photons is demonstrated. The right inset shows
the height of the first four satellite peaks as a function of the
microwave amplitude. Reprinted with permission from
\cite{VanRMP(03)}. \copyright 2003 American Physical Society.}
\label{QD_fig6}
\end{figure}

The basic idea when performing photon-assisted tunneling
spectroscopy is to measure the energy differences between states
in the two dots by using microwaves (typical frequencies ranging
from 0-75GHz)
\cite{FujSM(97),FujJJAP(97),OosNature(98),VanPhysB(99),BliSS(96)}.
Without microwaves, a resonant current flows through the double
quantum dot device provided that
$\mu_{left}>\varepsilon_1=\varepsilon_2>\mu_{right}$, where
$\mu_{left}$ and $\mu_{right}$ and $\varepsilon_1,\varepsilon_2$
are the chemical potentials and the discrete energy levels of the
quantum dots respectively. In the presence of microwaves,the
conditions for transport change because the external field can
induce inelastic events such that
$\Delta\varepsilon\equiv\varepsilon_1-\varepsilon_2=n\hbar\omega$.
Experimentally, this idea was first put forward by Blick et al in
Ref.~\cite{BliSS(96)} and by Fujisawa and Tarucha in
Ref.~\cite{FujSM(97)}.

To use photon-assisted tunneling as a spectroscopic tool for
double quantum dots, two different configurations can be used:
pumping \cite{StaPRL(96),BruPhysE(97)} and large bias voltage
\cite{StoPRB(96)}. The pumping configuration is operated at zero
bias voltage across the system, absorption of a photon with energy
$\Delta\varepsilon$ leads to a finite dc current which is positive
or negative depending on the sign of $\Delta\varepsilon$
(Fig.~\ref{configurations}, upper graph). photon-assisted
tunneling spectroscopy in double dots can also be investigated in
the large bias voltage case, where both absorption and emission
contribute to a positive dc current
(Fig.~\ref{configurations},lower graph).
\subsection{Photon-assisted tunneling in weakly coupled double quantum dots I: pumping configuration.
\label{pumping}} Photon-assisted tunneling in weakly coupled
double quantum dots in the pumping configuration was first
investigated theoretically by Stafford and Wingreen in Ref.~\cite{StaPRL(96)}.
By using Floquet theory they found that in the
strong localized eigenstates limit $|\Delta\varepsilon|>> t_C$
($t_C$ is the interdot hopping), at the N-photon resonance
$N\hbar\omega=\sqrt{(\Delta\varepsilon)^2+4|t_C|^2}\simeq
\Delta\varepsilon$, the electronic orbital on one dot hybridizes
with the {\it N}th sideband of the electronic orbital on the other
dot such that the quasienergy eigenstates become delocalized. This
results in a {\it renormalization} of the Rabi frequency which
becomes:
\begin{equation}
\Omega_R=2|t_C|J_N(\beta).
\end{equation}
By combining this Floquet theory with the Keldysh technique for
non-equilibrium Green's functions they were able to obtain a
general expression for the pumped current and found that this
current is maximized when the Rabi frequency $\Omega_R$ equals the
coupling to the leads $\Gamma_L=\Gamma_R=\Gamma$. At bias voltages
large compared to $\Gamma$, the current at the photon-assisted
tunneling peak is:
\begin{equation}
\bar{J}_{res}=\frac{e\Gamma}{2\hbar}(\frac{\Omega_R^2}{\Omega_R^2+\Gamma^2}).
\end{equation}
Transport is thus characterized by the ratio of the Rabi frequency
$\Omega_R$ to the tunneling rate to the leads $\Gamma$. If
$\Omega_R>>\Gamma$, the bottleneck for transport is the tunneling
to the leads and then the current is proportional to $\Gamma$ (in
fact, in this limit $\bar{J}_{res}=\frac{e\Gamma}{2\hbar}$ is the
largest current possible for this coupling to the leads). In the
opposite limit $\Gamma>>\Omega_R$,
$\bar{J}_{res}=\frac{e\Omega_R^2}{2\hbar\Gamma}$ and the
resonances are broadened in energy by $\Gamma$.
\subsection{Photon-assisted tunneling in weakly coupled double quantum dots II: Large bias voltage configuration.
\label{PAT-large-bias}} Photon-assisted tunneling in weakly
coupled double quantum dots in the large bias voltage
configuration was investigated theoretically by Stoof and Nazarov
in Ref.~\cite{StoPRB(96)} by using a density matrix approach.
Within this approach, the master equation for the reduced density
matrix elements can be written as \cite{Blum,Cohen}:
\begin{eqnarray}
\dot{\rho}(t)_{s's} &=& -i\omega_{s's}{\rho (t)}_{s's}
- \frac{i}{\hbar}\langle s'|[\mathcal{H}_T,\hat{\rho}(t)]|s\rangle\nonumber\\
&+& \left \{ \begin{array}{ll}
\sum_{m\neq s} W_{sm}\rho_{mm} - \sum_{k\neq s} W_{ks}\rho_{ss} & (s=s') \\
-\gamma_{s's}\rho_{s's} & (s\neq s')
\end{array}
\right. \label{eq-master}
\end{eqnarray}
where ${\rho (t)}_{s's}$ are the matrix elements of the density
operator $\hat{\rho}(t)$ in the basis defined by the many body
states $|s\rangle$ (with energy $E_s$) of each uncoupled quantum
dot. The first two terms in Eq.~(\ref{eq-master}) represent
reversible (coherent) dynamics between the quantum dots in terms
of the transition frequencies $\omega_{s',s}=(E_{s'}-E_{s})/\hbar$
and the interdot tunneling Hamiltonian $\mathcal{H}_T$ (see
below). The next two terms describe the irreversible dynamics due
to the coupling with the external leads. $W_{mn}$ are the
transition rates from a state $|n\rangle$ to a state $|m\rangle$.
$\gamma_{s's}$ accounts for the induced decoherence due to
interactions with the reservoirs: ${\rm Re}~\{\gamma_{s's}\} =
(\sum_{k\neq s} {W}_{ks}+ \sum_{k\neq s'} {W}_{ks'})/2$.

Within the two-level picture, the density matrix can be expressed
in an effective Hilbert space consisting of three states $|L
\rangle=|N+1,M \rangle$, $|R \rangle=|N,M+1\rangle$ and the
'empty' state $|0 \rangle=|N,M \rangle$, which describes a
situation with no extra electron in either of the dots. This
Hilbert space is defined by a pseudospin $\hat{\sigma}_z \equiv |L
\rangle \langle L|-|R \rangle \langle R|$ and
$\hat{\sigma}_x\equiv |L \rangle \langle R|+|R \rangle \langle
L|$. The effective Hamiltonian can be written in terms of the
pseudospin operators as:
\begin{eqnarray}\label{pseudospinHamiltonian}
{\mathcal H}(t)={\mathcal H}_0(t)+{\mathcal
H}_T=\frac{\varepsilon(t)}{2} \hat{\sigma}_z+ t_C\hat{\sigma}_x
\end{eqnarray}
with $\varepsilon(t)=\Delta\varepsilon+eV_{ac} cos\omega t$. The
coupling to external free electron reservoirs ${\mathcal
H}_{res}=\sum_{k_\alpha}\epsilon_{k_\alpha}
c_{k_\alpha}^{\dagger}c_{k_\alpha}$ is described by the usual
tunnel Hamiltonian
\begin{eqnarray}
  {\mathcal H}_c=\sum_{k_\alpha{}}
(V_k^\alpha{} c_{k_\alpha{}}^{\dagger}s_\alpha{}+H.c.),
\label{coupling_spin-boson}
\end{eqnarray}
with $\hat{s}_\alpha=|0 \rangle \langle \alpha|$ ($\alpha$=L,R).
In the limit of large bias voltage, Eqs.~(\ref{eq-master}) can be
written in the basis defined by $|L \rangle$, $|R \rangle$ and $|0
\rangle$ as \cite{StoPRB(96)}:
\begin{eqnarray}
  \frac{\partial}{\partial t}\rho_{LL}(t)&=&-it_C\left[
\rho_{RL}(t)-\rho_{LR}(t) \right]+ \Gamma_L \rho_{00}(t)
\nonumber\\
\frac{\partial}{\partial
t}\rho_{RR}(t)&=&it_C\left[\rho_{RL}(t)-\rho_{LR}(t)\right]
-{\Gamma}_R \rho_{RR}(t)\nonumber\\
\frac{\partial}{\partial t}\rho_{LR}(t)&=& -\frac{{\Gamma}_{R}}{2}
\rho_{LR}(t)+i\varepsilon(t)\rho_{LR}(t)+
it_C\left[\rho_{RR}(t)-\rho_{LL}(t)\right]\nonumber\\
\frac{\partial}{\partial t}\rho_{RL}(t)&=& -\frac{{\Gamma}_{R}}{2}
\rho_{RL}(t)-i\varepsilon(t)\rho_{RL}(t)-
it_C\left[\rho_{RR}(t)-\rho_{LL}(t)\right], \label{Stoof-Nazarov}
\end{eqnarray}
To lowest order in the interdot tunneling $t_C$, Stoof and Nazarov
showed that the photon-assisted tunneling current is given by:
\begin{equation}
\label{Stoof} I_{PAT}=et_C^2\sum_{n=-\infty}^{\infty}J_n^2(\beta)
\frac{\Gamma_R}{\frac{\Gamma_R^2}{4}+(n\hbar\omega-\Delta\varepsilon)^2}.
\end{equation}
Eq.~(\ref{Stoof}) thus predicts that the photon-assisted tunneling
current is composed of a number of photon sidebands, separated by
the photon energy $\hbar\omega$ and width $\Gamma_R$. This is in
good agreement with the experiments by van der Wiel et al
Fig.~\ref{QD_fig5}.
\begin{figure}
\begin{center}
\includegraphics[width=0.75\columnwidth,angle=0]{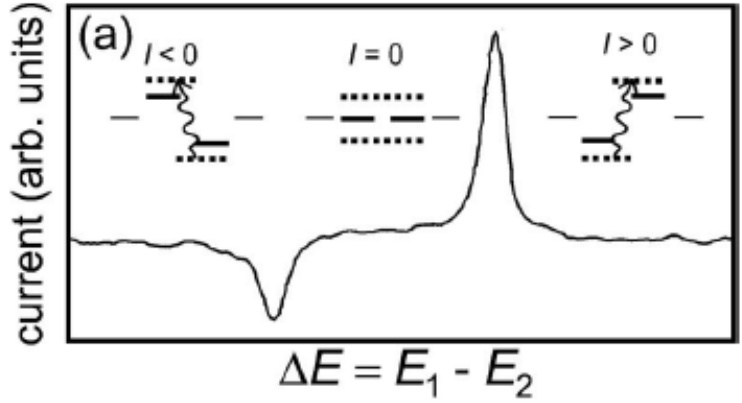}
\end{center}
\caption[]{Pumped photon-assisted tunneling current as a function
of the level separation $\Delta\varepsilon$. The positive peak
corresponds to pumping electrons from the left to the right dot
($\Delta\varepsilon <0$) while the negative one corresponds to the
opposite effect when $\Delta\varepsilon >0$. The central resonance
is absent because the bias voltage is zero. Reprinted with
permission from \cite{VanRMP(03)}. \copyright 2003 American
Physical Society.} \label{pump1}
\end{figure}
The relative position of the energy levels $\Delta\varepsilon$ is
shifted by the gate voltage $V_{g1}$ in the figure. The central
peak corresponds to elastic tunneling while the satellite
resonances involve the emission (left satellite) or absorption
(right satellite) of one photon. The slight asymmetry of the
central resonance at large negative gate voltages can be
understood in terms of relaxation processes due to emission of
phonons. In the large bias voltage configuration, these relaxation
processes contribute also to the current. This is the case even
for very low temperatures when spontaneous emission of phonons
always gives a contribution to the current for $\Delta\varepsilon
>0$ \cite{FujSci(98)} . By increasing the microwave frequency, the
distance between the main resonance and the satellites increases
linearly Fig.~\ref{QD_fig5}. b. This, as we will see below,
indicates the absence of quantum coherence between quantum dots.
By increasing the microwave power at fixed frequency (the
parameter $\beta$ in Eq. ~(\ref{Stoof})), it is possible to
measure multiphoton processes. This is shown in Fig.~\ref{QD_fig6}
where absorption of multiple photons (up to 11 photons) is demonstrated.\\
The above theory has been extended in Ref.~\cite{HazelPRB(01)} to
include a coherent pumping mechanism via inelastic cotunneling
processes.

Importantly, by performing photon-assisted tunneling spectroscopy
one is able to distinguish whether the two level system exhibits
quantum coherence or not, as we shall discuss in the next
subsection.
\subsection{Photon-assisted tunneling in strongly coupled double quantum dots}
An increase of the interdot tunneling coupling delocalizes the
electron wave function over the entire double dot structure.
Provided that the simplified two-level picture is correct,
elementary quantum mechanics tells us that the new eigenstates of
this problem are now the symmetric (bonding) and antisymmetric
(antibonding) combinations of the localized states. The new
eigenvalues are expressed in terms of the energies of the
uncoupled states as:
\begin{eqnarray}
\varepsilon_B&=&\frac{1}{2}\{(\varepsilon_1+\varepsilon_2)-\sqrt{(\Delta\varepsilon)^2+4t_C^2}\}\nonumber\\
\varepsilon_A&=&\frac{1}{2}\{(\varepsilon_1+\varepsilon_2)+\sqrt{(\Delta\varepsilon)^2+4t_C^2}\},
\end{eqnarray}
such that
$\varepsilon_A-\varepsilon_B=\sqrt{(\Delta\varepsilon)^2+4t_C^2}$.

We have described photon-assisted tunneling spectroscopy in weakly
coupled quantum dots. The next natural step would be to use this
technique to study strongly coupled dots in order to investigate
quantum coherence across the double dot system. This is not an
easy task though. As we mentioned, relaxation processes due to
spontaneous emission always contribute to the current for
$\Delta\varepsilon >0$ \cite{FujSci(98)} . With increasing the
interdot tunneling coupling between dots the spontaneous emission
rate also increases which renders the large bias voltage configuration
unapropriate to study the strong coupling regime. This difficulty
can be overcome by using the pumping configuration, the advantage
being that relaxation processes can lower the current but do not
contribute to it. An example of pumped current due to
photon-assisted tunneling in this configuration is given in
Fig.~\ref{pump1}. The positive peak corresponds to pumping
electrons from the left to the right dot ($\Delta\varepsilon <0$)
while the negative one corresponds to the opposite effect when
$\Delta\varepsilon >0$. The central resonance is absent because
the bias voltage is zero.

When the interdot tunneling coupling is strong enough, the
formation of the bonding and antibonding states results in a new
condition for observing a pumped photon-assisted tunneling
current: {\it to promote electrons from the low energy state to
the high energy state one needs now microwaves of frequency}:
\begin{equation}
\label{bonding-antibonding}
\hbar\omega=\varepsilon_A-\varepsilon_B=\sqrt{(\Delta\varepsilon)^2+4t_C^2}.
\end{equation}
Eq.~(\ref{bonding-antibonding}) can be rewritten as:
\begin{equation}
\label{bonding-antibonding2}
\Delta\varepsilon=\sqrt{(\hbar\omega)^2-4t_C^2}.
\end{equation}
This has to be compared with the condition for weak coupling
$t_C<<\Delta\varepsilon$ which is:
\begin{equation}
\label{linear} \hbar\omega=\Delta\varepsilon.
\end{equation}

From the previous reasoning, it is thus obvious that if one 
measures the pumped photon-assisted tunneling current
through a double quantum dot in the strong coupling regime the
position of the peak (antipeak) at positive (negative)
$\Delta\varepsilon$ should follow the hyperbolic form given by
Eq.~(\ref{bonding-antibonding}) instead of the linear relation in
Eq.~(\ref{linear}).
\begin{figure}
\begin{center}
\includegraphics[width=0.75\columnwidth,angle=0]{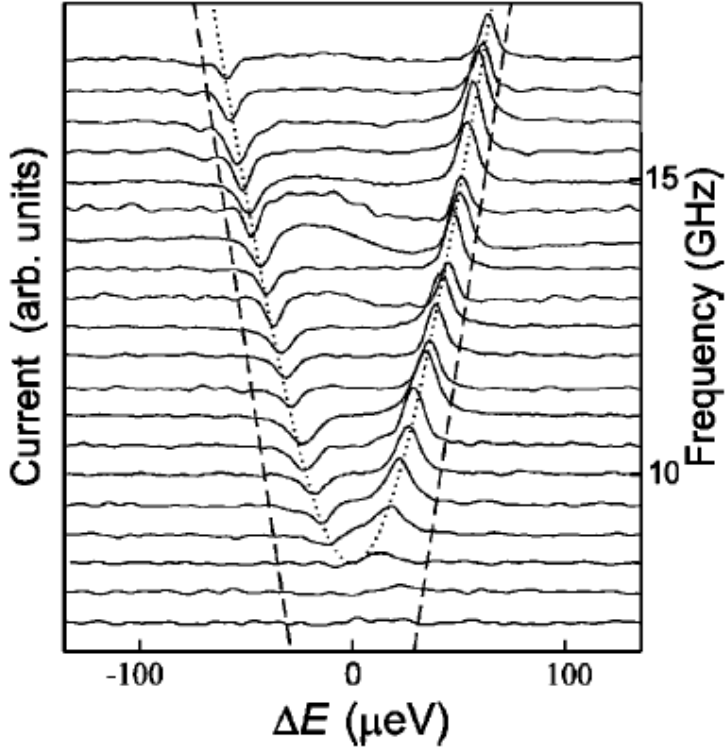}
\end{center}
\caption[]{Pumped photon-assisted tunneling current as a function
of the level separation $\Delta\varepsilon$ for different
microwave frequencies. The curves are offset such that the right
vertical axis gives the frequency. By using the interdot tunneling
coupling $t_C$ as a fitting parameter, the satellite peaks follow
the hyperbola $\sqrt{(\Delta\varepsilon)^2+4t_C^2}$ (dotted line).
The dashed line indicates the linear behavior expected for weak
coupling. Reprinted with permission from \cite{VanRMP(03)}.
\copyright 2003 American Physical Society.} \label{QD_fig8}
\end{figure}
This was experimentally demonstrated by Oosterkamp et al in
Ref.~\cite{OosNature(98)}. They measured the pumped current as a
function of the uncoupled energy difference $\Delta\varepsilon$
for different microwave frequencies and showed that indeed the
position of the resonances deviates from the linear relation in
Eq.~(\ref{linear}) when $t_C$ is fixed and
$\Delta\varepsilon\rightarrow 0$. These results are presented in
Fig.~\ref{QD_fig8}.

Similarly to the weak coupling case, the Rabi frequency becomes
renormalized by the microwave field. In this case,
$\hbar\omega>\Delta\varepsilon$ such that the interdot tunneling
coupling renormalizes as, see Eq.~(\ref{CDT}):
\begin{equation}
\tilde{t}_C=J_0(\beta)t_C.
\end{equation}
Namely, a strong microwave field reduces the tunnel coupling such
that the condition for pumping becomes:
\begin{equation}
\label{bonding-antibonding-tilde}
\Delta\varepsilon=\sqrt{(\hbar\omega)^2-4(J_0(\beta)t_C)^2}.
\end{equation}
\subsection{Spin-polarized pumps}
Cota et al recently followed up the ideas described in section
\ref{pumping} in order to investigate pumping of spin-polarized
electrons \cite{CotNano(03)}. Interestingly, the application of ac
voltages allows to control the degree of polarization of the
current flowing through a double quantum dot even in the case
where the contact leads are not spin polarized. This is of
importance, for understanding and controlling the behavior of
spins in nanostructures has become the subject of intense
investigation due to its relevance to quantum information
processing and spintronics \cite{Awschalom-book}.
\begin{figure}
\begin{center}
\includegraphics[width=0.6\columnwidth,angle=0]{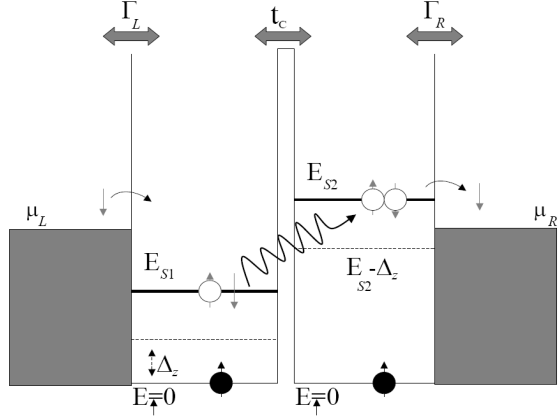}
\end{center}
\caption[]{Schematic representation of the double quantum dot in
the pumping configuration. $E_{S1}$ and $E_{S2}$ are the energies
of doubly occupied states in each dot. $E_{\uparrow}=0$ and
$E_{\downarrow}=\Delta_z$ are the energies of singly occupied
states ($\Delta_z$ is the Zeeman splitting). Dotted lines denote
chemical potentials. For $N \le 4$, the chemical potentials should
fulfill the conditions: $E_{S1}-E_{\uparrow}<\mu_L$,
$E_{S2}-E_{\uparrow}>\mu_R$ while $E_{S2}-\Delta_z<\mu_R$, in
order to obtain pumping of $\downarrow$ electrons.}
\label{CotaNano_03_fig1}
\end{figure}

An schematic diagram of the operation principle of the spin-pump
proposed in Ref.~\cite{CotNano(03)} is shown in
Fig.~\ref{CotaNano_03_fig1}. $E_{S1}$ and $E_{S2}$ are the
energies of the doubly occupied states in each dot, with a Zeeman
splitting $\Delta_z>kT$ on both dots (it is assumed that the leads
are unpolarized). The frequency of the ac field is tuned such that
$\hbar\omega\sim E_{S2}-E_{S1}$. Preparing the system initially in
the state $|\downarrow\uparrow,\uparrow\rangle$ (or in the state
$|\uparrow,\uparrow\rangle$ which is immediately filled by a
$\downarrow$ electron when $E_{S1}<\mu_L$), pumping of
$\downarrow$ spin is obtained in the regime where the chemical
potential for taking $\downarrow$ electrons out of the right dot
fulfils $E_{S2}>\mu_R$ while the chemical potential for taking
$\uparrow$ electrons out of the right dot fulfils
$E_{S2}-\Delta_z<\mu_R$. Then, a spin-polarized pump is realized
through the sequence:
$|\downarrow\uparrow,\uparrow\rangle\rightarrow
|\uparrow,\downarrow\uparrow\rangle \rightarrow
|\uparrow,\uparrow\rangle\rightarrow|\downarrow\uparrow,\uparrow\rangle$
or $|\downarrow\uparrow,\uparrow\rangle\rightarrow
|\uparrow,\downarrow\uparrow\rangle \rightarrow
|\downarrow\uparrow,\downarrow\uparrow\rangle\rightarrow|\downarrow\uparrow,\uparrow\rangle$
which involve states of double occupation on both dots.

The above qualitative explanation can be substantiated by studying
the problem with a reduced density matrix, see
Eq.~(\ref{eq-master}), fully taking into account the dynamics of a
Hilbert space comprising the sixteen states:
$|1\rangle=|0,0\rangle$, $|2\rangle=|\uparrow,0\rangle$
$|3\rangle=|\downarrow,0\rangle$,$|4\rangle=|0,\uparrow\rangle$
$|5\rangle=|0,\downarrow\rangle$,$|6\rangle=|\uparrow,\uparrow\rangle$,
$|7\rangle=|\downarrow,\downarrow\rangle$,
$|8\rangle=|\uparrow,\downarrow\rangle$,$|9\rangle=|\downarrow,\uparrow\rangle$,
$|10\rangle=|\uparrow\downarrow,0\rangle$,$|11\rangle=|0,\uparrow\downarrow\rangle$,
$|12\rangle=|\uparrow\downarrow,\uparrow\rangle$,
$|13\rangle=|\uparrow\downarrow,\downarrow\rangle$,
$|14\rangle=|\uparrow,\uparrow\downarrow\rangle$,
$|15\rangle=|\downarrow,\uparrow\downarrow\rangle$,
$|16\rangle=|\uparrow\downarrow,\uparrow\downarrow\rangle$. To
account for intrinsic decoherent processes acting even in the
isolated system, a term $T_2^{-1}$ is added to $\gamma_{s's}$ in
Eq.~(\ref{eq-master}) for terms involving spin-flips. Typically,
$T_2$ is at least an order of magnitude smaller than $T_1$, the
spin relaxation time \footnote{Recent experiments on vertical
quantum dots measured typical relaxation times $T_1\approx 200\mu
eV$ \cite{FujiNatu(02)}. The observed relaxation time can be
understood by inelastic cotunneling. Spin-orbit interactions are
also predicted to give an important contribution to spin
relaxation in GaAs quantum dots \cite{KhaetPRB(00)}.}. $T_1$ is
given by $(W _{\uparrow\downarrow}+W_{\downarrow\uparrow})^{-1}$,
where $W _{\uparrow\downarrow}$ and $W_{\downarrow\uparrow}$ are
spin-flip relaxation rates, such that $W
_{\uparrow\downarrow}/W_{\downarrow\uparrow}\approx
\exp(\Delta_z/kT)$. These spin-relaxation rate terms are taken
into account in the evolution equations (\ref{eq-master}) for the
diagonal elements of the reduced density matrix.
\begin{figure}
\begin{center}
\includegraphics[width=0.6\columnwidth,angle=0]{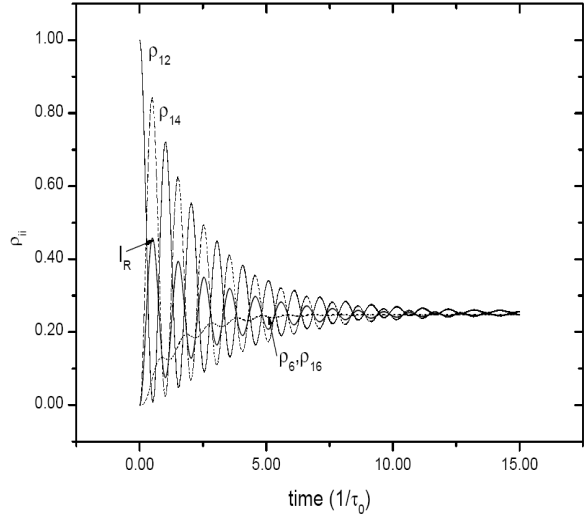}
\end{center}
\caption[]{Density matrix elements and current to the right lead
$I_R(t)$ as a function of time (in units of the period of Rabi
oscillations $\tau_0$). Parameters of the double quantum dot are:
$\Delta_z=6$ (Zeeman splitting), $U_L=6$, $U_R=12$ (on-site
interactions), $t_{c}=0.1$ (interdot tunneling),
$\Gamma_L=\Gamma_R=0.01$ (coupling to the leads),
$\mu_L=\mu_R=\mu=10$ (chemical potentials of the leads). When the
ac field (intensity $V_{ac}=\hbar\omega$) is tuned to resonance,
$\hbar\omega=\sqrt{((E_{14}-E_{12})^2 + 4t_C^2)}$, there is a
$(\downarrow)$ spin-dependent pumped current for unpolarized
leads. Initial state is $\rho_{12}=1$} \label{CotaNano_03_fig2}
\end{figure}

An example from this calculation is shown in
Fig.~\ref{CotaNano_03_fig2} where the dynamics of the relevant
density matrices $\rho_{12}=\langle
\downarrow\uparrow,\uparrow|\hat
\rho|\downarrow\uparrow,\uparrow\rangle$, $\rho_{14}=\langle
\uparrow,\uparrow\downarrow|\hat
\rho|\uparrow,\uparrow\downarrow\rangle$, $\rho_{6}=\langle
\uparrow,\uparrow|\hat \rho|\uparrow,\uparrow\rangle$ and
$\rho_{16}=\langle \uparrow\downarrow,\uparrow\downarrow|\hat
\rho|\uparrow\downarrow,\uparrow\downarrow\rangle$, (for
convenience we use the notation $\rho_i=\rho_{ii}$) is plotted.
The current to the right lead, which is given by the expression:
\begin{eqnarray}
I_R(t)&=& W_{1,4}\rho_4 + W_{1,5}\rho_5 + W_{2,6}\rho_6 +
W_{3,7}\rho_7
+ W_{2,8}\rho_8 + W_{3,9}\rho_9 \nonumber \\
&+& (W_{4,11} + W_{5,11})\rho_{11} + W_{10,12}\rho_{12} +
W_{10,13}\rho_{13} + (W_{6,14} +
W_{8,14})\rho_{14} \nonumber \\
&+& (W_{7,15} + W_{9,15})\rho_{15}
+ (W_{12,16} + W_{13,16})\rho_{16} \nonumber \\
&-& ( (W_{4,1}+W_{5,1})\rho_1 + (W_{6,2}+W_{8,2})\rho_2
+ (W_{7,3}+W_{9,3})\rho_3 \nonumber \\
&+& W_{11,4}\rho_4 + W_{11,5}\rho_5 + W_{14,6}\rho_6 +
W_{15,7}\rho_7
+ W_{14,8}\rho_8 + W_{15,9}\rho_9 \nonumber \\
&+& (W_{12,10}+W_{13,10})\rho_{10}+ W_{16,12}\rho_{12} +
W_{16,13}\rho_{13}), \label{eq-current}
\end{eqnarray}
is also plotted. After a few periods of the Rabi oscillation, both
a steady-state polarized current and a finite population of the
states $|6\rangle=|\uparrow,\uparrow\rangle$ and
$|12\rangle=|\uparrow\downarrow,\uparrow\rangle$ are reached
demonstrating the efficiency of the pump.
\begin{figure}
\begin{center}
\includegraphics[width=0.6\columnwidth,angle=0]{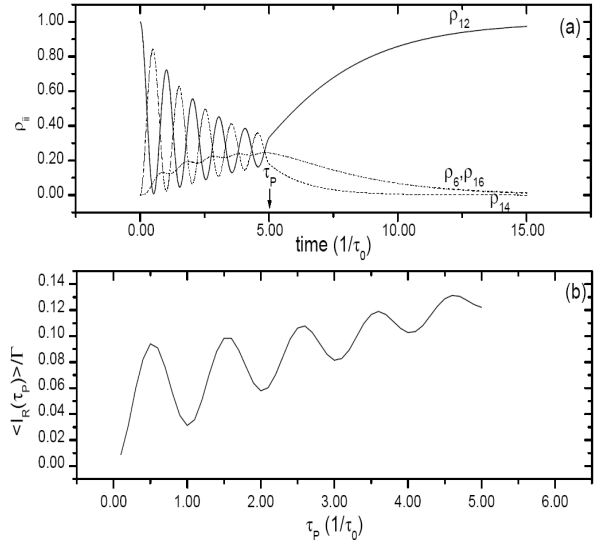}
\end{center}
\caption[]{(a) Time evolution of density matrix with a pulsed ac
field of duration $\tau_P$ tuned to resonance condition. (b) The
time-averaged spin-polarized current to lead $R$ in units of
$\Gamma$, as a function of pulse length $\tau_P$. The oscillations
in the current reflect Rabi oscillations of $\downarrow$ spins
within the double dot. Parameter values are the same as in
Fig.~\ref{CotaNano_03_fig2}.} \label{CotaNano_03_fig3}
\end{figure}

Further insight into the dynamics of the pump can be gained by
studying pulsed ac fields. As shown experimentally
\cite{NakNatu(99),Fujicondmat} and theoretically
\cite{HazelPRB(01),loss}, by applying a short pulse via a gate
electrode, Rabi oscillations can be resolved and observed through
current measurements. This is an important example of the
possibility of observation and control of coherent quantum state
time evolution. Importantly, it is crucial for the system to be
able to return to the initial (ground) state after the pulse has
been turned off. This is shown in Fig.~\ref{CotaNano_03_fig3}
where the system is initially prepared in state
$|12\rangle=|\uparrow\downarrow,\uparrow\rangle$, a pulsed ac
field (duration of the pulse $\tau_P$) is applied and then the
system is let to evolve for a time $\sim 3\tau_P$
(Fig.~\ref{CotaNano_03_fig3}(a)). We see, from the time evolution
of the density matrix elements  that the Rabi oscillations are
clearly resolved for $t < \tau_P$, and that the system eventually
regains the initial ground state ($\rho_{12}\rightarrow 1$) after
the field is turned off. Next, by applying pulses of different
length $\tau_P$ in sequence, one can calculate the time-averaged
current to the right lead $\langle I_R(\tau_P)\rangle$. The
results (Fig.~\ref{CotaNano_03_fig3}(b)) show that the Rabi
oscillations of $\downarrow$ spins between the two quantum dots
can be clearly resolved and observed through current measurements.

Note that in the above set-up, states of double occupancy in each
dot play a decisive role in obtaining the spin-polarized pumping
mechanism with unpolarized leads. Indeed, if the configuration is
such that states of doubly occupancy in both dots are above the
chemical potentials in the leads (such that only states with up to
two electrons in the double dot are relevant) the pump becomes
ineficcient: Let us assume that one starts from the state
$|\downarrow,\uparrow\rangle$, the ac field is tuned to resonance
and the pumping mechanism starts to populate the state
$|11\rangle=|0,\downarrow\uparrow\rangle$. Very rapidly, the
spin-triplet state $|6\rangle=|\uparrow, \uparrow\rangle$ will
dominate the dynamics because the state
$|12\rangle=|\uparrow\downarrow, \uparrow\rangle$ is unavailable.
As a consequence, the total current eventually goes to zero for
$\rho_{11}\rightarrow 0$ and $\rho_6\rightarrow 1$\footnote{This
spin blockade of the current due to the Pauli exclusion principle
has been observed in weakly coupled double dot systems by Ono et
al \cite{OnoSci(02)}.}. In other words, the appearance of the
triplet
$\rho_6=\langle\uparrow,\uparrow|\hat\rho|\uparrow,\uparrow\rangle$
blocks the pumping characteristics of the system, due to the Spin
Blockade effect. This result is general, with $\rho_6$ growing
with time more or less rapidly depending on the parameters of the
problem\footnote{A spin-pump with two electrons in the double dot
has been proposed very recently by Sun et al in
Ref.~\cite{SunPRL(03)}. Here a rather stringent condition, a
spatially {\it nonuniform} magnetic field in the double dot
system, is needed such that the ground state in the double dot
system is $|\uparrow, \downarrow\rangle$ instead of $|\uparrow,
\uparrow\rangle$.}.
\subsection{Photon-assisted tunneling in strongly dissipative double quantum dots}
As mentioned in section \ref{PAT-large-bias}, the current in the
large bias voltage configuration has always an inelastic
contribution that can be understood in terms of relaxation
processes due to emission of phonons. This is the case even for
very low temperatures when spontaneous emission of phonons always
gives a contribution to the current for $\Delta\varepsilon >0$
\cite{FujSci(98)}. The effect of a phonon bath on the transport
properties of double quantum dots in the low bias voltage regime
has been investigated experimentally by Qin et al
\cite{QinPRB(01)}.

In the presence of a generic dissipative bosonic bath (${\mathcal
H}_B= \sum_Q\omega_{Q} a^{\dagger}_Q a_Q$), the Hamiltonian of
Eq.~\ref{pseudospinHamiltonian} becomes:
\begin{eqnarray}\label{spinbosonHamiltonian}
{\mathcal H}_{SB}(t)&=& \Big[\frac{\varepsilon(t)}{2} +\sum_Q
\frac{g_{Q}}{2} \left(a_{-Q} + a^{\dagger}_Q \right)\Big]
\hat{\sigma}_z + t_C \hat{\sigma}_x +{\mathcal H}_B.
\end{eqnarray}
The effects of the bosonic bath are fully encapsulated in a
spectral density
\begin{eqnarray}\label{Jdef}
J({\omega})\equiv\sum_Q |g_{Q}|^2\delta(\omega-\omega_Q),
\end{eqnarray}
where $\omega_Q$ are the frequencies of the bosons and the $g_Q$
denote interaction constants. The Hamiltonian of
Eq.~(\ref{spinbosonHamiltonian}) is known in the literature as the
driven spin-boson Hamiltonian\footnote{The driven spin-boson
Hamiltonian has been studied extensively during the last years in
the context of quantum dissipative systems. For further details,
we refer the reader to the review by Grifoni and H\"anggi in
Ref.~\cite{GrifPR(98)}.}.

The coupling to external reservoirs is again described by
Eq.~(\ref{coupling_spin-boson}). In the presence of the bosonic
bath, Eqs.~(\ref{Stoof-Nazarov}) describing the reduced density
matrix are modified as follows:
\begin{eqnarray}\label{Brandes-Aguado-Platero}
\frac{\partial}{\partial t}\rho_{LL}(t)&=&-it_C\left\{
\rho_{RL}(t)-\rho_{LR}(t)\right\} 
+\Gamma_L\left[1-\rho_{LL}(t) - \rho_{RR}(t)\right]
\nonumber\\
\frac{\partial}{\partial t}\rho_{RR}(t) &=&it_C
\left\{\rho_{RL}(t)-\rho_{LR}(t)
\right\} 
-{\Gamma}_R\rho_{RR}(t)
\nonumber\\
\rho_{RL}(t)&=& -\int_0^tdt'   e^{i\int_{t'}^{t}ds\,\varepsilon(s)}\nonumber\\
&\times&\left[ \left( \frac{\Gamma_R}{2}\rho_{RL}(t') + it_C
\rho_{LL}(t')\right) C(t-t') - it_C \rho_{RR}(t')
C^*(t-t') \right] \nonumber\\
\rho_{LR}(t)&=& -\int_0^tdt'   e^{-i\int_{t'}^{t}ds\,\varepsilon(s)}\nonumber\\
&\times&\left[ \left( \frac{\Gamma_R}{2} \rho_{LR}(t') - it_C
\rho_{LL}(t')\right) C^*(t-t') + it_C
\rho_{RR}(t') C(t-t') \right].\nonumber\\
\end{eqnarray}
The boson correlation function for a harmonic bath with spectral
density $J(\omega)$, Eq.~(\ref{Jdef}), and at equilibrium
temperature $k_BT$ enters as,
\begin{eqnarray}
  C(t)&\equiv&e^{-Q(t)}\nonumber\\
Q(t)\!&\equiv&\!\int_0^{\infty}d\omega \frac{J(\omega)}{\omega^2}
\left[ \left(1- \cos \omega t\right) \coth
\left(\frac{\hbar\omega}{2k_BT}\right) + i \sin \omega t \right].
\end{eqnarray}
Without driving, $e^{i\int_{t'}^{t}ds\,\varepsilon(s)}\rightarrow
e^{i\Delta\varepsilon(t-t')}$, Eqs.~(\ref{Brandes-Aguado-Platero})
were solved by Brandes and Kramer \cite{BrandPRL(99)}. The
stationary current without ac reads:
\begin{eqnarray}\label{currentstat}
& &\overline{I} =et_C^2\frac{2\mbox{\rm
Re}(\hat{C}_{\Delta\varepsilon})+\Gamma_R|
\hat{C}_{\Delta\varepsilon}|^2}
{|1+\Gamma_R\hat{C}_{\Delta\varepsilon}/2|^2+2t_C^2B_{\Delta\varepsilon}},
\end{eqnarray}
with
\begin{eqnarray}
B_{\Delta\varepsilon}&\equiv& \mbox{\rm
Re}\left\{(1+\Gamma_R\hat{C}_{\Delta\varepsilon}/2)\left[
 \frac{\hat{C}_{-\Delta\varepsilon}}{\Gamma_R}+\frac{\hat{C}_{\Delta\varepsilon}^*}
{\Gamma_L}\left(1+\frac{\Gamma_L}{\Gamma_R}\right)\right]\right\}\nonumber,
\end{eqnarray}
and
$\hat{C}_{\Delta\varepsilon}\equiv\hat{C}(z=-i\Delta\varepsilon)$,
where $\hat{C}(z)$ denotes the Laplace transform:
\begin{eqnarray}
  \label{eq:Laplacedef}
  \hat{C}(z)=\int_{0}^{\infty}dt e^{-zt}C(t).
\end{eqnarray}
$\overline{I}^{(2)}$ is obtained by expanding
Eq.~(\ref{currentstat}) to lowest order in $t_C$, namely:
\begin{eqnarray}
\overline{I}^{(2)}=2e{\rm
Re}[t_C^2\hat{C}_{\Delta\varepsilon}/({1+\Gamma_R
\hat{C}_{\Delta\varepsilon}/2})].
\end{eqnarray}

In the driven case, the stationary current to lowest order
($t_C^2$) reads \cite{Brand-Aguado-Plat}:
\begin{eqnarray}
  \label{eq:currentstatnew}
  \overline{I}_{PAT}^{(2)}
&=&2et_C^2\sum_n
J_n^2\left(\frac{{eV_{ac}}}{\hbar\omega}\right)\mbox{\rm Re}
\left(\frac{\hat{C}_{\Delta\varepsilon+n\hbar\omega}}
{1+\frac{\Gamma_R}{2}\hat{C}_{\Delta\varepsilon+n\hbar\omega}}\right).
\end{eqnarray}
Remarkably, Eq.~(\ref{eq:currentstatnew}) is given by a
Tien-Gordon formula for {\it arbitrary electron-boson coupling}:
the current in the driven system is expressed by a sum over
current contributions (including the coupling to the dissipative
bosonic bath) from side-bands $\Delta\varepsilon+n\hbar\omega$,
weighted with squares of Bessel functions. Explicitly,
\begin{eqnarray}
  \label{eq:currentstatnew1}
  \overline{I}_{PAT}^{(2)}&\equiv&\sum_n J_n^2\left(\frac{{eV_{ac}}}{\hbar\omega}\right)
\left.\overline{I}^{(2)}\right|^{{V_{ac}} =
0}_{\Delta\varepsilon\to\Delta\varepsilon+n\hbar\omega};
\end{eqnarray}

Without bath, $\hat{C}_{\Delta\varepsilon}\to i/\Delta\varepsilon$
and Eq.~(\ref{eq:currentstatnew}) reduces to Eq.~(\ref{Stoof}). In
order to go beyond the Tien-Gordon approximation,
Eq.~(\ref{eq:currentstatnew1}), one has to perform a systematic
expansion of the current in $t_C$. The simplest way to do this is
by a numerical solution of Eqs.~(\ref{Brandes-Aguado-Platero})
\cite{Brand-Aguado-Plat}.
\subsection{Floquet theory for investigating ac-driven quantum
dots \label{ac-driven isolated double quantum dots}}
Ever since the pioneering
work of Anderson \cite{AndPR(58)}, it has been known that random
spatial disorder can cause electronic states to become localised
in quantum systems. As we have discussed in previous sections, it
has been found recently that an ac driving field can produce a
similar intriguing effect termed {\em dynamical localisation}, in
which the tunneling dynamics of a particle can be destroyed. One
of the first systems in which this effect was predicted is that of
a particle moving in a double-well potential \cite{GrosPRL(91)}. A
physical realization of this could consist of two coupled quantum dots
containing a single electron
--- the simplest type of artificial molecule possible.
If this system is prepared with the electron occupying one of the
quantum dots, one can expect it to tunnel across to the other quantum dot on a time
scale set by the Rabi frequency. However, if an ac field of the
correct strength and frequency is applied to the system, the
tunneling is destroyed, and the particle will remain trapped in
the initial well.

Weak time-dependent fields are generally treated as small
perturbations, which produce transitions between the eigenstates
of the unperturbed quantum system. This approach, however, is not
applicable to treat the strong driving fields required to produce
dynamical localisation, and instead the technique of Floquet
analysis \cite{GrifPR(98)}, which is valid in all regimes of driving,
has proven to be extremely effective. In this approach, the
important quantities to calculate are the {\em quasi-energies},
which play a similar role in driven systems to the eigenenergies
in the undriven case. In particular, dynamical localisation occurs
when two quasi-energies of states participating in the dynamics
approach each other, and become either degenerate (a crossing) or
close to degenerate (an avoided crossing). Using this formalism,
analytic and numerical studies of the double-well system have
shown \cite{holtzpb(92),hangepl(92),ShirPR(65)} that in the limit of high
frequencies, quasi-energy crossings occur when the ratio of the
field strength to the frequency is a root of the Bessel function
$J_0$.
Adding a second electron to the coupled quantum dot system, however,
introduces considerable complications. At the low electron
densities typically present in quantum dots, strong correlations produced
by the Coulomb interaction can significantly influence the
electronic structure. One of the most dramatic consequences of
this is the formation of {\em Wigner molecule} states \cite{JauEPL(93)}
that will be discussed in subsection \ref{Wigner}.
Understanding the interplay between electron correlations and the
driving field is, however, extremely desirable, as the ability to
rapidly control electrons using ac fields \cite{ColeNatu(01)} has
immediate applications to quantum metrology \cite{TamboPRL(99)}, where
a possible coherent turnstile device formed by a triple
well operating in a picosecond time
scale was proposed,
and quantum information processing. In particular, manipulating
entangled electrons on short timescales is of great importance to
the field of quantum computation \cite{BenNatu(00)}.
These kind of problems can be studied by applying the Floquet formalism to
systems of {\em interacting} particles. We illustrate this by
describing a system of two interacting electrons confined to a pair
of coupled quantum dots. A consequence of the interaction is that the
system only responds strongly to the field when the 
frequency is in resonance with the Coulomb interaction energy, namely $n\hbar\omega=U$.
When this condition is satisfied, CDT, which 
in this case is governed by the roots of higher-order
Bessel functions (order $n$), can occur.

\subsubsection{The driven double quantum-dot: a three-level system}

Here we describe a simplified model of a double quantum dot, in which each quantum dot is
replaced by a single site. Electrons are able to tunnel between
the sites, and the effect of interactions is included by means of a
Hubbard-$U$ term. This simple model captures all the main physics originating
from the interplay between strong ac driving and electronic correlations.
The Hamiltonian of this simplified system reads \cite{CrefPRB(02)a,CreSV03}:
\begin{equation}
H = t_{hub} \sum_{\sigma} \left( c_{1 \sigma}^{\dagger} c_{2
\sigma}^{ } + \mbox{H.c.} \right) + \sum_{i=1}^{2} \left(U_{hub}
n_{i \uparrow} n_{i \downarrow} + E_i(t) n_i \right).
\label{hubbard}
\end{equation}
Here $t_{hub}$ is the hopping parameter, and for the remainder of
this discussion we shall take $\hbar=1$, and measure all
energies in units of $t_{hub}$. $E_i(t)$ is the external electric
potential applied to site $i$. Clearly only the potential
difference, $E_1 - E_2$, is of physical importance, so one may
choose to take the symmetric parametrization:
\begin{equation}
E_1(t) = \frac{E}{2} \cos \omega t, \qquad E_2(t) = -\frac{E}{2}
\cos \omega t. \label{field}
\end{equation}
The Hilbert space of Hamiltonian (\ref{hubbard}) is
six-dimensional, comprising three singlet states and a three
dimensional triplet space. Measurements on semiconductor quantum dots have
shown that the spin-flip relaxation time is typically extremely
long \cite{FujiNatu(02)}, and so it is a good starting point to neglect
spin-flip terms in the Hamiltonian. Consequently the singlet and
triplet sectors are completely decoupled, and so if the initial
state possesses a definite parity this will be retained throughout
its time-evolution, and only states of the same parity need to be
included in the basis.

\begin{figure}
\begin{center}
\includegraphics[width=.8\textwidth]{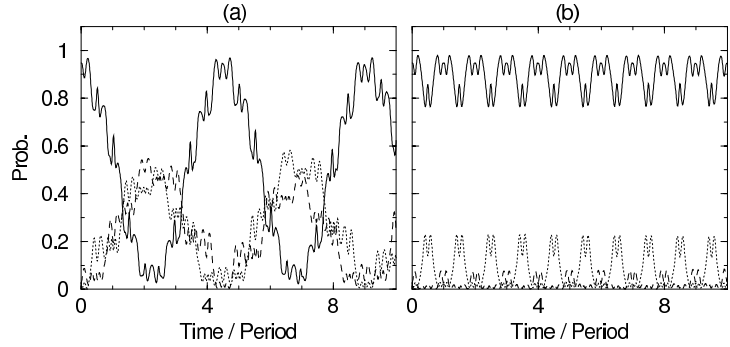}
\end{center}
\caption{Time evolution of the driven double quantum dot system for
$U_{hub}=8$ and $\omega=4$. (a) electric potential, $E = 30.0$; (b)
$E = 33.5$. Thick solid line = $p_{RL}(t)$, dotted line =
$p_{LL}(t)$, dashed line = $p_{RR}(t)$.} \label{two-site}
\end{figure}

The time evolution of this model system \cite{CrefPRB(02)a}
already contains the main physics obtained from more complicated
approaches like full simulations of detailed physical model of two
interacting electrons confined to a pair of coupled GaAs quantum dots
\cite{CrefPRB(02)a,TamboPRL(99),TamEPL(01)}. A great deal of
information can be extracted from the time evolution of the
probability functions $p_{LL}(t)$, $p_{RR}(t)$ and $p_{RL}(t)$,
which are respectively the probability that both electrons are in
the left quantum dot, both are in the right quantum dot, and that one electron is in
each of the quantum dots:
\begin{eqnarray}
p_{LL}(t)&=&\int_{L}dz_{1} \int_{L} dz_{2} |\Phi (z_1,z_2,t)|^2,\quad
p_{RR}(t)=\int_{R}dz_{1} \int_{R} dz_{2} |\Phi (z_1,z_2,t)|^2\nonumber\\
p_{RL}(t)&=&\int_{R}dz_{1} \int_{L} dz_{2} |\Phi (z_1,z_2,t)|^2
\end{eqnarray}
 The Coulomb interaction favors separating the
electrons, and thus for strong interactions the ground-state has a
large value of $p_{RL}$, and relatively small values of $p_{LL}$
and $p_{RR}$. In Fig.~\ref{two-site} the time evolution of these
quantities for $U_{hub}=8$ and $\omega = 4$, at two different
values of electric potential are plotted \cite{CrefPRB(02)a}. The
ground state of the static Hamiltonian is used as the initial
state. It is the dynamics of the ground state, i.e., the singlet
which is interesting because the electronic configuration will
oscillate between single and double occupation of the quantum
dots. In both cases the detailed form of the time-evolution is
highly complicated, but it is clear that the system behaves in two
distinct ways. In Fig.~\ref{two-site}a the value of $p_{RL}$
periodically cycles between its initial high value (indicating
that each dot holds approximately one electron) to nearly zero,
while the values of $p_{LL}$ and $p_{RR}$ correspondingly rise and
fall at its expense. This behavior is very different to that shown
in Fig.~\ref{two-site}b, where $p_{RL}$ never drops below a value
of 0.78, and the other two probabilities oscillate with a very
small amplitude. It thus appears that CDT is occurring in the
second case, and that the system's time evolution is essentially
frozen. If one terms the minimum value of $p_{RL}$ attained during
the time-evolution $p_{min}$, one can use this to quantify whether CDT
occurs, as a high value of $p_{min}$ signifies that tunneling has
been destroyed, while a low value indicates that the electrons are
free to move between the quantum dots.

\begin{figure}
\begin{center}
\includegraphics[width=.9\textwidth]{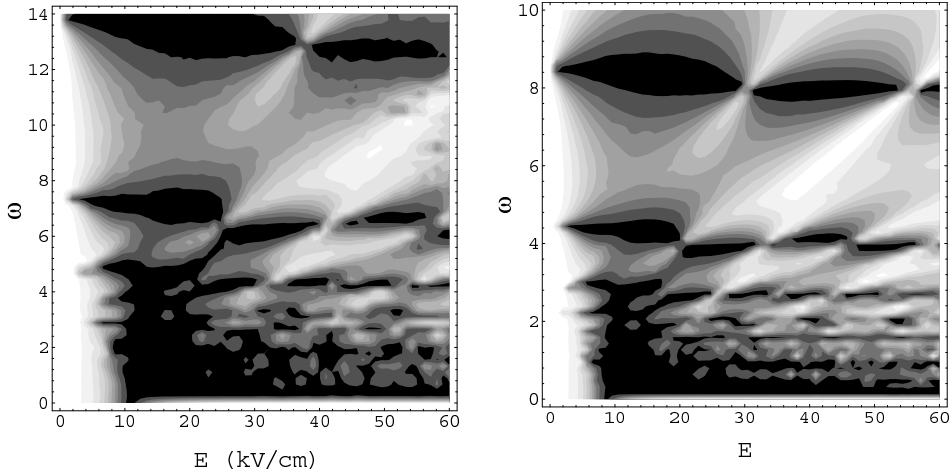}
\end{center}
\caption{$p_{min}$ as a function of the strength $E$ and energy $\hbar
\omega$ of the ac field: (a) for a full simulation of a quantum dot system
($\hbar \omega$ in units of meV) (b) for the two-site model with
$U_{hub}=8$ (both axes in units of $t_{hub}$). } \label{interact}
\end{figure}

\begin{figure}
\begin{center}
\includegraphics[width=.7\textwidth]{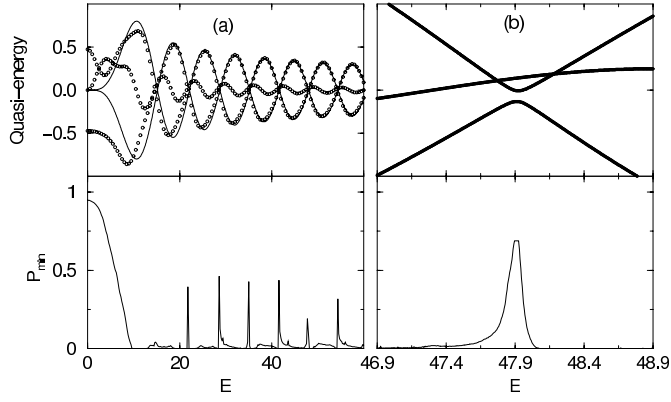}
\end{center}
\caption{(a) Quasi-energy spectrum for the two-site model for
$U_{hub} = 8$ and $\omega = 2$, circles = exact results, lines =
perturbation theory, (b) magnified view of exact results for a
single avoided crossing. Beneath are the corresponding plots of
$p_{min}$.} \label{quasi}
\end{figure}

This is illustrated in Fig.~\ref{interact}b where a contour plot of
$p_{min}$ as a function of both of the frequency and strength of
the ac field is presented. Dark areas correspond to low values of
$p_{min}$, and it can be seen that they form horizontal bands,
indicating that the system is excited strongly by the ac field
only at ``resonant'' values of $\omega$. Close examination of this
plot reveals that these bands occur at frequencies $\omega =
U_{hub},  U_{hub}/2,  U_{hub}/3 \dots$, at which the system can
absorb an integer number of photons to overcome the Coulomb
repulsion between electrons, thereby enabling tunneling processes
such as $| \uparrow, \downarrow \rangle \ \rightarrow \ | 0,
\uparrow \downarrow \rangle$ to occur. We can additionally observe
that these bands are punctuated by narrow zones in which CDT
occurs. Their form can be seen more clearly in the cross-section
of $p_{min}$ given in Fig.~\ref{quasi}a, which reveals them to be
narrow peaks. These peaks are approximately equally spaced along
each resonance, the spacing increasing with $\omega$. Another
contour plot of $p_{min}$ is shown in Fig.~\ref{interact}a, this
time obtained from a full simulation of two interacting electrons
confined to a pair of coupled GaAs quantum dots \cite{CrefPRB(02)a}. The
striking similarity between these results clearly indicates that
the simple, effective model (\ref{hubbard}) indeed captures the
essential processes occurring in the full system. \\
Tamborenea et
al \cite{TamEPL(01)} performed a full numerical simulation
of the two particle Schr\"odinger equation considering the configuration
interaction. Their results were similar to those presented in
Fig.~\ref{interact}a, which correspond to a numerical integration of
the Schr\"odinger equation by Creffield et al \cite{CrefPRB(02)a}.
However there is an important difference which comes from the
larger sample size considered in \cite{TamEPL(01)}. The narrower structure
studied in \cite{CrefPRB(02)a} (Fig.~\ref{interact}a)
 allows finer detail to be shown, and
 to resolve the punctuated regions of the bands in which CDT occurs
 and in which the system remains localized in its initial state.
 These regions are fundamental
 to describe the two electron dynamics for different intensities
 and frequencies of the ac field.

These results are radically different to those
obtained for non-interacting particles. In this case an analogous
plot of delocalisation shows a fan-like structure \cite{hangepl(92)},
in which localisation occurs along lines given
by $\omega = E / x_j$, where $x_j$ is the $j$-th root of the
Bessel function $J_0(x)$.

\begin{figure}
\begin{center}
\includegraphics[width=.7\textwidth]{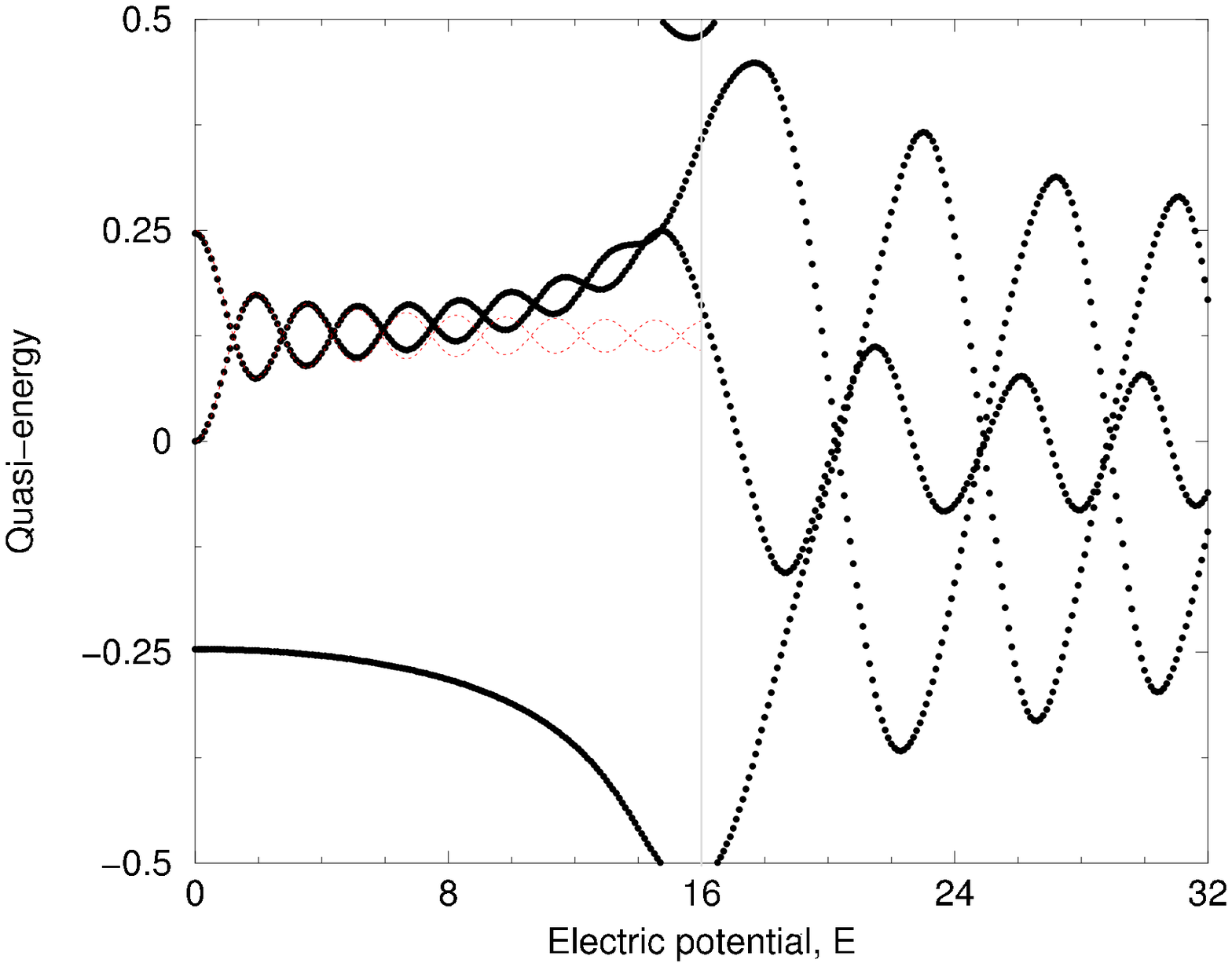}
\end{center}
\caption{(a) Quasi-energy spectrum for the two-site model for
$U_{hub} = 16$ and $\omega = 2$, circles = exact results, solid line
= two-level approximation, $\epsilon_{\pm} = \pm (\Delta/2) J_0(2
E/\omega)$.} 
\label{transition}
\end{figure}

In Fig.~\ref{quasi}a the Floquet quasi-energies as a
function of the field strength for $\omega = 2$, one of the
resonant frequencies visible in Fig.~\ref{interact}b, are shown. We see that
the system possesses {\em two} distinct regimes of behavior,
depending on whether the driving potential is weaker of stronger
than $U_{hub}$. For weak fields $E < U_{hub}$, as studied previously
in Ref.~\cite{ZhaPLA(00)}, the Floquet spectrum consists of one isolated
state (which evolves from the ground state) and two states which
make a set of exact crossings. Although in this regime $p_{min}$ shows
little structure, these crossings do in fact influence the
system's dynamics. This is demonstrated in
Fig.~\ref{transition} where the Floquet quasi-energies in the weak-field
regime for the case of $U_{hub} = 16$ are plotted. Beneath, the
minimum value of $p_{LL}$ attained during the time-evolution is shown
(where this time the state $|\uparrow \downarrow, 0 \rangle$ has
been used as the initial state). It can be seen that for this
choice of initial condition, the crossings of the quasi-energies
again produce CDT and freeze the initial state
--- despite the Coulomb repulsion between the electrons.
This surprising result may be understood as follows. For large
values of $U_{hub}$, the singlet eigenstates of the undriven system
consist of the ground state, separated by the Hubbard gap $U_{hub}$
from two almost degenerate excited states. For small values of the
driving potential, the two excited states remain isolated from the
ground state, and constitute an effective two-level system with a
level-splitting of $\Delta \simeq 4 t_{hub}^2/U_{hub}$. Thus if the
system is prepared in an initial state which projects mainly onto
the excited states, its dynamics will be governed by the two-level
approximation \cite{holtzpb(92),hangepl(92),ShirPR(65)}, and CDT will
occur at the roots of $J_0$. We show in Fig.~\ref{transition}a the
quasi-energies obtained from the two-level approximation, which
give excellent agreement with the actual results with {\em no}
adjustable parameters. As $E$ becomes comparable to the Hubbard
gap, however, the two excited states are no longer isolated from
the ground state, and all three levels must be taken into account.
This can be seen in the progressive deviation of the
quasi-energies from the two-level approximation as the electric
potential approaches $U_{hub}$ \cite{ChaMJ(03)}.

When the electric potential exceeds $U_{hub}$, the system displays a
very different behavior, in which $p_{min}$ remains close to zero
except at a series of narrow peaks, corresponding to the close
approaches of two of the quasi-energies. A detailed examination of
these approaches (see Fig.~\ref{quasi}b) reveals them to be {\em
avoided crossings} between the Floquet states which evolve from
the ground state and the higher excited state, and have the same
generalized parity. The remaining state, of opposite parity, makes
small oscillations around zero, but its exact crossings with the
other two states do not correlate with any structure in $p_{min}$.

To interpret this behavior in the strong-field regime,
one can obtain analytic expressions for the quasi-energies via the
perturbation theory described in Subsection \ref{method}. The first
step is to solve the eigenvalue equation, see Eq.~(\ref{floqeq}), in the
absence of the tunneling component $H_t$. In a real-space
representation the interaction terms are diagonal, and so it can
be readily shown that an orthonormal set of eigenvectors is given
by:
\begin{eqnarray}
|\epsilon_0(t)\rangle &=& \left( \exp \left[i \epsilon_0 t
\right], \ 0, \ 0 \right)
\nonumber \\
|\epsilon_+(t)\rangle &=& \left( 0, \ \exp \left[-i (U_{hub} -
\epsilon_+) t + i \frac{E}{\omega} \sin \omega t \right], \ 0
\right)
\nonumber \\
|\epsilon_-(t)\rangle &=& \left( 0, \ 0, \ \exp \left[-i (U_{hub} -
\epsilon_-) t - i \frac{E}{\omega} \sin \omega t \right] \right)
\end{eqnarray}
Imposing $T$-periodic boundary conditions reveals the
corresponding eigenvalues (modulo $\omega$) to be $\epsilon_0 = 0$
and $\epsilon_{\pm} = U_{hub}$. These eigenvalues represent the
zeroth-order approximation to the Floquet quasi-energies, and for
frequencies such that $U_{hub} = n \ \omega$ all three eigenvalues
are degenerate. This degeneracy is lifted by the perturbation
$H_t$, and to first-order, the quasi-energies are obtained by
diagonalizing the perturbing operator $P_{ij} = \langle \langle
\epsilon_i | H_t | \epsilon_j \rangle \rangle_T$. By using
$\exp\left[-i \beta \sin \omega t \right] =
\sum_{m=-\infty}^{\infty} J_m (\beta) \exp \left[-i m \omega t \right]$
to rewrite the form of $|\epsilon_{\pm}(t)\rangle$, the matrix
elements of $P$ can be obtained straightforwardly: 
\begin{equation}
P=\left(\begin{array}{ccc} 
0&0&- \sqrt{2}J_n(E/\omega)\\ 
0&0&- \sqrt{2}J_n(E/\omega)\\
- \sqrt{2}J_n(E/\omega)&- \sqrt{2}J_n(E/\omega)&0
\end{array}\right),
\end{equation}           
and its eigenvalues
subsequently found to be $\epsilon_0 = 0$ and $\epsilon_{\pm} =
\pm 2J_n(E/\omega)$ (the matrix elements and eigenvalues are given
in units of the interdot hopping t), where $n=
\frac{U}{\omega}$. For the non-interacting case U=0, the solution for independent electrons is recovered:
$\epsilon_{\pm}=\pm 2J_0(E/\omega)$.\\
Fig.~\ref{quasi}a demonstrates the excellent agreement between this
result (with $n = 4$) and the exact quasi-energies for strong and
moderate fields, which allows the position of the peaks in
$p_{min}$ to be found by locating the roots of $J_n$. Similar
excellent agreement occurs at the other resonances. For weak
fields, however, the interaction terms do not dominate the
tunneling terms and the perturbation theory breaks down, although
we are still able to treat the system phenomenologically by using
an effective two-level approximation.

Zhang et al., considered as well a Hubbard type model to study the dynamics
of two interacting electrons in double quantum dots \cite{ZhaPLA(00),ZhaJPC(01),ZhaJPC(00)}.
They numerically analyzed the character exchange
of the Floquet states at avoided triple crossings which appears
in a three level system as  a function of the field
intensity
in the neighborhood of an avoided crossing.
They observed that the degree of CDT at the avoided crossings depends
on the exchange of character between the three participating states.
\begin{figure}
\begin{center}
\includegraphics[width=.7\textwidth]{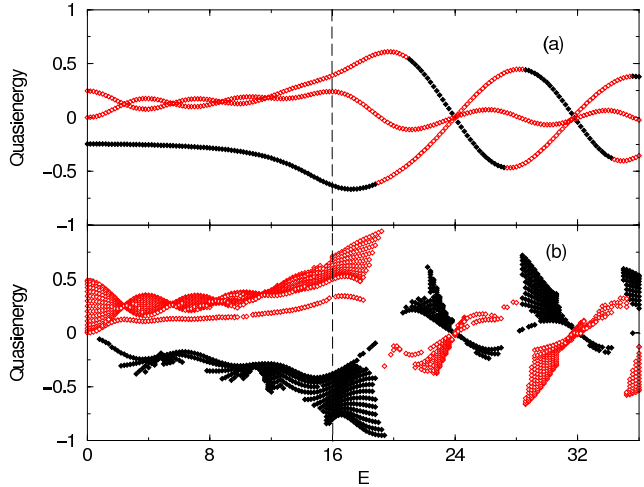}
\end{center}
\caption{Quasienergy spectrum for (a) a two-site system, and (b) a sixteen-site system for
$U_{hub}=16$ and $\omega=2$. Symbols indicate the characteristic of the corresponding 
Floquet state: hollow diamonds = doubly-occupied states, solid diamonds = neighbor-states, 
points= states with wide separation between electrons. The widely-separated states (points)
exhibit a series of miniband collapses over the whole range of $E$. For weak driving 
the doubly-occupied states show a similar set of collapses but with {\it half} the period. 
The vertical dashed line marks the boundary between the weak and strong driving regimes.}
\label{quasi-array}
\end{figure}
                      
 Recent calculations \cite{array} study the dynamics of two interacting
 electrons in quantum dot arrays driven by ac-fields. In this
 system also two different regimes are found as a function of
 the ratio between the strength of the field and the inter-electron Coulomb
 repulsion. When the ac field dominates, CDT occurs at certain frequencies
 , in which transport along the array is suppresed. In the other limit: weak
 driving regime, an interesting result is found:
 the two electrons can bind into a single composite particle
 - despite of the strong Coulomb repulsion between them, which can then be
 controlled by the ac field. These results can be explained in terms of
 the quasienergy spectrum \cite{array}.
In particular, these two regimes of weak and strong driving are a generic effect and
the effects seen in a two-site system arise in an analogous way such as the two intercrossing 
quasienergies in the weak driving regime are replaced by a miniband of states. 
This is illustrated in Fig.~\ref{quasi-array} where the quasienergy spectrum 
for a two-site system and a sixteen-site 
system are compared. 
\section{Photon assisted tunneling in quantum dots II: strongly correlated quantum dots \label{quantumdotsII}}
\subsection{Beyond the Coulomb blockade: Kondo effect \label{Kondo}}
\subsubsection{Basics}
So far, we have restricted ourselves to describe transport in the
sequential regime, namely transport to lowest order in the
coupling to the reservoirs. This is not the only contribution
though: under certain conditions  higher-order tunneling processes
become more and more relevant as the resistance of the tunneling
barriers approaches the quantum of resistance
$R_t=h/e^2=25.813k\Omega$.
\begin{figure}
\begin{center}
\includegraphics[width=0.75\columnwidth,angle=0]{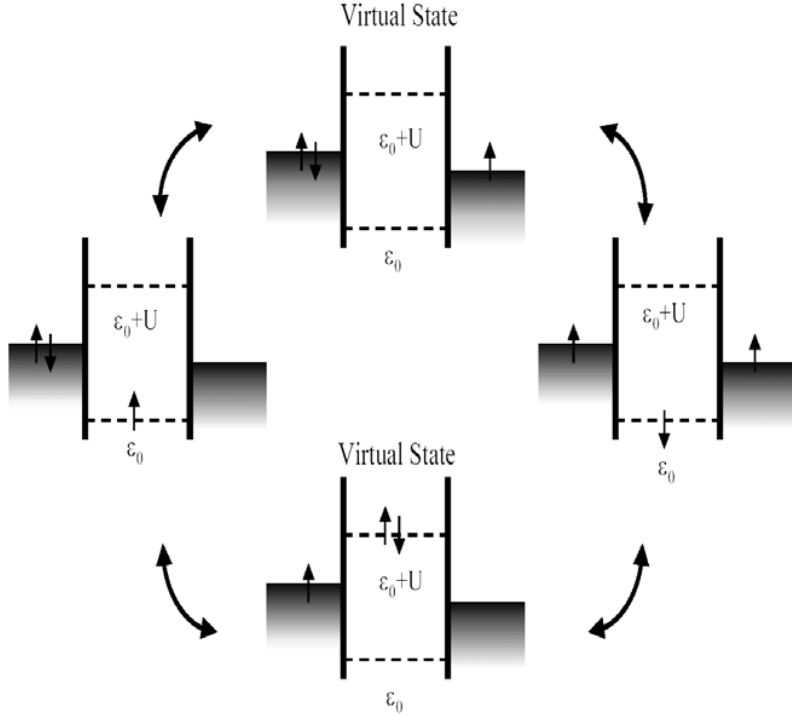}
\end{center}
\caption[]{Schematic diagram of the energy profile of a 
single-level quantum dot. This system is an artificial realization of 
the Anderson impurity model. By fluctuating through virtual states, 
empty and doubly occupied dot, the spin
of the artificial impurity is effectively flipped.}
\label{scheme-kondo}
\end{figure}
In this situation, quantum fluctuations dominate transport because
electrons are allowed to tunnel via intermediate virtual states
where first order tunneling would be supressed. Thus, the
intrinsic width of the energy levels of the quantum dot $\Gamma$
does not only include contributions from direct elastic tunneling
but also tunneling via virtual states. These higher-order
tunneling events are referred to as cotunneling processes.
Experimental results on cotunneling in semiconductor quantum dots
have been reported in
Refs.~\cite{GlaZPB(91),PasPRL(93),CronPRL(97),FranPRL(01)}. Photon
assisted tunneling in the cotunneling regime has been studied
theoretically by Flensberg in Ref.~\cite{FlePRB(97)}, but, to our
knowledge, no experiments in this regime exist to date.

Higher-order tunneling events lead to spectacular effects when the
spin of the electrons is also involved: a quantum dot with a net
spin coupled to electron reservoirs resembles a magnetic impurity
coupled to itinerant electrons in a metal and, thus, can exhibit
Kondo effect. The Kondo effect appears in dilute alloys containing
localized moments as a crossover from weak to strong coupling
between itinerant electrons of the host non-magnetic metal and the
unpaired localized electron of the magnetic impurity as the
temperature is reduced well below the Kondo temperature ($T_K$)
\cite{Hew}. Due to spin exchange interaction, see below, the
conduction electrons tend to screen the non-zero spin of the
magnetic impurity such that a many-body spin singlet state,
consisting of the impurity spin and the itinerant electrons
surrounding the impurity, forms.

As we mentioned, a quantum dot with a net spin and coupled to
reservoirs mimics the above situation. In the simplest model of a
magnetic impurity, the Anderson model, there is only a single spin
degenerate level coupled to itinerant electrons. The same
situation applies for quantum dots with an odd number of electrons
(only the topmost level which can accomodate a spin up or down is
considered) \footnote{This is not the only situation where Kondo
effect can appear in quantum dots. For other examples where the
Kondo effect appears for an even number of electrons in the
quantum dot see
Refs.~\cite{Integ1,Integ2,Integ3,Integ4,Integ5,Integ6}.}. The role
of the itinerant electrons in the usual Anderson model is played
here by the electron reservoirs to which the quantum dot is
coupled to. Fig.~\ref{scheme-kondo} (left) shows a schematic
diagram of the energy profile of a quantum dot using the language
of the Anderson model. A quantum dot with an odd number of
electrons can be represented by the topmost occupied level
$\varepsilon_0$ which is below the Fermi energy of the leads
$\varepsilon_F$ and is occupied by a spin (up in the figure).
Adding another electron to the quantum dot costs a charging energy
$U$ such that the double occupancy state has an energy
$\varepsilon_0+U$ well above the Fermi energy. On the other hand,
it would at least cost $|\varepsilon_0|$ to remove the electron
from the dot. The Hamiltonian which describes this system is:
\begin{equation}\label{chap4equa1}
\mathcal{H}  =  \mathcal{H}_{\rm leads} + \mathcal{H}_{\rm qd} +
\mathcal{H}_{\rm T}\,,
\end{equation}
where each term of the total Hamiltonian is defined as follows
\begin{eqnarray}
\label{chap4equa2} &&\mathcal{H}_{\rm leads}  = \sum_{\alpha
k\sigma} \epsilon_{\alpha k} c^\dagger_{\alpha k\sigma}c_{\alpha
k\sigma}
 \\
&&\mathcal{H}_{\rm qd} =  \sum _\sigma \epsilon _{0\sigma}
d^\dagger_\sigma d_\sigma + U d^\dagger_\uparrow d_\uparrow
d^\dagger_\downarrow d_\downarrow
 \\
&&\mathcal{H}_{\rm T} =  \sum _{\alpha k\sigma}\left (V_{\alpha
k}c^\dagger_{\alpha k\sigma} d_\sigma +V^*_{\alpha k}
d^\dagger_\sigma c_{\alpha k\sigma}\right)
\end{eqnarray}
The operator $d^\dagger_\sigma$ creates an electron with spin
$\sigma =\uparrow,\downarrow$ in the quantum dot, while $c^\dagger
_{\alpha k\sigma}$ creates an electron in the reservoir $\alpha
=L,R$ with energy $\epsilon_k $ ($k$ labels the rest of quantum
numbers). $V_{\alpha k}$ is the coupling between the quantum dot
and the reservoirs, which contributes to the intrinsic width of
the energy levels of the quantum dot $\Gamma=\Gamma_L+\Gamma_R$
with
\begin{equation}\label{chap4equa3}
\Gamma_{L(R)}(\epsilon)=-2\mathcal{I}m \left [ \Sigma_{L(R)}^{\rm
sp,r}(\epsilon +i \eta )\right]=2 \pi\sum_{k\in L(R)}
|V_{k}|^2\delta \left(\epsilon-\epsilon_{k}\right)\,,
\end{equation}
where $\Sigma_{L(R)}^{\rm sp,r}(\epsilon)$ is the hybridization
single-particle retarded self-energy, see below. In the simplest
case (wideband limit) one neglects the principal value of the
hybridization self-energy and consider the imaginary part to be an
energy independent constant, i.e., $\Sigma_{L(R)}^{\rm
sp,r}(\epsilon)=\Lambda_{L(R)}(\epsilon)-i\Gamma_{L(R)}(\epsilon)/2\approx
-i\Gamma_{L(R)}/2=-i\pi\rho_0 V^2$, where $\rho_0$ is the electron
density of states in the leads.

To lowest order, the transport in the Coulomb blockade region is
inhibited because $|\varepsilon_F-\varepsilon_0|>>\Gamma$.
Nonetheless, quantum uncertainty allows the system to visit
classically forbidden virtual states (empty or doubly occupied)
for a short period of time $\Delta t\sim h/|\varepsilon_0|$ or
$\Delta t\sim h/U$, respectively (see middle graphs in
Fig.~\ref{scheme-kondo}). Within the short timescale $\Delta t$
another electron must tunnel to the dot (if the virtual state is
the empty one) or out of the dot (if the virtual state is the
doubly occupied one). However, the initial and final states (left
and right graphs in Fig.~\ref{scheme-kondo}, respectively) may
have opposite spins, namely the spin has flip. These spin flip
processes can be described rigurously by an effective exchange
Hamiltonian
\begin{eqnarray}
\label{kondoHamiltonian} \mathcal{H}_{\rm K}=\mathcal{H}_{\rm
leads}&+&\overrightarrow{S}.\sum_{\sigma\sigma'}\{J_{LL}c^\dagger
_{L0\sigma}
\frac{\overrightarrow{\sigma}_{\sigma\sigma'}}{2}c_{L0\sigma'}
+J_{LR}c^\dagger_{L0\sigma}
\frac{\overrightarrow{\sigma}_{\sigma\sigma'}}{2}c_{R0\sigma'}\nonumber\\
&+&J_{RR}c^\dagger_{R0\sigma}
\frac{\overrightarrow{\sigma}_{\sigma\sigma'}}{2}c_{R0\sigma'}
+J_{RL}c^\dagger
_{R0\sigma}
\frac{\overrightarrow{\sigma}_{\sigma\sigma'}}{2}c_{L0\sigma'}\}
\end{eqnarray}
The local degree of freedom in the quantum dot is a spin
($S^2=3/4$)
$\overrightarrow{S}=1/2\sum_{\sigma\sigma'}d^\dagger_\sigma
\overrightarrow{\sigma}_{\sigma\sigma'} d_{\sigma'}$, where the
components of $\overrightarrow{\sigma}$ are the Pauli matrices,
and $c_{L(R)0\sigma}\equiv\sum_k c_{L(R)k\sigma}$. The exchange
Hamiltonian in Eq.~(\ref{kondoHamiltonian}) (together with a
scattering term not shown here) is derived from the Anderson
Hamiltonian by means of a canonical transformation
(Schrieffer-Wolff transformation \cite{Hew}) which integrates out
the aforementioned virtual excited states. The exchange constant
in this effective Hamiltonian can be written in terms of the
original parameters as:
\begin{equation}
\label{exchange} J_{\alpha\alpha
'}=\frac{\sqrt{\Gamma_\alpha\Gamma_{\alpha '}}}{\pi\rho_0}
\,\biggr (\frac{1}{U+\varepsilon_0}-\frac{1}{\varepsilon_0}\biggr
)\,.
\end{equation}
For symmetrical coupling to the leads one has $J\equiv
J_{LL}=J_{RR}=J_{RL}=J_{LR}=\frac{\Gamma}{\pi\rho_0}\biggr
(\frac{U}{|\varepsilon_0||U+\varepsilon_0|}\biggr)$ which for
$U>>\varepsilon_0$ is $J=\frac{2 V^2}{|\varepsilon_0|}$. Many spin
flip events mediated by the exchange interaction in
Eq.~(\ref{kondoHamiltonian}) lead to the formation of a many-body
spin singlet state, consisting of the localized spin and the spins
of the reservoirs. The energy scale for this singlet state is the
Kondo temperature $T_K=D\sqrt{\rho_0J}e^{-1/(2\rho_0J)}$, where
$D$ is a high energy cutoff.  In the language of the Anderson
Hamiltonian, the Kondo temperature reads $T_K\sim
D'e^{-\pi|\varepsilon_0||U+\varepsilon_0|/(2\Gamma U)}$ This
singlet is reflected in the local density of states (DOS) of the
quantum dot as a narrow peak around $\varepsilon_F$: the
Abrikosov-Suhl (AS) or Kondo resonance.

The Kondo effect leads to many remarkable properties and has been
the subject of extensive research for decades in the context of
metals with magnetic impurities \cite{Hew}. In recent years,
spectacular advances in nanotechnology have made it possible to
experimentally study Kondo physics in quantum dots
\cite{Gold1,CronSci(98),Stutt,Gold2,Blick,WielSci(00)}. These
experiments confirm early theoretical predictions \cite{Kon} that
low-temperature transport through quantum dots in the Coulomb
blockade regime should exhibit Kondo physics as described above.
Kondo physics in quantum dots manifests as an increase of the
linear conductance ($\mathcal{G}$) as one lowers the temperature
in regions with an odd number of electrons, which, again in the
simplest case, corresponds to a net spin S=1/2.
\begin{figure}
\begin{center}
\includegraphics[width=0.75\columnwidth,angle=0]{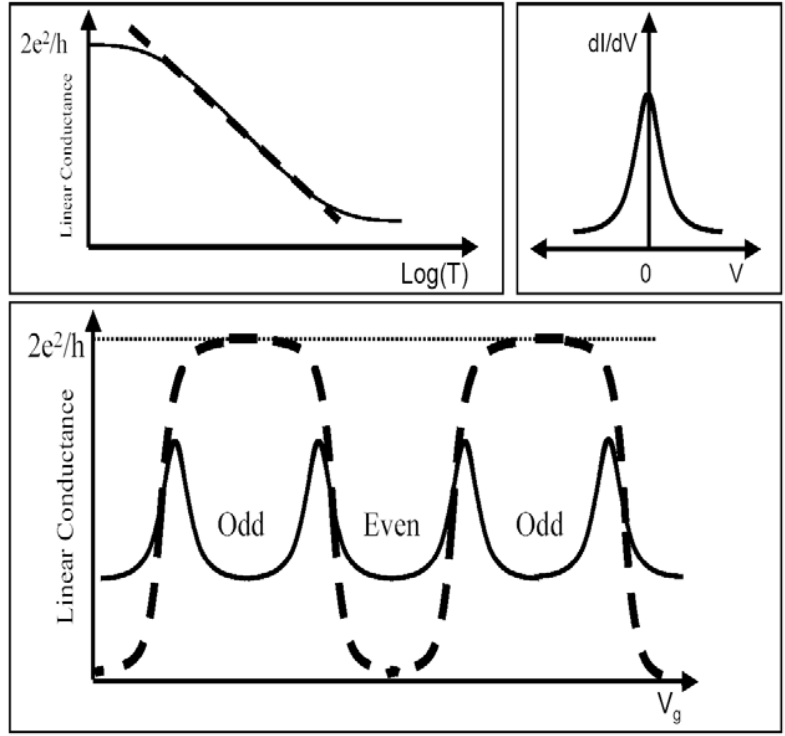}
\end{center}
\caption[]{Main transport characteristics of a quantum dot in the
Kondo regime (schematic).} \label{scheme-kondo2}
\end{figure}
This linear conductance increase can be explained in terms of the
increasing DOS around $\varepsilon_F$ as one lowers the
temperature, namely the Kondo resonance. For $T<<T_K$, the Kondo
effect increases the linear conductance to {\it its highest
possible value $2e^2/h$}. In other words, the spin-flip processes
leading to the Kondo effect are able to make an otherwise Coulomb
blockaded dot {\it perfectly transparent}. This limit of perfect
conductance $\mathcal{G}=2e^2/h$ is called the unitary limit.
Furthermore, the conductance, divided by its value at absolute
zero, depends only on the temperature divided by $T_K$, namely
$\mathcal{G}/\mathcal{G}_0=f(T/T_K)$. Importantly, $f(T/T_K)$ is
an universal function such that the behavior of a system with
parameters $\varepsilon_0$, $U$, etc, depends only on $T_K$:
different systems with the same $T_K$ behave in an universal
fashion. In bulk metals, the Kondo effect produces the opposite
behavior, it {\it decreases} the conductance because in this case
the scattering from magnetic impurities mixes electron states with
different momenta which increases the resistance.
\begin{figure}
\begin{center}
\includegraphics[width=0.75\columnwidth,angle=0]{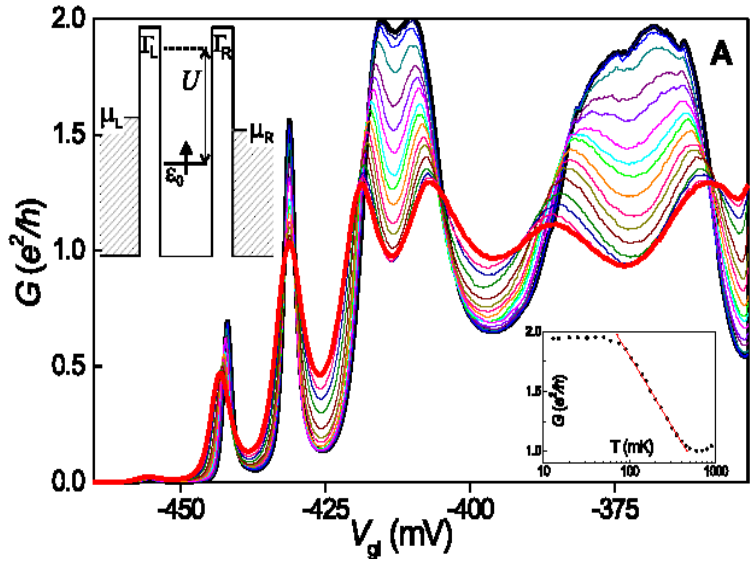}
\end{center}
\caption[]{Linear conductance versus gate voltage for different
temperatures. At the largest temperature, T=800mK, the conductance
exhibits Coulomb oscillations. At the lowest temperature, T=15mK,
the valley conductance around $V_{gl}=-413mV$ reaches the {\it
unitary limit} (see right inset). Reprinted with permission from
\cite{WielSci(00)}. \copyright 2000 American Association for the
Advancement of Science.} \label{QD_fig9}
\end{figure}
The main theoretical predictions of linear transport through
quantum dots in the Kondo regime are schematically depicted in
Fig.~\ref{scheme-kondo2}. Experimentally, linear transport through
quantum dots in the Kondo regime was first studied
in Refs.~\cite{Gold1,CronSci(98),Stutt} and later in
Refs.~\cite{Gold2,Blick,WielSci(00)}. We show
an example of these kind of experiments in Fig.~\ref{QD_fig9}
where we show the first demonstration of the unitary limit by van
der Wiel et al \cite{WielSci(00)}. In the nonlinear regime, the
hallmark of the Kondo effect is a zero bias anomaly in the
differential conductance as shown schematically in Fig.~\ref{scheme-kondo2}; an example 
from the experiments by
Cronenwett {\it et al} \cite{CronSci(98)} is shown in Fig.~\ref{QD_fig10}. Remarkably,
quantum dots provide the possibility to control and modify the
Kondo effect experimentally: the continuous tuning of the relevant
parameters governing the Kondo effect \cite{Hew} as well as the
possibility of studying Kondo physics when the system is driven
out of equilibrium, either by dc
\cite{HersPRL(91),WinPRL(93),LevyPRL(93),WinPRB(94),SivPRB(96),KonPRB(96)}
or ac voltages
\cite{HetPRL(95),NgPRL(96),SchiPRL(96),LopPRL(98),AvisPRL(98),KamPRL(99),KamPRB(00),NordPRB(00),LopPRB(01)},
which we shall discuss in the next subsection.
\begin{figure}
\begin{center}
\includegraphics[width=0.75\columnwidth,angle=0]{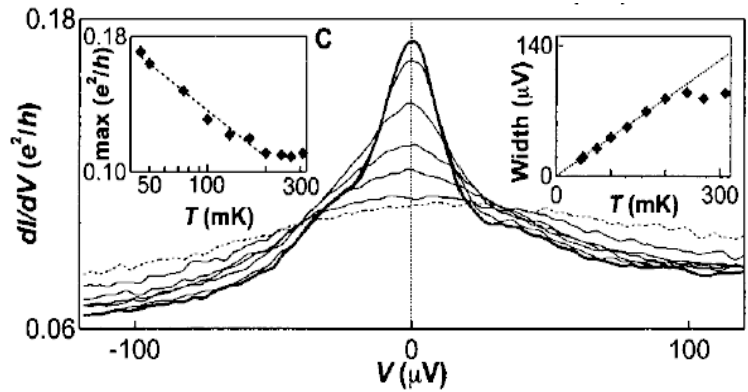}
\end{center}
\caption[]{Differential conductance $dI/dV$ versus bias voltage
$V$ for different temperatures (from T=45mK to T=270mK). The gate
center is set in the middle of a Kondo valley. The peak maximum
(left inset) is logarithmic in T. Reprinted with permission from
\cite{CronSci(98)}. \copyright 1998 American Association for the
Advancement of Science.} \label{QD_fig10}
\end{figure}
These kind of studies pave the way for the study of strongly
correlated electron physics in artificial systems. Moreover, they
provide a unique testing ground in which to investigate the
interplay of strongly correlated electron physics, quantum
coherence and non-equilibrium physics.

\subsubsection{Kondo physics in quantum dots with ac driving: Introduction}
Even before the first experimental demonstrations of Kondo effect
in quantum dots, some papers addressed theoretically different
aspects of the transport through ac-driven quantum dots in the
Kondo regime \cite{HetPRL(95),NgPRL(96),SchiPRL(96)}. These kind
of studies are motivated by the possibility of studying Kondo
physics in nonequilibrium situations not available in bulk metals.
An ac potential can be applied to the central gate, $\epsilon_0\to
\epsilon_0(t)\equiv \epsilon_0+ eV_{ac}cos\omega t$, thus
modulating the position of the energy levels of the quantum dot
with respect to the leads. In this way, the ac potential can be
used to periodically modify the Kondo temperature or to alternate
between situations with strong spin fluctuations (Kondo regime) or
charge fluctuations (mixed-valence regime). Alternatively, one may
apply an ac bias to the leads.

As described in section \ref{Keldysh}, the linear transport
through an ac driven quantum dot can be characterized by a
time-averaged spectral density such that the linear conductance is
given by~\cite{JauPRB(94)}
\begin{eqnarray}
{\mathcal G}_{0}=\frac{e^2}{\hbar} \int d\epsilon
\frac{\Gamma_L\Gamma_R}{\Gamma_L+\Gamma_R} \left( -\frac{\partial
f(\epsilon)}{\partial \epsilon} \right) \sum_\sigma
\rho_\sigma^{\rm dot} (\epsilon) \label{averageconductance}
\end{eqnarray}
where the time-averaged spectral density $\rho_\sigma^{\rm
dot}(\epsilon)$ is defined as
\begin{equation}\label{averageDOS}
\rho_\sigma^{\rm dot} (\epsilon)\equiv\langle
A_{\sigma}(\epsilon)\rangle
=\frac{\omega_0}{2\pi}\int_{0}^{\frac{2\pi} {\omega_0}}d\bar{t}
\,\rho_\sigma^{\rm dot}(\epsilon,\bar{ t})\,
\end{equation}
and $\frac{\partial f(\epsilon)}{\partial \epsilon}$ is the
derivative of the Fermi function. The time dependent spectral
density $\rho_{\sigma}^{\rm dot}(\epsilon,\bar{t})$, with
$\bar{t}=\left(t+t'\right)/2$, is defined as the imaginary part of
the Fourier transform with respect to $\tau =t-t'$ of the retarded
quantum dot Green's function
\begin{equation}\label{timeDOS}
\rho_\sigma^{\rm dot}(\epsilon,\bar{t})= -\frac{1}{\pi}
\mathcal{I}m\int_{-\infty}^\infty {\mathbf
G}^{r}_{d,\sigma}\left(\bar{t}+\frac{\tau}{2},\bar{t}-\frac{\tau}{2}\right)e^{i\epsilon\tau}d\tau\,.
\end{equation}

Despite a considerable amount of work, the physical picture of the
influence of microwaves on the Kondo conductance is still
controversial. In the following, we mention some specific examples
of theoretical work focusing on different aspects of the problem.

Goldin and Avishai~\cite{AvisPRL(98)} considered the case of a
very strong ac bias with the help of third-order perturbation
theory in the exchange constant. They concluded that the zero bias
anomaly is supressed by the ac field and contains sidebands at
multiples of the applied frequency. Furthermore, the zeroth and
the first harmonics of the ac current are strongly enhanced by the
Kondo effect while the other harmonics are small.

Nordlander {\it et al.,} analyzed in Ref.~\cite{NordPRB(00)} the
effects of an ac voltage applied to the central gate by using a
selfconsistent nonperturbative approach called non-crossing
approximation (NCA) for calculating $\rho_\sigma^{\rm
dot}(\epsilon,\bar{t})$. They found a rich behavior of the
conductance on the driving frequency and amplitude. At low
frequencies a strong ac potential produces sidebands of the Kondo
peak and a slow, roughly logarithmic, decrease of the linear
conductance over several decades of frequency. The strength of the
sidebands can be obtained analytically for the special case where
a perturbative treatment in the tunneling coupling is appropriate.
This limit can be better understood in terms of a time-dependent
Kondo model which, with respect to properties near the Fermi
level,  is equivalent to the Anderson Hamiltonian. In this limit
the dot can be replaced simply by a dynamical Heisenberg spin
$\vec{S}$ ($S^2=3/4$), which scatters electrons both within and
between reservoirs:
\begin{equation}
\sum_{kk'\sigma\sigma'}J_{kk'}(t)
\left(\vec{S}\cdot\vec{\sigma}_{\sigma\sigma'}+
\frac{1}{2}\delta_{\sigma\sigma'}\right)
c^\dagger_{k\sigma}c_{k'\sigma'}, \label{hamsd}
\end{equation}
where the components of $\vec{\sigma}$ are the Pauli spin
matrices. For near Fermi level properties, the relationship
between the Kondo and Anderson Hamiltonians is
$J(t)=|V^2/\epsilon_0(t)|$ for $U = \infty$. Near the Fermi level,
$w_{\rm leads}(\epsilon)/\hbar$, which is the total rate at which
lead electrons of energy $\epsilon$ undergo intralead and
interlead scattering by the dot, has a Kondo peak. Furthermore, if
$J$ is modulated as $J(t)= {\langle{J}\rangle}(1+\alpha \cos
\omega t)$, then an electron scattered by the dot will be able to
absorb or emit multiple quanta of energy $\hbar \omega$, leading
to satellites of the Kondo peak in $w_{\rm leads}(\epsilon)$. One
can then obtain $\langle\rho_\sigma^{\rm dot}(\epsilon,t)\rangle$
through the exact Anderson model relation
\begin{equation}
w_{\rm leads}(\epsilon) = \Gamma_{\rm dot}(\epsilon)
\langle\rho_\sigma^{\rm dot}(\epsilon,t)\rangle / \rho_{\rm
leads}(\epsilon), \label{wleads}
\end{equation}
where $\rho_{\rm leads}(\epsilon)$ is the state density per spin
in the leads. The above can be illustrated explicitly using
perturbation theory in $J$. Keeping all terms of order $J^2$ and
logarithmic terms to order $J^3$ one gets,
\begin{equation}
w_{\rm leads}(\epsilon) = 2\pi \langle J^2\rangle \rho\bigg[
1+3\langle{J}\rangle\rho\sum_{n=-1}^1 a_n
g(\epsilon+n\hbar\omega)\bigg], \label{GammaLeads}
\end{equation}
where $\rho=\rho_{\rm leads}(0)$, $a_0=1$, $a_{\pm 1} =
\alpha^2/(2+\alpha^2)$, $\langle
J^2\rangle=(1+\frac{1}{2}\alpha^2){\langle{J}\rangle}^2$, and
\begin{equation}
g(\epsilon)= \frac{1}{2}\int_{-D}^{D}\!d\epsilon'\,
        \frac{1-2f(\epsilon')}{\epsilon'-\epsilon}
\rightarrow \ln\left|\frac{D}{\epsilon}\right|, \label{gdef}
\end{equation}
the last limit being approached when $ T \ll |\epsilon|$. The
coefficients $a_{\pm 1} =V_{ac}^2/(2\epsilon_0^2+V_{ac}^2)$ are
the strengths of the first satellites above and below a central
peak of unit strength.

At high frequencies, photon-assisted tunneling processes result in
a effective temperature
$T_{effec}=T+\Gamma_{PAT}=T+J_1^2(\beta)\Gamma(E+n\hbar\omega)$
such that even at zero temperature photon-assisted tunneling
processes provides a cutoff for the Kondo singularity and reduce
the conductance. Later, Kaminski {\it et al} pointed out in
Refs.~\cite{KamPRL(99),KamPRB(00)} that even in the absence of dot
ionization, low frequency microwaves are able to flip the spin of
the dot, thus producing decoherence in the Kondo state. In
Refs.~\cite{KamPRL(99),KamPRB(00)} only small frequencies and
intensities were considered. In the following we describe a model
by L\'opez  {\it et al} \cite{LopPRB(01),Lopeztesis} where the Kondo effect
in quantum dots with ac driving is studied for all ranges of ac
parameters.
\subsubsection{$U^2$ perturbative solution}
The static Anderson model is exactly solvable \cite{Hew} but a
reliable method to obtain dynamical properties in the whole range
of $U/\Gamma$ is not available. Some approximation is thus needed
to evaluate the quantum dot retarded Green's function in
Eq.~(\ref{averageDOS}). For $U\rightarrow\infty$ the NCA allows to
study transport properties at intermediate temperatures $T\lesssim
T_K$. As we have described in the previous section, this is the
route followed by Nordlander et al in Ref.~\cite{NordPRB(00)}.
Furthermore, the NCA method can be formulated in a fully
time-dependent form such that non-equilibrium time-dependent
properties of quantum dots in the Kondo regime can be studied
\cite{NordPRL(99)}. NCA, however, breaks down as $T\rightarrow 0$
and does not recover properly the Fermi liquid $T=0$ regime. Other
methods are thus called for. Second order finite U perturbation
theory gives reliable qualitative results in the symmetric case
$\varepsilon_0=-U/2$ but exhibits anomalies away from this special
point. These anomalies can be circumvented by interpolating the
selfenergy of order $U^2$ in order to achieve a proper behavior in
the limits $U/\Gamma\rightarrow 0$ and $\Gamma/U\rightarrow 0$ and
good analytic properties both, in
$\varepsilon\rightarrow\varepsilon_F$ and $\varepsilon \rightarrow
\pm \infty$ limits. In addition, charge conservation is obtained
by introducing a selfconsistent parameter in this interpolative
self-energy in order to fulfill the Friedel sum rule
\cite{LevyPRL(93),MartSSC(82)}. This interpolative scheme was
generalized to the ac case in Ref.~\cite{LopPRL(98)}. In this
work, the following ansatz for the quantum dot Green's function in
the presence of an ac potential is proposed
\begin{eqnarray}
G^{r}_{d,\sigma}(t,t')=\exp\left[-i\frac{V_{\rm ac}}{\omega_0}
\left (\sin\omega_0 t-\sin\omega_0
t'\right)\right]\,\tilde{G}^{r}_{d,\sigma}(t-t')\,. \label{anstaz}
\end{eqnarray}
where $\tilde{G}^{r}_{d,\sigma}(t-t')$ is the static quantum dot
retarded Green's function. In Eq.~(\ref{anstaz}) the ac potential
breaks the symmetry under temporal translation only by introducing
a global phase in the total quantum dot Green's function. The
physics of this ansatz is to assume that the only effect of the ac
potential in the many-body state consists of flipping the quantum
dot spin in a coherent way. Thereby using this ansatz one only
accounts for coherent tunneling processes involving the absorption
or emission of photons. As we shall see below, the neglect
of inelastic tunneling processes, via multiphotonic events, 
is not a good approximation in many cases. From Eq.~(\ref{anstaz}) one can obtain
the time averaged Green's function $\langle
G^{r}_{d\sigma}(\epsilon)\rangle $ by Fourier transforming with
respect to $\tau =t-t'$ and performing the time average in the
time variable $\bar{t}=\left(t+t'\right)/2$
\begin{eqnarray}
\langle
G^{r}_{d\sigma}(\epsilon)\rangle=\sum_{m=-\infty}^{\infty}\,
J^2_{m}(\beta) \tilde{G}^{r}_{d\sigma}(\epsilon+m\omega_{0})\,,
\end{eqnarray}
Here, $\tilde{G}^{r}_{d\sigma}(\epsilon+m\omega_{0})$ is given by
\begin{eqnarray}
\tilde{G}^{r}_{d\sigma}(\epsilon+m\omega_{0})= \frac{1}{\left
[\epsilon+m\omega_0-\epsilon_{0\sigma}-\Sigma(\epsilon+m\omega_0)+i\Gamma\right
]} \,,
\end{eqnarray}
where $\Sigma(\epsilon+m\omega_0)$ is the correlation self-energy
obtained by the interpolative method with a Friedel sum rule which
is generalized to the ac case \cite{LopPRL(98)}.
$\Sigma(\epsilon+m\omega_0)$ depends on the time-averaged quantum
dot occupation, $\langle n_{d\sigma}^{ac}\rangle$, in the presence
of ac potential
\begin{eqnarray}
\langle n_{d\sigma}^{ac}\rangle=-\frac{1}{\pi}\int \, d\epsilon \,
\mathcal{I} m \langle G^{r}_{d\sigma}(\epsilon)\rangle
f(\epsilon)\,,
\end{eqnarray}

Using this model one obtains a DOS consisting of a Kondo peak at
$\varepsilon_{F}$ roughly weighted by $J^2_{0}(\beta)$ and
satellites at $\varepsilon\pm m\omega_0$ with weights
$J^2_{m}(\beta)$. As a consequence the linear conductance departs
from the unitary limit. As we shall see in the following,
inelastic photon-assisted tunneling processes lead, in many cases,
to a strong reduction of the central peak in the Kondo spectrum
and therefore the linear conductance is strongly suppressed. Only
in cases where the absorption and emission probability of photons
is small (small $\beta$) do not contribute and the ansatz of
Eq.~(\ref{anstaz}) is a good approximation.

Within the context of perturbation theory in $U$, the simplest
extension that goes beyond the description given by the ansatz in
Eq.~(\ref{anstaz}) is just to calculate selfenergies to second
order in $U$ without any assumption about how the breakdown of
time-translational invariance modifies the propagators. Using the
language of non-equilibrium Green's functions (see Section \ref{Keldysh})
the second order self-energies read
\begin{eqnarray}
\Sigma^{r,(2)}_{d,\sigma}(t,t')=\theta(t-t')[\Sigma^{<,(2)}_{d,\sigma}(t,t')-\Sigma^{>,(2)}_{d,\sigma}(t,t')]\,,
\label{sigma2retarded}
\end{eqnarray}
and
\begin{eqnarray}
\Sigma^{>,(2)}_{d,\sigma}(t,t')=iU^2G^>_{d,\sigma}(t,t')G^<_{d,\bar{\sigma}}(t',t)G^>_{d,\bar\sigma}(t,t')\,,
\label{sigma2greater}
\end{eqnarray}
\begin{eqnarray}
\Sigma^{<,(2)}_{d,\sigma}(t,t')=-iU^2
G^<_{d,\sigma}(t,t')G^>_{d,\bar{\sigma}}(t',t)G^<_{d,\bar\sigma}(t,t')\,.
\label{sigma2lesser}
\end{eqnarray}

The bare propagators in
Eqs.~(\ref{sigma2greater},\ref{sigma2lesser}) already include a
Hartree correction, the coupling to the leads and the ac
potential. In a first step, one calculates these propagators,
without ac, including the Hartree contribution (given by $U\langle
n_{d,\bar{\sigma}}\rangle$, where $\langle
n_{d,\sigma}\rangle=\langle d_\sigma^\dagger(t)d_\sigma(t)\rangle$
is the quantum dot occupation) and the coupling to the leads:
\begin{eqnarray}
\label{hartreepropagator} g^{r,a}_{d,\sigma}(t-t')= \mp
i\theta(\pm t \mp t')
\exp^{\left[-i\int_{t'}^{t}dt_1\left(\epsilon_{0\sigma}+U\langle
n_{d,\bar{\sigma}}\rangle \mp i\sum_{\alpha\in
\{L,R\}}\frac{\Gamma_\alpha}{2}\right)\right]}.
\end{eqnarray}

Including the time modulation of the quantum dot level, the
retarded and advanced quantum dot Green's functions read:
\begin{eqnarray}
\label{barepropagator} G^{r,a}_{d,\sigma}(t,t')=
\exp\left[-i\frac{V_{\rm ac}}{\omega_0} \left (\sin\omega_0
t-\sin\omega_0 t'\right)\right]\,g^{r,a}_{d,\sigma}(t-t')\,.
\end{eqnarray}
Finally, the lesser and greater bare propagators can be obtained
using
\begin{equation}
G^{<,>}_{d,\sigma}(t,t')=\int dt_1 \int dt_2
G^r_{d,\sigma}(t,t_1)\Sigma^{<,>}_{\rm
hp}(t_1,t_2)G^a_{d,\sigma}(t_2,t')\,, \label{bare}
\end{equation}
where $\Sigma^{<,>}_{\rm hp}(t_1,t_2)$ are the lesser and greater
coupling self-energies~\cite{JauPRB(94)}. Finally, one can obtain
the retarded Green's function (to second order in $U$) solving the
Dyson equation:
\begin{eqnarray}
\label{dyson} &&\left [i\frac{\partial}{\partial
t}-\bar{\epsilon}_{0\sigma}(t)+i\sum_{\alpha\in
L,R}\Gamma_\alpha/2\,\right] {\mathbf G}^{r,(2)}_{d,\sigma}(t,t')
=\delta(t-t')\nonumber\\
&&+\int dt_1 \Sigma^{r,(2)}_{d,\sigma}(t,t_1)\,{\mathbf
G}^{r,(2)}_{d,\sigma}(t_1,t'),
\end{eqnarray}
where $\bar{\epsilon}_{0\sigma}(t)$=$\epsilon_{0\sigma}+Un_
{d,\bar{\sigma}}(t)+V_{\rm ac}\cos \omega_0t$. This retarded
Green's function can be used in
Eqs.~(\ref{averageconductance},\ref{averageDOS},\ref{timeDOS}) to
calculate the linear conductance through the quantum
dot\cite{LopPRB(01)}.
\subsubsection{Spin-flip cotunneling rate and average conductance}
The main effect of the ac potential consists in a reduction of the
time-averaged DOS at $\varepsilon_F$. This reduction can be
interpreted as decoherence induced by ac excitations, either by
real photon-assisted induced excitations at large ac
frequencies~\cite{NordPRB(00)} or virtual spin-flip cotunneling
processes at small ac frequencies~\cite{KamPRL(99),KamPRB(00)}.
These processes introduce a quenching of the Kondo peak causing a
deviation of the linear conductance from the unitary limit. It is
difficult to extract the magnitude of this lifetime induced by the
ac potential from the analytical expressions above. Following
Refs.~\cite{KamPRL(99),KamPRB(00)} a simple estimate for the
lifetime can be obtained from the rate of spin-flip cotunneling.
In the case of spin-flip cotunneling the simplest process involves
the hopping of one electron out of the dot to a state above the
Fermi level while another electron in the reservoirs, with
opposite spin, enters into the dot. The rate of virtual spin-flip
cotunneling which takes into account one photon processes is
restricted to the case of very low ac frequencies and amplitudes,
i.e., $\omega_0, V_{\rm ac}\ll\epsilon_0, \epsilon_0+U$. Under
these conditions the rate of spin-flip cotunneling was derived in
Ref.~\cite{KamPRL(99),KamPRB(00)}. In the symmetric case the rate
obtained there is zero. Without restrictions, the expression for
the rate can be generalized quite easily \cite{LopPRB(01)}. By
means of a modified Schrieffer-Wolff
transformation~\cite{KamPRL(99),KamPRB(00)} one can obtain a Kondo
Hamiltonian with a time dependent exchange constant,
\begin{eqnarray}
J_{\alpha\alpha '}(t)=\frac{\sqrt{\Gamma_\alpha\Gamma_{\alpha
'}}}{4\pi\rho_{\rm leads}}
\sum_{n,m}J_n(\beta)J_m(\beta)\exp \left[i\,\left(n-m\right)\omega_0t\right]\nonumber\\
 \times\,\biggr (\frac{1}{\epsilon_0+n\omega_0}-\frac{1}{\epsilon_0+U+n\omega_0}
+\frac{1}{\epsilon_0+m\omega_0}-\frac{1}{\epsilon_0+U+m\omega_0}
\biggr )\,. \label{time-exchange}
\end{eqnarray}
To second order in $J_{\alpha\alpha '}(t)$ the rate of spin-flip
cotunneling can be found as
\begin{eqnarray}
\gamma=\frac{1}{2\pi}\sum_{\alpha\alpha',nm}{{\bf
J}_{\alpha\alpha'}^{nm}}^2\,|n-m|\,\omega_0\,.
\label{spin-flip-cot}
\end{eqnarray}
with
\begin{eqnarray}\label{chap4equa22}
{\bf
J}_{\alpha\alpha'}^{nm}&=&\frac{\sqrt{\Gamma_\alpha\Gamma_{\alpha
'}}J_n(\beta)J_m(\beta)}
{4}\nonumber\\
&\times&\, \biggr (\frac{1}{\epsilon_0+n\omega_0}-\frac{1}
{\epsilon_0+U+n\omega_0}+\frac{1}{\epsilon_0+m\omega_0}-\frac{1}{\epsilon_0+U+m\omega_0}\biggr)\,.
\end{eqnarray}
In the limit of very low ac frequencies and taking into account
one photon processes Eq.~(\ref{chap4equa22}) reduces to the
expression for the rate obtained in Ref.~\cite{KamPRB(00)} by
Kaminski {\it et al}:
\begin{eqnarray}\label{chap4equa23}
\gamma=\frac{\omega_0}{8\pi} \left
[\frac{\left(\Gamma_L+\Gamma_R\right)U}{U+\epsilon_0}\right ]^2
\left[\frac{V_{\rm
ac}\left(U+2\epsilon_0\right)}{\left(\epsilon_0+U\right)\epsilon_0}\right
]^2\,.
\end{eqnarray}
Equation~(\ref{time-exchange}) shows that the rate of spin-flip
cotunneling depends on the absorption or emission probability of
photons through the Bessel functions, the energy denominators and
the window of energy given by $|n-m|\omega_0$, its behavior as a
function of the ac frequency depends on two opposite effects. On
one hand, by increasing $\omega_0$ the window of allowed states
becomes larger but on the other hand the absorption or emission
probability diminishes. The competition of these two opposite
effects produces a maximal rate at certain frequency $\omega_{t}$.

These results can be connected with the results for the
conductance obtained from
Eqs.~(\ref{averageconductance},\ref{averageDOS},\ref{timeDOS},\ref{dyson}),
by using an exact Anderson model relation for the scattering rate.
In particular, the finite lifetime induced by the ac field reduces
the scattering rate by introducing a finite lifetime (even at
$T=0$). One can define an effective time-averaged self-energy
$\langle\Sigma^{\rm ac}_{\rm int}(E_F)\rangle$ in the presence of
irradiation such that the time-averaged DOS at $E_F$ can be
written as
\begin{equation}\label{rate_ac}
 \pi\Gamma\langle A(E_F)\rangle=\frac{\Gamma^2}
{\left [\Gamma-\mathcal{I}m\langle\Sigma^{\rm ac}_{\rm
int}(E_F)\rangle\right]^2}\,.
\end{equation}
The imaginary part of this effective self-energy can be identified
as the total rate of decoherence induced by the ac potential,
including the spin-flip cotunneling derived above, i.e.,
\begin{equation}\label{nu}
\nu=-\mathcal{I}m\langle\Sigma^{\rm ac}_{\rm int}(E_F)\rangle.
\end{equation}
Both, the rate of spin-flip cotunneling and the total decoherence
rate obtained from Eq.~(\ref{nu}),  present a non-monotonous
behavior as a function of the external frequency with a maximum at
$\omega_{t}$.
\begin{figure}
\begin{center}
\includegraphics[width=0.75\columnwidth,angle=0]{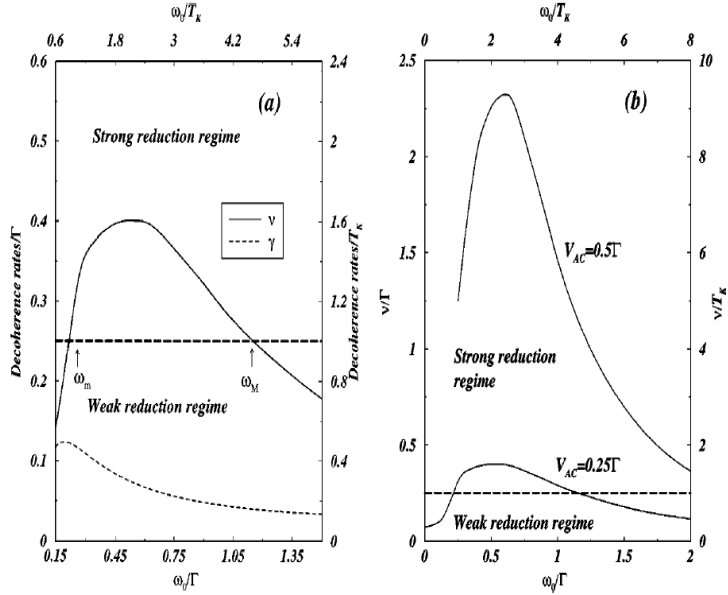}
\end{center}
\caption[]{(a) Rates of decoherence vs $\omega_0$ for $V_{\rm
ac}=0.25\Gamma\approx 2T_K$. Solid line shows the total rate of
decoherence obtained from Eq.(\ref{nu}), dashed line depicts the
rate of spin-flip cotunneling derived from
Eq.~(\ref{spin-flip-cot}). (b) Total rate of decoherence vs
$\omega_0$ for two intensities: $V_{\rm ac}=0.25\Gamma\approx T_K$
and $V_{\rm ac}=0.5\Gamma\approx 2T_K$. Strong and weak reduction
regimes (SRR and WRR, respectively) are separated by the
horizontal line $\nu=T_{K}$. For $\omega_0<\omega_{m}$ and
$\omega_0>\omega_{M}$ the system is in the weak reduction regime
whereas the strong reduction regime is achieved for
$\omega_{m}<\omega_0<\omega_{M}$. Reprinted with permission from
\cite{LopPRB(01)}. \copyright 2003 American Physical Society.}
\label{QD_fig11}
\end{figure}
One example of this behavior is shown in Fig.~(\ref{QD_fig11})
where the both rates as a function of $\omega_0$ are plotted. This
dependence defines two different regimes for the problem: {\it (i)
weak reduction} regime, which occurs when $\nu/T_K<1$. In this
case, the formation time for the Kondo state (given by
$1/T_K$~\cite{NordPRL(99)}) is shorter than the necessary time to
destroy it, which is given by the inverse of the decoherence rate,
and the system spends most of the time in a Kondo state without or
with little decoherence.
\begin{figure}
\begin{center}
\includegraphics[width=0.75\columnwidth,angle=0]{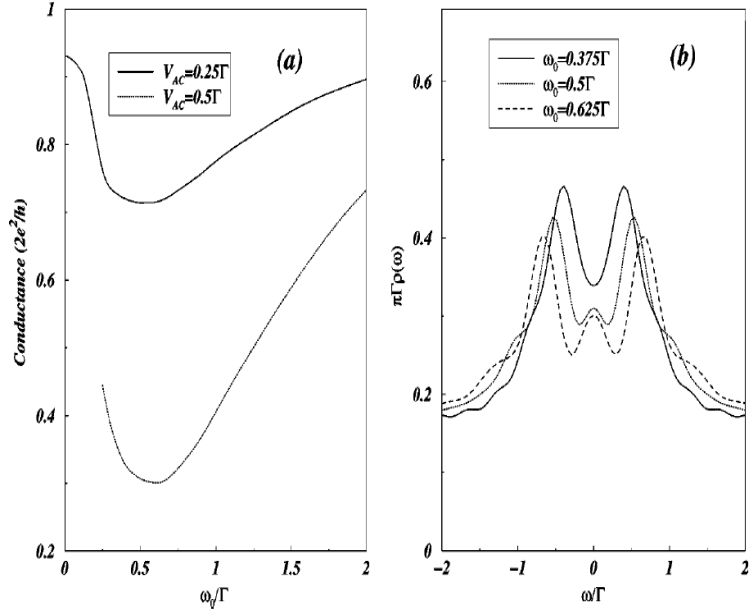}
\end{center}
\caption[]{ (a) Theoretical results for the linear conductance
(\ref{averageconductance}) as a function of $\omega_{0}$ for two
intensities $V_{\rm ac}=0.25\Gamma\approx T_K$ (solid line) and
$V_{\rm ac}=0.5\Gamma\approx 2T_K$ (dotted line) (b) Time-averaged
DOS for $\omega_0=0.375\Gamma\approx 3T_K/2$ (solid line),
$\omega_0=\Gamma/2\approx 2 T_K$ (dotted line),
$\omega_0=0.625\Gamma\approx 5 T_K/2$ (dashed line) and ac
amplitude $V_{\rm ac}=0.5\Gamma\approx 2T_K$. Reprinted with
permission from \cite{LopPRB(01)}. \copyright 2003 American
Physical Society.} \label{QD_fig12}
\end{figure}

On average, this translates into a high linear conductance
independently of the applied ac parameters. As long as the photon
absorption or emission rate is negligible, and therefore the ac is
not effective for inducing decoherence, it is irrelevant whether
or not the frequency is larger or smaller than $T_K$. {\it (ii)
strong reduction} regime which is found when $\nu/T_K>1$. In this
case, the decoherence time is shorter than $1/T_{K}$ and the
system spends most of the time in a state with a strong reduction
of the Kondo effect.\cite{LopPRB(01)}

The results for the conductance are presented in
Fig.~\ref{QD_fig12}a. As expected, the linear conductance behaves
non-monotonously as a function of the external frequency with a
minimum at $\omega_{t}$. As a function of intensity, the linear
conductance decays monotonously. At fixed $eV_{ac}/\hbar\omega_0$,
the maximum reduction of conductance occurs for the largest
frequency (Fig.~\ref{QD_fig12}a, dotted curve).

The above results can be understood in terms of the average
density of states (\ref{averageDOS}). One example of such DOS is
shown in Fig.~\ref{QD_fig12}b. Remarkably, the average DOS
exhibits photon sidebands, namely replicas of the Kondo resonance,
at $\hbar\omega_0\lesssim\Gamma$. This result is in qualitative
agreement with
Refs.~\cite{HetPRL(95),LopPRL(98),AvisPRL(98),NordPRB(00)}. Hence
in an experiment one may also expect replicas of the zero bias
anomaly, spaced by $\hbar\omega$, in the differential conductance
provided that the microwaves do not ionize the quantum dot.
However, this is still an open question as we discuss in the next
section.
\subsubsection{Experiments}
Experimentally, the effect of microwaves on the transport
properties through quantum dots in the Kondo regime was studied by
Elzerman {\it et al} \cite{ElzJLTP(00)}. In the entire frequency
range studied (10-50 GHz) the Kondo resonance vanishes by
increasing the microwave power such that no evidence of sideband
formation is found.
\begin{figure}
\begin{center}
\includegraphics[width=0.75\columnwidth,angle=0]{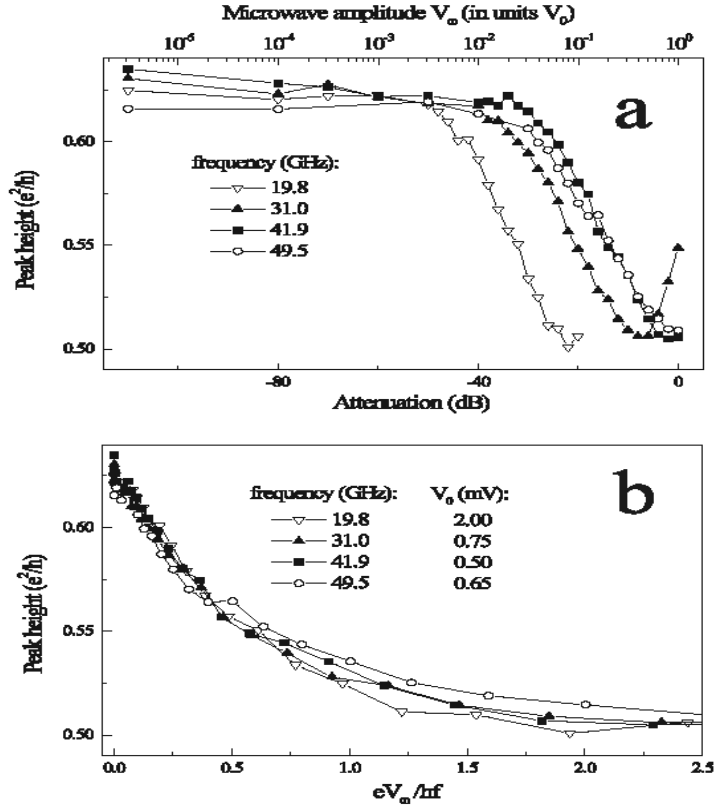}
\end{center}
\caption[]{a) Zero-bias conductance versus microwave attenuation
(lower scale). The upper scale gives the amplitude $V_\omega$ of
the microwaves in units of the (microwave-dependent) amplitude
$V_0$ without attenuation. b) Same data versus
$eV_\omega/\hbar\omega$. Reprinted with permission from
\cite{ElzJLTP(00)}. \copyright 2000 Plenum Academic Publishers.}
\label{QD_fig13}
\end{figure}
A possible explanation for the absence of sidebands is the extra
source of decoherence induced by the finite dc voltage applied to
measure the differential conductance $dI/dV_{\rm sd}|_{eV_{\rm
sd}=\hbar\omega_0}$. This extra source of decoherence which
reduces the Kondo effect \cite{KamPRL(99),KamPRB(00)} is not
included in the calculations of Refs.
~\cite{HetPRL(95),LopPRL(98),AvisPRL(98),NordPRB(00)}. Concerning
the linear conductance, the authors of the experiment in
Ref.~\cite{ElzJLTP(00)} conclude that the Kondo effect is reduced
more effectively by microwave-induced dephasing than by
temperature. Interestingly, they also find that the reduction of
the linear conductance shows an scaling behavior, independently of
$\omega$, as a function of $eV_\omega/\hbar\omega$ (Fig.
\ref{QD_fig13}), where $V_\omega$ is the estimated amplitude of
the ac signal relative to the amplitude without attenuation $V_0$.
For each frequency, $V_0$ is estimated by comparing the onset of
supression by microwaves, $eV_\omega=\alpha V_0$, with the thermal
fluctuations $eV_\omega\sim k_BT_{eff}$.
\subsection{Photon-assisted tunneling in one-dimensional quantum dots \label{1D}}
The experimental realization of one-dimensional (1D) quantum wires
has opened new possibilities to investigate electron transport in
the presence of strong electron-electron interactions and
impurities. For example, it has been possible to study Coulomb
blockade through 1D islands created by random impurities in
cleaved-edge-overgrotwth GaAs/AlGaAs quantum wires with low
electron densities. The temperature dependence of the conductance
through these systems showed, for the first time, clear evidence
of non-Fermi liquid behavior. Contrary to the dc-transport case,
where several theoretical works exist \cite{dc-transport1D}, the
investigations of transport in 1D systems in the presence of ac
fields are scarce. Furthermore, most of the theoretical work on
ac-transport in Luttinger liquids focuses on the single barrier
case \cite{SasSSC(96),FecEPL(99),FecPRB(01),CunEPL(99)}. The
photoconductance of a 1D quantum dot embedded in a non-fermi
liquid has been studied by Vicari {\it et al} \cite{VicEPJ01} by
means of the bosonization technique. Motivated by the spin-charge
separation in a Luttinger liquid, they consider spinless electrons
such that only the collective low-energy charge density modes are
relevant. These modes are described by the Hamiltonian:
\begin{equation}
H_0=\frac{v_F}{2}\int_{-\infty}^{\infty}dx[\Pi^2(x)+\frac{1}{g^2}(\partial\Theta(x))^2].
\end{equation}
Their quantization is described by the field operator $\Theta(x)$
and its conjugate $\Pi(x)$. $\Theta(x)$ represents the
long-wavelength part of the electron density
$\rho(x)=\rho_0+\sqrt{\frac{1}{\pi}}\partial_x\Theta(x)$, where
$\rho_0=k_F/\pi$ is the mean electron density. $1/g$ is the
constant that renormalizes the velocity of the charge modes
$v=v_F/g$ due to interactions:
\begin{equation}
\frac{1}{g}=\sqrt{1+\frac{V(q\rightarrow 0)}{\pi v_F}},
\end{equation}
where $V(q)$ is the Fourier transform of the 3D Coulomb
interaction projected along the wire and $v_F$ is the Fermi
velocity. The quantum dot is described by two symmetric delta-like
barriers (amplitude $V_B$ and positions $x_{1}$ and $x_{2}$,
respectively), their presence induces $2k_F$-backscattering
between left and right moving electrons such that the quantum dot
is described in bosonized form as:
\begin{equation}
H_D=\rho_0 V_B cos[\pi N_+]cos[\pi(n_0+N_-)],
\end{equation}
with $N_\pm=[\Theta(x_2)\pm\Theta(x_1)]/\sqrt{\pi}$. The
unbalanced particles between left and right leads are described by
$N_+/2$, while $N_-$ describes the fluctuations of the particle
number in the dot with respect to the mean electron number
$n_0=\rho_0(x_2-x_1)$, such that the coupling to a time dependent
gate voltage $V_g(t)=V_g+V_{ac}cos\omega t$ occurs through
$H_g=-eV_g(t)N_-$.

Vicari {\it et al} find that the photoconductance is strongly
influenced by the strong electron interaction. At fixed
temperature, the position of the sidebands does not depend on the
interactions but their intensity is strongly reduced by decreasing
$g$ (namely, increasing the interactions) from the non-interacting
limit ($g=1$).

As a function of temperature, the maxima of the sidebands peaks
scale according to a non-Fermi liquid power law similar to the dc
case:
\begin{equation}
\mathcal{G}_{max}\propto (k_BT)^{1/g-2},
\label{photoconductance1D}
\end{equation}
Eq.~(\ref{photoconductance1D}) defines two regimes for the
conductance. In the weak interacting regime $1/2<g\leq 1$, the
conductance has a peak-like behavior which can be enhanced by
decreasing the temperature. On the other hand, in the strong
interactions regime $g\leq 1/2$ the ac field is no longer able to
split the dc conductance in a series of sidebands. Note that this
is the characteristic behavior of metallic systems, namely for
strong interactions $g\leq 1/2$ the quantum dot behaves as if it
had a continuous density of states
instead of a discrete one.\\
Finally, let us mention that although these theoretical
predictions have not been tested experimentally, the recent
experimental advances in the study of electron transport through
carbon nanotubes or cleaved-edge-overgrotwth GaAs/AlGaAs quantum
wires suggest that the observation of the above effects is not far
from reach. Importantly, these kind of experiments would allow to
study in a well controlled way high-frequency effects and ac
transport in non-Fermi liquid systems.
\subsection{Wigner molecule regime
in ac-driven quantum dots \label{Wigner}} As we have remarked
during this chapter, the Coulomb interaction between the electrons
can significantly affect the properties (transport, dynamics, etc)
of a quantum dot. Such strongly correlated problems are
notoriously difficult to treat, and the addition of a
time-dependent field complicates the problem even further. When
the mean inter-electron separation exceeds a certain critical
value, however, a surprising simplification occurs, as the Coulomb
interaction dominates the kinetic energy and drives a transition
to a quasi-crystalline arrangement which minimizes the total
electrostatic energy. In analogy to the phenomenon of Wigner
crystallization in bulk two-dimensional systems
\cite{WigPR(34),TanPRB(89)} such a state is termed a {\em Wigner
molecule} \cite{JauEPL(93)}. As the electrons in the Wigner state
are sharply localised in space, the system can be naturally and
efficiently discretized by placing lattice points just at these
spatial locations. A many-particle basis can then be constructed
by taking Slater determinants of single-particle states defined on
these lattice sites, from which an effective Hamiltonian of
Hubbard-type can be generated to describe the low-energy dynamics
of the system \cite{JeffPRB(96)}. A major advantage of this
technique over standard discretization
\cite{AkbPRB(01),SchuPRB(02)} schemes, in which a very large
number of lattice points is taken to approximate the continuum
limit, is that the dimension of the effective Hamiltonian is much
smaller (typically by many orders of magnitude), which permits the
investigation of systems which would otherwise be prohibitively
complex. This approach has proven to be extremely successful in
treating a variety of static problems, including one-dimensional
quantum dots \cite{JeffPRB(96)}, two-dimensional quantum dots with
polygonal boundaries \cite{CreffPRB(99),CreffPRB(00)}, and
electrons confined to quantum rings
\cite{HausPhysB(96),KoskPRB(01)}. We further develop this method
here by including a time-dependent electric field, and study the
temporal dynamics of the system as it is driven out of
equilibrium.

\begin{figure}
\begin{center}
\includegraphics[width=.5\textwidth]{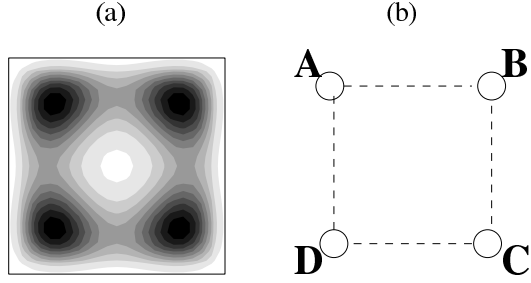}
\end{center}
\caption{(a) Ground-state charge-density for a two-electron square
quantum dot. GaAs material parameters are used, and the
side-length of the quantum dot is 800 nm, placing it in the Wigner
regime. The dark areas indicate peaks in the charge-density. (b)
Lattice points used for the effective lattice-Hamiltonian.}
\label{states}
\end{figure}

Let us consider a system of two electrons confined to a square
quantum dot with a hard-wall confining potential --- a simple
representation of a two-dimensional semiconductor quantum dot.
Such a system can be produced by gating a two-dimensional electron
gas confined at a heterojunction interface, and by placing a gate
split into four quadrants over the heterostructure
\cite{AustinSST(97)}, the potentials at the corners of the quantum
dot can be individually regulated. In Fig.\ref{states}a we show
the ground-state charge-density obtained from the exact
diagonalization of a square quantum dot \cite{CreffPRB(99)}, for
device parameters placing it deep in the Wigner molecule regime.
It can be seen that the charge-density is sharply peaked at four
points, located close to the vertices of the quantum dot. This
structure arises from the Coulomb interaction between the
electrons, which tends to force them apart into diagonally
opposite corners of the dot. As there are two such diagonal
states, degenerate in energy, we can understand the form of the
ground-state by considering it to be essentially a superposition
of these two states (with a small admixture of higher energy
states). The four points at which the peaks occur define the sites
on which the effective lattice-Hamiltonian operates, as shown in
Fig.\ref{states}b.

Following Creffield {\it et al} in Ref.~\cite{CrefPRB(02)b}, one
thus reduces the original problem to an effective
lattice-Hamiltonian of the form:
\begin{eqnarray}
&&H =  \sum_{\langle i, j \rangle, \sigma} \big[ t_{hub}( c_{i
\sigma}^{\dagger} c_{j \sigma}^{ } + \mbox{H.c.})
+V_{hub} n_i n_j \big] + \sum_{i} \big[ U_{hub}  n_{i \uparrow}
n_{i \downarrow} + E_i(t) n_i \big] . \label{Hamiltonianwigner}
\end{eqnarray}
Here $V_{hub}$ represents the Coulomb repulsion between electrons
occupying neighboring sites, and $U_{hub}$ is the standard
Hubbard-$U$ term, giving the energy cost for double-occupation of
a site.

$E_i(t)$ denotes the electric potential at site $i$, which in
general can have a static and a time-dependent component. In
experiment, static offsets can arise either from deviations of the
confining potential of the quantum dot from the ideal geometry, or
by the application of gating voltages to the corners of the
quantum dot. Applying corner potentials in this way would
substantially enhance the stability of the Wigner molecule state,
and could also be used to ensure that the multiplet of states
included in the effective lattice-model is well-separated from the
other excited states of the quantum dot system. In this discussion
the effects of static gates are not explicitly considered. Also,
the influence of small, accidental offsets encountered in
experiments is neglected as they are expected to have only minor
effects, and indeed may even stabilize CDT \cite{StoPRE(99)}. For
convenience, we consider applying an ac field aligned with the
$x$-axis of the quantum dot, which can be parameterized as:
\begin{equation}
E_A = E_D  = \frac{E}{2} \cos \omega t, \qquad E_B = E_C = -
\frac{E}{2} \cos \omega t
\end{equation}
where A,B,C,D label the sites as shown in Fig.\ref{states}b. We
emphasize that although we have the specific system of a
semiconductor quantum dot in mind, the effective-Hamiltonian we
are using can describe a wide range of physical systems, including
$2 \times 2$ arrays of connected quantum dots \cite{StaffPRB(97)},
and our results are thus of general applicability.

\begin{figure}
\begin{center}
\includegraphics[width=.5\textwidth]{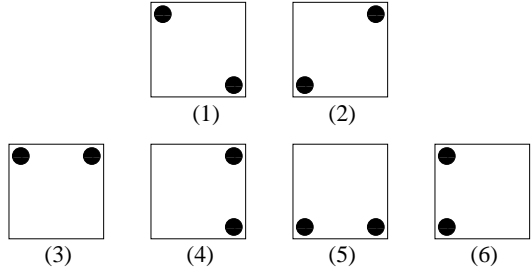}
\end{center}
\caption{Schematic representation of the two-particle basis states
for the singlet sector of the Hamiltonian. The ground state of the
quantum dot is approximately a superposition of states (1) and
(2).} \label{basis}
\end{figure}

As for the case of the double quantum dot in Section
\ref{ac-driven isolated double quantum dots}, no spin-flip terms
in (\ref{Hamiltonianwigner}) are included and so the singlet and
triplet sectors are again decoupled. Initial states with singlet
symmetry are chosen, which corresponds to the symmetry of the
system's ground-state. Simple state counting reveals that the
singlet sector has a dimension of ten, and can be spanned by the
six states shown schematically in Fig.\ref{basis}, together with
the four states in which each site is doubly-occupied.

\subsubsection{Interacting electrons, double occupancy excluded
\label{sec_Uinf}}

\begin{figure}
\begin{center}
\includegraphics[width=.5\textwidth]{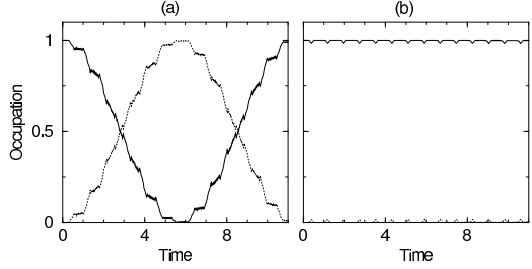}
\end{center}
\caption{Time development of the system for $U_{hub}$ infinite,
$V_{hub}=80$ and $\omega=8$: (a) electric potential, E = 100.0 (b)
E = 115.7. Solid line indicates the occupation of sites A and D,
the dotted line the occupation of sites B and C.} \label{Uinf}
\end{figure}

If the Hubbard-$U$ term is taken to be infinitely large one works
in the sub-space of states with no double occupation. The Hilbert
space is thus six-dimensional, and one can use the states shown in
Fig.\ref{basis} as a basis. We show in Fig.\ref{Uinf} the
time-dependent number occupation of the four sites at two
different values of $E$, in both cases using state $(6)$ as the
initial state, and setting the ac frequency to $\omega = 8$. In
Fig.\ref{Uinf}a $E$ has a value of 100.0, and it can be clearly
seen that the electrons perform driven Rabi oscillations between
the left side of the quantum dot and the right. Accordingly, the
occupation number of the sites varies continuously between zero
and one. In Fig.\ref{Uinf}b, however, we see that changing the
electric potential to a value of $E=115.7$ produces dramatically
different behavior. The occupations of sites A and D only vary
slightly from unity, while sites B and C remain essentially empty
throughout the time-evolution. Only a small amount of charge can
transfer per period of the driving field between the left and
right sides of the system, producing the small spikes visible in
this figure. The amplitude of these features is extremely small,
however, indicating that the tunneling between left and right
sides has been almost totally destroyed.

\begin{figure}
\begin{center}
\includegraphics[width=.5\textwidth]{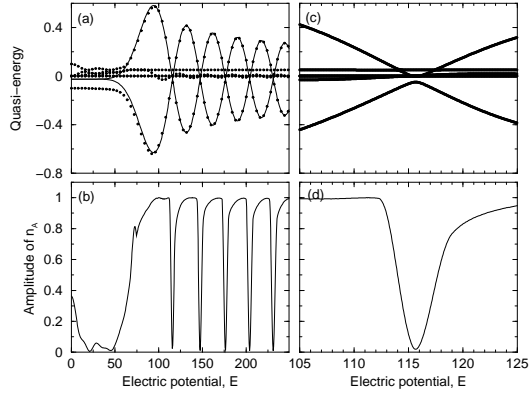}
\end{center}
\caption{(a) Quasi-energies of the system for $U_{hub}$ infinite,
$V_{hub}=80$ and $\omega=8$: circles = exact results,
lines=perturbative solution [$\pm 2 J_{10}(E/\omega)$]. (b)
Amplitude of oscillation of the occupation of site A. (c) Detail
of quasi-energy spectrum, showing an avoided crossing. (d) Detail
of amplitude of oscillations.} \label{Uinf_floq}
\end{figure}

A comparison of the amplitude of the oscillations of $n_A$ with
the quasi-energy spectrum, as a function of the electric potential
$E$, indeed confirms that CDT is occurring, see
Fig.\ref{Uinf_floq}. Similarly to the double quantum dot system,
we can see in Fig.\ref{Uinf_floq}a that the quasi-energies have
two different regimes of behavior. The first of these is the
weak-field regime, $E < V_{hub}$, at which the driving field does
not dominate the dynamics. In this regime the quasi-energy
spectrum, and correspondingly, the amplitude of oscillations shows
little structure. The second regime occurs at strong values of
potential, $E > V_{hub}$, for which the quasi-energy spectrum
clearly shows a sequence of close approaches. In
Fig.\ref{Uinf_floq}c an enlargement of one of these approaches is
presented which reveals it to be an {\em avoided crossing}.
Employing the perturbative method described in Section
\ref{method} demonstrates that the two quasi-energies involved in
these avoided crossings are described by $\pm 2J_n(E/\omega)$,
where $n$ is equal to $V_{hub} / \omega$. We may thus again think
of $n$ as signifying the number of photons the system needs to
absorb to overcome the Coulomb repulsion between the electrons
occupying neighboring sites. The results in Fig.\ref{Uinf_floq}b
and Fig.\ref{Uinf_floq}d clearly show that the locations of the
avoided crossings correspond exactly to quenching of the
oscillations in $n_A$, and so confirm that CDT indeed occurs at
these points.

\subsubsection{Interacting electrons, double-occupancy permitted}

In the most general case one has to consider the competition
between the $U_{hub}$ and $V_{hub}$ terms. Setting $U_{hub}$ to a
finite value means that the four doubly-occupied basis states are
no longer energetically excluded from the dynamics, and
accordingly the full ten-dimensional basis set has to be included.

\begin{figure}
\begin{center}
\includegraphics[width=.5\textwidth]{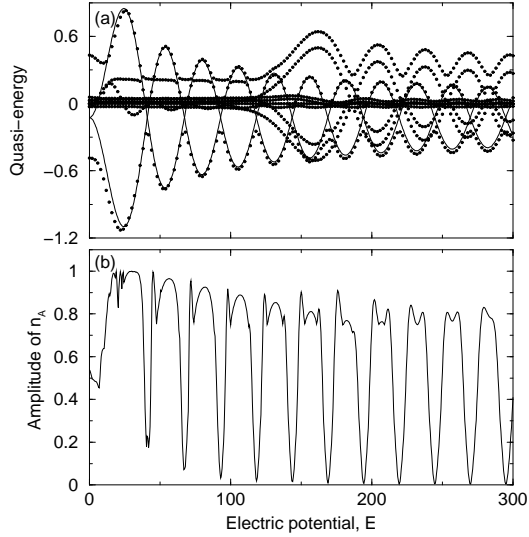}
\end{center}
\caption{(a) Quasi-energies of the system for  $U_{hub}=160$ and
$V_{hub}=16$, $\omega=8$: circles = exact results,
lines=perturbative solution [$\pm 2 J_{2}(E/\omega)$]. (b)
Amplitude of oscillation of the occupation of site A, with (6) as
the initial state.} \label{U160_1}
\end{figure}

Although it is difficult to obtain precise estimates for the
values of parameters of the effective Hamiltonian, it is clear
that in general $U_{hub} > V_{hub}$. Accordingly, the parameters
$U_{hub} = 160,  V_{hub}=16$ are chosen. Again, the frequency of
the ac field is set to $\omega=8$, and the quasi-energy spectrum
obtained by sweeping over the field strength is studied
\cite{CrefPRB(02)b}, see Fig.\ref{U160_1}a. It is immediately
clear from this figure that for electric potentials $E < U_{hub}$
the form of the spectrum is extremely similar to the
infinite-$U_{hub}$ case. Performing perturbation theory confirms
that, as in the previous case, the behavior of the quasi-energies
is given by $\pm 2 J_n(E/\omega)$ where $n=V_{hub} / \omega$. The
amplitude of the oscillations of $n_A$ when the system is
initialized in state $(6)$ is shown in Fig.\ref{U160_1}b which
demonstrates that at the locations of the avoided crossings the
tunneling parallel to the field is again quenched.

When the electric potential exceeds the value of $U_{hub}$,
however, new structure appears in the quasi-energy spectrum. A
group of four quasi-energies, that for weaker fields cluster
around zero, become ``excited'' and make a sequence of avoided
crossings as the field strength is increased. Perturbation theory
predicts that these high-field quasi-energies are given by $\pm 2
J_m(E/\omega)$, where $m=(U_{hub}-V_{hub})/\omega$, and thus these
avoided crossings arise when the absorption of $m$ photons equates
to the electrostatic energy difference between the two electrons
being on neighboring sites, and doubly-occupying one site. This
then indicates that this structure arises from the coupling of the
ac field to the doubly-occupied states.

\begin{figure}
\begin{center}
\includegraphics[width=.5\textwidth]{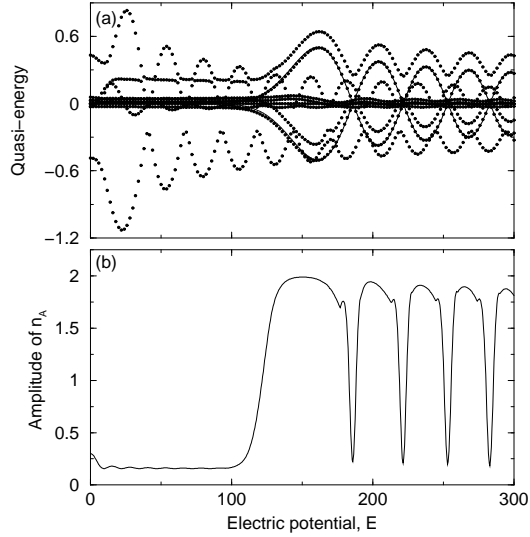}
\end{center}
\caption{(a) Quasi-energies of the system for  $U_{hub}=160$ and
$V_{hub}=16$, $\omega=8$: circles = exact results,
lines=perturbative solution [$\pm 2 J_{18}(E/\omega)$]. (b)
Amplitude of oscillation of the occupation of site A, with site A
doubly-occupied as the initial state.} \label{U160_2}
\end{figure}

To probe this phenomenon, the time evolution of the system from an
initial state consisting of {\em two} electrons occupying site A
is studied in Fig.\ref{U160_2}b. It can be seen that for electric
potentials weaker than $U_{hub}$ the amplitude of the oscillations
in $n_A$ remains small, and shows little dependence on the field.
As the potential exceeds $U_{hub}$, this picture changes, and the
ac field drives large oscillations in $n_A$, and in fact mainly
forces charge to oscillate between sites A and B. At the
high-field avoided crossings, however, the tunneling between A and
B is suppressed, which shuts down this process. Instead, the only
time-evolution that the system can perform consists of {\em
undriven} Rabi oscillations between sites A and D, perpendicular
to the field. As these oscillations are undriven they have a much
longer time-scale than the forced dynamics, and thus during the
interval over which we evolve the system the occupation of A only
changes by a small amount, producing the very sharp minima visible
in Fig.\ref{U160_2}b, centered on the roots of $J_m(E/\omega)$.

As the tunneling perpendicular to the field is undriven, it is
straightforward to evaluate the time evolution of the initial
state, if we assume that the left side of the quantum dot is
completely decoupled from the right side: The occupation of sites
A and D is then given by \cite{CrefPRB(02)b}:
\begin{equation}
n_A(t) = 1 + \cos \Omega_R t, \quad n_D = 1 - \cos \Omega_R t
\label{rabi}
\end{equation}
where $\Omega_R = 4 t_{hub}^2/(U_{hub} - V_{hub})$. For field
intensities such that CDT occurs, the electron dynamics is
restricted to the direction perpendicular to the applied field
where Rabi oscillations take place. The decay of the amplitude of
the Rabi oscillations as a function of time will indicate the
degree of dynamical localization achieved for the particular field
parameters considered \cite{CrefPRB(02)b}.
If  Rabi oscillations are damped, it will  indicate that the
isolation between the left and right sides of the quantum dot is
not perfect.
 Tuning the parameters of the
driving field therefore gives a simple and controllable way to
investigate how a two-electron wavefunction can decohere in a
quantum dot.

These results, together with those of Section \ref{ac-driven
isolated double quantum dots}, show that ac fields may not only be
used as a spectroscopic tools to probe the electronic structure of
a quantum dot systems, but can also be used to dynamically control
the time-evolution of the system. The tunability of the CDT
effect, and its ability to discriminate between doubly-occupied
and singly-occupied states, make it an excellent means for rapid
manipulation of the dynamics of strongly correlated electrons in
mesoscopic systems.
\section{Photon assisted shot noise \label{shot-noise}}
In a quantum conductor out of equilibrium, electronic current
noise originates from the dynamical fluctuations of the current
away from its average:
\begin{equation}
\Delta\hat{I}(t)\equiv \hat{I}(t)-\langle \hat{I}(t)\rangle.
\end{equation}
Shot noise, defined as the zero frequency limit of the power
spectral density
\begin{equation}
\mathcal {S}_{I}(\omega)\equiv \int_{-\infty}^{\infty} d\tau
e^{i\omega\tau}\langle \{\Delta\hat{I}(\tau),\Delta\hat{I}(0)\}
\rangle,
\end{equation}
provides us with a sensitive tool to study correlations between
carriers. In particular, shot noise experiments reveal the charge
and statistics of the quasiparticles relevant for electronic
transport. Also, information about internal energy scales can be
extracted from noise experiments\footnote{For a detailed review
about shot noise see Ref.~\cite{BlaPR(00)}.}. For uncorrelated
carriers with charge $q$, $\mathcal{S}_I(0)=2qI$ ({\it full} shot
noise or Poissonian noise). The Fano factor
($\gamma\equiv\frac{\mathcal{S}_I(0)}{2qI}$) quantifies deviations
from the Poissonian noise.

\begin{figure}
\begin{center}
\includegraphics[width=0.75\columnwidth,angle=0]{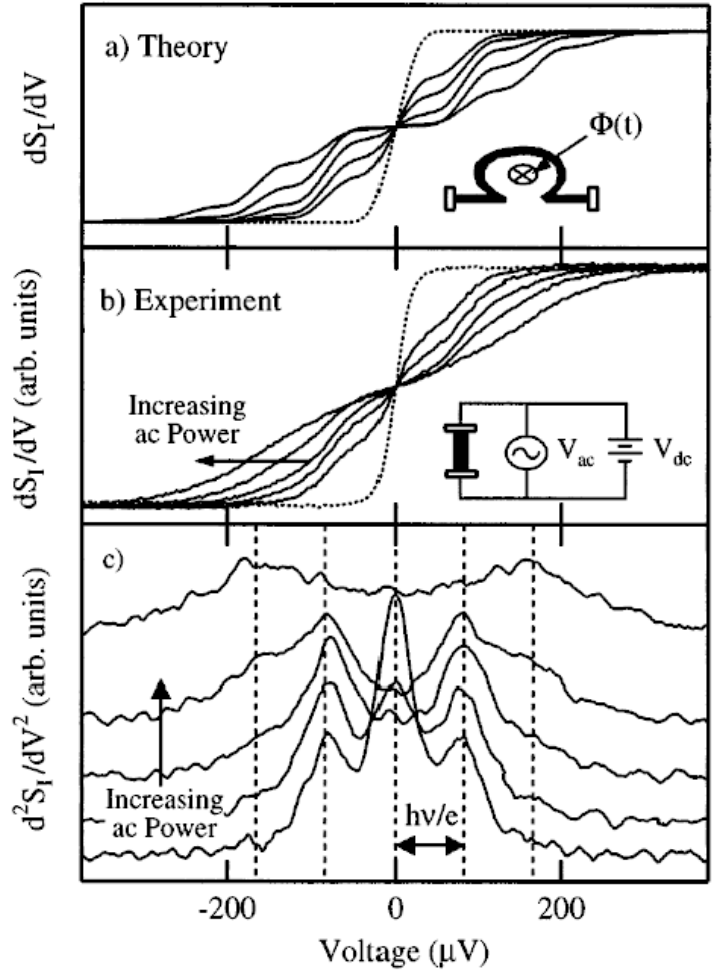}
\end{center}
\caption[]{a) Theoretical results from Eq.~(\ref{PATshotnoise1})
for different ac amplitudes. b) and c) Experimental results of
Schoelkopf {\it et al} \cite{SchoPRL(98)}. Reprinted with
permission from \cite{SchoPRL(98)}. \copyright 1998 American
Physical Society.} \label{shotnoise1}
\end{figure}

Photon assisted shot noise was first observed by Schoelkopf {\it
et al} \cite{SchoPRL(98)}. They measured both dc transport and
noise in a diffusive metallic conductor (namely, shorter than the
electron phase-breaking length) irradiated by microwaves
($f_{ac}$=2-40GHz). Interestingly, their experiment demonstrates the first
observation of photon assisted transport in a {\it linear system}:
the dc conductance remains completely unafected by the microwaves
in this linear mesoscopic system but the shot noise develops clear
features associated with photon assisted transport. In particular,
the differential shot noise $d{S}_{I}(0)/dV$ shows steps at
voltages corresponding to the photon energies $V=nhf_{ac}/e$. This
behavior can be easily understood within the framework of
scattering theory by noting that the shot noise of a coherent
conductor can be written as
\cite{SchoPRL(98),LesPRL(94),PedPRB(98)}:
\begin{eqnarray}
{S}_{I}(0)&=&\frac{2e^2}{h} \sum_i D_i^24k_BT+\frac{2e^2}{h}
\sum_{n=-\infty}^{\infty}J_n^2(\beta)\sum_i D_i(1-D_i)\nonumber\\
&\times&\{(nhf_{ac}+eV)coth[\frac{nhf_{ac}+eV}{2k_BT}]\nonumber\\
&+&(nhf_{ac}-eV)coth[\frac{nhf_{ac}-eV}{2k_BT}]\},
\label{PATshotnoise1}
\end{eqnarray}
where the $D_i$'s are the transmission probabilities of the
different conduction channels of the conductor. For zero
temperature, Eq.~(\ref{PATshotnoise1}) develops singularities at
voltages $V=nhf_{ac}/e$. The results of the experiments of
Ref.~\cite{SchoPRL(98)} are presented in Fig.~(\ref{shotnoise1}).

Very recently, a spectacular proof of photon assisted noise has
been demonstrated. Reydellet {\it et al} \cite{ReyPRL(03)} show
experimentally that photon-assisted processes do generate shot
noise {\it even in the absence of net dc electron transport}. This
noise can be interpreted as generated by photon-created
electron-hole pair partitioning. Without microwaves, noise in this
mesoscopic system can be understood as follows: the left reservoir
emits electrons at a frequency $eV/h$ such that the incoming
current is $I_0=e(eV/h)$. Asuming a single mode with transmission
probability $D$, the transmitted current is $I=DI_0$ and the
conductance $\mathcal{G}=e^2/hD$ (Landauer formula). Shot noise
originates from the quantum partition noise generated by electrons
either transmitted or reflected. Quantum partition results in
current fluctuations, bandwidth $\Delta f$, $\Delta
I^2=2eI_0D(1-D)\Delta f$. The binomial statistics of the
partitioning is reflected in the factor $D(1-D)$ \cite{BlaPR(00)}.
The microwaves change the frequencies of the emitted electrons (as
well as the probabilities of being emitted). In the limit
$hf_{ac}>>k_BT$ one can define an effective noise temperature
$T_N=S_{I}(0)/4\mathcal{G}k_B$ such that:
\begin{eqnarray}
T_N=T\Big(J_0^2(\beta)+\frac{\sum_iD_i^2}{\sum_iD_i}[1-J_0^2(\beta)]\Big)
+\sum_{n=1}^{\infty}\frac{nhf_{ac}}{k_B}J_n^2(\beta)\frac{\sum_i
D_i(1-D_i)}{\sum_iD_i}\nonumber\\. \label{PATshotnoise2}
\end{eqnarray}
The first term represents thermal noise (Johnson-Nyquist) while
the second one originates from photo-excited electron-hole pairs.
When the modes are either fully transmitting or reflecting
($D_i=1$ or $0$) there is no partition noise and only thermal
noise contributes to Eq.~(\ref{PATshotnoise2}).
\begin{figure}
\begin{center}
\includegraphics[width=0.75\columnwidth,angle=0]{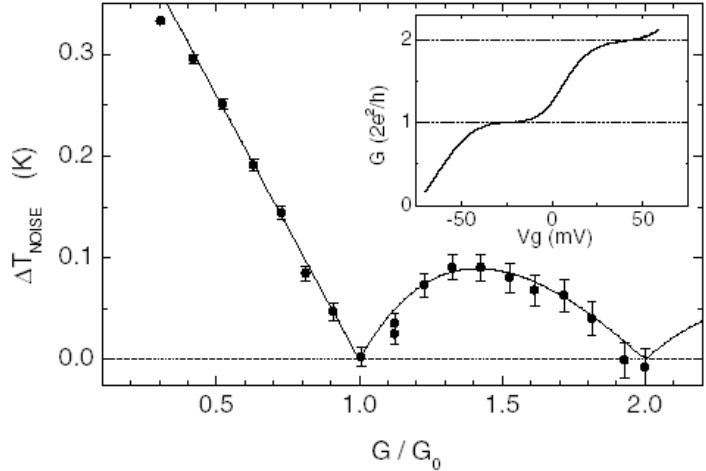}
\end{center}
\caption[]{Noise temperature (thermal noise substracted) as a
function of the transmission $\mathcal{G}/\mathcal{G}_0$ for a
quantum point contact irradiated with microwaves at 17.32 GHz
($\beta=2.3$). The solid line is a fit to the quantum supression
of the noise $\sum_i D_i(1-D_i)/\sum_iD_i$. Inset: conductance
versus ate voltage. Reprinted with permission from
\cite{ReyPRL(03)}. \copyright 2003 American Physical Society.}
\label{shotnoise2}
\end{figure}
In intermediate situations, the noise temperature is proportional
to the Fano factor $\sum_i D_i(1-D_i)/\sum_iD_i$ which
unambiguosly demonstrates photon-assisted partition noise as shown
in the experiments of Ref.~\cite{ReyPRL(03)} (see
Fig.~(\ref{shotnoise2})).

Theoretically, photon assisted shot noise has been also studied in
the context of quantum dots by Sun and coworkers in
Ref.~\cite{SunPRB(00)} where they consider a single resonant state
(without electron-electron interactions) coupled to leads (with
time dependent voltages). When the broadening of the resonant
state $\Gamma$ is smaller than $hf_{ac}$, they conclude that the
differential shot noise (at fixed dc voltage) versus gate voltage
shows a resonant structure reminiscent of the sideband structure
in the current versus gate voltage curves. When $\Gamma>>hf_{ac}$
the system resembles the single-channel conductor studied by
Schoelkopf {\it et al} \cite{SchoPRL(98)} and the differential
shot noise versus dc voltage shows a step-like behavior.

Recently, an extended theory of photon assisted shot noise has
been put forward by Camalet et al \cite{CamPRL(03)}. They
consider a generic situation where a multisite nanoscale conductor
(without dissipation and electron-electron interactions) is
coupled to leads. The Hamiltonian reads in a tight-binding
approximation with $N$ orbitals $|n\rangle$
\begin{equation}
H_{\rm wire}(t)= \sum_{n,n'} H_{nn'}(t) c^{\dag}_n
c^{\phantom{\dag}}_{n'}\;.
\end{equation}
As usual, the leads are modeled by ideal electron gases,
$H_\mathrm{leads}=\sum_q \epsilon_{q} (c^{\dag}_{Lq}
c^{\phantom{\dag}}_{Lq} + c^{\dag}_{Rq} c^{\phantom{\dag}}_{Rq})$,
where $c_{Lq}^{\dag}$ ($c_{Rq}^{\dag}$) creates an electron in the
state $|Lq \rangle$ ($|Rq \rangle$) in the left (right) lead.  The
tunneling Hamiltonian, $H_{T} = \sum_{q} \left( V_{Lq}
c^{\dag}_{Lq} c^{\phantom{\dag}}_1 + V_{Rq} c^{\dag}_{Rq}
c^{\phantom{\dag}}_N \right) + \mathrm{h.c.}$, establishes the
contact between the sites $|1\rangle$, $|N\rangle$ and the
respective lead. The influence of an applied ac-field 
of frequency $\omega=2\pi/{\mathcal T}$ 
results in a periodic
time-dependence of the Hamiltonian: $H_{nn'}(t+{\mathcal
T})=H_{nn'}(t)$ such that a generalized Floquet approach for the
evaluation of correlation functions can be developed. Here, we
sketch the derivation.

The Heisenberg equations for the wire operators read
\begin{eqnarray}
\dot c_{1(N)} &=& -\frac{i}{\hbar} \sum_{n'} H_{1(N),n'}(t)\, c_{n'}
             -\frac{\Gamma_{L(R)}}{2\hbar}c_{1(N)} + \xi_{L(R)}(t), \nonumber \\
\dot c_n &=& -\frac{i}{\hbar} \sum_{n'} H_{nn'}(t)\, c_{n'}\,,
\quad n=2,\ldots,N-1. \label{camaleteq1}
\end{eqnarray}
Within the wide-band limit approximation, the dissipative terms
are memory free and the Gaussian noise
$\xi_{L(R)}(t)=-(i/\hbar)\sum_q V^*_{L(R)
q}e^{-i\epsilon_q(t-t_0)/\hbar}c_{L(R) q}(t_0)$, with
$\langle\xi_{L(R)}(t)\rangle = 0$, obeys
\begin{eqnarray}
\langle\xi^\dagger_\alpha(t)\,\xi_{\alpha '}(t')\rangle
 &=& \delta_{\alpha,\alpha '}\frac{\Gamma_\alpha}{2\pi\hbar^2}
\int\!\! d\epsilon\, e^{i\epsilon(t-t')/\hbar}f_\alpha(\epsilon),\quad\alpha\in L,R
\end{eqnarray}
where $f_{L(R)}(\epsilon)$
denotes the Fermi function at temperature $T$ and chemical
potential $\mu_{L(R)}$. Without the inhomogeneity,
Eqs.~(\ref{camaleteq1}) are linear with $\mathcal{T}$-periodic
coefficients. Thus, it is
possible to construct a complete solution with the help of a
Floquet ansatz which in this case reads 
\begin{equation}
|\psi_\alpha(t)\rangle =
\exp[(-i\epsilon_\alpha/\hbar-\gamma_\alpha)\, t]\,
|u_\alpha(t)\rangle.  
\end{equation}
The Floquet states
$|u_{\alpha}(t)\rangle=\sum_k |u_{\alpha k}\rangle\exp(-ik\omega
t)$ obey the eigenvalue equation
\begin{equation}
\Big(\mathcal{H}(t) - i\Sigma
-i\hbar\frac{d}{dt}\Big)|u_{\alpha}(t)\rangle =( \epsilon_{\alpha}
-  i\hbar\gamma_{\alpha}) |u_{\alpha}(t)\rangle , \label{eq:Fs}
\end{equation}
where $\mathcal{H}(t)=\sum_{n,n'} |n \rangle H_{nn'}(t) \langle n'
|$ and $2 \Sigma =|1\rangle\Gamma_L\langle 1 | + |N
\rangle\Gamma_R\langle N |$.
Note that the eigenvalue
equation (\ref{eq:Fs}) is non-Hermitian (compare with Eq.~(\ref{floqeq})) in Section \ref{Floquet}), 
its eigenvalues
$\epsilon_{\alpha} - i\hbar\gamma_{\alpha}$ are generally complex
valued and the (right) eigenvectors are not mutually orthogonal.
Therefore, one needs to solve also the adjoint Floquet equation
yielding again the same eigenvalues but providing the adjoint
eigenvectors $|u_\alpha^+(t)\rangle$. It can be shown that the
the Floquet states $|u_\alpha(t)\rangle$ together with the adjoint
states $|u_\alpha^+(t)\rangle$ form at equal times a complete
bi-orthogonal basis: $\langle u^+_{\alpha}(t)|u_{\beta}(t)\rangle
= \delta_{\alpha\beta}$ and $\sum_{\alpha} |u_{\alpha}(t)\rangle
\langle u^+_{\alpha} (t)|= \mathbf{1}$.
For $\Gamma_{L/R}=0$, both $|u_\alpha(t)\rangle$ and
$|u_\alpha^+(t)\rangle$ reduce to the usual Floquet states.

The Floquet states $|u_\alpha(t)\rangle$ allow to write the
general solution of Eq.~(\ref{camaleteq1}) in closed form.  In the
asymptotic limit $t_0\to -\infty$, it reads
\begin{eqnarray}
c_n(t)=\sum_{\alpha}\int\limits_0^\infty & d\tau\,
  \langle n|u_\alpha(t)\rangle
  e^{(-i\epsilon_\alpha/\hbar-\gamma_\alpha)\tau}
  \langle u_\alpha^+(t-\tau)|\nonumber\\
&\times\big\{
  |1\rangle\xi_L(t-\tau)+|N\rangle\xi_R(t-\tau)\big\} .
\label{camaleteq2}
\end{eqnarray}
The average current and the shot noise can be obtained from the
closed form of the wire operators in Eq.~(\ref{camaleteq2}). In
particular, the shot noise reads:
\begin{eqnarray}
S_I(0) &=& \frac{e^2}{2\pi\hbar} \Gamma_L \Gamma_{R} \sum_k
\int\!\! d\epsilon\, \Big\{
  \Gamma_L \Gamma_{R} \Big| \sum_{k'} G_{N1}^{(k'-k)}(\epsilon+k\hbar\omega)
   G_{N1}^{(k')}(\epsilon)^* \, \Big|^2 \nonumber\\
&\times& f_L(\epsilon) [1-f_L(\epsilon+k\hbar\omega)]
\nonumber\\
&+& \Big| G_{1N}^{(-k)}(\epsilon+k\hbar\omega) + i \Gamma_L
\sum_{k'}
        G_{1N}^{(k'-k)}(\epsilon+k\hbar\omega)G_{11}^{(k')}(\epsilon)^* \Big|^2\nonumber\\
&\times& f_{L}(\epsilon) [1-f_{R}(\epsilon+k\hbar\omega)] \Big\}
+(L,1)\leftrightarrow(R,N). \label{camaletnoise}
\end{eqnarray}
The retarded Green's functions
\begin{equation}
G_{nn'}^{(k)}(\epsilon) =\sum_{\alpha,k'}
 \frac{\langle n|u_{\alpha,k'+k}\rangle\langle u_{\alpha,k'}^+|n'\rangle}
      {\epsilon-(\epsilon_\alpha+k'\hbar\omega-i\hbar\gamma_\alpha)}
\end{equation}
describe the propagation of an electron from orbital $|n'\rangle$
to orbital $|n\rangle$. 

Using Eq.~(\ref{camaletnoise}), Camalet
and coworkers study the noise properties of a simple wire with N=3
sites with equal energies and coupled to each other by a hopping
matrix element $\Delta$. The on-site energies are modulated by an ac dipole field
as $\epsilon_n(t)=\epsilon_n-A(N+1-2n)/2cos(\omega t)$, $n=1,2,3$. $A$ is thus the electric 
field strength multiplied by the electron charge and the distance bewteen neighboring sites.
Remarkably, they find that when
$\Delta_{eff}=J_0(\frac{A}{\hbar\omega})\Delta\rightarrow 0$ {\it both} the dc
current and the shot noise vanish (note that typically, as
demonstrated experimentally by Reydellet and coworkers
\cite{ReyPRL(03)}, a system with zero dc current is {\it not}
noiseless). At current supression, the Fano factor exhibits a
sharp maximum and two pronounced minima nearby. These results
suggest that external ac fields could be used to obtain nanoscale
devices with controllable noise levels.
Note, however, that the above derivation neglects important effects like dissipation 
in the wire and electron-electron interactions. 

\section{Conclusions}
In this review, we have attempted to give an overview of the physics of photon assisted 
tunneling in semiconductor nanostructures. Along the review, we have shown how the interplay of 
nonlinearity, time-dependent fields, electron-electron interactions and quantum confinement 
leads to new transport phenomena 
in nanostructures. During the last few decades, the study of these phenomena has lead to
important developments in the fields of mesoscopic physics and nanoscience with a 
a wide scope ranging from the study of very basic concepts
of quantum theory, like the demonstration of quantum coherence in artificial two level systems 
and the possibility of manipulating the dynamics of electrons in man-made structures,  
to engineering questions concerning ultimate
speed limits of nanoelectronic devices.

A great deal of information can be extracted from simple models like the Tien-Gordon model. 
In particular, the key concept of photo side-bands and their physical meaning, 
namely that photon absorption ($m>0$) and emission ($m<0$)
can be viewed as creating an effective electron density of states at energies $E\pm m\hbar\omega$
with a probability given by $J_{m}^2(\frac{eV_{ac}}{\hbar\omega})$, is already present in this simple description
of photon assisted tunneling. 

In many cases, more sophisticated theoretical tools are called for. 
Some of these tools, like The Floquet formalism, the non-equilibrium Green's function technique or 
the density matrix technique, to name just a few, have been described in the review together with 
concrete applications to the study of the available experimental information. We have seen along the review how by
using these theoretical techniques one can sed light on intriguing experimental observations like, for instance, 
the absolute negative conductance observed in THz irradiated semiconductor superlattices or the nontrivial 
features of photon-assisted transport through quantum dots in the Coulomb Blockade regime.

The use of these theoretical tools allows not only to explain experimental evidence but also to predict 
new effects which are not yet tested experimentally. Among these predictions 
we have described the possibility of realizing different electron pumps, including spin-polarized ones, 
by using ac fields, 
the coherent destruction of tunneling in artificial molecules, ac-induced 
sidebands of the Kondo resonance in the density of states of a strongly correlated quantum dot or the possibility
of obtaining nanoscale devices with controllable shot noise levels by using ac fields, just to mention a few.

Although we have tried to present a review as exhaustive as possible, we are conscious that some of the topics covered
along these pages would surely deserve a more in-depth treatment. We hope that the biased treatment of some topics,
which obviously reflects the author's views and partiality on some subjects,
will be compensated by the reader's desire of going deeply into little covered aspects (or even uncovered aspects
we may have overlooked) of photon-assisted transport through semiconductor nanostructures. 

We have provided many examples along the review which illustrate how one can obtain nontrivial physics 
by ac-driving a few electron system.
Many important questions remain open, though.
For instance, most of the calculations presented here study the response of the system to a field 
that is assumed to be known, but
the actual field inside a few electron system can be rather different than the applied one due to interactions. 
Developments along this line are thus extremely desirable. 

This is just an instance which demonstrates that, despite being rather mature, 
photon-assisted transport is still a very dynamic research area.
We exemplify this with the last
Sections of the review, photon-assisted tunneling in strongly correlated quantum dots
and photon-assisted shot noise: two areas which, doubtlessly, will bring us 
new exciting developments in the coming years.

\section{Acknowledgements}
We greatly acknowledge the support of the
Ministerio de Ciencia y Tecnolog\'{\i}a of Spain through the grant
MAT2002-02465 (R. A. and G. P.) and the "Ram\'on y Cajal" program (R. A.). 
We also thank the support of the EU through the RTN "Nanoscale and Dynamics" HPRN-CT-2000-00144.
During the last years we have benefited from collaborations and discussions with many colleagues, 
in particular, among others, Luis Bonilla, Tobias Brandes, Markus B\"uttiker, 
Valmir Chitta, Ernesto Cota, Charles Creffield, 
Silvano De Franceschi, Toshimasa Fujisawa, Leonid Glazman, Peter H\"anggi, Jes\'us I\~narrea,
Antti-Pekka Jauho, Sigmund Kohler, Leo Kouwenhoven, Bernhard Kramer, Rosa L\'opez, Jan Kees Maan, 
Miguel Moscoso, Tjerk Oosterkamp, David S\'anchez, 
Carlos Tejedor, Sergio Ulloa, Mathias Wagner, J\"urgen Weis and Wilfred van der Wiel.
We also thank Markus B\"uttiker, Charles Creffield and Antti-Pekka Jauho for a critical reading of the manuscript.

\end{document}